\documentclass[10pt, notitlepage, twoside, openright]{report}
\usepackage[applemac]{inputenc}
\usepackage{a4, fullpage}
\usepackage{cite}
\usepackage{amssymb, amsmath}
\usepackage{pstricks, pst-node, pst-tree}
\usepackage[figuresleft]{rotating}
\usepackage{rotate}
\usepackage{tocbibind}
\usepackage{nextpage}
\usepackage{graphicx}
\usepackage{relate}
\usepackage{titling}
\newcommand{\ceil}[1]{\left\lceil{#1}\right\rceil}

\newtheorem{lemma}{Lemma}
\newtheorem{theorem}{Theorem}
\newtheorem{corollary}{Corollary}

\newtheorem{prop}{Proposition}

\newcommand{\qed}{\hfill\ensuremath{\Box}\medskip\\\noindent}
\newenvironment{proof}{\noindent\emph{Proof. }}

\newcommand{\ignore}[1]{}

\newcommand{\depth}{\ensuremath{\mathrm{depth}}}
\newcommand{\cdepth}{\ensuremath{\mathrm{cdepth}}}
\newcommand{\ldepth}{\ensuremath{\mathrm{ldepth}}}
\newcommand{\size}{\ensuremath{\mathrm{size}}}
\newcommand{\heavy}{\ensuremath{\mathrm{heavy}}}
\newcommand{\leaves}{\ensuremath{\mathrm{leaves}}}
\newcommand{\roots}{\ensuremath{\mathrm{root}}}
\newcommand{\keyroots}{\ensuremath{\mathrm{keyroots}}}
\newcommand{\parent}{\ensuremath{\mathrm{parent}}}
\newcommand{\child}{\ensuremath{\mathrm{child}}}
\newcommand{\nca}{\ensuremath{\mathrm{nca}}}
\newcommand{\lab}{\ensuremath{\mathrm{label}}}
\newcommand{\post}{\ensuremath{\mathrm{post}}}
\newcommand{\pre}{\ensuremath{\mathrm{pre}}}
\newcommand{\rr}{\ensuremath{\mathrm{rr}}}
\newcommand{\lr}{\ensuremath{\mathrm{lr}}}

\newcommand{\norm}[1]{\ensuremath{| #1 |}}
\newcommand{\true}{\ensuremath{\mathsf{true}}}

\newcommand{\match}{\ensuremath{\textsc{match}}}
\newcommand{\mopc}{\ensuremath{\textsc{mop}}}
\newcommand{\mc}[1]{\ensuremath{\mathcal{#1}}}

\newcommand{\rn}{\ensuremath{\textsc{right}}}
\newcommand{\leftn}{\ensuremath{\textsc{left}}}
\newcommand{\restrict}[2]{\ensuremath{\mathop{#1|}_{#2}}}
\newcommand{\leftof}{\ensuremath{\textsc{leftof}}}
\newcommand{\Pred}{\ensuremath{\mathsf{Pred}}}
\newcommand{\Successor}{\ensuremath{\mathsf{Succ}}}
\newcommand{\Next}{\ensuremath{\mathsf{Next}}}

\newcommand{\Insertt}{\ensuremath{\textsc{Insert}}}
\newcommand{\Eq}{\ensuremath{\textsc{Eq}}}
\newcommand{\Ancestor}{\ensuremath{\textsc{Ancestor}}}
\newcommand{\mop}{\ensuremath{\mathrm{mop}}}
\newcommand{\fl}{\ensuremath{\mathrm{fl}}}

\newcommand{\first}{\ensuremath{\mathrm{first}}}
\newcommand{\emb}{\ensuremath{\mathrm{emb}}}

\newcommand{\Nca}{\ensuremath{\textsc{Nca}}}

\newcommand{\Fl}{\ensuremath{\textsc{Fl}}}
\newcommand{\Mop}{\ensuremath{\textsc{Mop}}}
\newcommand{\Mopsim}{\ensuremath{\textsc{MopSim}}}
\newcommand{\Match}{\ensuremath{\textsc{Match}}}

\newcommand{\Rightof}{\ensuremath{\textsc{Rightof}}}

\newcommand{\Deep}{\ensuremath{\textsc{Deep}}}
\newcommand{\Parent}{\ensuremath{\textsc{Parent}}}
\newcommand{\Child}{\ensuremath{\textsc{Child}}}
\newcommand{\Emb}{\ensuremath{\textsc{Emb}}}

\newcommand{\Cluster}{\ensuremath{\textsc{Cluster}}}

\newcommand{\Down}{\ensuremath{\textsc{Down}}}
\newcommand{\Up}{\ensuremath{\textsc{Up}}}

\newcommand{\Visit}{\ensuremath{\textsc{Visit}}}

\newcommand{\path}{\ensuremath{\mathrm{path}}}

\newcommand{\pqtime}[2]{\ensuremath{\left\langle{#1},{#2}\right\rangle}}
\newcommand{\Pre}{\ensuremath{\mathrm{Pre}}}
\newcommand{\prcfont}[1]{{\ensuremath{\mathsf{#1}}}}
\newcommand{\Succ}{\prcfont{Succ}}
\newcommand{\Equal}{\prcfont{Eq}}
\newcommand{\Move}{\prcfont{Move}}
\newcommand{\Close}{\prcfont{Close}}
\newcommand{\Nextt}{\prcfont{Next}}

\newcommand{\Insert}{\prcfont{Insert}}

\newcommand{\Member}{\prcfont{Member}}

\newcommand{\phrase}{\ensuremath{\mathrm{phrase}}}

\newcommand{\occ}{\ensuremath{\mathit{occ}}}
\newcommand{\reference}{\ensuremath{\mathrm{reference}}}

\newcommand{\rpre}{\ensuremath{\mathrm{rpre}}}
\newcommand{\rsuf}{\ensuremath{\mathrm{rsuf}}}
\newcommand{\lastmatch}{\ensuremath{\mathrm{lastmatch}}}
\newcommand{\zlw}{\rm \texttt{ZLW}}
\newcommand{\zla}{\rm \texttt{ZL78}}
\newcommand{\zlb}{\rm \texttt{ZL77}}
\newcommand{\deltab}{\ensuremath{\bar{\delta}}}

\begin{document}

\thispagestyle{empty}

\begin{titlepage}
{
\hfill\includegraphics[scale=.5]{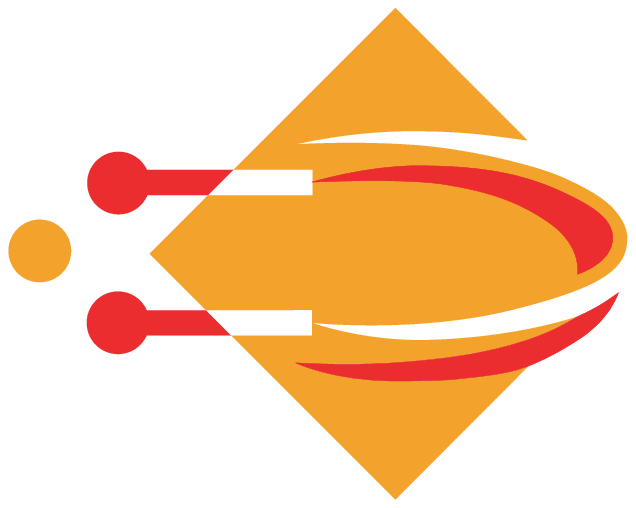}
\sf
\begin{flushright}
\vspace*{2cm}
{\huge {Pattern Matching in Trees and Strings}}\\
\vspace*{.2cm}    
{\large Philip Bille}\\
\rule{\textwidth}{2pt}
\vspace{2cm}
\end{flushright}
\begin{center}
A PhD Dissertation \\[0.1cm]
Presented to the Faculty of the IT University of Copenhagen \\[0.1cm]
in Partial Fulfilment of the Requirements for the PhD Degree
\end{center}
\vfill
\begin{flushleft}
\rule{\textwidth}{2pt}\\
\hfill June 2007 
\end{flushleft}
}
\end{titlepage}
\cleartooddpage[\thispagestyle{empty}]

\pagenumbering{roman}
\setcounter{page}{1}
\chapter*{Preface}
\addcontentsline{toc}{chapter}{Preface}
The work presented in this dissertation was done while I was enrolled as a PhD student at the IT University of Copenhagen in the 4-year PhD program. My work was funded by the EU-project "Deep Structure, Singularities, and Computer Vision" (IST Programme of the European Union (IST-2001-35443)). During the summer of 2003 my advisors Stephen Alstrup and Theis Rauhe left to start their own company and  my advisors then became Lars Birkedal  and Anna {\"O}stlin Pagh. In the period from March 2003 to September 2003 I was on paternity leave. I received my Masters Degree in January 2005. In Spring 2005 I visited Martin Farach-Colton at Rutgers University twice for a total period of two months. In the period from October 2006 to April 2007 I was on another 6 months of paternity leave. Finally, in the remaining period I came back to finish the present dissertation. 

I want to thank all of the inspiring people that I have worked with during my PhD. In particular, I want to thank Stephen Alstrup and Theis Rauhe for introducing me to their unique perspective on algorithms. I also want to thank Lars Birkedal, Anna {\"O}stlin Pagh, and Rasmus Pagh for tons of guidance. I am grateful to Martin Farach-Colton for the very pleasant stay at Rutgers University. Thanks to all of my co-authors: Stephen Alstrup, Theis Rauhe, Inge Li Gørtz, Martin Farach-Colton, Rolf Fagerberg, Arjan Kuijper, Ole Fogh Olsen, Peter Giblin, and Mads Nielsen. Thanks to the people who read and commented on earlier drafts of the dissertation: Inge Li Gørtz, Søren Debois, Rasmus Pagh, and Theis Rauhe. Finally, thanks to all of my colleagues at the IT University for creating a fun and interesting work environment.

\chapter*{Abstract}
\addcontentsline{toc}{chapter}{Abstract}
We study the design of efficient algorithms for combinatorial pattern matching. More concretely, we study algorithms for tree matching, string matching, and string matching in compressed texts. 

\paragraph{Tree Matching Survey} We begin with a survey on tree matching problems for labeled trees based on deleting, inserting, and relabeling nodes. We review the known results for the tree edit distance problem, the tree alignment distance problem, and the tree inclusion problem. The survey covers both ordered and unordered trees. For each of the problems we present one or more of the central algorithms for each of the problems in detail.
 

\paragraph{Tree Inclusion}
Given rooted, ordered, and labeled trees $P$ and $T$ the \emph{tree inclusion problem} is to determine if $P$ can be obtained from $T$ by deleting nodes in $T$. We show that the tree inclusion problem can be solved in $O(n_T)$ space with the following running times: 
\begin{equation*}
\min
\begin{cases}
      O(l_Pn_T), \\
      O(n_Pl_T\log \log n_T + n_T), \\
      O(\frac{n_Pn_T}{\log n_T} + n_{T}\log n_{T}).
\end{cases}
\end{equation*}
Here $n_S$ and $l_S$ denotes the number of nodes and leaves in tree $S \in \{P, T\}$, respectively, and we assume that $n_P \leq n_T$. Our results matches or improves the previous time complexities while using only $O(n_T)$ space. All previous algorithms required $\Omega(n_Pn_T)$ space in worst-case.
%

\paragraph{Tree Path Subsequence}
Given rooted and labeled trees $P$ and $T$ the \emph{tree path subsequence problem} is to report which paths in $P$ are
subsequences of which paths in $T$. Here a path begins at the root and ends at a leaf. We show that the tree path subsequence problem can be solved in $O(n_T)$ space with the following running times:
\begin{equation*}
\min
\begin{cases}
    O(l_{P}n_{T} + n_P) , \\
     O(n_{P}l_{T}+n_T) , \\
     O(\frac{n_{P}n_{T}}{\log n_{T}}+ n_T + n_P \log n_P).
\end{cases}
\end{equation*}
As our results for the tree inclusion problem this matches or improves the previous time complexities while using only $O(n_T)$ space. All previous algorithms required $\Omega(n_Pn_T)$ space in worst-case.

\paragraph{Regular Expression Matching Using the Four Russian Technique}
Given a regular expression $R$ and a string $Q$ the \emph{regular expression matching problem} is to determine if $Q$ matches any of the strings specified by $R$. We give an algorithm for regular expression matching using $O(nm/\log n + n + m\log m)$ and $O(n)$ space, where $m$ and $n$ are the lengths of $R$ and $Q$, respectively. This matches the running time of the fastest known algorithm for the problem while improving the space from $O(nm/\log n)$ to $O(n)$. Our algorithm is based on the Four Russian Technique. We extend our ideas to improve the results for the \emph{approximate regular expression matching problem}, the \emph{string edit distance problem}, and the \emph{subsequence indexing problem}.

\paragraph{Regular Expression Matching Using Word-Level Parallelism} We revisit the regular expression matching problem and develop new algorithms based on word-level parallel techniques. On a RAM with a standard instruction set and word length $w \geq \log n$, we show that the problem can be solved in $O(m)$ space with the following running times:
\begin{equation*}
\begin{cases}
      O(n\frac{m \log w}{w} + m \log w) & \text{ if $m > w$} \\
      O(n\log m + m\log m) & \text{ if $\sqrt{w} < m \leq  w$} \\
      O(\min(n+ m^2, n\log m + m\log m)) & \text{ if $m \leq \sqrt{w}$.}
\end{cases}
\end{equation*} 
This improves the best known time bound among algorithms using $O(m)$ space. Whenever $w \geq \log^2 n$ it improves all known time bounds regardless of how much space is used. 

\paragraph{Approximate String Matching and Regular Expression Matching on Compressed Texts}
Given strings $P$ and $Q$ and an \emph{error threshold} $k$, the \emph{approximate string matching problem} is to find all ending positions of substrings in $Q$ whose \emph{unit-cost string edit distance} to $P$ is at most $k$. The unit-cost string edit distance is the minimum number of insertions, deletions, and substitutions needed to convert one string to the other. We study the approximate string matching problem when $Q$ is given in compressed form using Ziv-Lempel compression schemes (more precisely, the {\zla } or {\zlw } schemes). We present a time-space trade-off for the problem. In particular, we show that the problem can be solved in $O(nmk + \occ)$ time and $O(n/mk + m + \occ)$ space, where $n$ is the length of the compressed version of $Q$, $m$ is the length of $P$, and $\occ$ is the number of matches of $P$ in $Q$. This matches the best known bound while improving the space by a factor $\Theta(m^2k^2)$. We extend our techniques to improve the results for regular expression matching on Ziv-Lempel compressed strings.

\tableofcontents

\cleartooddpage
\pagenumbering{arabic}
\setcounter{page}{1}
\chapter{Introduction}
In this dissertation we study the design of efficient algorithms for combinatorial pattern matching. More concretely, we study algorithms for tree matching, string matching, and string matching in compressed strings.

The dissertation consists of this introduction and the following (revised) papers.

\begin{description}
  \item[Chapter \ref{chap:tree1}] A Survey on Tree Edit Distance and Related Problems. Philip Bille. \emph{Theoretical Computer Science}, volume 337(1-3), 2005, pages 217--239.
  \item[Chapter \ref{chap:tree2}] The Tree Inclusion Problem: In Optimal Space and Faster. Philip Bille and Inge Li Gørtz. 
In \emph{Proceedings of the 32nd International Colloquium on Automata, Languages and Programming}, Lecture Notes in Computer Science, volume 3580, 2005,  pages 66--77. 
  \item[Chapter \ref{chap:tree3}] Matching Subsequences in Trees. Philip Bille and Inge Li Gørtz. 
In \emph{Proceedings of the 6th Italian Conference on Algorithms and Complexity}, Lecture Notes in Computer Science, volume 3998, 2006, pages 248--259. 
  \item[Chapter \ref{chap:string1}] Fast and Compact Regular Expression Matching. Philip Bille and Martin Farach-Colton. Submitted to a journal, 2005.
  \item[Chapter \ref{chap:string2}] New Algorithms for Regular Expression Matching. Philip Bille. In \emph{Proceedings of the 33rd International Colloquium on Automata, Languages and Programming}, Lecture Notes in Computer Science, volume 4051, 2006, pages 643--654. 
  \item[Chapter \ref{chap:string3}] Improved Approximate String Matching and Regular Expression Matching on Ziv-Lempel Compressed Texts. Philip Bille and, Rolf Fagerberg, and Inge Li Gørtz. In \emph{Proceedings of the 18th Annual Symposium on Combinatorial Pattern Matching}, 2007, to appear.
\end{description} 

In addition to the above papers I have coauthored the following 3 papers during my PhD that are not included in the dissertation:
\begin{itemize}
  \item Labeling Schemes for Small Distances in Trees. Stephen Alstrup, Philip Bille, and Theis Rauhe.
\emph{SIAM Journal of Discrete Mathematics}, volume 19(2), pages 448 - 462.
  \item From a 2D Shape to a String Structure using the Symmetry Set. Arjan Kuijper, Ole Fogh Olsen, Peter Giblin, Philip Bille, and Mads Nielsen. In \emph{Proceedings of the 8th European Conference on Computer Vision}, Lecture Notes in Computer Science, Volume 3022, 2004, pages 313 - 325.
  \item Matching 2D Shapes using their Symmetry Sets. Arjan Kuijper, Ole Fogh Olsen, Peter Giblin, and Philip Bille.
In \emph{Proceedings of the 18th International Conference on Pattern Recognition}, 2006, pages 179-182.
\end{itemize}
Of these three papers, the first paper studies compact distributed data structures for trees. The other two are papers on image analysis are related to our work on tree matching. The tree matching papers in the dissertation and the related image analysis papers are all part of the EU-project ``Deep Structure, Singularities, and Computer Vision'' that funded my studies. The project was a collaboration of 15 researchers from Denmark, The United Kingdom, and The Netherlands working in Mathematics, Computer Vision, and Algorithms. The overall objective of the project was to develop methods for matching images and shapes based on multi-scale singularity trees and symmetry sets. The algorithms researchers (Stephen Alstrup, Theis Rauhe, and myself) worked on algorithmic issues in tree matching problems. 

\section{Chapter Outline}
The remaining introduction is structured as follows. In Section~\ref{i:model} we define the model of computation. In Section~\ref{i:treematching} we summarize our contributions for tree matching and their relationship to previous work. We do the same for  string matching and compressed string matching in Sections~\ref{i:stringmatching}, and~\ref{i:compressed}, respectively. In Section~\ref{i:techniques} we give an overview of the central techniques used in this dissertation to achieve our results and in Section~\ref{i:conclusion} we conclude the introduction.

\section{Computational Model}\label{i:model}
Before presenting our work, we briefly define our model of computation. The \emph{Random Access Machine} model (RAM), formalized by Cook and Reckhow~\cite{CR1972}, captures many of the properties of a typical computer. We will consider the \emph{word-RAM} model variant as defined by Hagerup~\cite{Hagerup1998}. Let $w$ be a positive integer parameter called the \emph{word length}. The memory of the word-RAM is an infinite array of \emph{cells} each capable of storing a $w$-bit integer called a \emph{word}. We adopt the usual assumption that $w\geq \log n$, where $n$ is the size of the input, i.e., an index or pointer to the input fits in a word. Most of the problems in this dissertation are defined according to a set of \emph{characters} or \emph{labels} called an \emph{alphabet}. We assume that each input element from alphabet is encoded as a w-bit integer in a word. 

The instruction set includes operations on words such as addition, subtraction, bitwise shifting, bitwise and, bitwise or and bitwise xor, multiplication, and division. Each operation can be computed in unit time. The space complexity of an algorithm is the maximum number of cells used at any time beside the input, which is considered read-only. The time to access a cell at index $i$ is $O(\ceil{(\log i)/w})$, i.e., the access time is proportional to the number of words needed to write the index in binary. In particular, any data structure of size $2^{O(w)}$ can be accessed in constant time. We will only encounter super-constant access time in our discussion of the regular expression matching problem where very large data structures appear.

Word-RAM algorithms can be \emph{weakly non-uniform}, that is, the algorithm has access to a fixed number of word-size constants that depend on $w$. These constants may be thought of a being computed at ``compile time''. For several of our results, we use a deterministic dictionary data structure of Hagerup et al.~\cite{HMP2001} that requires weak non-uniformity. However, in all cases our results can easily be converted to work without weak non-uniformity (see Section~\ref{i:datastructures} for details).

\section{Tree Matching}\label{i:treematching}
The problem of comparing trees occurs in areas as diverse as structured text data bases (XML), computational biology, compiler optimization, natural language processing, and image analysis \cite{KTSK2000, CO1982, KM1995, RR1992, Tai1979, ZS1989}. For example, within computational biology the secondary structure of RNA is naturally represented as a tree~\cite{Waterman1995, Gusfield1997}. Comparing the secondary structure of RNA helps to determine the functional similarities between these molecules.

In this dissertation we primarily consider comparing trees based on simple \emph{tree edit operations} consisting of deleting, inserting, and relabeling nodes. Based on these operations researcher have derived several interesting problems such as the \emph{tree edit distance problem}, the \emph{tree alignment distance problem}, and the \emph{tree inclusion problem}. Chapter~\ref{chap:tree1} contains a detailed survey of each of these problems. The survey covers both \emph{ordered trees}, with a left-to-right order among siblings, and \emph{unordered trees}. For each problem one or more of the central algorithms are presented in detail in order to illustrate the techniques and ideas used for solving the problem. 

The survey is presented in the original published form except for minor typographical corrections. However, significant progress has been made on many of the problems since publication. To account for these, we give a brief introduction to each of the problems and discuss recent developments, focusing on our own contributions to the tree inclusion problem and the tree path subsequence problem. 

\subsection{Tree Edit Operations}\label{i:editoperation}
Let $T$ be a rooted tree. We call $T$ a \emph{labeled tree} if each node is a assigned a symbol from a finite alphabet $\Sigma$. We say that $T$ is an \emph{ordered tree} if a left-to-right order among siblings in $T$ is given. If  $T$ is an ordered tree the \emph{tree edit operations} are defined as follows:
 \begin{description}
  \item[relabel] Change the label of a node $v$ in $T$.
  \item[delete] Delete a non-root node $v$ in $T$ with parent $v'$, making the children of $v$ become the children of $v'$. The children are inserted in the place of $v$ as a subsequence in the left-to-right order of the children of $v'$.
   \item[insert] The complement of delete. Insert a node $v$ as a child of $v'$ in $T$ making $v$ the parent of a consecutive subsequence of the children of $v'$.   
\end{description}
For unordered trees the operations can be defined similarly. In this case, the insert and delete operations works on a \emph{subset} instead of a subsequence. Figure~\ref{t1:operationexample} on page \pageref{t1:operationexample} illustrates the operations. 

\subsection{Tree Edit Distance}\label{i:treeeditdistance}
Let $P$ and $T$ be two rooted and labeled trees called the \emph{pattern} and the \emph{target}, respectively. The \emph{tree edit distance} between $P$ and $T$ is the minimum cost  of transforming $P$ to $T$ by sequence of tree edit operations called an \emph{edit script}. The cost of each tree edit operation is given by metric \emph{cost function} assigning a real value to each operation depending on the labels of the nodes involved. The cost of a sequence of edit operations is the sum of the costs of the operations in the sequence. The \emph{tree edit distance problem} is to compute the tree edit distance and a corresponding minimum cost edit script. 

To state the complexities for the problem let $n_P$, $l_P$, $d_P$, and $i_P$ denote the number of nodes, number of leaves, the maximum depth, and the maximum in-degree of $P$, respectively. Similarly, define $n_T$, $l_T$, $d_T$, and $i_T$ for $T$. For simplicity in our bounds we will assume w.l.o.g. that $n_P \leq n_T$. 

The \emph{ordered} version of the tree edit distance problem was originally introduced by Tai~\cite{Tai1979}, who gave an algorithm using $O(n_Pn_Tl^2_Pl^2_T)$ time and space. In worst-case this is $O(n_P^3n_T^3) = O(n_T^6)$. Zhang and Shasha~\cite{ZS1989} gave an improved algorithm using $O(n_Pn_T\min(l_P, d_P)\min(l_T,d_T))$ time and $O(n_Pn_T)$ space. Note that in worst-case this is $O(n^2_Pn^2_T) = O(n^4_T)$ time.  Klein~\cite{Klein1998} showed how to improve the worst-case running time to $O(n^2_Pn_T\log n_T) = O(n_T^3\log n_T)$. The latter two algorithms are both based on \emph{dynamic programming} and may be viewed as different ways of computing a subset of the same  dynamic programming table. The basic dynamic programming idea is presented in Section~\ref{t1:simpledynamic} and a detailed presentation of Zhang and Shasha's and Klein's algorithms is given in Section~\ref{t1:zhangshasha} and~\ref{t1:klein}.

Using fast matrix multiplication Chen~\cite{Chen2001} gave an algorithm using $O(n_Pn_T + l^2_Pn_T + l^{2.5}_Pl_T)$ time and $O((n_P + l_P^2)\min(l_T, d_T) + n_T)$ space. In worst-case this algorithm runs in $O(n_Pn_T^{2.5}) = O(n_T^{3.5})$ time.

In~\cite{DT2005} Dulucq and Touzet introduced the concept of \emph{decomposition strategies} as a framework for algorithms based on the same type of dynamic program as \cite{ZS1989, Klein1998}. They proved a lower bound of $\Omega(n_Pn_T\log n_P \log n_T)$ for any such strategy. Very recently, Demaine et al.~\cite{DMRW2007} gave a new algorithm for tree edit distance within the decomposition strategy framework. In worst-case this algorithms uses $O(n^2_Pn_T(1+\log \frac{n_T}{n_P})) = O(n^3_T)$ time and $O(n_Pn_T)$ space. They also proved a matching worst-case lower bound for all algorithms within the decomposition strategy framework.

An interesting special case of the problem is the \emph{unit-cost tree edit distance problem}, where the goal is to  compute the \emph{number} of edit operations needed to transform $P$ to $T$.  Inspired by techniques from string matching~\cite{Ukkonen1985, LV1989}, Zhang and Shasha~\cite{SZ1990} proposed an algorithm for the ordered unit-cost tree edit distance problem. If $u$ is the number of tree edit operations needed to transform $P$ into $T$ their algorithm runs in $O(u^2 \min\{n_P,n_T\} \min\{l_P,l_T\})$ time. Hence, if the distance between $P$ and $T$ is small this algorithm significantly improves the bounds for the general tree edit distance problem. In a recent paper, Akutsu et al.~\cite{AFT2006} gave an approximation algorithm for the unit-cost tree edit distance problem. They gave an algorithm using $O(n_Pn_T)$ time that approximates the unit-cost tree edit distance for bounded degree trees to within a factor of $O(n^{3/4}_T)$. The idea in their algorithm is to extract modified \emph{Euler strings} (the sequence of labels obtained by visiting the tree in a depth-first left-to-right order) and subsequently compute the \emph{string edit distance} (see Section~\ref{i:stringeditdistance}) between these. This algorithm is based on earlier work on the relationship between the unit-cost tree edit distance and string edit distance of the corresponding Euler strings~\cite{Akutsu2006}.

Zhang et al.~\cite{ZSS1992} showed that the \emph{unordered} tree edit distance problem (recast as a decision problem) is NP-complete even for binary trees with an alphabet of size 2. Later, Zhang and Jiang~\cite{ZJ1994} showed that the problem is MAX-SNP hard.

\subsection{Constrained Tree Edit Distance}\label{i:constrainedtreeeditdistance}
Given that unordered tree edit distance is NP-complete and the algorithms for ordered tree edit distance are not practical for large trees, several authors have proposed restricted forms and variations of the problem. Selkow~\cite{Selkow1977} introduced the \emph{degree-1 edit distance}, where insertions and deletions are restricted to the leaves of the trees. Zhang et al.~\cite{Zhang1996a, ZWS1996} introduced the \emph{degree-2 edit distance}, where insertions and deletions are restricted to nodes with zero or one child. Zhang~\cite{Zhang1995, Zhang1996} introduced the \emph{constrained edit distance} that generalizes the degree-2 edit distance. Informally, constrained edit scripts must transform disjoint subtrees to disjoint subtrees (see Section~\ref{t1:constrainededitdistance}). In~\cite{Zhang1995, Zhang1996} Zhang presented algorithms for the constrained edit distance problem. For the ordered case he obtained $O(n_Pn_T)$ time and for the unordered case he obtained $O(n_Pn_T(i_P + i_T)\log (i_P + i_T))$ time. Both use space $O(n_Pn_T)$. Richter~\cite{Richter1997} presented an algorithm for the ordered version of the problem using $O(n_Pn_Ti_Pi_T)$ time and $O(n_Pd_Ti_T)$. Hence, for small degree and low depth trees this is a space improvement of Zhang's algorithm. Recently, Wang and Zhang~\cite{WZ2006} showed how to achieve $O(n_Pn_T)$ and $O(n_P \log n_T)$ space. The key idea is to process subtrees of $T$ according to a \emph{heavy-path decomposition} of $T$ (see Section~\ref{i:treedecompositions}).  

For other variations and analysis of the tree edit distance problem see Section~\ref{t1:othervariants} and also the recent work in~\cite{Touzet2003, DT2003, GK2005, Touzet2005, JP2006}.

\subsection{Tree Alignment Distance}\label{i:treealignmentdistance}
An \emph{alignment} of $P$ and $T$ is obtained by inserting specially labeled nodes (called \emph{spaces}) into $P$ and $T$ so they become isomorphic when labels are ignored. The resulting trees are then \emph{overlayed} on top of each other giving the alignment $A$. The cost of the alignment is the cost of all pairs of opposing labels in $A$ and the optimal alignment is the alignment of minimum cost. The \emph{tree alignment distance problem} is to compute a minimum cost alignment of $P$ and $T$. 

For strings the alignment distance and edit distance are equivalent notions. More precisely, for any two strings $A$ and $B$ the edit distance between $A$ and $B$ equals the value of an optimal alignment of $A$ and $B$~\cite{Gusfield1997}. However, for trees edit distance and alignment distance can be different (see the discussion in Section~\ref{t1:treealignmentdistance}).

The tree alignment distance problem was introduced by Jiang et al.~\cite{JWZ1995} who gave algorithms for both the ordered and unordered version of the problem. For the ordered version they gave an algorithm using $O(n_Pn_T(i_P + i_T)^2)$ time and $O(n_Pn_T(i_P + i_T))$ space. Hence, if $P$ and $T$ have small degrees this algorithm outperforms the known algorithms for ordered tree edit distance. For the unordered version Jiang et al.~\cite{JWZ1995} show how to modify their algorithm such that it still runs in $O(n_Pn_T)$ time for bounded degree trees. On the other hand, if one of the trees is allowed to have arbitrary degree the problem becomes MAX SNP-hard. Recall that the unordered tree edit distance problem  is MAX SNP-hard even if both tree have bounded degree. The algorithm by Jiang et al.~\cite{JWZ1995} for ordered tree alignment distance is discussed in detail in Section~\ref{t1:jiangwangzhang}. 

For \emph{similar} trees Jansson and Lingas~\cite{JL2003} presented a fast algorithm for ordered tree alignment. More precisely, if an optimal alignment requires at most $s$ spaces their algorithm computes the alignment in $O((n_P + n_T)\log (n_P + n_T)(i_P + i_T)^3s^2)$ time\footnote{Note that the result reported in Chapter~\ref{chap:tree1} is the slightly weaker bound from the conference version of their paper~\cite{JL2001}.}. Their algorithm may be viewed as a generalization of the fast algorithms for comparing similar sequences, see e.g., Section 3.3.4 in \cite{SM1997}. The recent techniques for space-efficient computation of constrained edit distances of Wang and Zhang~\cite{WZ2006}  also also apply to alignment of trees. Specifically, Wang and Zhang gave an algorithm for the tree alignment distance problem using $O(n_Pn_T(i_P + i_T)^2)$ time and $O(n_P i_T \log n_T (i_P + i_T))$ space. Hence, they match the running time of Jiang et al.~\cite{JWZ1995} and whenever $i_T \log n_T = o(n_T)$ they improve the space. This result improves an earlier space-efficient but slow algorithm by Wang and Zhao~\cite{WZ2003}.

Variations for more complicated cost functions for the tree alignment distance problem can be found in~\cite{HTGK2003, JHS2006}.

\subsection{Tree Inclusion}\label{i:treeinclusion}
The tree inclusion problem is defined as follows. We say that $P$ is \emph{included} in $T$ if $P$ can be obtained from $T$ by deleting nodes in $T$. The tree inclusion problem is to determine if $P$ can be included in $T$ and if so report all subtrees of $T$ that include $P$. 

The tree inclusion problem has recently been recognized as a query primitive for XML databases, see~\cite{SM2002, YLH2003,YLH2004, ZADR03, SN2000, TRS2002}. The basic idea is that an XML database can be viewed as a labeled and ordered tree, such that queries correspond to solving a tree inclusion problem (see Figure~\ref{t2:inclusionexample} on page~\pageref{t2:inclusionexample}).

The tree inclusion problem was introduced by Knuth \cite[exercise 2.3.2-22]{Knuth1969} who gave a sufficient
condition for testing inclusion. Kilpel\"{a}inen and Mannila \cite{KM1995} studied both the ordered and unordered version of the problem. For unordered trees they showed that the problem is NP-complete. The same result was obtained independently by Matou\v{s}ek and Thomas~\cite{MT1992}. For ordered trees Kilpel\"{a}inen and Mannila~\cite{KM1995} gave a simple dynamic programming algorithm using $O(n_Pn_T)$ time and space. This algorithm is presented in detail in Section~\ref{t1:kilpelainenmannila}. 

Several authors have improved the original dynamic programming algorithm. Kilpel\"{a}inen~\cite{Kilpelainen1992} gave a more space efficient version of the above algorithm using $O(n_Pd_T)$ space. Richter~\cite{Richter1997a} gave an algorithm using $O(\sigma_P n_T + m_{P,T}d_T)$ time, where $\sigma_P$ is the size of the alphabet of the labels in $P$ and $m_{P,T}$ is the set of \emph{matches}, defined as the number of pairs of nodes in $P$ and $T$ that have the same label. Hence, if the number of matches is small the time complexity of this algorithm improves the $O(n_Pn_T)$ time bound. The space complexity of the algorithm is $O(\sigma_P n_T + m_{P,T})$. Chen~\cite{Chen1998} presented a more complex algorithm using $O(l_Pn_T)$  time and $O(L_1l_P \min(d_T, l_T))$ space. Notice that the time and space complexity is still  $\Omega(n_Pn_T)$ in worst-case. 

A variation of the problem was studied by Valiente~\cite{Valiente2005} and Alonso and Schott~\cite{AS2001} gave an efficient average case algorithm. 

\paragraph{Our Results and Techniques}
In Chapter~\ref{chap:tree2} we give three new algorithms for the tree inclusion problem that together improve all the previous time and space bounds. More precisely, we show that the tree inclusion problem can be solved in $O(n_T)$ space with the following running time (Theorem~\ref{t2:thm:main}): 
\begin{equation*}
\min
\begin{cases}
      O(l_Pn_T), \\
      O(n_Pl_T\log \log n_T + n_T), \\
      O(\frac{n_Pn_T}{\log n_T} + n_{T}\log n_{T}).
\end{cases}
\end{equation*}
Hence, when either $P$ or $T$ has few leaves we obtain fast algorithms. When both trees have many leaves and $n_{P} = \Omega (\log^{2} n_{T})$, we instead improve the previous quadratic time bound by a logarithmic factor. In particular, we significantly improve the space bounds which in practical situations is a likely bottleneck. 

Our new algorithms are based on a different approach than the previous dynamic programming algorithms. The key idea is to construct a data structure on $T$ supporting a small number of procedures, called the \emph{set procedures}, on subsets of nodes of $T$. We show that any such data structure implies an
algorithm for the tree inclusion problem. We consider various implementations of this data structure all of which use linear space. The first one gives an algorithm with $O(l_Pn_T)$ running time. Secondly, we show that the running time
depends on a well-studied problem known as the \emph{tree color problem}. We give a connection between the tree color problem and the tree inclusion problem and using a data structure of Dietz \cite{Die89} we immediately obtain an algorithm with $O(n_Pl_T\log \log n_T + n_T)$ running time (see also Section~\ref{i:datastructures}).

Based on the simple algorithms above we show how to improve the worst-case running time of the set procedures by a logarithmic factor. The general idea is to divide $T$ into small trees called \emph{clusters} of logarithmic size, each of which overlap with other clusters on at most $2$ nodes. Each cluster is represented by a constant number of nodes in a \emph{macro tree}. The nodes in the macro tree are then connected according to the overlap of the cluster they represent. We show how to efficiently preprocess the clusters and the macro tree such that the set procedures use constant time for each cluster. Hence, the worst-case quadratic running time is improved by a logarithmic factor (see also Section~\ref{i:treedecompositions}).

\subsection{Tree Path Subsequence}\label{i:treepathsubsequence}
In Chapter~\ref{chap:tree3} we study the \emph{tree path subsequence problem} defined as follows. Given two sequences of labeled nodes $p$ and $t$, we say that $p$ is a \emph{subsequence} of $t$ if $p$ can be obtained by removing nodes from $t$. Given two rooted, labeled trees $P$ and $T$ the \emph{tree path subsequence problem} is to determine which paths in $P$ are subsequences of which paths in $T$. Here a path begins at the root and ends at a leaf. That is, for each path $p$ in $P$, we must report all paths $t$ in $T$ such that $p$ is a subsequence of $t$. 

In the tree path subsequence problem each path is considered \emph{individually}, in the sense that removing a node from a path do not affect any of the other paths that the node lies on. This should be seen in contrast to the tree inclusion problem where each node deletion affects \emph{all} of these paths. By the definition tree path subsequence does not fit into tree edit operations framework and whether or not the trees are ordered does not matter as long as the paths can be uniquely identified. 

A necessary condition for $P$ to be included in $T$ is that all paths in $P$ are subsequences of paths in $T$. As we will see shortly, the tree path subsequence problem can be solved in polynomial time and therefore we can use algorithms for tree path subsequence as a fast heuristic for unordered tree inclusion (recall that unordered tree inclusion is NP-complete). Section~\ref{t3:applications} contains a detailed discussion of applications.

Tree path subsequence can be solved trivially in polynomial time using basic techniques. Given two strings (or labeled paths) $a$ and $b$, it is straightforward to determine if $a$ is a subsequence of $b$ in $O(|a|+|b|)$ time. It follows that we can solve tree path subsequence in worst-case $O(n_Pn_T(n_P + n_T))$ time. Alternatively, Baeza-Yates~\cite{BaezaYates1991} gave a data structure using $O(|b|\log |b|)$ preprocessing time such that testing whether $a$ is a subsequence of $b$ can be done in $O(|a|\log |b|)$ time. Using this data structure on each path in $T$ we obtain solution to the tree path subsequence problem using $O(n^2_T\log n_T + n^2_P\log n_T)$ time. The data structure for subsequences can be improved as discussed in Section~\ref{i:subsequenceindexing}. However, a specialized and more efficient solution was discovered by Chen~\cite{Chen2000} who showed how to solve the tree path subsequence problem in $O(\min(l_{P}n_{T} + n_P, n_{P}l_{T}+n_T))$ time and $O(l_{P}d_{T} + n_P + n_T)$ space. Note that in worst-case this is $\Omega(n_Pn_T)$ time and space.

\paragraph{Our Results and Techniques}
In Chapter~\ref{chap:tree3} we give three new algorithms for the tree path subsequence problem improving the previous time and space bounds. Concretely, we show that the problem can be solved in $O(n_T)$ space with the following time complexity (Theorem~\ref{t3:main}):
\begin{equation*}
\min
\begin{cases}
    O(l_{P}n_{T} + n_P) , \\
     O(n_{P}l_{T}+n_T) , \\
     O(\frac{n_{P}n_{T}}{\log n_{T}}+ n_T + n_P \log n_P).
\end{cases}
\end{equation*}
The first two bounds in Theorem~\ref{t3:main} match the previous time bounds while improving the space to linear. The latter bound improves the worst-case $O(n_Pn_T)$ running time whenever $\log n_{P} = O(n_{T}/ \log n_{T})$. Note that -- in worst-case -- the number of pairs consisting of a path from $P$ and a path $T$ is $\Omega(n_{P}n_{T})$, and therefore we need at least as many bits to report the solution to TPS. Hence, on a RAM with logarithmic word size our worst-case bound is optimal.

The two first bounds are achieved using an algorithm that resembles the algorithm of Chen~\cite{Chen2000}. At a high level, the algorithms are essentially identical and therefore the bounds should be regarded as an improved analysis of Chen's algorithm. The latter bound is achieved using an entirely new algorithm that improves the worst-case $O(n_Pn_T)$ time. Specifically, whenever $\log n_{P} = O(n_{T}/ \log n_{T})$ the running time is improved by a logarithmic factor. 

Our results are based on a simple framework for solving tree path subsequence. The main idea is to traverse $T$ while maintaining a subset of nodes in $P$, called the \emph{state}. When reaching a leaf $z$ in $T$ the
state represents the paths in $P$ that are a subsequences of the path from the root to $z$. At each step the state is updated using a simple procedure defined on subset of nodes. The result of Theorem~\ref{t3:main} is obtained by taking the best of two algorithms based on our framework: The first one uses a simple data structure to maintain the state. This leads to an algorithm using $O(\min(l_{P}n_{T} + n_P, n_{P}l_{T}+n_T))$ time. At a high level this algorithm resembles the algorithm of Chen~\cite{Chen2000} and achieves the same running time. However, we improve the analysis of the algorithm and show a space bound of  $O(n_T)$. Our second algorithm combines several 
techniques. Starting with a simple quadratic time and space algorithm, we show how to reduce the space to $O(n_P \log n_T)$ using a \emph{heavy-path decomposition} of $T$. We then divide $P$ into small subtrees of size $\Theta(\log n_T)$ called \emph{micro trees}. The micro trees are then preprocessed such that subsets of nodes in a micro tree can be maintained in constant time and space leading to a logarithmic improvement of the time and space bound (see also Section~\ref{i:treedecompositions}).

\section{String Matching}\label{i:stringmatching}
String matching is a classical core area within theoretical and practical algorithms, with numerous applications in areas such as computational biology, search engines, data compression, and compilers, see~\cite{Gusfield1997}. 

In this dissertation we consider the string edit distance problem, approximate string matching problem, regular expression matching problem, approximate regular expression matching problem, and the subsequence indexing problem. In the following sections we present the known results and our contributions for each of these problems. 

\subsection{String Edit Distance and Approximate String Matching}\label{i:stringeditdistance}
Let $P$ and $Q$ be two strings. The \emph{string edit distance} between $P$ and $T$ is the minimum cost of transforming $P$ to $Q$ by a sequence of insertions, deletions, and substitutions of characters called the \emph{edit script}. The cost of each edit operation is given by a metric cost function. The \emph{string edit distance problem} is to compute the string edit distance between $P$ and $Q$ and a corresponding minimum cost edit script. Note that the string edit distance is identical to the tree edit distance if the trees are paths. 

The string edit distance problem has numerous applications. For instance, algorithms for it and its variants are widely used within computational biology to search for gene sequences in biological data bases. Implementations are available in the popular Basic Local Alignment Search Tool (BLAST)~\cite{AGMML1990}. 

To state the complexities for the problem, let $m$ and $n$ be the lengths of $P$ and $Q$, respectively, and assume w.l.o.g. that $m \leq n$. The standard textbook solution to the problem, due to Wagner and Fischer~\cite{WF1974}, fills in an $m+1 \times n+1$ size \emph{distance matrix} $D$ such that $D_{i,j}$ is the edit distance between the $i$th prefix of $P$ and the $j$th prefix of $Q$. Hence, the string edit distance between $P$ and $T$ can be found in $D_{m,n}$. Using dynamic programming each entry in $D$ can be computed in constant time leading to an algorithm using $O(mn)$ time and space. Using a classic divide and conquer technique of Hirschberg~\cite{Hirschberg1975} the space can be improved to $O(m)$. More details of the dynamic programming algorithm can be found in Section~\ref{s1:sec:stringedit}. 

For general cost functions Crochemore et al.~\cite{CLZ2003} recently improved the running time for the string edit distance problem to $O(nm/\log m + n)$ time and space for a constant sized alphabet. The result is achieved using a partition of the distance matrix based on a Ziv-Lempel factoring~\cite{ZL1978} of the strings. 

For the \emph{unit-cost string edit distance problem} faster algorithms are known. Masek and Paterson~\cite{MP1980} showed how to encode and compactly represent small submatrices of the dynamic programming table. The space needed for the encoded submatrices is $\Omega(n)$ but the dynamic programming algorithm can now be simulated in $O(mn/\log^2 n + m + n)$ time\footnote{Note that the result stated by the authors is a $\log n$ factor slower. This is because they assumed a computational model where operations take time proportional to their length in bits. To be consistent we have restated their result in the uniform cost model.}. This encoding and tabulating idea in this algorithm is often referred to as the \emph{Four Russian technique} after Arlazarov et al.~\cite{ADKF1970} who introduced the idea for boolean matrix multiplication. The algorithm of Masek and Paterson assumes a constant sized alphabet and this restriction cannot be trivially removed. The details of their algorithm is given i Section~\ref{s1:sec:stringedit}. 

Instead of encoding submatrices of the dynamic programming table using large tables, several algorithms based on simulating the dynamic programming algorithm using the arithmetic and logical operations of the word RAM have been suggested~\cite{BYG1992, WM1992b, Wright1994, BYN1996, Myers1999, HN2005}. We will refer to this technique as \emph{word-level parallelism} (see also Section~\ref{i:wordlevelparallelism}). Myers~\cite{Myers1999} gave a very practical $O(nm/w + n + m\sigma)$ time and $O(m\sigma/w + n+ m)$ space algorithm based on word-level parallelism. The algorithm can be modified in a straightforward fashion to handle arbitrary alphabets in $O(nm/w + n + m\log m)$ time and $O(m)$ space by using \emph{deterministic dictionaries}~\cite{HMP2001}. 

A close relative of the string edit distance problem is the \emph{approximate string matching problem}. Given strings $P$ and $Q$ and an \emph{error threshold} $k$, the goal is to find all ending positions of substrings in $Q$ whose unit-cost string edit distance to $P$ is at most $k$. Sellers~\cite{Sellers1980} showed how a simple modification of the dynamic programming algorithm for string edit distance can be used solve approximate string matching. Consequently, all of the bounds listed above for string edit distance also hold for approximate string matching. 

For more variations of the string edit distance and approximate string matching problems and algorithms optimized for various properties of the input strings see, e.g.,~\cite{Gotoh1982, Ukkonen1985b, Myers1986, MM1988, EGG1988, LV1989, EGCI1992, LMS1998,  MNU2004, ALP2004, CM2007, CH2002}. For surveys see \cite{Myers1991, Navarro2001a, Gusfield1997}.

\paragraph{Our Results and Techniques}
In Section~\ref{s1:sec:stringedit} we revisit the Four Russian algorithm of Masek and Paterson~\cite{MP1980} and the assumption that the alphabet size must be constant. We present an algorithm using $O(nm\log \log n/\log^2 n + m + n)$ time and $O(n)$ space that works for any alphabet (Theorem~\ref{s1:thm:stringeditdistance}). Thus, we remove the alphabet assumption at the cost of a factor $\log \log n$ in the running time. Compared with Myers' algorithm~\cite{Myers1999} (modified to work for any alphabet) that uses $O(nm/w + n + m\log m)$ time our algorithm is faster when $\frac{\log n}{\log \log n} = o(w)$ (assuming that the first terms of the complexities dominate). Our result immediately generalizes to approximate string matching.

The key idea to achieve our result is a more sophisticated encoding of submatrices of the distance matrix that maps input characters corresponding to the submatrix into a small range of integers. However, computing this encoding directly requires too much time for our result. Therefore we construct a two-level decomposition of the distance matrix such that multiple submatrices can be efficiently encoded simultaneously. Combined these ideas lead to the stated result. 

\subsection{Regular Expression Matching}\label{i:regularexpressionmatching}
Regular expressions are a simple and flexible way to recursively describe a set of strings composed from simple characters using union, concatenation, and Kleene star. Given a regular expression $R$ and a string $Q$ the \emph{regular expression matching problem} is to decide if $Q$ matches one of the string denoted by $R$. 

Regular expression are frequently used in the lexical analysis phase of compilers to specify and distinguish tokens to be passed to the syntax analysis phase. Standard programs such as Grep, the programming languages Perl~\cite{Wall1994} and Awk~\cite{AKW1988}, and most text editors, have mechanisms to deal with regular expressions. Recently, regular expression have also found applications in computational biology for protein searching~\cite{NR2003}.

Before discussing the known complexity results for regular expression matching we briefly present  some of the basic concepts. More details can be found in Aho et al.~\cite{ASU1986}. 
 
The set of \emph{regular expressions} over an alphabet $\Sigma$ is defined recursively as follows: A character $\alpha \in \Sigma$ is a regular expression, and if $S$ and $T$ are regular expressions then so is the
  \emph{concatenation}, $(S)\cdot(T)$, the \emph{union}, $(S)|(T)$, and the \emph{star}, $(S)^*$. The \emph{language} $L(R)$ generated by $R$ is defined as follows: $L(\alpha) = \{\alpha\}$, $L(S \cdot T) = L(S)\cdot L(T)$, that is, any string formed by the concatenation of a string in $L(S)$ with a string in $L(T)$, $L(S)|L(T) = L(S) \cup L(T)$, and $L(S^*) = \bigcup_{i \geq 0} L(S)^i$, where $L(S)^0 = \{\epsilon\}$ and $L(S)^i = L(S)^{i-1} \cdot L(S)$, for $i > 0$.     
Given a regular expression $R$ and a string $Q$ the \emph{regular expression matching problem} is to decide if $Q \in L(R)$.  
       
A \emph{finite automaton} is a tuple $A = (V, E, \Sigma, \theta, \Phi)$, where $V$ is a set of nodes called \emph{states}, $E$ is set of directed edges between states called \emph{transitions} each labeled by a character from $\Sigma \cup \{\epsilon\}$, $\theta \in V$ is a \emph{start state}, and $\Phi \subseteq V$ is a set of \emph{final states}. In short, $A$ is an edge-labeled directed graph with a special start node and a set of accepting nodes. 
$A$ is a \emph{deterministic finite automaton} (DFA) if $A$ does not contain any $\epsilon$-transitions, and all outgoing transitions of any state have different labels. Otherwise, $A$ is a \emph{non-deterministic automaton} (NFA). 
We say that $A$ \emph{accepts} a string $Q$ if there is a path from the start state to an accepting state such that the concatenation of labels on the path spells out $Q$. Otherwise, $A$ \emph{rejects} $Q$.

Let $R$ be a regular expression of length $m$ and let $Q$ be a string of length $n$. The classic solution to regular expression matching is to first construct a NFA $A$ accepting all strings in $L(R)$. There are several NFA constructions with this property~\cite{MY1960, Glushkov1961, Thomp1968}. Secondly, we \emph{simulate} $A$ on $Q$ by producing a sequence of state-sets $S_0, \ldots, S_n$  such that $S_i$ consists of all states in $A$ for which there is a path from the start state of $A$ that spells out the $i$th prefix of $Q$. Finally, $S_n$ contains an accepting state of $A$ if and only if $A$ accepts $Q$ and hence we can determine if $Q$ matches a string in $L(R)$ by inspecting $S_n$.

Thompson~\cite{Thomp1968} gave a simple well-known NFA construction for regular expressions that we will call a \emph{Thompson-NFA} (TNFA). For $R$ the TNFA $A$ has at most $2m$ states and $4m$ transitions, a single accepting state, and can be computed in $O(m)$ time. Each of the state-set in the simulation of $A$ on $Q$ can be computed in $O(m)$ time using a breadth-first search of $A$. This implies an algorithm for regular expression matching using $O(nm)$ time. Each of the state-sets only depends on the previous one and therefore the space is $O(m)$. The full details of Thompson's construction is given in Section~\ref{s2:sec:regex}.

We note that the regular expression matching problem is sometimes defined as reporting all of the ending positions of substrings of $Q$ matching $R$. Thompson's algorithm can easily be adapted without loss of efficiency for this problem. Simply add the start state to the current state-set before computing the next and inspect the accepting state of the state-sets at each step. All of the algorithms presented in this section can be adapted in a similar fashion such that the bounds listed below also hold for this variation.

In practical implementations regular expression matching is often solved by converting the NFA accepting the regular expression into a DFA before simulation. However, in worst-case the standard DFA-construction needs $O(2^{2m}\ceil{m/w}\sigma)$ space. With a more succinct representation of the DFA the space can be reduced to $O((2^m + \sigma)\ceil{m/w})$~\cite{NR2004, WM1992b}. Note that the space complexity is still exponential in the length of the regular expression. Normally, it is reported that the time complexity for simulating the DFA is $O(n)$, however, this analysis does not account for the limited word size of the word RAM. In particular, since there are $2^m$ states in the DFA each state requires $\Omega(m)$ bits to be addressed. Therefore we may need $\Omega(\ceil{m/w})$ time to identify the next state and thus the total time to simulate the DFA becomes $\Omega(n\ceil{m/w})$. This bound is matched by Navarro and Raffinot~\cite{NR2004} who showed how to solve the problem in $O((n + 2^m)\ceil{m/w})$ time and $O((2^m + \sigma)\ceil{m/w})$ space. Navarro and Raffinot~\cite{NR2004} suggested using a \emph{table splitting technique} to improve the space complexity of the DFA algorithm for regular expression matching. For any $s$ this technique gives an algorithm using $O((n + 2^{m/s})s\ceil{m/w})$ time and $O((2^{m/s}s + \sigma)\ceil{m/w})$ space. 

The DFA-based algorithms are primarily interesting for sufficiently small regular expressions. For instance, if $m = O(\log n)$ it follows that regular expression matching can be solved in $O(n)$ time and $O(n + \sigma)$ space.
Several heuristics can be applied to further improve the DFA based algorithms, i.e., we often do not need to fill in all entries in the table, the DFA can be stored in an adjacency-list representation and minimized, etc. None of these improve the above worst-case complexities of the DFA based algorithms.

Myers~\cite{Myers1992} showed how to efficiently combine the benefits of NFAs with DFA. The key idea in Myers' algorithm is to decompose the TNFA built from $R$ into $O(\ceil{m/\log n})$ subautomata each consisting of $\Theta(\log n)$ states. Using the Four Russian technique~\cite{ADKF1970} each subautomaton is converted into a DFA using $O(2^m) = O(n)$ space giving a total space complexity of $O(nm/\log n)$. The subautomata can then be simulated in constant time leading to an algorithm using $O(nm/\log n + (n+m)\log m)$ time. The details of Myers' algorithm can be found in Sections~\ref{s1:sec:regex} and~\ref{s2:sec:simul}. 

For variants and extension of the regular expression matching problem see~\cite{KMb1995, MOG1998, Yamamoto2001, NR2003, YM2003, ISY2003}.

\paragraph{Our Results and Techniques}
In Section~\ref{s1:sec:regex} we improve the space complexity of Myers' Four Russian algorithm. We present an algorithm using $O(nm/\log n + n + m\log m)$ time and $O(n)$ space (Theorem~\ref{s1:thm:regex}). Hence, we match or improve the running time of Myers' algorithm while we significantly improve the space complexity from $O(nm/\log n)$ to $O(n)$.

As in Myers' algorithm, our new result is achieved using a decomposition of the TNFA into small subautomata of  $\Theta(\log n)$ states. To improve the space complexity we give a more efficient encoding. First, we represent the labels of transitions in each subautomaton  using deterministic dictionaries~\cite{HMP2001}. Secondly, we bound the number of distinct TNFAs without labels on transitions. Using this bound we show that it is possible to encode \emph{all}  TNFAs with $x = \Theta(\log n)$ states in total space $O(n)$, thereby obtaining our result.

Our space-efficient Four Russian algorithm for regular expression matching is faster than Thompson's algorithm which uses $O(nm)$ time. However, to achieve the speedup we use $\Omega(n)$ space, which may still be significantly larger than the $O(m)$ space used by Thompson's algorithm. In Chapter~\ref{chap:string2} we study a different and more space-efficient approach to regular expression matching. Specifically, we show that regular expression matching can be solved in $O(m)$ space with the following running times (Theorem~\ref{s2:thm:main}): 
\begin{equation*}
\begin{cases}
      O(n\frac{m \log w}{w} + m \log w) & \text{ if $m > w$} \\
      O(n\log m + m\log m) & \text{ if $\sqrt{w} < m \leq  w$} \\
      O(\min(n+ m^2, n\log m + m\log m)) & \text{ if $m \leq \sqrt{w}$.}
\end{cases}
\end{equation*} 
To compare these bounds with previous results, let us assume a conservative word length of $w = \log n$. When the regular expression is ``large'', e.g., if  $m > \log n$, we achieve an $O(\frac{\log n}{\log \log n})$ factor speedup over Thompson's algorithm using $O(m)$ space. In this case we simultaneously match the best known time and space bounds for the problem, with the exception of an $O(\log \log n)$ factor in time. Next, consider the case when the regular expression is ``small'', e.g., $m = O(\log n)$.  In this case, we get an algorithm using $O(n\log \log n)$ time and $O(\log n)$ space. Hence, the space is improved exponentially at the cost of an $O(\log \log n)$ factor in time. In the case of an even smaller regular expression, e.g., $m = O(\sqrt{\log n})$, the slowdown can be eliminated and we achieve optimal $O(n)$ time. For larger word lengths, our time bounds improve. In particular, when $w > \log n \log \log n$ the bound is better in all cases, except for $\sqrt{w} \leq m \leq w$, and when $w > \log^2n$ it improves the time bound of Myers' algorithm.

As in Myers' and our previous algorithms for regular expression matching, this algorithm is based on a decomposition of the TNFA. However, for this result, a slightly more general decomposition is needed to handle different sizes of subautomata. We provide this by showing how any ``black-box'' algorithm for simulating small TNFAs can efficiently converted into an algorithm for simulating larger TNFAs (see Section~\ref{s2:sec:simul} and Lemma~\ref{s2:lem:simulation}). To achieve $O(m)$ space we cannot afford to encode the subautomata as in the Four Russian algorithms. Instead we present two algorithms that simulate the subautomata using word-level parallelism. The main problem in doing so is the complicated dependencies among states in TNFAs. A state may be connected via long paths of $\epsilon$-transitions to number of other states, all of which have to be traversed in parallel to simulate the TNFA. Our first algorithm, presented in Section~\ref{s2:sec:simple}, simulates TNFAs with $O(\sqrt{w})$ states in constant time for each step. The main idea is to explicitly represent the transitive closure of the $\epsilon$-paths compactly  in a constant number of words. Combined with a number of simple word operations to we show how to compute the next state-set in constant time. Our second, more complicated algorithm, presented in Section~\ref{s2:sec:notsimple} simulates TNFAs of with $O(w)$ states in $O(\log w)$ time for each step. Instead of representing the transitive closure of the of the $\epsilon$-paths this algorithm recursively decomposes the TNFA into $O(\log w)$ levels that represent increasingly smaller subautomata. Using this decomposition we then show to traverse all of the $\epsilon$-paths in constant time for each level. We combine the two algorithms with our black-box simulation of large TNFAs, and choose the best algorithm in the various cases to get the stated result.

\subsection{Approximate Regular Expression Matching}
Given a regular expression $R$, a string $Q$, and an error threshold $k$ the \emph{approximate regular expression matching problem} is to determine if the minimum unit-cost string edit distance between $Q$ and a string in $L(R)$ is at most $k$. As in the above let $m$ and $n$ be the lengths of $R$ and $Q$, respectively. 

Myers and Miller~\cite{MM1989} introduced the problem and gave an $O(nm)$ time and $O(m)$ space algorithm. Their algorithm is an extension of the standard dynamic programming algorithm for approximate string matching adapted to handle regular expressions. Note that the time and space complexities are the same as in the simple case of strings. Assuming a constant sized alphabet,  Wu et al.~\cite{WMM1995} proposed a Four Russian algorithm using $O(\frac{mn\log(k+2)}{\log n} + n + m)$ time and $O(\frac{m\sqrt{n}\log(k+2)}{\log n} + n + m)$ space. This algorithm combines decomposition of TNFAs into subautomata as the earlier algorithm of Myers for regular expression matching~\cite{Myers1992} and the dynamic programming idea of Myers and Miller~\cite{MM1989} for approximate regular expression matching. Recently, Navarro~\cite{Navarro2004} proposed a practical DFA based solution for small regular expressions. 

Variants of approximate regular expression matching including extensions to more complex cost functions can be found in~\cite{MM1989, KMa1995, Myers1996, MOG1998, NR2003}.  

\paragraph{Our Results and Techniques}
In Section~\ref{s1:sec:appregexmatching} we present an algorithm for approximate regular expression matching using $O(\frac{mn\log(k+2)}{\log n} + n + m\log m)$ time and $O(n)$ space that works for any alphabet (Theorem~\ref{s1:thm:approxregex}). Hence, we match the running time of Wu et al.~\cite{WMM1995} while improving the space complexity from  $O(\frac{m\sqrt{n}\log(k+2)}{\log n} + n + m)$ to $O(n)$. 

We obtain the result as a simple combination and extension of the techniques used in our Four Russian algorithm for regular expression matching and the algorithm of Wu et al.~\cite{WMM1995}.

\subsection{Subsequence Indexing}\label{i:subsequenceindexing}
Recall that a subsequence of a string $Q$ is a string that can be obtained from $Q$ by deleting zero or more characters. The \emph{subsequence indexing problem} is to preprocess a string $Q$ into a data structure efficiently supporting queries of the form: ``Is $P$ a subsequence of $Q$?'' for any string $P$.  

Baeza-Yates~\cite{BaezaYates1991} introduced the problem and gave several algorithms. Let $m$ and $n$ denote the length of $P$ and $Q$, respectively, let $\sigma$ be the size of the alphabet. Baeza-Yates showed that the subsequence indexing problem can either be solved using $O(n\sigma)$ space and $O(m)$ query time, $O(n\log \sigma)$ space and $O(m\log \sigma)$ query time, or $O(n)$ space and $O(m\log n)$ query time. For these algorithm the preprocessing time matches the space bounds. 

The key component in Baeza-Yates' solutions is a DFA called the \emph{directed acyclic subsequence graph} (DASG).  Baeza-Yates obtains the first trade-off listed above by explicitly constructing the DASG and using it to answer queries. The second trade-off follows from an encoded version of the DASG and the third trade-off is based on simulating the DASG using predecessor data structures.

Several variants of subsequence indexing have been studied, see~\cite{DFGGK1997, BCGM1999} and the surveys~\cite{Tronicek2001, CMT2003}.

%

\paragraph{Our Results and Techniques}
In Section~\ref{s1:sec:subseq} we improve the bounds for subsequence indexing. We show how to solve the problem  using $O(n\sigma^{1/2^l})$ space and preprocessing time and $O(m(l+1))$ time for queries, for $0 \leq l \leq \log \log \sigma$ (Theorem~\ref{s1:subseqthm}). In particular, for constant $l$ we get a data structure using $O(n\sigma^\epsilon)$ space and preprocessing time and $O(m)$ query time and for $l = \log \log \sigma$ we get a data structure using $O(n)$ space and preprocessing time and $O(m\log \log \sigma)$ for queries.

The key idea is a simple two-level decomposition of the DASG that efficiently combines the explicit DASG with a fast predecessor structure. Using the classical van Emde Boas data structure~\cite{BKZ1977} leads to $O(n)$ space and preprocessing time with $O(m\log \log \sigma)$ query time. To get the full trade-off, we replace this data structure with a recent one by Thorup~\cite{Thorup2003}.

\section{Compressed String Matching}\label{i:compressed}
\emph{Compressed string matching} covers problems that involve searching for an (uncompressed) pattern in a compressed target text without decompressing it. The goal is to search more efficiently than the obvious approach of decompressing the target and then performing the matching. Modern text data bases, e.g., for biological data and World Wide Web data, are huge. To save time and space the data must be kept in compressed form while allowing searching. Therefore, efficient algorithms for compressed string matching are needed. 

Amir and Benson~\cite{AB1992,AB1992a} initiated the study of compressed string matching. Subsequently, several researchers have proposed algorithms for various types of string matching problems and compression methods~\cite{AB1992, FT1998, KTSMA1998, KNU2003, Navarro2003, MUN2003}. For instance, given a string $Q$ of length $u$ compressed with the Ziv-Lempel-Welch scheme~\cite{Welch1984} into a string of length $n$, Amir et al.~\cite{ABF1996} gave an algorithm for finding all exact occurrences of a pattern string of length $m$ in $O(n + m^2)$ time and space. Algorithms for \emph{fully compressed pattern matching}, where both the pattern and the target are compressed have also been studied (see the survey by Rytter~\cite{Rytter1999}).

In Chapter~\ref{chap:string3}, we study approximate string matching and regular expression matching problems in the context of compressed texts. As in previous work on these problems~\cite{KNU2003, Navarro2003} we focus on the popular \zla\  and \zlw\  compression schemes~\cite{ZL1978, Welch1984}. These compression schemes adaptively divide the input into substrings, called \emph{phrases}, which can be compactly encoded using references to other phrases. During encoding and decoding with the \zla/\zlw\  compression schemes the phrases are typically stored in a \emph{dictionary trie} for fast access. Details of Ziv-Lempel compression can be found in Section~\ref{s3:zlc}.

\subsection{Compressed Approximate String Matching}\label{i:approximatestringcompressed}
Recall that given strings $P$ and $Q$ and an error threshold $k$, the approximate string matching problem is to find all ending positions of substrings of $Q$ whose unit-cost string edit distance to $P$ is at most $k$. Let $m$ and $u$ denote the length of $P$ and $Q$, respectively. For our purposes we are particularly interesting in the fast algorithms for small values of $k$, namely, the $O(uk)$ time algorithm by Landau and Vishkin~\cite{LV1989} and the more recent $O(uk^4/m + u)$ time algorithm due to Cole and Hariharan~\cite{CH2002} (we assume w.l.o.g. that $k < m$). Both of these can be implemented in $O(m)$ space.

K{\"a}rkk{\"a}inen et al.~\cite{KNU2003} initiated the study of compressed approximate string matching with the \zla/\zlw\  compression schemes. If $n$ is the length of the compressed text, their algorithm achieves $O(nmk + \occ)$ time and $O(nmk)$ space, where $\occ$ is the number of occurrences of the pattern. For special cases and restricted versions of compressed approximate string matching, other algorithms have been proposed~\cite{MKTSA2000, NR1998}.  An experimental study of the problem and an optimized practical implementation can be found in~\cite{NKTSA01}. Crochemore et al.~\cite{CLZ2003} gave an algorithm for the fully compressed version of the problem. If $m'$ is the length of the compressed pattern their algorithm runs in $O(um' + nm)$ time and space.   

\paragraph{Our Results and Techniques}
In Section~\ref{s3:approx} we show how to efficiently use algorithms for the uncompressed approximate string matching problem to achieve a simple time-space trade-off. Specifically, let $t(m,u, k)$ and $s(m,u,k)$ denote the time and space, respectively, needed by any algorithm to solve the (uncompressed) approximate string matching problem with error threshold $k$ for pattern and text of length $m$ and $u$, respectively. We show that if $Q$ is compressed using {\zla } then given a parameter $\tau \geq 1$ we can solve compressed approximate string matching in $O(n(\tau + m + t(m, 2m+2k,k)) + \occ)$ expected time and $O(n/\tau + m + s(m,2m+2k,k) + \occ)$ space (Theorem~\ref{s3:thm:approx}). The expectation is due to hashing and can be removed at an additional $O(n)$ space cost. In this case the bound also hold for \zlw\  compressed strings. We assume that the algorithm for the uncompressed problem produces the matches in sorted order (as is the case for all algorithms that we are aware of). Otherwise, additional time for sorting must be included in the bounds. 

To compare our result with the algorithm of K{\"a}rkk{\"a}inen et al.~\cite{KNU2003}, plug in the Landau-Vishkin algorithm and set $\tau = mk$. This gives an algorithm using $O(nmk + \occ)$ time and $O(n/mk + m + \occ)$ space. These bounds matches the best known time bound while improving the space by a factor $\Theta(m^2k^2)$. Alternatively, if we plug in the Cole-Hariharan algorithm and set $\tau = k^4 + m$ we get an algorithm using $O(nk^4 + nm + \occ)$ time and $O(n/(k^4 + m) + m + \occ)$ space. Whenever $k = O(m^{1/4})$ this is $O(nm + \occ)$ time and $O(n/m + m + \occ)$ space. 

The key idea for our result is a simple $o(n)$ space data structure for {\zla } compressed texts. This data structures compactly represents a subset of the dynamic dictionary trie whose size depends on the parameter $\tau$. Combined with the compressed text the data structure enables fast access to relevant parts of the trie, thereby allowing algorithms to solve compressed string matching problems in $o(n)$ space. To the best of our knowledge, all previous non-trivial compressed string matching algorithm for \zla/\zlw\  compressed text, with the exception of a very slow algorithm for exact string matching by Amir et al.~\cite{ABF1996}, explicitly construct the trie and therefore use $\Omega(n)$ space. 

Our bound depends on the special nature of {\zla } compression scheme and do not in general hold for \zlw\ compressed texts. However, whenever we use $\Omega(n)$ space in the trade-off we have sufficient space to explicitly construct the trie and therefore do not need our $o(n)$ space data structure. In this case the bound holds for \zlw\ compressed texts and hashing is not needed. Note that even with $\Omega(n)$ space we significantly improve the previous bounds.

\subsection{Compressed Regular Expression Matching}
Let $R$ be a regular expression and let $Q$ be string. Recall that deciding if $Q\in L(R)$ and finding all occurrences of substrings of $Q$ matching $L(R)$ was the same problem for all of the finite automaton-based algorithms discussed in Section~\ref{i:regularexpressionmatching}. In the compressed setting this is not the case since the complexities we obtain for the substring variant of the problem may be dominated by the number of reported occurrences. In this section we therefore define regular expression matching as follows: Given a regular expression $R$ and a string $Q$, the \emph{regular expression matching problem} is to find all ending positions of substrings in $Q$ matching a string in $L(R)$.

The only solution to the compressed problem is due to Navarro~\cite{Navarro2003}, who studied the problem on \zla/\zlw\  compressed strings. This algorithms depends on a complicated mix of Four Russian techniques and word-level parallelism. As a similar improvement is straightforward to obtain for our algorithm we ignore these factors in the bounds presented here. With this simplification Navarro's algorithm uses $O(nm^2 + \occ \cdot m\log m)$ time and $O(nm^2)$ space, where $m$ and $n$ are the lengths of the regular expression and the compressed text, respectively. 

\paragraph{Our Results and Techniques}
We show that if $Q$ is compressed using {\zla } or {\zlw } then given a parameter $\tau \geq 1$ we can solve compressed regular expression matching in $O(nm(m + \tau) + \occ\cdot m \log m)$ time and $O(nm^2/\tau + nm)$ space (Theorem~\ref{s3:thm:regularex}). If we choose $\tau = m$ we obtain an algorithm using $O(nm^2 + \occ\cdot m \log m)$ time and $O(nm)$ space. This matches the best known time bound while improving the space by a factor $\Theta(m)$. With word-parallel techniques these bounds can be improved slightly. The full details are given in Section~\ref{s3:sec:wordparallel}.

As in the previous section we obtain this result by representing information at a subset of the nodes in dictionary trie depending on the parameter $\tau$. In this case the total space used is always $\Omega(n)$ and therefore we have sufficient space to store the trie.

\section{Core Techniques}\label{i:techniques}
In this section we identify the core techniques used in this dissertation. 

\subsection{Data Structures}\label{i:datastructures}
The basic goal of data structures is to organize information compactly and support fast queries. Hence, it is not surprising that using the proper data structures in the design of pattern matching algorithms is important. A good example is the tree data structures used in our algorithms for the tree inclusion problem (Chapter~\ref{chap:tree2}). Let $T$ be a rooted and labeled tree. A node $z$ is a \emph{common ancestor} of nodes $v$ and $w$ if it is an ancestor of both $v$ and $w$. The \emph{nearest common ancestor} of $v$ and $w$ is the common ancestor of $v$ and $w$ of maximum depth. The \emph{nearest common ancestor problem} is to preprocess $T$ into a data structure supporting nearest common ancestor queries. Several linear-space data structures for the nearest common ancestor problem that supports queries in constant time are known~\cite{HT1984, BFC2000, AGKR2004}. The \emph{first ancestor of $w$ labeled $\alpha$} is the ancestor of $w$ of maximum depth labeled $\alpha$. The \emph{tree color problem} is to preprocess $T$ into a data structure supporting first label queries. This is well-studied problem~\cite{Die89,MM1996, FM1996, AHR1998}. In particular, Dietz~\cite{Die89} gave a linear space solution supporting queries in $O(\log \log n_T)$ time. We use data structures for both the nearest common ancestor problem and the tree color problem extensively in our algorithms for the tree inclusion problem. More precisely, let $v$ be a node in $P$ with children $v_1,\ldots, v_k$. After computing which subtrees of $T$ that include each of the subtrees of $P$ rooted at $v_1,\ldots, v_k$ we find the subtrees of $T$ that include the subtree of $P$ rooted at $v$ using a series of nearest common ancestor and first label queries. Much of the design of our algorithms for tree inclusion was directly influenced by our knowledge of these data structures.

We use \emph{dictionaries} in many of our results to handle large alphabets efficiently. Given a subset $S$ of elements from a universe $U$ a dictionary preprocesses $S$ into a data structure supporting membership queries of the form: ``Is $x\in S$?'' for any $x \in U$. The  dictionary also supports retrieval of \emph{satellite data} associated with $x$. In many of our results we rely on a dictionary construction due to Hagerup et al.~\cite{HMP2001}. They show how to preprocess a subset $S$ of $n$ elements from the universe $U = \{0,1\}^w$ in $O(n\log n)$ time into an $O(n)$ space data structure supporting membership queries in constant time. The preprocessing  makes heavy uses of weak non-uniformity to obtain an \emph{error correcting code}. A suitable code can be computed in $O(w2^w)$ time and no better algorithm than brute force search is known. In nearly all of our algorithms that use this dictionary data structure we only work with polynomial sized universes. In this case, the dictionary can be constructed in the above stated bound without the need for weak non-uniformity. The only algorithms in this dissertation that use larger universes are our algorithms for regular expression matching in Chapter~\ref{chap:string2}. Both of these algorithms construct a deterministic dictionary for $m$ elements in $O(m\log m)$ time. However, in the first algorithm (Section~\ref{s2:sec:simple}) we may replace the dictionary with another dictionary data structure by Ru\v{z}i\'{c}~\cite{Ruzic2004} that runs in $O(m^{1+\epsilon})$ preprocessing time and does not use weak non-uniformity. Since the total running time of the algorithm is $\Omega(m^2)$ this does not affect our result. Our second algorithm (Section~\ref{s2:sec:notsimple}) uses $\Omega(\log m)$ time in each step of the simulation and therefore we may simply use a sorted array and binary searches to perform the lookup.

A key component in our result for compressed string matching (Chapter~\ref{chap:string3}) is an efficient dictionary for sets that dynamically change under insertions of elements. This is needed to maintain our sublinear space data structure for representing a subset of the trie while the trie is dynamically growing through additions of leaves (see Section \ref{s3:selectingcompression}).  For this purpose we use the dynamic perfect hashing data structure by Dietzfelbinger et al.~\cite{DKMMRT1994} that supports constant time membership queries and constant amortized expected time insertions and deletions.

Finally, for the subsequence indexing problem, presented in Section~\ref{s1:sec:subseq}, we use the van Emde Boas predecessor data structure(vEB)~\cite{Boas1977, BKZ1977}. For $x$ integers in the range $[1,X]$ a vEB answers queries in $O(\log \log X)$ time and combined with perfect hashing the space complexity is $O(x)$~\cite{MN1990}. To get the full trade-off we replace the vEB with a more recent data structure by Thorup~\cite[Thm. 2]{Thorup2003}. This data structure supports successor queries of $x$ integers in the range $[1,X]$ using $O(xX^{1/2^l})$ preprocessing time and space with query time $O(l+1)$, for $0\leq l \leq \log \log X$. P\v{a}tra\c{s}cu and Thorup~\cite{PT2006} recently showed that in linear space the time bounds for the van Emde Boas data structures are optimal. Since predecessor searches is the computational bottleneck in our algorithms for subsequence queries we cannot hope to get an $O(n)$ space data and $O(m)$ query time using the techniques presented in Section~\ref{s1:sec:subseq}.

\subsection{Tree Techniques}\label{i:treedecompositions}
Several combinatorial properties of trees are used extensively throughout the dissertation. The simplest one is the \emph{heavy-path decomposition}~\cite{HT1984}. The technique partitions a tree into disjoint \emph{heavy-paths}, such that at most a logarithmic number of distinct heavy-paths are encountered on any root-to-leaf path (see Section~\ref{t3:heavy} for more details). The heavy-path decomposition is used in Klein's algorithm~\cite{Klein1998} (presented in Section~\ref{t1:klein}) to achieve a worst-case efficient algorithm for tree edit distance. To improve the space of the constrained tree edit distance problem and tree alignment Wang and Zhang~\cite{WZ2006} order the computation of children of nodes according to a heavy path decomposition. In our worst-case algorithm for the tree path subsequence problem (Section~\ref{t3:worstcase}) we traverse the target tree according to a heavy-path traversal to reduce the space of an algorithm from $O(n_Pn_T)$ to $O(n_P\log n_T)$.

Various forms of grouping or clustering of nodes in trees is used extensively. Often the relationship between the clusters is represented as another tree called a \emph{macro tree}. In particular, in our third algorithm for the tree inclusion problem (Section~\ref{t2:micromacro}) we cluster the target tree into small logarithmic sized subtrees overlapping in at most two nodes. A macro tree is used to represent the overlap between the clusters and internal properties of the clusters. This type of clustering is well-known from several tree data structures see e.g.,~\cite{AHT2000, AHLT1997, Frederickson1997}, and the macro-tree representation is inspired by a related construction of Alstrup and Rauhe~\cite{AR2002c}. 

In our worst-case algorithm for tree path subsequence (Section~\ref{t3:worstcase}), a simpler tree clustering due to Gabow and Tarjan~\cite{GT1989} is used. Here we cluster the pattern tree into logarithmic sized subtrees that may overlap only in their roots. We also construct a macro tree from these overlaps. Note that to achieve our worst-case bound for the tree path subsequence we are both clustering the pattern tree and using a heavy-path decomposition of  the target tree.

For the regular expression matching and approximate regular expression problem we cluster TNFAs into small subautomata of varying sizes (see Sections~\ref{s1:sec:clustering} and~\ref{s2:sec:clustering}). This clustering is based on a clustering of the \emph{parse tree} of the regular expression and is similar to the one by Gabow and Tarjan~\cite{GT1989}. Our second word-level parallel algorithm for regular expression matching (Section~\ref{s2:sec:notsimple}) uses a recursive form of this clustering on subautomata of TNFAs to efficiently traverse paths of $\epsilon$-transitions in parallel.

For the subsequence indexing problem we cluster the DASG according to the size of the alphabet. The clusters are represented in a macro DASG.  

Finally, for compressed string matching we show how to efficiently select a small subset of nodes in the dynamic dictionary trie such that the minimum distance from a node to a node is bounded by a given parameter.

\subsection{Word-RAM Techniques}\label{i:wordlevelparallelism}
The Four Russian technique~\cite{ADKF1970} is used in several algorithms to achieve speedup. The basic idea is to tabulate and encode solutions to all inputs of small subproblems, and use this to achieve a speedup. Combined with tree clustering we use the Four Russian technique in our worst-case efficient algorithms for tree inclusion and tree path subsequence (Sections~\ref{t2:micromacro} and~\ref{t3:worstcase}) to achieve logarithmic speedups. Our results for string edit distance, regular expression matching, and approximate regular expression matching are improvements of previously known Four Russian techniques for these problems (see Sections~\ref{s1:sec:stringedit}, \ref{s1:sec:regex}, and~\ref{s1:sec:appregexmatching}). 

Four Russian techniques have been widely used. For instance, many of the recent subcubic algorithms for the all-pairs-shortest-path problem make heavy use of this technique~\cite{Takaoka2004, Zwick2004, Han2004, Chan2006, Han2006, Chan2007}.

Our latest results for regular expression matching (Chapter~\ref{chap:string2}) does not use the Four Russian technique. Instead of simulating the automata using table-lookups we simulate them using the instruction set of the word RAM. This kind of technique is often called \emph{word-level parallelism}. Compared to our Four Russian algorithm this more space-efficient since the large tables are avoided. Furthermore, the speedup depends on the word length rather than the available space for tables and therefore our algorithm can take advantage of machines with long word length.

Word-level parallelism has been used in many areas of algorithms. For instance, in the fast algorithms for sorting integers
\cite{Boas1977, FredmanWillardFussion, AH1997, AHNR1998, HT2002}. Within the area of string matching many of the fastest practical algorithms are based on word-level parallel techniques, see e.g.,~\cite{BYG1992, Myers1999, Navarro2001a}. In string matching, the term \emph{bit-parallelism}, introduced by Baeza-Yates~\cite{BaezaYates1989}, is often used instead of the term word-level parallelism.

\section{Discussion}\label{i:conclusion}
I will conclude this introduction by discussing which of the contributions in this dissertation I find the most interesting. 

First, I want to mention our results for the tree inclusion problem. As the volume of tree structured data is growing rapidly in areas such as biology and image analysis I believe algorithms for querying of trees will become very important in the near future. Our work shows how to obtain a fast and space-efficient algorithm for a very simple tree query problem, but the ideas may be useful to obtain improved results for more sophisticated tree query problems.

Secondly, I want to mention our results for the regular expression matching using word-level parallelism. Modern computers have large word lengths and support a sophisticated set of instructions, see e.g.,~\cite{PWW1997, TONH1996, TH1999, OFW1999, DDHS2000}. Taking advantage of such features is a major challenge for the string matching community. Some of the steps used in our regular expression matching algorithms resemble some of these sophisticated instructions, and therefore it is likely possible to implement a fast practical version of the algorithm. We believe that some of the ideas may also be useful to improve other string matching problems. 

Finally, I want to mention our results for compressed string matching. Almost all of the available algorithms for compressed string matching problems require space at least linear in the size of the compressed text. Since space is a likely bottleneck in practical situations more space-efficient algorithms are needed. Our work solves approximate string matching efficiently using sublinear space, and we believe that the techniques may be useful in other compressed string matching problems.

\emptythanks
\chapter{A Survey on Tree Edit Distance and Related Problems}\label{chap:tree1}

\title{A Survey on Tree Edit Distance and Related Problems}
\author{Philip Bille \\ IT University of Copenhagen \\ \texttt{beetle@itu.dk}}
\date{}
\cleartooddpage
\maketitle

\begin{abstract}
We survey the problem of comparing labeled trees based on simple local operations of deleting, inserting, and relabeling nodes. These operations lead to the tree edit distance, alignment distance, and inclusion problem. For each problem we review the results available and present, in detail, one or more of the central algorithms for solving the problem. 
\end{abstract}

\section{Introduction}
Trees are among the most common and well-studied combinatorial structures in computer science. In particular, the problem of comparing trees occurs in several diverse areas such as computational biology, structured text databases, image analysis, automatic theorem proving, and compiler optimization \cite{Tai1979, ZS1989,  KM1995, KTSK2000, CO1982, RR1992, ZSW1994}. For example, in computational biology, computing the similarity between trees under various distance measures is used in the comparison of RNA secondary structures \cite{ZS1989, JWZ1995}. 

\begin{figure}[t]
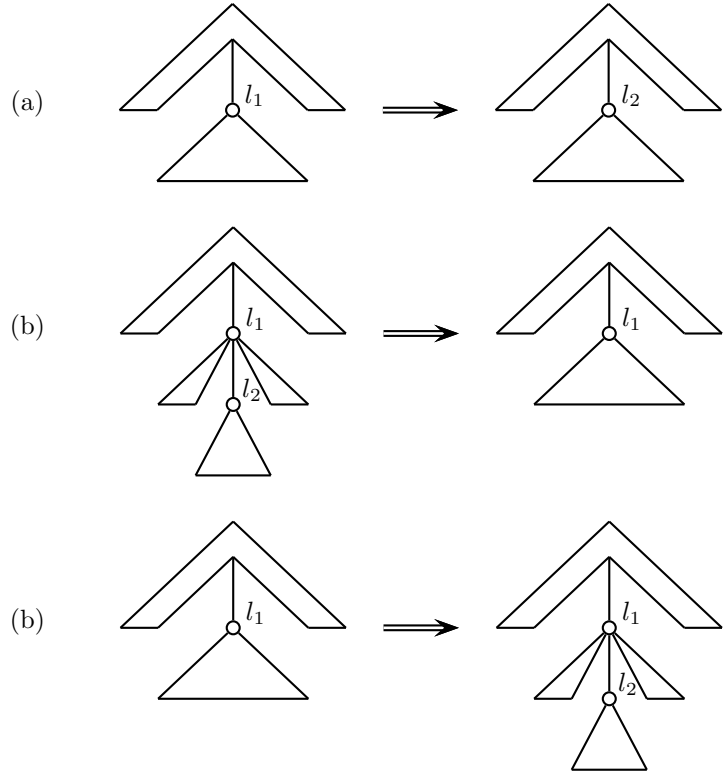

\begin{center}
  	
  \begin{psmatrix}[colsep=0.5cm,rowsep=0.05cm,labelsep=1pt]
  && &&&  \pnode{root1} &&&&&&&&&& \pnode{root2} \\
  &&&& &  \pnode{c1} &&&&&&&&&& \pnode{c2}  \\\\
  (a) && \pnode{l1}&\pnode{ll1} &&\cnode{.1}{cc1} \rput(0.3,0.2){$l_1$} && \pnode{rr1} &\pnode{r1} 
  &\pnode{a1}&&\pnode{a2}& 
  \pnode{l2} &\pnode{ll2} &&\cnode{.1}{cc2}\rput(0.3,0.2){$l_2$}&&\pnode{rr2} & \pnode{r2} \\ \\
  &&&\pnode{lll1} &&&& \pnode{rrr1} &&&&&& \pnode{lll2} &&&& \pnode{rrr2} \\
  \ncline{l1}{root1}
  \ncline{r1}{root1}
  \ncline{ll1}{l1}
  \ncline{rr1}{r1}
  \ncline{ll1}{c1}
  \ncline{rr1}{c1}
  
  \ncline{cc1}{c1}
  \ncline{lll1}{cc1}
  \ncline{rrr1}{cc1}
  \ncline{rrr1}{lll1}
  
  \ncline{l2}{root2}
  \ncline{r2}{root2}
  \ncline{ll2}{l2}
  \ncline{rr2}{r2}
  \ncline{ll2}{c2}
  \ncline{rr2}{c2}
  
  \ncline{cc2}{c2}
  \ncline{lll2}{cc2}
  \ncline{rrr2}{cc2}
  \ncline{rrr2}{lll2}
  
  \ncline[doubleline=true]{->}{a1}{a2}
   \end{psmatrix} 
  \begin{psmatrix}[colsep=0.5cm,rowsep=0.05cm,labelsep=1pt]
  && &&&  \pnode{root1} &&&&&&&&&& \pnode{root2} && \\
  &&&& &  \pnode{c1} &&&&&&&&&& \pnode{c2} && \\\\
  (b) && \pnode{l1}&\pnode{ll1} &&\cnode{.1}{cc1} \rput(0.3,0.2){$l_1$} && \pnode{rr1} &\pnode{r1} 
  &\pnode{a1}&&\pnode{a2}& 
  \pnode{l2} &\pnode{ll2} &&\cnode{.1}{cc2}\rput(0.3,0.2){$l_1$}&&\pnode{rr2} & \pnode{r2} \\ \\
  &&&\pnode{lll1}& \pnode{llr1} &\cnode{.1}{ccc1}\rput(0.25,0.2){$l_2$} & \pnode{rrl1} & \pnode{rrr1} &&&&&& 
  \pnode{lll2}&&&&\pnode{rrr2} \\ \\
  &&&&\pnode{cl1} & & \pnode{cr1} \\
  \ncline{l1}{root1}
  \ncline{r1}{root1}
  \ncline{ll1}{l1}
  \ncline{rr1}{r1}
  \ncline{ll1}{c1}
  \ncline{rr1}{c1}
  
  \ncline{cc1}{c1}
  \ncline{lll1}{cc1}
  \ncline{rrr1}{cc1}
  \ncline{lll1}{llr1}
  \ncline{rrr1}{rrl1}
  \ncline{llr1}{cc1}
  \ncline{rrl1}{cc1}
  
  \ncline{ccc1}{cc1}
  \ncline{cl1}{ccc1}
  \ncline{cr1}{ccc1}
  \ncline{cr1}{cl1}

  \ncline{l2}{root2}
  \ncline{r2}{root2}
  \ncline{ll2}{l2}
  \ncline{rr2}{r2}
  \ncline{ll2}{c2}
  \ncline{rr2}{c2}
  
  \ncline{cc2}{c2}
  \ncline{lll2}{cc2}
  \ncline{rrr2}{cc2}
  \ncline{rrr2}{lll2}

  \ncline[doubleline=true]{->}{a1}{a2}
   \end{psmatrix} 
     \begin{psmatrix}[colsep=0.5cm,rowsep=0.05cm,labelsep=1pt]
  && &&&  \pnode{root1} &&&&&&&&&& \pnode{root2} && \\
  &&&& &  \pnode{c1} &&&&&&&&&& \pnode{c2} && \\\\
  (b) && \pnode{l1}&\pnode{ll1} &&\cnode{.1}{cc1} \rput(0.3,0.2){$l_1$} && \pnode{rr1} &\pnode{r1} 
  &\pnode{a1}&&\pnode{a2}& 
  \pnode{l2} &\pnode{ll2} &&\cnode{.1}{cc2}\rput(0.3,0.2){$l_1$}&&\pnode{rr2} & \pnode{r2} \\ \\
  &&&\pnode{lll1} &&&&\pnode{rrr1} &&&&&& 
  \pnode{lll2}  & \pnode{llr2} &\cnode{.1}{ccc2}\rput(0.25,0.2){$l_2$} & \pnode{rrl2} &  \pnode{rrr2} \\ \\
  &&&&&&&&&&&&&&\pnode{cl2} & & \pnode{cr2} \\
  \ncline{l1}{root1}
  \ncline{r1}{root1}
  \ncline{ll1}{l1}
  \ncline{rr1}{r1}
  \ncline{ll1}{c1}
  \ncline{rr1}{c1}
  
  \ncline{cc1}{c1}
  \ncline{lll1}{cc1}
  \ncline{rrr1}{cc1}
  \ncline{rrr1}{lll1}

  \ncline{l2}{root2}
  \ncline{r2}{root2}
  \ncline{ll2}{l2}
  \ncline{rr2}{r2}
  \ncline{ll2}{c2}
  \ncline{rr2}{c2}
  
  \ncline{cc2}{c2}
  \ncline{lll2}{cc2}
  \ncline{rrr2}{cc2}
  \ncline{lll2}{llr2}
  \ncline{rrr2}{rrl2}
  \ncline{llr2}{cc2}
  \ncline{rrl2}{cc2}

  \ncline{ccc2}{cc2}
  \ncline{cl2}{ccc2}
  \ncline{cr2}{ccc2}
  \ncline{cr2}{cl2}

  \ncline[doubleline=true]{->}{a1}{a2}
   \end{psmatrix} 

    \caption{(a) A relabeling of the node label $l_1$ to $l_2$. (b) Deleting the node labeled $l_2$. (c) Inserting a node labeled $l_2$ as the child of the node labeled $l_1$.}
  \label{t1:operationexample}
\end{center}
\end{figure}

Let $T$ be a rooted tree. We call $T$ a \emph{labeled tree} if each node is a assigned a symbol from a fixed finite alphabet $\Sigma$. We call $T$ an \emph{ordered tree} if a left-to-right order among siblings in $T$ is given. In this paper we consider matching problems based on simple primitive operations applied to labeled trees. If $T$ is an ordered tree these operations are defined as follows:
 \begin{description}
  \item[relabel] Change the label of a node $v$ in $T$.
  \item[delete] Delete a non-root node $v$ in $T$ with parent $v'$, making the children of $v$ become the children of $v'$. The children are inserted in the place of $v$ as a subsequence in the left-to-right order of the children of $v'$.
   \item[insert] The complement of delete. Insert a node $v$ as a child of $v'$ in $T$ making $v$ the parent of a consecutive subsequence of the children of $v'$.   
\end{description}
Figure \ref{t1:operationexample} illustrates the operations. 
For unordered trees the operations can be defined similarly. In this case, the insert and delete operations works on a \emph{subset} instead of a subsequence. We define three problems based on the edit operations. Let $T_1$ and $T_2$ be labeled trees (ordered or unordered). 

\paragraph{Tree edit distance}
Assume that we are given a \emph{cost function} defined on each edit operation. An \emph{edit script} $S$ between $T_1$ and $T_2$ is a sequence of edit operations turning $T_1$ into $T_2$. The cost of  $S$ is the sum of the costs of the operations in $S$. An \emph{optimal edit script} between $T_1$ and $T_2$ is an edit script between $T_1$ and $T_2$ of minimum cost and this cost is the \emph{tree edit distance}. The \emph{tree edit distance problem} is to compute the edit distance and a corresponding edit script.

\paragraph{Tree alignment distance}
Assume that we are given a cost function defined on pair of labels. An \emph{alignment} $A$ of $T_1$ and $T_2$ is obtained as follows. First we insert nodes labeled with \emph{spaces} into $T_1$ and $T_2$ so that they become isomorphic when labels are ignored. The resulting trees are then \emph{overlayed} on top of each other giving the alignment $A$, which is a tree where each node is labeled by a pair of labels. The \emph{cost} of $A$ is the sum of costs of all pairs of opposing labels in $A$. An \emph{optimal alignment} of $T_1$ and $T_2$ is an alignment of minimum cost and this cost is called the \emph{alignment distance} of $T_1$ and $T_2$. The \emph{alignment distance problem} is to compute the alignment distance and a corresponding alignment.

\paragraph{Tree inclusion}
$T_1$ is \emph{included} in $T_2$ if and only if  $T_1$ can be obtained by deleting nodes from $T_2$. The \emph{tree inclusion problem} is to determine if $T_1$ is included in $T_2$.
\bigskip\par

In this paper we survey each of these problems and discuss the results obtained for them. For reference, Table \ref{t1:results} on page \pageref{t1:results} summarizes most of the available results. All of these and a few others are covered in the text. The tree edit distance  problem is the most general of the problems. The alignment distance corresponds to a kind of restricted edit distance, while tree inclusion is a special case of both the edit and alignment distance problem. Apart from these simple relationships, interesting variations on the edit distance problem has been studied leading to a more complex picture. 

\begin{sidewaystable}
  \centering 
  \begin{tabular}{|c|c|c|c|c|}
\multicolumn{5}{c}{\textbf{Tree edit distance}} \\ \hline
   variant 	& 	type 	& 	time 	                                                               & space              &   reference   \\ \hline
   general & O            &    $O(|T_1||T_2|D_1^2D_2^2)$ &$O(|T_1||T_2|D_1^2D_2^2)$&  \cite{Tai1979}\\\hline
   general & O 		&  	$O(|T_1||T_2|\min(L_1,D_1)\min(L_2,D_2))$ & $O(|T_1||T_2|)$  & \cite{ZS1989} \\\hline
   general & O            &    $O(|T_1|^2|T_2|\log |T_2|)$                                & $O(|T_1||T_2|)$  & \cite{Klein1998} \\\hline
   general & O            &$O(|T_1||T_2| + L_1^2|T_2| + L_1^{2.5}L_2)$&$O((|T_1| + L_1^2)\min(L_2,D_2)+|T_2|)$ &\cite{Chen2001}\\ \hline
   general & U             & \multicolumn{2}{|c|}{MAX SNP-hard}                           & \cite{ZJ1994} \\\hline
   constrained& O     &   $O(|T_1||T_2|)$                                                     &  $O(|T_1||T_2|)$ & \cite{Zhang1995} \\\hline
   constrained& O     &  $O(|T_1||T_2|I_1I_2)$                                           & $O(|T_1||D_2I_2)$  & \cite{Richter1997} \\\hline
   constrained& U      & $O(|T_1||T_2|(I_1+I_2)\log(I_1 + I_2))$            & $O(|T_1||T_2|)$ & \cite{Zhang1996} \\\hline
   less-constrained & O &   $O(|T_1||T_2|I_1^3I_2^3(I_1+I_2))$                   &$O(|T_1||T_2|I_1^3I_2^3(I_1+I_2))$                            &\cite{LST2001} \\\hline
   less-constrained & U &   \multicolumn{2}{|c|}{MAX SNP-hard}                   & \cite{LST2001} \\\hline
   unit-cost      & O & $O(u^2 \min(|T_1|,|T_2|)\min( L_1, L_2))$         & $O(|T_1||T_2|)$&\cite{SZ1990} \\ \hline
   $1$-degree & O & $O(|T_1||T_2|)$                                                          &  $O(|T_1||T_2|)$ &  \cite{Selkow1977} \\ \hline
   \multicolumn{5}{c}{ } \\ 
\multicolumn{5}{c}{\textbf{Tree alignment distance}} \\ \hline
   general   & O       & $O(|T_1||T_2|(I_1 + I_2)^2)$                                   &  $O(|T_1||T_2|(I_1 + I_2))$ & \cite{JWZ1995} \\ \hline
   general   & U       & \multicolumn{2}{|c|}{MAX SNP-hard}                 &   \cite{JWZ1995} \\ \hline
   similar     & O       & $O((|T_1| + |T_2|)\log(|T_1| + |T_2|)(I_1+I_2)^4s^2)$ & $O((|T_1| + |T_2|)\log(|T_1| + |T_2|)(I_1+I_2)^4s^2)$            & \cite{JL2001} \\ \hline
\multicolumn{5}{c}{ } \\
\multicolumn{5}{c}{\textbf{Tree inclusion}} \\\hline
   general   & O        & $O(|T_1||T_2|)$                                         & $O(|T_1|\min(D_2L_2))$ &\cite{Kilpelainen1992}\\ \hline
   general   & O        & $O(|\Sigma_{T_1}||T_2| + m_{T_1,T_2}D_2)$ & $O(|\Sigma_{T_1}||T_2| + m_{T_1,T_2})$ &\cite{Richter1997a} \\ \hline
   general   & O        & $O(L_1|T_2|)$                                         & $O(L_1\min(D_2L_2))$ &\cite{Chen1998} \\ \hline  
   general   & U        & \multicolumn{2}{|c|}{NP-hard}         & \cite{KM1995, MT1992} \\ \hline
\end{tabular}
  \caption{Results for the tree edit distance, alignment distance, and inclusion problem listed according to variant. $D_i$, $L_i$, and $I_i$ denotes the depth, the number of leaves, and the maximum degree respectively of $T_i$, $i=1,2$. The type is either O for ordered or U for unordered. The value $u$ is the unit cost edit distance between $T_1$ and $T_2$ and the value $s$ is the number of spaces in the optimal alignment of $T_1$ and $T_2$. The value $\Sigma_{T_1}$ is set of labels used in $T_1$ and $m_{T_1,T_2}$ is the number of pairs of nodes in $T_1$ and $T_2$ which have the same label.}\label{t1:results}
\end{sidewaystable}

Both the ordered and unordered version of the problems are reviewed. For the unordered case, it turns out that all of the problems in general are NP-hard. Indeed, the tree edit distance and alignment distance problems are even MAX SNP-hard \cite{ALMSS1992}. However, under various interesting restrictions, or for special cases,   polynomial time algorithms are available. For instance, if we impose a \emph{structure preserving} restriction on the unordered tree edit distance problem, such that disjoint subtrees are mapped to disjoint subtrees, it can be solved in polynomial time. Also, unordered alignment for constant degree trees can be solved efficiently. 

For the ordered version of the problems polynomial time algorithms exists. These are all based on the classic technique of \emph{dynamic programming} (see, e.g., \cite[Chapter 15]{CLRS2001}) and most of them are simple  combinatorial algorithms. Recently however, more advanced techniques such as fast matrix multiplication have been applied to the tree edit distance problem \cite{Chen2001}. 

The survey covers the problems in the following way. For each problem and variations of it we review results for both the ordered and unordered version. This will in most cases include a formal definition of the problem, a comparison of the available results and a description of the techniques used to obtain the results. More importantly, we will also pick one or more of the central algorithms for each of the problems and present it in almost full detail. Specifically, we will describe the algorithm, prove that it is correct, and analyze its time  complexity. For brevity, we will omit the proofs of a few lemmas and skip over some less important details. Common for the algorithms presented in detail is that, in most cases, they are the basis for more advanced algorithms. Typically, most of the algorithms for one of the above problems are refinements of the same dynamic programming algorithm. 

The main technical contribution of this survey is to present the problems and algorithms in a common framework. Hopefully, this will enable the reader to gain a better overview and deeper understanding of the problems and how they relate to each other. In the literature, there are some discrepancies in the presentations of the problems. For instance, the ordered edit distance problem was considered by Klein \cite{Klein1998} who used edit operations on edges. He presented an algorithm using a reduction to a problem defined on balanced parenthesis strings. In contrast, Zhang and Shasha \cite{ZS1989} gave an algorithm based on the postorder numbering on trees. In fact, these algorithms share many features which become apparent if considered in the right setting. In this paper we present these algorithms in a new framework bridging the gap between the two descriptions. 

Another problem in the literature is the lack of an agreement on a definition of the edit distance problem. The definition given here is by far the most studied and in our opinion the most natural. However, several alternatives ending in very different distance measures have been considered \cite{Lu1979, TT1988, Selkow1977, Lu1989}. In this paper we review these other variants and compare them to our definition. We should note that the edit distance problem defined here is sometimes referred to as the \emph{tree-to-tree correction problem}.

This survey adopts a \emph{theoretical} point of view. However, the problems above are not only interesting mathematical problems but they also occur in many practical situations and it is important to develop algorithms that perform well on \emph{real-life} problems. For practical issues see, e.g., \cite{WZJS1994,TSKK1998, SWSZ2002}. 

We restrict our attention to \emph{sequential} algorithms. However, there has been some research in parallel algorithms for the edit distance problem, e.g., \cite{ZS1989, Zhang1996a, SZ1990}.

This summarizes the contents of this paper. Due to the fundamental nature of comparing trees and its many applications several other ways to compare trees have been devised. In this paper, we have chosen to limit ourselves to a handful of problems which we describe in detail. Other problems include \emph{tree pattern matching} \cite{Kosaraju1989, DGM1990} and \cite{CO1982, RR1992, ZSW1994},  \emph{maximum agreement subtree} \cite{KA1994, FT1994}, \emph{largest common subtree} \cite{AH1994, KMY1995}, and \emph{smallest common supertree} \cite{NRT2000, GN1998}.

\subsection{Outline}
In Section \ref{t1:preliminaries} we give some preliminaries. In Sections \ref{t1:treeeditdistance}, \ref{t1:treealignmentdistance}, and \ref{t1:treeinclusion} we survey the tree edit distance, alignment distance, and inclusion problems respectively. We conclude in Section \ref{t1:conclusion} with some open problems.

\section{Preliminaries and Notation}\label{t1:preliminaries}
In this section we define notations and definitions we will use throughout the paper. For a graph $G$ we denote the set of nodes and edges by $V(G)$ and $E(G)$ respectively. Let $T$ be a rooted tree. The root of $T$ is denoted by $\roots(T)$. The \emph{size} of $T$, denoted by $|T|$, is $|V(T)|$. The \emph{depth} of a node $v\in V(T)$, $\depth(v)$, is the number of edges on the path from $v$ to $\roots(T)$. The \emph{in-degree} of a node $v$, $\deg(v)$ is the number of children of $v$. We extend these definitions such that $\depth(T)$ and $\deg(T)$ denotes the maximum depth and degree respectively of any node in $T$. A node with no children is a leaf and otherwise an internal node. The number of leaves of $T$ is denoted by $\leaves(T)$. We denote the parent of node $v$ by $\parent(v)$. Two nodes are siblings if they have the same parent. For two trees $T_1$ and $T_2$, we will frequently refer to $\leaves(T_i)$, $\depth(T_i)$, and $\deg(T_i)$ by $L_i$, $D_i$, and $I_i$, $i=1,2$.

Let $\theta$ denote the empty tree and let $T(v)$ denote the subtree of $T$ rooted at a node $v \in V(T)$. If $w\in V(T(v))$ then $v$ is an ancestor of $w$, and if $w\in V(T(v))\backslash \{v\}$ then $v$ is a proper ancestor of $w$. If $v$ is a (proper) ancestor of $w$ then $w$ is a (proper) descendant of $v$. A tree $T$ is \emph{ordered} if a left-to-right order among the siblings is given. For an ordered tree $T$ with root $v$ and children $v_1, \ldots ,v_i$, the \emph{preorder traversal} of $T(v)$ is obtained by visiting $v$ and then recursively visiting $T(v_k)$, $1 \leq k \leq i$, in order. Similarly, the \emph{postorder traversal} is obtained by first visiting $T(v_k)$, $1 \leq k \leq i$, and then $v$. The \emph{preorder number} and \emph{postorder number} of a node $w \in T(v)$, denoted by $\pre(w)$ and $\post(w)$, is the number of nodes preceding $w$ in the preorder and postorder traversal of $T$ respectively. The nodes to the \emph{left} of $w$ in $T$ is the set of nodes $u \in V(T)$ such that $\pre(u) < \pre(w)$ and $\post(u) < \post(w)$. If $u$ is to the left of $w$ then $w$ is to the \emph{right} of $u$.

A forest is a set of trees. A forest $F$ is ordered if a left-to-right order among the trees is given and each tree is ordered. Let $T$ be an ordered tree and let $v\in V(T)$. If $v$ has children $v_1,\ldots, v_i$ define $F(v_s, v_t)$, where $1 \leq s \leq t \leq i$, as the forest $T(v_s), \ldots , T(v_r)$. For convenience, we set $F(v) = F(v_1,v_i)$. 

We assume throughout the paper that labels assigned to nodes are chosen from a finite alphabet $\Sigma$. Let $\lambda \not \in \Sigma$  denote a special \emph{blank} symbol and define $\Sigma_{\lambda} = \Sigma \cup \lambda$. We often define a \emph{cost function}, $\gamma : (\Sigma_{\lambda} \times \Sigma_{\lambda})  \backslash (\lambda,\lambda) \rightarrow \mathbb{R}$, on pairs of labels. We will always assume that $\gamma$ is a distance metric. That is, for any $l_1$,$l_2$,$l_3 \in \Sigma_{\lambda}$ the following conditions are satisfied:
\begin{enumerate}
  \item $\gamma(l_1,l_2) \geq 0,\, \gamma(l_1, l_1) = 0$.
  \item $\gamma(l_1, l_2) = \gamma(l_2, l_1)$. 
  \item $\gamma(l_1, l_3)\leq \gamma(l_1, l_2) + \gamma(l_2, l_3)$.
\end{enumerate}

\section{Tree Edit Distance}\label{t1:treeeditdistance}
\begin{figure}[t]
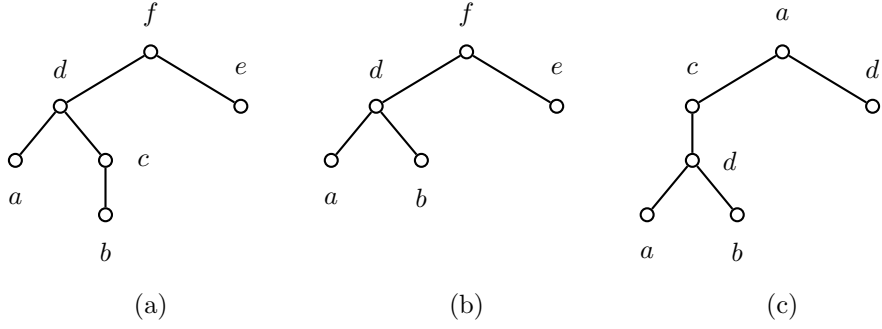

\begin{center}
  \begin{psmatrix}[colsep=0.6cm,rowsep=0.3cm,labelsep=1pt]
  & & & \cnode{.1}{root1}\rput(0,.5){$f$} & & & & & & &
  \cnode{.1}{root2}\rput(0,.5){$f$} & & & & & & & 
  \cnode{.1}{root3}\rput(0,.5){$a$}\\
  & \cnode{.1}{l1}\rput(0,.5){$d$} & & & & \cnode{.1}{r1}\rput(0,.5){$e$} &&& 
  \cnode{.1}{l2}\rput(0,.5){$d$}  &&&& \cnode{.1}{r2}\rput(0,.5){$e$}  &&&
  \cnode{.1}{l3}\rput(0,.5){$c$} &&&& \cnode{.1}{r3}\rput(0,.5){$d$} \\
  \cnode{.1}{ll1}\rput(0,-.5){$a$} & & \cnode{.1}{lr1}\rput(.5,0){$c$} &&&&&
  \cnode{.1}{ll2}\rput(0,-.5){$a$} & & \cnode{.1}{lr2}\rput(0,-.5){$b$} &&&&&&
  \cnode{.1}{ll3}\rput(.5,0){$d$} \\
  && \cnode{.1}{lrr1}\rput(0,-.5){$b$} & & &&&&&&&&&& 
  \cnode{.1}{lll3}\rput(0,-.5){$a$} && \cnode{.1}{llr3}\rput(0,-.5){$b$} \\
  &&& \rput(0,-.5){(a)} &&&&&&& \rput(0,-.5){(b)} &&&&&&& \rput(0,-.5){(c)} \\

  \ncline{l1}{root1} \ncline{r1}{root1}
  \ncline{ll1}{l1} \ncline{lr1}{l1} \ncline{lrr1}{lr1}
  \ncline{l2}{root2} \ncline{r2}{root2}
  \ncline{ll2}{l2} \ncline{lr2}{l2} 
  \ncline{l3}{root3} \ncline{r3}{root3}
  \ncline{ll3}{l3} 
   \ncline{lll3}{ll3} \ncline{llr3}{ll3}	
  \end{psmatrix} 
   \caption{Transforming (a) into (c) via editing operations. (a) A tree. (b) The tree after deleting the node labeled $c$. (c) The tree after inserting the node labeled $c$ and relabeling $f$ to $a$ and $e$ to $d$.}
  \label{t1:editexample}
\end{center}
\end{figure}

In this section we survey the tree edit distance problem. Assume that we are given a \emph{cost function} defined on each edit operation. An \emph{edit script} $S$ between two trees $T_1$ and $T_2$ is a sequence of edit operations turning $T_1$ into $T_2$. The cost of  $S$ is the sum of the costs of the operations in $S$. An \emph{optimal edit script} between $T_1$ and $T_2$ is an edit script between $T_1$ and $T_2$ of minimum cost. This cost is called the \emph{tree edit distance}, denoted by $\delta(T_1,T_2)$. An example of an edit script is shown in Figure \ref{t1:editexample}. 

The rest of the section is organized as follows. First, in Section \ref{t1:editmappings}, we  present some preliminaries and formally define the problem. In Section \ref{t1:generalorderededitdistance} we survey the results obtained for the ordered edit distance problem and present two of the currently best algorithms for the problem. The unordered version of the problem is reviewed in Section \ref{t1:generalunorderededitdistance}.
In Section \ref{t1:constrainededitdistance} we review results on the edit distance problem when various \emph{structure-preserving} constraints are imposed. Finally, in Section \ref{t1:othervariants} we consider some other variants of the problem.

\subsection{Edit Operations and Edit Mappings}\label{t1:editmappings}
Let $T_1$ and $T_2$ be labeled trees. Following \cite{Tai1979} we represent each edit operation by $(l_1 \rightarrow l_2)$, where $(l_1,l_2) \in (\Sigma_{\lambda} \times \Sigma_{\lambda})  \backslash (\lambda,\lambda)$. The operation is a relabeling if $l_1 \neq \lambda$ and $l_2 \neq \lambda$, a deletion if  $l_2 = \lambda$, and an insertion if $l_1 = \lambda$. We extend the notation such that $(v \rightarrow w)$ for nodes $v$ and $w$ denotes $(\lab(v)\rightarrow \lab(w))$. Here, as with the labels, $v$ or $w$ may be $\lambda$. Given a metric cost function $\gamma$ defined on pairs of labels we define the cost of an edit operation by setting $\gamma(l_1 \rightarrow l_2) = \gamma(l_1, l_2)$. The cost of a sequence $S = s_1,\ldots, s_k$ of operations is given by $\gamma(S) = \sum_{i=1}^k \gamma(s_i)$. The edit distance, $\delta(T_1,T_2)$, between $T_1$ and $T_2$ is formally defined as:
\begin{equation*}
\delta(T_1,T_2) = \min\{\gamma(S)\mid S \text{ is a sequence of operations transforming $T_1$ into $T_2$}\}.
\end{equation*}
Since $\gamma$ is a distance metric $\delta$ becomes a distance metric too. 

\begin{figure}[t]
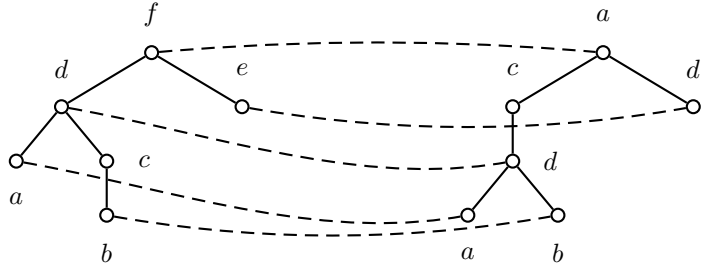

\begin{center}
  \begin{psmatrix}[colsep=0.6cm,rowsep=0.3cm,labelsep=1pt]
  & & & \cnode{.1}{root1}\rput(0,.5){$f$} & & & & & & &
   & & &  
  \cnode{.1}{root3}\rput(0,.5){$a$}\\
  & \cnode{.1}{l1}\rput(0,.5){$d$} & & & & \cnode{.1}{r1}\rput(0,.5){$e$} &&& 
      &&&
  \cnode{.1}{l3}\rput(0,.5){$c$} &&&& \cnode{.1}{r3}\rput(0,.5){$d$} && \\
  \cnode{.1}{ll1}\rput(0,-.5){$a$} & & \cnode{.1}{lr1}\rput(.5,0){$c$} &&&&&
   &&&&
  \cnode{.1}{ll3}\rput(.5,0){$d$} \\
  && \cnode{.1}{lrr1}\rput(0,-.5){$b$} & & &&&&&& 
  \cnode{.1}{lll3}\rput(0,-.5){$a$} && \cnode{.1}{llr3}\rput(0,-.5){$b$} \\
  \ncline{l1}{root1} \ncline{r1}{root1}
  \ncline{ll1}{l1} \ncline{lr1}{l1} \ncline{lrr1}{lr1}
  \ncline{l3}{root3} \ncline{r3}{root3}
  \ncline{ll3}{l3} 
   \ncline{lll3}{ll3} \ncline{llr3}{ll3}	
  \psset{linestyle=dashed}
  \nccurve[angleA=5,angleB=175]{root1}{root3}
  \nccurve[angleA=350,angleB=190]{r1}{r3}
  \nccurve[angleA=350,angleB=190]{l1}{ll3}
  \nccurve[angleA=350,angleB=190]{ll1}{lll3}
  \nccurve[angleA=350,angleB=190]{lrr1}{llr3}
  \end{psmatrix} 
   \caption{The mapping corresponding to the edit script in Figure \ref{t1:editexample}.}
  \label{t1:mappingexample}
  \end{center}
\end{figure}

An \emph{edit distance mapping} (or just a \emph{mapping}) between $T_1$ and $T_2$ is a representation of the edit operations, which is used in many of the algorithms for the tree edit distance problem. Formally, define the triple $(M, T_1, T_2)$ to be an \emph{ordered edit distance mapping} from $T_1$ to $T_2$, if $M \subseteq V(T_1)\times V(T_2)$ and for any pair $(v_1,w_1), (v_2,w_2) \in M$:
\begin{enumerate}
  \item $v_1 = v_2$ iff $w_1 = w_2$. (one-to-one condition)
  \item $v_1$ is an ancestor of $v_2$ iff $w_1$ is an ancestor of $w_2$. (ancestor condition)
  \item $v_1$ is to the left of $v_2$ iff $w_1$ is to the left of $w_2$. (sibling condition)
\end{enumerate}
Figure \ref{t1:mappingexample} illustrates a mapping that corresponds to the edit script in Figure \ref{t1:editexample}.
We define the \emph{unordered edit distance mapping} between two unordered trees as the same, but without the sibling condition. We will use $M$ instead of $(M,T_1,T_2)$ when there is no confusion. Let $(M,T_1,T_2)$ be a mapping. We say that a node $v$ in $T_1$ or $T_2$ is \emph{touched by a line} in M if $v$ occurs in some pair in $M$. Let $N_1$ and $N_2$ be the set of nodes in $T_1$ and $T_2$ respectively not touched by any line in $M$. The cost of $M$ is given by:
\begin{equation*}
\gamma(M) = \sum_{(v,w) \in M} \gamma(v \rightarrow w) + \sum_{v \in N_1} \gamma(v \rightarrow \lambda) + \sum_{w \in N_2} \gamma(\lambda \rightarrow w)  
\end{equation*}
Mappings can be composed. Let $T_1$, $T_2$, and $T_3$ be labeled trees. Let $M_1$ and $M_2$ be a mapping from  $T_1$ to $T_2$ and $T_2$ to $T_3$ respectively. Define
\begin{equation*}
M_1 \circ M_2 = \{(v,w) \mid \exists u \in V(T_2) \text{ such that $(v,u) \in M_1$ and $(u,w) \in M_2$}\}
\end{equation*}
With this definition it follows easily that $M_1 \circ M_2$ itself becomes a mapping from $T_1$ to $T_3$. Since $\gamma$ is a metric, it is not hard to show that a minimum cost mapping is equivalent to the edit distance:
\begin{equation*}
\delta(T_1,T_2) = \min\{\gamma(M) \mid (M,T_1,T_2) \text{ is an edit distance mapping}\}. 
\end{equation*}

Hence, to compute the edit distance we can compute the minimum cost mapping. We extend the definition of edit distance to forests. That is, for two forests $F_1$ and $F_2$, $\delta(F_1, F_2)$ denotes the edit distance between $F_1$ and $F_2$. The operations are defined as in the case of trees, however, roots of the trees in the forest may now be deleted and trees can be merged by inserting a new root. The definition of a mapping is extended in the same way.

\subsection{General Ordered Edit Distance}\label{t1:generalorderededitdistance}
The ordered edit distance problem was introduced by Tai \cite{Tai1979} as a generalization of the well-known \emph{string edit distance problem} \cite{WF1974}. Tai presented an algorithm for the ordered version using $O(|T_1|| T_2|| L_1|^2| L_2|^2)$ time and space. Subsequently, Zhang and Shasha \cite{ZS1989} gave a simple algorithm improving the bounds to $O(|T_1||T_2|\min(L_1,D_1)\min(L_2,D_2))$ time and $O(|T_1||T_2|)$ space. This algorithm was modified by Klein \cite{Klein1998} to get a better worst case time bound of $O(|T_1|^2|T_2|\log |T_2|)$\footnote{Since the edit distance is symmetric this bound is in fact $O(\min(|T_1|^2|T_2|\log |T_2|, |T_2|^2|T_1|\log |T_1|)$. For brevity we will use the short version.}
under the same space bounds. We present the latter two algorithms in detail below. Recently, Chen \cite{Chen2001} has presented an algorithm using $O(|T_1||T_2| + L_1^2|T_2| + L_1^{2.5}L_2)$ time and $O((|T_1| + L_1^2)\min(L_2,D_2)+|T_2|)$ space.  Hence, for certain kinds of trees the algorithm improves the previous bounds. This algorithm is more complex than all of the above and uses results on fast matrix multiplication. Note that in the above bounds we can exchange $T_1$ with $T_2$ since the distance is symmetric. 

\subsubsection{A Simple Algorithm}\label{t1:simpledynamic}
We first present a simple recursion which will form the basis for the two dynamic programming algorithms we present in the next two sections. We will only show how to compute the edit distance. The corresponding edit script can be easily obtained within the same time and space bounds. The algorithm is due to Klein \cite{Klein1998}. However, we should note that the presentation given here is somewhat different. We believe that our framework is more simple and provides a better connection to previous work. 

Let $F$ be a forest and $v$ be a node in $F$. We denote by $F - v$ the forest obtained by deleting $v$ from $F$. Furthermore, define $F - T(v)$ as the forest obtained by deleting $v$ and all descendants of $v$. The following lemma provides a way to compute edit distances for the general case of forests. 

\begin{lemma}\label{t1:EDrecursion}
Let $F_1$ and $F_2$ be ordered forests and $\gamma$ be a metric cost function defined on labels. Let $v$ and $w$ be the rightmost (if any) roots of the trees in $F_1$ and $F_2$ respectively. We have, 
\begin{equation*}
\begin{aligned}
  	\delta(\theta, \theta) &= 0\\
  	\delta(F_1,\theta)  &= \delta(F_1-v,\theta) +  \gamma (v \rightarrow\lambda)\\ 
  	\delta(\theta, F_2) &= \delta(\theta,F_2 - w) + \gamma(\lambda \rightarrow w)\\
	\delta(F_1,F_2) &= \min 
		\begin{cases}
			\delta(F_1 - v,F_2) + \gamma (v \rightarrow \lambda)  \\
      		\delta(F_1,F_2 - w) + \gamma(\lambda \rightarrow w)  \\
      		\delta(F_1(v),F_2(w)) + \delta(F_1 - T_1(v), F_2 - T_2(w)) + \gamma(v \rightarrow w) 
		\end{cases}
\end{aligned}
\end{equation*}
\end{lemma}
\begin{proof}
The first three equations are trivially true. To show the last equation consider a minimum cost mapping $M$ between $F_1$ and $F_2$. There are three possibilities for $v$ and $w$: 
\begin{description}
  \item[Case 1:] $v$ is not touched by a line. Then $(v,\lambda) \in M$ and the first case of the last equation applies.
  \item[Case 2:] $w$ is not touched by a line. Then $(\lambda, w) \in M$ and the second case of the last equation applies.
  \item[Case 3:] $v$ and $w$ are both touched by lines. We show that this implies $(v,w) \in M$. Suppose $(v,h)$ and $(k,w)$ are in $M$. If $v$ is to the right of $k$ then $h$ must be to right of $w$ by the sibling condition. If $v$ is a proper ancestor of $k$ then $h$ must be a proper ancestor of $w$ by the ancestor condition. Both of these cases are impossible since $v$ and $w$ are the rightmost roots and hence $(v,w) \in M$. By the definition of mappings the equation follows. {\hfill$\Box$\\\noindent}
\end{description}\end{proof}
Lemma \ref{t1:EDrecursion} suggests a dynamic programming algorithm. The value of $\delta(F_1, F_2)$ depends on a constant number of subproblems of smaller size. Hence, we can compute $\delta(F_1, F_2)$ by computing $\delta(S_1,S_2)$ for all pairs of subproblems $S_1$ and $S_2$ in order of increasing size. Each new subproblem can be computed in constant time. Hence, the time complexity is bounded by the number of subproblems of $F_1$ times the number of subproblems of $F_2$. 

To count the number of subproblems, define for a rooted, ordered forest $F$ the $(i,j)$-\emph{deleted subforest}, $0\leq i+j \leq |F|$, as the forest obtained from $F$ by  first deleting the rightmost root repeatedly $j$ times and then, similarly, deleting the leftmost root $i$ times. We call the $(0,j)$-deleted and  $(i,0)$-deleted subforests, for $0\leq j \leq |F|$, the \emph{prefixes} and the \emph{suffixes} of $F$ respectively.  The number of $(i,j)$-deleted subforests of $F$ is $\sum_{k=0}^{|F|} k= O(|F|^2)$, since for each $i$ there are $|F|-i$ choices for $j$. 

It is not hard to show that all the pairs of subproblems $S_1$ and $S_2$ that can be obtained by the recursion of Lemma \ref{t1:EDrecursion} are deleted subforests of $F_1$ and $F_2$. Hence, by the above discussion the time complexity is bounded by $O(|F_1|^2|F_2|^2)$. In fact, fewer subproblems are needed, which we will show in the next sections.
\subsubsection{Zhang and Shasha's Algorithm}\label{t1:zhangshasha}
The following algorithm is due to Zhang and Shasha~\cite{ZS1989}. Define the \emph{keyroots} of  a rooted, ordered tree $T$ as follows:
\begin{equation*}
\keyroots(T) = \{\roots(T)\} \cup \{v\in V(T) \mid v \text{ has a left sibling}\}
\end{equation*}
The \emph{special} subforests of $T$ is the forests $F(v)$, where $v \in \keyroots(T)$. The \emph{relevant subproblems of $T$ with respect to the keyroots} is the prefixes of all special subforests $F(v)$. In this section we refer to these as the \emph{relevant subproblems}.
\begin{lemma}\label{t1:keyrootlemma}
For each node $v\in V(T)$, $F(v)$ is a relevant subproblem.
\end{lemma}
It is easy to see that, in fact, the subproblems that can occur in the above recursion are either subforests of the form $F(v)$, where $v\in V(T)$, or prefixes of a special subforest of $T$. Hence, it follows by Lemma \ref{t1:keyrootlemma} and the definition of a relevant subproblem, that to compute $\delta(F_1,F_2)$ it is sufficient to compute $\delta(S_1,S_2)$ for all relevant subproblems $S_1$ and $S_2$ of $T_1$ and $T_2$ respectively. 

The relevant subproblems of a tree $T$ can be counted as follows. For a node $v \in V(T)$ define the \emph{collapsed depth} of $v$, $\cdepth(v)$, as the number of keyroot ancestors of $v$. Also, define $\cdepth(T)$ as the maximum collapsed depth of all nodes $v\in V(T)$.
\begin{lemma}\label{t1:cdepthlemma}
For an ordered tree $T$ the number of relevant subproblems, with respect to the keyroots is bounded by $O(|T|\cdepth(T))$.
\end{lemma}
\begin{proof}
The relevant subproblems can be counted using the following expression:
\begin{equation*}
\sum_{v\in \keyroots(T)}|F(v)| < \sum_{v\in \keyroots(T)}|T(v)| = \sum_{v\in V(T)}\cdepth(v) \leq |T|\cdepth(T)
\end{equation*}
Since the number prefixes of a subforest $F(v)$ is $|F(v)|$ the first sum counts the number of relevant subproblems of $F(v)$. To prove the first equality note that for each node $v$ the number of special subforests containing $v$ is the collapsed depth of $v$. Hence, $v$ contributes the same amount to the left and right side. The other equalities/inequalities follow immediately. {\hfill$\Box$\\\noindent}
\end{proof}
\begin{lemma}
For a tree $T$, $\cdepth(T) \leq \min\{\depth(T),\leaves(T)\}$
\end{lemma}
Thus, using dynamic programming the problem can be solved in $O(|T_1||T_2|\min\{ D_1,L_1\}\min\{ D_2,L_2\})$ time and space. To improve the space complexity we carefully compute the subproblems in a specific order and discard some of the intermediate results. Throughout the algorithm we maintain a table called the \emph{permanent table} storing the distances $\delta(F_1(v), F_2(w))$, $v_1 \in V(F_1)$ and $w_2 \in V(F_2)$, as they are computed. This uses $O(|F_1||F_2|)$ space. When the distances of all special subforests of $F_1$ and $F_2$ are availiable in the permanent table, we compute the distance between all prefixes of $F_1$ and $F_2$ in order of increasing size and store these in a table called the \emph{temporary table}. The values of the temporary table that are distances between special subforests are copied to the permanent table and the rest of the values are discarded. Hence, the temporary table also uses at most $O(|F_1||F_2|)$ space. By Lemma~\ref{t1:EDrecursion} it is easy to see that all values needed to compute $\delta(F_1, F_2)$ are availiable. Hence, 
\begin{theorem}[\cite{ZS1989}]
For ordered trees $T_1$ and $T_2$ the tree edit distance problem can be solved in time $O(|T_1||T_2|\min\{D_1,L_1\} \min \{D_2,L_2\})$ and space $O(|T_1||T_2|)$.
\end{theorem}
\subsubsection{Klein's Algorithm}\label{t1:klein}
In the worst case, that is for trees with linear depth and a linear number of leaves, Zhang and Shasha's algorithm of the previous section still requires $O(|T_1|^2|T_2|^2)$ time as the simple algorithm. In \cite{Klein1998} Klein obtained a better worst case time bound of $O(|T_1|^2|T_2|\log |T_2|)$. The reported space complexity of the algorithm is $O(|T_1|^2|T_2|\log |T_2|)$ which is significantly worse than the algorithm of Zhang and Shasha. However, according to Klein \cite{Klein2002} this algorithm can also be improved to $O(|T_1||T_2|)$.

The algorithm is based on an extension of the recursion in Lemma \ref{t1:EDrecursion}. The main idea is to consider all of the $O(|T_1|^2)$ deleted subforests of $T_1$ but only $O(|T_2|\log |T_2|)$ deleted subforests of $T_2$. In total the worst case number of subproblems is thus reduced to the desired bound above.

A key concept in the algorithm is the decomposition of a rooted tree $T$ into disjoint paths called \emph{heavy paths}. This technique was introduced by Harel and Tarjan \cite{HT1984}. We define the \emph{size} a node $v\in V(T)$ as $|T(v)|$. We classify each node of $T$ as either \emph{heavy} or \emph{light} as follows. The root is light. For each internal node $v$ we pick a child $u$ of $v$ of maximum size among the children of $v$ and classify $u$ as heavy. The remaining children are light. We call an edge to a light child a \emph{light edge}, and an edge to a heavy child a \emph{heavy edge}.  The \emph{light depth} of a node $v$, $\ldepth(v)$, is the number of light edges on the path from $v$ to the root.

\begin{lemma}[\cite{HT1984}]\label{t1:lightdepthlemma}
  For any tree $T$ and any $v \in V(T)$, $\ldepth(v) \leq \log |T| + O(1)$.
\end{lemma}
By removing the light edges $T$ is partitioned into heavy paths.

We define the \emph{relevant subproblems of $T$ with respect to the light nodes} below.  We will refer to these as \emph{relevant subproblems} in this section. First fix a heavy path decomposition of $T$. For a node $v$ in $T$ we recursively define the relevant subproblems of $F(v)$ as follows: $F(v)$ is relevant. If $v$ is not a leaf, let $u$ be the heavy child of $v$ and let $l$ and $r$ be the number of nodes to the left and to the right of $u$ in $F(v)$ respectively. Then, the $(i,0)$-deleted subforests of $F(v)$, $0\leq i\leq l$, and the $(l,j)$-deleted subforests of $F(v)$, $0\leq j\leq r$ are relevant subproblems. Recursively, all relevant subproblems of $F(u)$ are relevant.  

The relevant subproblems of $T$ with respect to the light nodes is the union of all relevant subproblems of $F(v)$ where $v \in V(T)$ is a light node.
\begin{lemma}
For an ordered tree $T$ the number of relevant subproblems with respect to the light nodes is bounded by $O(|T|\,\ldepth(T))$.
\end{lemma}
\begin{proof}
Follows by the same calculation as in the proof of Lemma \ref{t1:cdepthlemma}.
\end{proof}{\hfill$\Box$\\\noindent}

Also note that Lemma \ref{t1:keyrootlemma} still holds with this new definition of relevant subproblems. 
Let $S$ be a relevant subproblem of $T$ and let $v_l$ and $v_r$ denote the leftmost and rightmost root of $S$ respectively. The \emph{difference node} of $S$ is either $v_r$ if $S-v_r$ is relevant or $v_l$ if $S-v_l$ is relevant.  The recursion of Lemma \ref{t1:EDrecursion} compares the rightmost roots. Clearly, we can also choose to compare the leftmost roots resulting in a new recursion, which we will refer to as the \emph{dual} of Lemma \ref{t1:EDrecursion}. Depending on which recursion we use, different subproblems occur. We now give a modified dynamic programming algorithm for calculating the tree edit distance. 
Let $S_1$ be a deleted tree of $T_1$ and let $S_2$ be a relevant subproblem of $T_2$. Let $d$ be the difference node of $S_2$. We compute $\delta(S_1,S_2)$ as follows. There are two cases to consider:
\begin{enumerate}
  \item If $d$ is the rightmost root of $S_2$ compare the rightmost roots of $S_1$ and $S_2$ using Lemma~\ref{t1:EDrecursion}.
  \item If $d$ is the leftmost root of $S_2$ compare the leftmost roots of $S_1$ and $S_2$ using the dual of Lemma~\ref{t1:EDrecursion}.
\end{enumerate}
 
It is easy to show that in both cases the resulting smaller subproblems of $S_1$ will all be deleted subforests of $T_1$ and the smaller subproblems of $S_2$ will all be relevant subproblems of $T_2$. Using a similar dynamic programming technique as in the algorithm of Zhang and Shasha we obtain the following:
\begin{theorem}[\cite{Klein1998}]
For ordered trees $T_1$ and $T_2$ the tree edit distance problem can be solved in time and space $O(|T_1|^2|T_2|\log |T_2|)$.
\end{theorem}

Klein \cite{Klein1998} also showed that his algorithm can be extended within the same time and space bounds to the \emph{unrooted ordered edit distance problem} between $T_1$ and $T_2$, defined as the minimum edit distance between $T_1$ and $T_2$ over all possible roots of $T_1$ and $T_2$. 

\subsection{General Unordered Edit Distance}\label{t1:generalunorderededitdistance}
In the following section we survey the unordered edit distance problem. This problem has been shown to be NP-complete \cite{ZSS1992, Zhang1989, ZSS1991} even for binary trees with a label alphabet of size $2$. The reduction is from the Exact Cover by $3$-set problem \cite{GJ1979}. Subsequently, the problem was shown to be MAX-SNP hard \cite{ZJ1994}. Hence, unless P=NP there is no PTAS for the problem \cite{ALMSS1992}. 
It was shown in \cite{ZSS1992} that for special cases of the problem polynomial time algorithms exists. If $T_2$ has one leaf, i.e., $T_2$ is a sequence, the problem can be solved in $O(|T_1|| T_2|)$ time. More generally, there is an algorithm running in time $O(|T_1|| T_2| + L_2! 3^{L_2} (L_2^3 +D_1^2)| T_1|)$. Hence, if the number of leaves in $T_2$ is logarithmic the problem can be solved in polynomial time.

\subsection{Constrained Edit Distance}\label{t1:constrainededitdistance}
The fact that the general edit distance problem is difficult to solve has led to the study of restricted versions of the problem. In \cite{Zhang1995, Zhang1996} Zhang introduced the \emph{constrained edit distance}, denoted by $\delta_c$, which is defined as an edit distance under the restriction that disjoint subtrees should be mapped to disjoint subtrees. Formally, $\delta_c(T_1, T_2)$ is defined as a minimum cost mapping $(M_c, T_1, T_2)$ satisfying the additional constraint, that for all $(v_1,w_1), (v_2,w_2), (v_3,w_3) \in M_c$:
\begin{itemize}
\item $\nca(v_1,v_2)$ is a proper ancestor of $v_3$ iff $\nca(w_1, w_2)$ is a proper ancestor of $w_3$.
\end{itemize}

\begin{figure}[t]
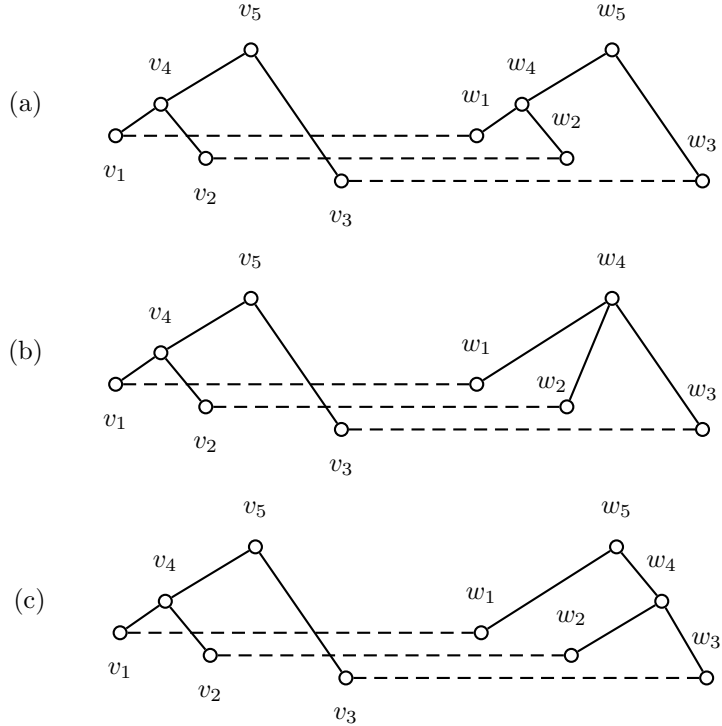

\begin{center}
  	
  \begin{psmatrix}[colsep=0.6cm,rowsep=0.3cm,labelsep=1pt]
  & & &&& \cnode{.1}{root1}\rput(0,.5){$v_5$} & & & &  & & & &
  \cnode{.1}{root2}\rput(0,.5){$w_5$} &&\\
  \rput(0,0){(a)} &&& \cnode{.1}{l1}\rput(0,.5){$v_4$} & & & & &&&& \cnode{.1}{l2}\rput(0,.5){$w_4$} &&& \\
  &&\rput(0,.3){\cnode{.1}{ll1}}\rput(0,-.2){$v_1$} && 
  \rput(0,0){\cnode{.1}{lr1}} \rput(0,-.5){$v_2$} &&&
  \rput(0,-.3){\cnode{.1}{r1}} \rput(0,-.8){$v_3$}&&& 
  
  \rput(0,.3){\cnode{.1}{ll2}}\rput(0,.8){$w_1$} &&
  \rput(0,0){\cnode{.1}{lr2}}\rput(0,.5){$w_2$}&&&
  \rput(0,-.3){\cnode{.1}{r2}}\rput(0,.2){$w_3$}&&& \\
  \psset{linestyle=solid}
  \ncline{l1}{root1}
  \ncline{ll1}{l1}
  \ncline{lr1}{l1}
  \ncline{r1}{root1}
  
  \ncline{l2}{root2}
  \ncline{ll2}{l2}
  \ncline{lr2}{l2}
  \ncline{r2}{root2}
  
 \psset{linestyle=dashed}
  \nccurve[angleA=0,angleB=180]{ll1}{ll2}
  \nccurve[angleA=0,angleB=180]{lr1}{lr2}
  \nccurve[angleA=0,angleB=180]{r1}{r2}
  \end{psmatrix} 
  \vspace{1cm}
  
\begin{psmatrix}[colsep=0.6cm,rowsep=0.3cm,labelsep=1pt]
  & & &&& \cnode{.1}{root1}\rput(0,.5){$v_5$} & & & &  & & & &
  \cnode{.1}{root2}\rput(0,.5){$w_4$} &&\\
  \rput(0,0){(b)} &&& \cnode{.1}{l1}\rput(0,.5){$v_4$} & & & & &&&& &&& \\
  &&\rput(0,.3){\cnode{.1}{ll1}}\rput(0,-.2){$v_1$} && 
  \rput(0,0){\cnode{.1}{lr1}} \rput(0,-.5){$v_2$} &&&
  \rput(0,-.3){\cnode{.1}{r1}} \rput(0,-.8){$v_3$}&&& 
  
  \rput(0,.3){\cnode{.1}{ll2}}\rput(0,.8){$w_1$} &&
  \rput(0,0){\cnode{.1}{lr2}}\rput(-.2,.3){$w_2$}&&&
  \rput(0,-.3){\cnode{.1}{r2}}\rput(0,.2){$w_3$}&&& \\
  \psset{linestyle=solid}
  \ncline{l1}{root1}
  \ncline{ll1}{l1}
  \ncline{lr1}{l1}
  \ncline{r1}{root1}
  
  \ncline{ll2}{root2}
  \ncline{lr2}{root2}
  \ncline{r2}{root2}
  
 \psset{linestyle=dashed}
  \nccurve[angleA=0,angleB=180]{ll1}{ll2}
  \nccurve[angleA=0,angleB=180]{lr1}{lr2}
  \nccurve[angleA=0,angleB=180]{r1}{r2}
  \end{psmatrix} 
  \vspace{1cm}
  
\begin{psmatrix}[colsep=0.6cm,rowsep=0.3cm,labelsep=1pt]
  & & &&& \cnode{.1}{root1}\rput(0,.5){$v_5$} & & & &  & & & &
  \cnode{.1}{root2}\rput(0,.5){$w_5$} &&\\
  \rput(0,0){(c)} &&& \cnode{.1}{l1}\rput(0,.5){$v_4$} & & & & &&&& &&&\cnode{.1}{l2}\rput(0,.5){$w_ 4$} \\
  &&\rput(0,.3){\cnode{.1}{ll1}}\rput(0,-.2){$v_1$} && 
  \rput(0,0){\cnode{.1}{lr1}} \rput(0,-.5){$v_2$} &&&
  \rput(0,-.3){\cnode{.1}{r1}} \rput(0,-.8){$v_3$}&&& 
  
  \rput(0,.3){\cnode{.1}{ll2}}\rput(0,.8){$w_1$} &&
  \rput(0,0){\cnode{.1}{lr2}}\rput(0,.5){$w_2$}&&&
  \rput(0,-.3){\cnode{.1}{r2}}\rput(0,.2){$w_3$}&&& \\
  \psset{linestyle=solid}
  \ncline{l1}{root1}
  \ncline{ll1}{l1}
  \ncline{lr1}{l1}
  \ncline{r1}{root1}
  
  \ncline{ll2}{root2}
  \ncline{l2}{root2}
  \ncline{lr2}{l2}
  \ncline{r2}{l2}
   
 \psset{linestyle=dashed}
  \nccurve[angleA=0,angleB=180]{ll1}{ll2}
  \nccurve[angleA=0,angleB=180]{lr1}{lr2}
  \nccurve[angleA=0,angleB=180]{r1}{r2}
  \end{psmatrix} 

   \caption{(a) A mapping which is constrained and less-constrained. (b) A mapping which is less-constrained but not constrained. (c) A mapping which is neither constrained nor less-constrained. }
  \label{t1:constrainedmappingexample}
  
  \end{center}
\end{figure}  
  
According to \cite{LST2001}, Richter \cite{Richter1997} independently introduced the \emph{structure respecting  edit distance} $\delta_s$. Similar to the constrained edit distance, $\delta_s(T_1,T_2)$ is defined as a minimum cost mapping $(M_s, T_1, T_2)$ satisfying the additional constraint, that for all $(v_1,w_1), (v_2,w_2), (v_3,w_3) \in M_s$ such that none of $v_1$, $v_2$, and $v_3$ is an ancestor of the others,
\begin{itemize}
\item $\nca(v_1,v_2) = \nca(v_1,v_3)$ iff $\nca(w_1,w_2) = \nca(w_1,w_3)$.
\end{itemize}
 
It is straightforward to show that both of these notions of edit distance are equivalent. Henceforth, we will refer to them simply as the constrained edit distance.  As an example consider the mappings of Figure \ref{t1:constrainedmappingexample}. (a) is a constrained mapping since $\nca(v_1, v_2) \neq \nca(v_1,v_3)$ and $\nca(w_1, w_2) \neq \nca(w_1, w_3)$. (b) is not constrained since $\nca(v_1, v_2) = v_4 \neq \nca(v_1,v_3) = v_5$, while $\nca(w_1,w_2) = w_4 = \nca(w_1,w_3)$. (c) is not constrained since $\nca(v_1, v_3) = v_5 \neq \nca(v_2,v_3)$, while $\nca(w_1, w_3) = v_5 \neq \nca(w_2,w_3) = w_4$.

In \cite{Zhang1995, Zhang1996} Zhang presents algorithms for computing minimum cost constrained mappings. For the ordered case he gives an algorithm using $O(|T_1||T_2|)$ time and for the unordered case he obtains a running time of $O(|T_1||T_2|(I_1 + I_2)\log(I_1 + I_2))$. Both use space $O(|T_1||T_2|)$. The idea in both algorithms is similar. Due to the restriction on the mappings fewer subproblem need to be considered and a faster dynamic programming algorithm is obtained. In the ordered case the key observation is a reduction to the string edit distance problem. For the unordered case the corresponding reduction is to a maximum matching problem. Using an efficient algorithm for computing a minimum cost maximum flow Zhang obtains the time complexity above. Richter presented an algorithm for the ordered constrained edit distance problem, which uses $O(|T_1||T_2|I_1I_2)$ time and $O(|T_1|D_2I_2)$ space. Hence, for small degree, low depth trees this algorithm gives a space improvement over the algorithm of Zhang. 

Recently, Lu et al.~\cite{LST2001} introduced the \emph{less-constrained edit distance}, $\delta_l$, which relaxes the constrained mapping. The requirement here is that for all $(v_1,w_1), (v_2,w_2), (v_3,w_3) \in M_l$ such that none of $v_1$, $v_2$, and $v_3$ is an ancestor of the others, $\depth(\nca(v_1,v_2)) \geq \depth(\nca(v_1,v_3))$, and $\nca(v_1,v_3) = \nca(v_2,v_3)$ if and only if $\depth(\nca(w_1,w_2)) \geq \depth(\nca(w_1,w_3))$ and $\nca(w_1,w_3) = \nca(w_2,w_3)$.

For example, consider the mappings in Figure \ref{t1:constrainedmappingexample}. (a) is less-constrained because it is constrained. (b) is not a constrained mapping, however the mapping is less-constrained since $\depth(\nca(v_1,v_2)) > \depth(\nca(v_1, v_3))$, $\nca(v_1,v_3) = \nca(v_2,v_3)$, $\nca(w_1, w_2) = \nca(w_1, w_3)$, and   $\nca(w_1, w_3) = \nca(w_2,w_3)$. (c) is not a less-constrained mapping since $\depth(\nca(v_1,v_2)) > \depth(\nca(v_1, v_3))$ and $\nca(v_1,v_3) = \nca(v_2,v_3)$, while $\nca(w_1, w_3) \neq \nca(w_2, w_3)$

In the paper \cite{LST2001} an algorithm for the ordered version of the less-constrained edit distance problem using $O(|T_1||T_2|I_1^3I_2^3(I_1+I_2))$ time and space is presented. For the unordered version, unlike the constrained edit distance problem, it is shown that the problem is NP-complete. The reduction used is similar to the one for the unordered edit distance problem. It is also reported that the problem is MAX SNP-hard. Furthermore, it is shown that there is no absolute approximation algorithm\footnote{An approximation algorithm $A$ is \emph{absolute} if there exists a constant $c>0$ such that for every instance $I$, $|A(I) - OPT(I)| \leq c$, where $A(I)$ and $OPT(I)$ are the approximate and optimal solutions of $I$ respectively \cite{Motwani1992}.} for the unordered less-constrained edit distance problem unless P=NP.

\subsection{Other Variants}\label{t1:othervariants}
In this section we survey results for other variants of edit distance. Let $T_1$ and $T_2$ be rooted trees. The \emph{unit cost edit distance} between  $T_1$ and $T_2$ is defined as the number of edit operations needed to turn $T_1$ into $T_2$.  In \cite{SZ1990} the ordered version of this problem is considered and a fast algorithm is presented. If $u$ is the unit cost edit distance between $T_1$ and $T_2$ the algorithm runs in $O(u^2 \min\{|T_1|,|T_2|\} \min\{L_1,L_2\})$ time. The algorithm uses techniques from Ukkonen \cite{Ukkonen1985} and Landau and Vishkin \cite{LV1989}. 

In \cite{Selkow1977} Selkow considered an edit distance problem where insertions and deletions are restricted to leaves of the trees. This edit distance is sometimes referred to as the \emph{$1$-degree edit distance}. He gave a simple algorithm using $O(|T_1||T_2|)$ time and space. Another edit distance measure where edit operations work on subtrees instead of nodes was given by Lu \cite{Lu1979}. A similar edit distance was given by Tanaka in \cite{TT1988, TT1995}. A short description of Lu's algorithm can be found in \cite{DZ1997}.

\section{Tree Alignment Distance}\label{t1:treealignmentdistance}
\begin{figure}[h]
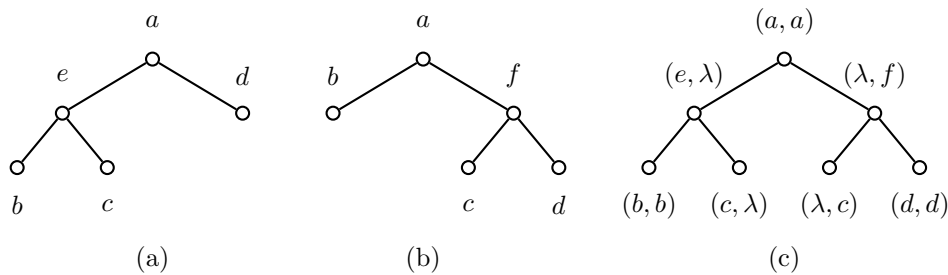

  \begin{center} 
  \begin{psmatrix}[colsep=0.6cm,rowsep=0.3cm,labelsep=1pt]
  \\
  & & & \cnode{.1}{root1}\rput(0,.5){$a$} & & & & & & 
  \cnode{.1}{root2}\rput(0,.5){$a$} & & & & & & & &
  \cnode{.1}{root3}\rput(0,.5){$(a,a)$} &&&\\
  & \cnode{.1}{l1}\rput(0,.5){$e$} & & & & \cnode{.1}{r1}\rput(0,.5){$d$} && 
  \cnode{.1}{l2}\rput(0,.5){$b$}  &&&& \cnode{.1}{r2}\rput(0,.5){$f$}  &&&& 
  \cnode{.1}{l3}\rput(0,.5){$(e,\lambda)$} &&&& \cnode{.1}{r3}\rput(0,.5){$(\lambda,f)$} && \\
  \cnode{.1}{ll1}\rput(0,-.5){$b$} & & \cnode{.1}{lr1}\rput(0,-.5){$c$} &&&&&&&&
  \cnode{.1}{rl2}\rput(0,-.5){$c$} & & \cnode{.1}{rr2}\rput(0,-.5){$d$} &&
  \cnode{.1}{ll3}\rput(0,-.5){$(b,b)$} & & \cnode{.1}{lr3}\rput(0,-.5){$(c,\lambda)$} &&
  \cnode{.1}{rl3}\rput(0,-.5){$(\lambda,c)$} & & \cnode{.1}{rr3}\rput(0,-.5){$(d,d)$} \\
  &&& \rput(0,-.5){(a)} &&&&&& \rput(0,-.5){(b)} &&&&&&&& \rput(0,-.5){(c)}& \\
  \ncline{root1}{l1}
  \ncline{root1}{r1}
  \ncline{ll1}{l1}
  \ncline{lr1}{l1}
  
  \ncline{root2}{l2}
  \ncline{root2}{r2}
  \ncline{rl2}{r2}
  \ncline{rr2}{r2}
  
  \ncline{root3}{l3}
  \ncline{root3}{r3}
  \ncline{ll3}{l3}
  \ncline{lr3}{l3}
  \ncline{rl3}{r3}
  \ncline{rr3}{r3}
  \end{psmatrix}	
  
    \caption{(a) Tree $T_1$. (b) Tree $T_2$. (c) An alignment of $T_1$ and $T_2$.}
  \label{t1:alignmentexample}
  \end{center}
\end{figure}

In this section we consider the alignment distance problem. Let $T_1$ and $T_2$ be rooted, labeled trees and let $\gamma$ be a metric cost function on pairs of labels as defined in Section \ref{t1:preliminaries}. An alignment $A$ of $T_1$ and $T_2$ is obtained by first inserting nodes labeled with $\lambda$ (called \emph{spaces}) into $T_1$ and $T_2$ so that they become isomorphic when labels are ignored, and then \emph{overlaying} the first augmented tree on the other one. The \emph{cost} of a pair of opposing labels in $A$ is given by $\gamma$. The cost of $A$ is the sum of costs of all opposing labels in $A$. An \emph{optimal alignment} of $T_1$ and $T_2$,  is an alignment of $T_1$ and $T_2$ of minimum cost. We denote this cost by $\alpha(T_1,T_2)$. Figure \ref{t1:alignmentexample} shows an example (from \cite{JWZ1995}) of an ordered alignment.

The tree alignment distance problem is a special case of the tree editing problem. In fact, it corresponds to a  restricted edit distance where all insertions must be performed before any deletions. Hence, $\delta(T_1,T_2) \leq \alpha(T_1,T_2)$. For instance, assume that all edit operations have cost $1$ and consider the example in Figure~\ref{t1:alignmentexample}. The optimal sequence of edit operations is achieved by deleting the node labeled $e$ and then inserting the node labeled $f$. Hence, the edit distance is $2$. The optimal alignment, however, is the tree depicted in (c) with a value of $4$. Additionally, it also follows that the alignment distance does not satisfy the triangle inequality and hence it is not a distance metric. For instance, in Figure~\ref{t1:alignmentexample} if $T_3$ is $T_1$ where the node labeled $e$ is deleted, then $\alpha(T_1, T_3) + \alpha(T_3, T_2) = 2 > 4 = \alpha(T_1, T_2)$. 

It is a well known fact that edit and alignment distance are equivalent in terms of complexity for sequences,  see, e.g., Gusfield \cite{Gusfield1997}. However, for trees this is not true which we will show in the following sections. In Section \ref{t1:orderedalignment} and Section \ref{t1:unorderedalignment} we survey the results for the ordered and unordered tree alignment distance problem respectively.

\subsection{Ordered Tree Alignment Distance}\label{t1:orderedalignment}
In this section we consider the ordered tree alignment distance problem. Let $T_1$ and $T_2$ be two rooted, ordered and labeled trees. The ordered tree alignment distance problem was introduced by Jiang et al. in \cite{JWZ1995}. The algorithm presented there uses $O(|T_1|| T_2| (I_1 + I_2)^2)$ time and $O(|T_1||T_2|(I_1 + I_2))$ space. Hence, for small degree trees, this algorithm is in general faster than the best known algorithm for the edit distance. We present this algorithm in detail in the next section. Recently, in \cite{JL2001}, a new algorithm was proposed designed for \emph{similar} trees. Specifically, if there is an optimal alignment of $T_1$ and $T_2$ using at most $s$ spaces the algorithm computes the alignment in time $O((|T_1| + |T_2|)\log(|T_1| + |T_2|)(I_1+I_2)^4s^2)$. This algorithm works in a way similar to the fast algorithms for comparing similar sequences, see, e.g., Section 3.3.4 in \cite{SM1997}. The main idea is to speedup the algorithm of Jiang et al. by only considering subtrees of $T_1$ and $T_2$ whose sizes differ by at most $O(s)$. 

\subsubsection{Jiang, Wang, and Zhang's Algorithm}\label{t1:jiangwangzhang}
In this section we present the algorithm of Jiang et al. \cite{JWZ1995}. We only show how to compute the alignment distance. The corresponding alignment can easily be constructed within the same complexity bounds. 
Let $\gamma$ be a metric cost function on the labels. For simplicity, we will refer to nodes instead of labels, that is, we will use $(v,w)$ for nodes $v$ and $w$ to mean $(\lab(v),\lab(w))$. Here, $v$ or $w$ may be $\lambda$. We extend the definition of $\alpha$ to include alignments of forests, that is, $\alpha(F_1,F_2)$ denotes the cost of an optimal alignment of forest $F_1$ and $F_2$.
\begin{lemma}\label{t1:ATbasic}
Let $v\in V(T_1)$ and $w\in V(T_2)$ with children $v_1,\ldots,v_i$ and $w_1,\ldots, w_j$ respectively. Then,  
\begin{equation*}
\begin{aligned}
\alpha(\theta, \theta) &= 0 \\
\alpha(T_1(v), \theta) &= \alpha(F_1(v), \theta) + \gamma(v, \lambda) \\
\alpha(\theta, T_2(w)) &= \alpha(\theta, F_2(w)) + \gamma(\lambda, w)\\
\alpha(F_1(v), \theta) &= \sum_{k=1}^i \alpha(T_1(v_k), \theta) \\
\alpha(\theta, F_2(w)) &= \sum_{k=1}^j \alpha(\theta, T_2(w_k)) \\
\end{aligned}
\end{equation*}
\end{lemma}
\begin{lemma}\label{t1:ATrecursion}
Let $v\in V(T_1)$ and $w\in V(T_2)$ with children $v_1,\ldots,v_i$ and $w_1,\ldots, w_j$ respectively. Then,  
\begin{equation*}
\alpha(T_1(v), T_2(w)) = \min
\begin{cases}
      \alpha(F_1(v), F_2(w)) + \gamma(v,w) \\
      \alpha(\theta, T_2(w)) + \min_{1\leq r \leq j} \{\alpha(T_1(v), T_2(w_r)) - \alpha(\theta, T_2(w_r)\}  \\
      \alpha(T_1(v), \theta)  + \min_{1\leq r \leq i} \{\alpha(T_1(v_r), T_2(w)) - \alpha(T_1(v_r), \theta)\}  
\end{cases}
\end{equation*}
\end{lemma}
\begin{proof}
Consider an optimal alignment $A$ of $T_1(v)$ and $T_2(w)$. There are four cases: (1) $(v,w)$ is a label in $A$, (2) $(v,\lambda)$ and $(k,w)$ are labels in $A$ for some $k\in V(T_1)$, (3)  $(\lambda,w)$ and $(v,h)$ are labels in $A$ for some $h\in V(T_2)$ or (4) $(v,\lambda)$ and $(\lambda, w)$ are in $A$. Case (4) need not be considered since the two nodes can be deleted and replaced by the single node $(v,w)$ as the new root. The cost of the resulting alignment is by the triangle inequality at least as small.  
\begin{description}
  \item[Case 1:] The root of $A$ is labeled by $(v,w)$. Hence,
  \begin{equation*}
  \alpha(T_1(v), T_2(w)) =   \alpha(F_1(v), F_2(w)) + \gamma(v,w)
  \end{equation*}
  \item[Case 2:] The root of $A$ is labeled by $(v,\lambda)$. Hence, $k \in V(T_1(w_s))$ for some $1\leq r \leq i$. It follows that,
  \begin{equation*}
   \alpha(T_1(v), T_2(w)) = \alpha(T_1(v), \theta)  + \min_{1\leq r \leq i} \{\alpha(T_1(v_r), T_2(w)) - \alpha(T_1(v_r), \theta)\}  
  \end{equation*}
  \item[Case 3:] Symmetric to case $2$. {\hfill$\Box$\\\noindent}
\end{description}
\end{proof}

\begin{lemma}\label{t1:AFrecursion}
Let $v\in V(T_1)$ and $w\in V(T_2)$ with children $v_1,\ldots,v_i$ and $w_1,\ldots, w_j$ respectively. For any $s$, $t$ such that $1\leq s \leq i$ and $1\leq t \leq j$,
\begin{equation*}
\alpha(F_1(v_1,v_s), F_2(w_1,w_t))  = \min
\begin{cases}
      \alpha(F_1(v_1, v_{s-1}), F_2(w_1, w_{t-1})) + \alpha(T_1(v_s),T_2(w_t)) \\
      \alpha(F_1(v_1, v_{s-1}), F_2(w_1, w_t)) + \alpha(T_1(v_s),\theta) \\
      \alpha(F_1(v_1, v_s), F_2(w_1, w_{t-1})) + \alpha(\theta, T_2(w_t)) \\
      \begin{aligned}
      \gamma(\lambda, w_t) +\min_{1\leq k <s}\{&\alpha(F_1(v_1, v_{k-1}), F_2(w_1, w_{t-1})) \\
								   			  & + \alpha(F_1(v_k,v_s), F_2(w_k))\}  
	\end{aligned}\\
	\begin{aligned}						
      \gamma(v_s,\lambda) + \min_{1\leq k <t}\{&\alpha(F_1(v_1, v_{s-1}), F_2(w_1, w_{k-1})) \\
      											  &	+\alpha(F_1(v_s), F_2(w_k, w_{t}))\}  
	\end{aligned}							  
      \end{cases}    
\end{equation*}
\end{lemma}

\begin{proof}
Consider an optimal alignment $A$ of $F_1(v_1, v_s)$ and  $F_2(w_1, w_t)$. The root of the rightmost tree in $A$ is labeled either $(v_s, w_t)$, $(v_s,\lambda)$ or $(\lambda, w_t)$. 
\begin{description}
  \item[Case 1:] The label is $(v_s, w_t)$. Then the rightmost tree of $A$ must be an optimal alignment of $T_1(v_s)$ and $T_2(w_t)$. Hence,  
  \begin{equation*}
	  \alpha(F_1(v_1,v_s), F_2(w_1,w_t)) = \alpha(F_1(v_1, v_{s-1}), F_2(w_1, w_{t-1})) + \alpha(T_1(v_s),T_2(w_t)).
  \end{equation*}
  \item[Case 2:] The label is $(v_s, \lambda)$. Then $T_1(v_s)$ is a aligned with a subforest $F_2(w_{t-k+1},w_t)$, where $0 \leq k \leq t$. The following subcases can occur:
  \begin{description}
	  \item[2.1 $(k=0)$.] $T_1(v_s)$ is aligned with $F_2(w_{t-k+1},w_t) = \theta$. Hence, 
	  \begin{equation*}
	  \alpha(F_1(v_1,v_s), F_2(w_1,w_t)) = \alpha(F_1(v_1, v_{s-1}), F_2(w_1, w_t)) + \alpha(T_1(v_s),\theta).
	  \end{equation*}
	  \item[2.2 $(k=1)$.]  $T_1(v_s)$ is aligned with $F_2(w_{t-k+1},w_t) = T_2(w_t)$. Similar to case $1$.
	  \item[2.3 $(k\geq 2)$.] The most general case. It is easy to see that:
	  \begin{equation*}
	  \begin{split}
	  \alpha(F_1(v_1,v_s), F_2(w_1,w_t)) = \gamma(v_s,\lambda) + \min_{1\leq r < t}\{&\alpha(F_1(v_1, v_{s-1}), F_2(w_1, w_{k-1}))) \\ &+  \alpha(F_1(v_s), F_2(w_k, w_{t})).  
	  \end{split}
	  \end{equation*}	   
  \end{description}
  \item[Case 3:] The label is $(\lambda, w_t)$. Symmetric to case $2$. {\hfill$\Box$\\\noindent}
\end{description}
\end{proof}

This recursion can be used to construct a bottom-up dynamic programming algorithm. Consider a fixed pair of nodes $v$ and $w$ with children $v_1,\ldots,v_i$ and $w_1,\ldots, w_j$ respectively. We need to compute the values $\alpha(F_1(v_h,v_k), F_2(w))$ for all $1\leq h \leq k \leq i$, and $\alpha(F_1(v), F_2(w_h,w_k))$ for all $1\leq h \leq k \leq j$. That is, we need to compute the optimal alignment of $F_1(v)$ with each subforest of $F_2(w)$ and, on the other hand, compute the optimal alignment of $F_2(w)$ with each subforest of $F_1(v)$. 
For any $s$ and $t$, $1\leq s \leq i$ and $1\leq t \leq j$, define the set:
\begin{gather*} 
A_{s,t} = \{\alpha(F_1(v_s, v_p), F_2(w_t, w_q)) \mid s \leq p \leq i, t \leq q \leq j\} \\
\end{gather*} 
To compute the alignments described above we need to compute $A_{s,1}$ and $A_{1,t}$ for all $1\leq s \leq i$ and $1\leq t \leq j$. Assuming that values for smaller subproblems are known it is not hard to show that $A_{s,t}$ can be computed, using Lemma \ref{t1:AFrecursion}, in time $O((i-s)\cdot (j-t) \cdot (i-s + j-t)) = O(ij(i+j))$. Hence,  the time to compute the $(i+j)$ subproblems, $A_{s,1}$ and $A_{1,t}$, $1\leq s \leq i$ and $1\leq t \leq j$, is bounded by $O(ij(i+j)^2)$. It follows that the total time needed for all nodes $v$ and $w$ is bounded by:
\begin{equation*}
\begin{split}
\sum_{v\in V(T_1)}\sum_{w\in V(T_2)} & O(\deg(v)\deg(w)(\deg(v) + \deg(w))^2)  \\
	& \leq  \sum_{v\in V(T_1)}\sum_{w\in V(T_2)} O(\deg(v)\deg(w)(\deg(T_1) + \deg(T_2))^2)   \\
      &  \leq  O( (I_1+ I_2)^2 \sum_{v\in V(T_1)}\sum_{w\in V(T_2)} \deg(v)\deg(w))  \\
      &  \leq  O(|T_1||T_2|(I_1 + I_2)^2)
\end{split}
\end{equation*}
In summary, we have shown the following theorem.
\begin{theorem}[\cite{JWZ1995}]
For ordered trees $T_1$ and $T_2$, the tree alignment distance problem can be solved in $O(|T_1||T_2|(I_1 + I_2)^2)$ time and $O(|T_1||T_2|(I_1 + I_2))$ space. 
\end{theorem}

\subsection{Unordered Tree Alignment Distance}\label{t1:unorderedalignment}
The algorithm presented above can be modified to handle the unordered version of the problem in a straightforward way \cite{JWZ1995}. If the trees have bounded degrees the algorithm still runs in $O(|T_1|T_2|)$ time. This should be seen in contrast to the edit distance problem which is MAX SNP-hard even if the trees have bounded degree. If one tree has arbitrary degree unordered alignment becomes NP-hard \cite{JWZ1995}. The reduction is, as for the edit distance problem, from the Exact Cover by 3-Sets problem \cite{GJ1979}.

\section{Tree Inclusion}\label{t1:treeinclusion}
\begin{figure}[t]
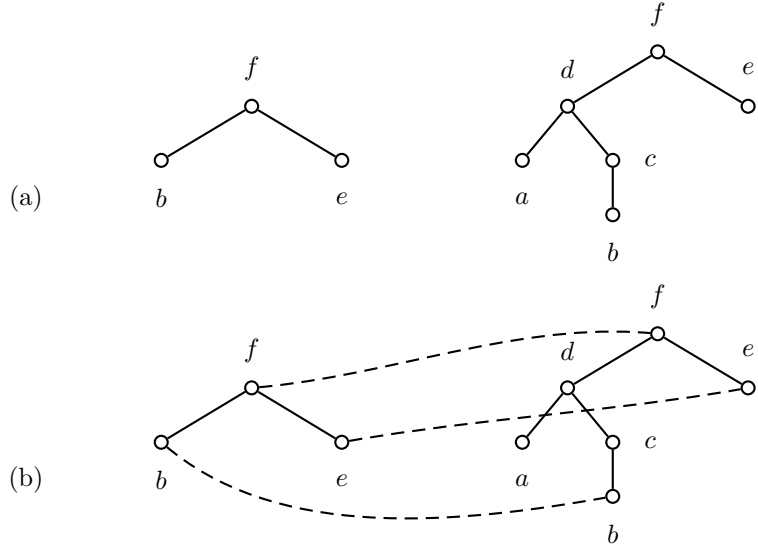

\begin{center}
  	
  \begin{psmatrix}[colsep=0.6cm,rowsep=0.3cm,labelsep=1pt]
  & & & & & & & & &  &&&&  & \cnode{.1}{root3}\rput(0,.5){$f$} &&\\
  &&&&& \cnode{.1}{root1}\rput(0,.5){$f$} &&&&&&& 
  \cnode{.1}{l3}\rput(0,.5){$d$} &&&& \cnode{.1}{r3}\rput(0,.5){$e$}\\
  \rput(0,-.5){(a)}&&& \cnode{.1}{l1}\rput(0,-.5){$b$} &&&& \cnode{.1}{r1}\rput(0,-.5){$e$} &&&&
  \cnode{.1}{ll3}\rput(0,-.5){$a$} && \cnode{.1}{lr3}\rput(.5,0){$c$} \\
  &&&&&&&&&&&&& \cnode{.1}{lrr3}\rput(0,-.5){$b$} \\ \\
  \ncline{l1}{root1} \ncline{r1}{root1}
  \ncline{l3}{root3} \ncline{r3}{root3}
  \ncline{ll3}{l3} 
   \ncline{lr3}{l3} 
   \ncline{lrr3}{lr3}	
  \end{psmatrix} 
  \begin{psmatrix}[colsep=0.6cm,rowsep=0.3cm,labelsep=1pt]
  & & & & & & & & &  &&&&  & \cnode{.1}{root3}\rput(0,.5){$f$} &&\\
  &&&&& \cnode{.1}{root1}\rput(0,.5){$f$} &&&&&&& 
  \cnode{.1}{l3}\rput(0,.5){$d$} &&&& \cnode{.1}{r3}\rput(0,.5){$e$}\\
  \rput(0,-.5){(b)}&&& \cnode{.1}{l1}\rput(0,-.5){$b$} &&&& \cnode{.1}{r1}\rput(0,-.5){$e$} &&&&
  \cnode{.1}{ll3}\rput(0,-.5){$a$} && \cnode{.1}{lr3}\rput(.5,0){$c$} \\
  &&&&&&&&&&&&& \cnode{.1}{lrr3}\rput(0,-.5){$b$} \\
  \ncline{l1}{root1} \ncline{r1}{root1}
  \ncline{l3}{root3} \ncline{r3}{root3}
  \ncline{ll3}{l3} 
   \ncline{lr3}{l3} 
   \ncline{lrr3}{lr3}	
  \psset{linestyle=dashed}
  \nccurve[angleA=5,angleB=175]{root1}{root3}
  \nccurve[angleA=10,angleB=190]{r1}{r3}
  \nccurve[angleA=320,angleB=190]{l1}{lrr3}
  \end{psmatrix} 
   \caption{(a) The tree on the left is included in the tree on the right by deleting the nodes labeled $d$, $a$ and $c$. (b) The embedding corresponding to (a). }
  \label{t1:inclusionexample}
  \end{center}
\end{figure}

In this section we survey the tree inclusion problem. Let $T_1$ and $T_2$ be rooted, labeled trees. We say that $T_1$ is \emph{included} in $T_2$ if there is a sequence of delete operations performed on $T_2$ which makes $T_2$ isomorphic to $T_1$. The \emph{tree inclusion problem} is to decide if $T_1$ is included in $T_2$. Figure \ref{t1:inclusionexample}(a) shows an example of an ordered inclusion. The tree inclusion problem is a special case of the tree edit distance problem: If insertions all have cost $0$ and all other operations have cost $1$, then $T_1$ can be included in $T_2$ if and only if $\delta(T_1,T_2) = 0$. According to \cite{Chen1998} the tree inclusion problem was initially introduced by Knuth \cite{Knuth1969}[exercise 2.3.2-22]. 

The rest of the section is organized as follows. In Section \ref{t1:orderingsandembeddings} we give some preliminaries and in Section \ref{t1:orderedtreeinclusion} and \ref{t1:unorderedtreeinclusion} we survey the known results on ordered and unordered tree inclusion respectively.

\subsection{Orderings and Embeddings}\label{t1:orderingsandembeddings}
Let $T$ be a labeled, ordered, and rooted tree. We define an ordering of the nodes of $T$ given by $v\prec v'$ iff $\post(v) <\post(v')$. Also, $v\preceq v'$ iff $v\prec v'$ or $v=v'$. Furthermore, we extend this ordering with two special nodes $\bot$ and $\top$ such that for all nodes $v \in V(T)$, $\bot \prec v \prec \top$. The \emph{left relatives}, $\lr(v)$, of a node $v\in V(T)$ is the set of nodes that are to the left of $v$ and similarly the \emph{right relatives}, $\rr(v)$, are the set of nodes that are to the right of $v$. 

Let $T_1$ and $T_2$ be rooted labeled trees. We define an \emph{ordered embedding} $(f, T_1, T_2)$ as an injective function $f : V(T_1) \rightarrow V(T_2)$ such that for all  nodes $v,u \in V(T_1)$, 
\begin{itemize}
  \item $\lab(v) = \lab(f(v))$. (label preservation condition)
  \item $v$ is an ancestor of $u$  iff $f(v)$ is an ancestor of $f(u)$. (ancestor condition)
  \item $v$ is to the left of $u$ iff $f(v)$ is to the left of $f(u)$. (sibling condition)
\end{itemize}
Hence, embeddings are special cases of mappings (see Section \ref{t1:editmappings}). An \emph{unordered embedding} is defined as above, but without the sibling condition. An embedding $(f, T_1, T_2)$ is \emph{root preserving} if $f(\roots(T_1)) = \roots(T_2)$. Figure \ref{t1:inclusionexample}(b) shows an example of a root preserving embedding.

\subsection{Ordered Tree Inclusion}\label{t1:orderedtreeinclusion}
Let $T_1$ and $T_2$ be rooted, ordered and labeled trees. The ordered tree inclusion problem has been the attention of much research. Kilpel\"{a}inen and Mannila \cite{KM1995} (see also \cite{Kilpelainen1992}) presented the first polynomial time algorithm using $O(|T_1|| T_2|)$ time and space. Most of the later improvements are refinements of this algorithm. We present this algorithm in detail in the next section. In \cite{Kilpelainen1992} a more space efficient version of the above was given using $O(|T_1|D_2)$ space. In \cite{Richter1997a} Richter gave an algorithm using $O(|\Sigma_{T_1}| |T_2| + m_{T_1,T_2}D_2)$ time, where $\Sigma_{T_1}$ is the alphabet of the labels of $T_1$ and $m_{T_1,T_2}$ is the set \emph{matches}, defined as the number of pairs $(v,w) \in T_1 \times T_2$ such that $\lab(v) = \lab(w)$. Hence, if the number of matches is small the time complexity of this algorithm improves the $(|T_1|| T_2|)$ algorithm. The space complexity of the algorithm is $O(|\Sigma_{T_1}| |T_2| + m_{T_1,T_2})$.  In \cite{Chen1998} a more complex algorithm was presented using $O(L_1|T_2|)$ time and $O(L_1 \min\{D_2, L_2\})$ space. In \cite{AS2001} an efficient average case algorithm was given.

\subsubsection{Kilpel\"{a}inen and Mannila's Algorithm}\label{t1:kilpelainenmannila}
In this section we present the algorithm of Kilpel\"{a}inen and Mannila \cite{KM1995} for the ordered tree inclusion problem. Let $T_1$ and $T_2$ be ordered labeled trees. Define $R(T_1, T_2)$ as the set of root-preserving embeddings of $T_1$ into $T_2$. 
We define $\rho(v,w)$, where $v\in V(T_1)$ and $w\in V(T_2)$:

\begin{equation*}
\rho(v, w) = \min_{\prec} \left(\{w' \in rr(w) \mid \exists f \in R(T_1(v), T_2(w')) \} \cup \{\top\} \right)
\end{equation*}

Hence, $\rho(v,w)$ is the closest right relative of $w$ which has a root-preserving embedding of $T_1(v)$.
Furthermore, if no such embedding exists $\rho(v,w)$ is $\top$. It is easy to see that, by definition, $T_1$ can be included in $T_2$ if and only if $\rho(v, \bot) \neq \top$. The following lemma shows how to search for root preserving embeddings.

\begin{lemma}\label{t1:Ilemma}
Let  $v$  be a node in $T_1$ with children $v_1, \ldots , v_i$. For a node $w$ in $T_2$, define a sequence $p_1, \ldots, p_i$ by setting $p_1 = \rho(v_1,\max_{\prec}\lr(w))$ and $p_k = \rho(v_k,p_{k-1})$, for $2\leq k \leq i$. There is a root preserving embedding $f$ of $T_1(v)$ in $T_2(v)$ if and only if $\lab(v) = \lab(w)$ and $p_i \in T_2(w)$, for all $1\leq k \leq i$. 
\end{lemma}
\begin{proof}
If there is a root preserving embedding between $T_1(v)$ and $T_2(w)$ it is straightforward to check that there is a sequence $p_i$, $1\leq i \leq k$ such that the conditions are satisfied. Conversely, assume that $p_k \in T_2(w)$ for all $1\leq k \leq i$ and $\lab(v) = \lab(w)$.  We construct a root-preserving embedding $f$ of $T_1(v)$ into $T_2(w)$ as follows. Let $f(v) = w$. By definition of $\rho$ there must be a root preserving embedding $f^k$, $1\leq k \leq i$, of $T_1(v_k)$ in $T_2(p_k)$. For a node $u$ in $T_1(v_k)$, $1\leq k \leq i$, we set $f(u) = f^k(u)$. Since $p_k\in \rr(p_{k-1})$, $2\leq k \leq i$, and  $p_k \in T_2(w)$ for all $k$, $1\leq k \leq i$, it follows that $f$ is indeed a root-preserving embedding. {\hfill$\Box$\\\noindent}
\end{proof}

Using dynamic programming it is now straightforward to compute $\rho(v, w)$ for all $v \in V(T_1)$ and $w \in V(T_2)$. For a fixed node $v$ we traverse $T_2$ in reverse postorder. At each node $w\in V(T_2)$ we check if there is a root preserving embedding of $T_1(v)$ in $T_2(w)$. If so we set $\rho(v,q) = w$, for all $q\in lr(w)$ such that $x\preceq q$, where $x$ is the next root-preserving embedding of $T_1(v)$ in $T_2(w)$. 

For a pair of nodes $v\in V(T_1)$ and $w\in V(T_2)$ we test for a root-preserving embedding using Lemma \ref{t1:Ilemma}. Assuming that values for smaller subproblems has been computed, the time used is $O(\deg(v))$. Hence, the  contribution to the total time for the node $w$ is $\sum_{v\in V(T_1)} O(\deg(v)) = O(|T_1|)$. It follows that the time complexity of the algorithm is bounded by $O(|T_1|| T_2|)$. Clearly, only $O(|T_1|| T_2|)$ space is needed to store $\rho$. Hence, we have the following theorem, 

\begin{theorem}[\cite{KM1995}]
For any pair of rooted, labeled, and ordered trees $T_1$ and $T_2$, the tree inclusion problem can be solved in $O(|T_1|| T_2|)$ time and space.
\end{theorem}

\subsection{Unordered Tree Inclusion}\label{t1:unorderedtreeinclusion}
In \cite{KM1995} it is shown that the unordered tree inclusion problem is NP-complete. The reduction used is from the Satisfiability problem \cite{GJ1979}. Independently, Matou\v{s}ek and Thomas \cite{MT1992} gave another proof of NP-completeness. 

An algorithm for the unordered tree inclusion problem is presented in \cite{KM1995} using $O(|T_1|I_12^{2I_1}|T_2|)$ time. Hence, if $I_1$ is constant the algorithm runs in $O(|T_1|| T_2|)$ time and if $I_1 = \log |T_2|$ the algorithm runs in $O(|T_1|\log |T_2| |T_2|^3)$.

\section{Conclusion}\label{t1:conclusion}
We have surveyed the tree edit distance, alignment distance, and inclusion problems. Furthermore, we have presented, in our opinion, the central algorithms for each of the problems. There are several open problems, which may be the topic of further research. We conclude this paper with a short list proposing some directions.

\begin{itemize}
  \item For the unordered versions of the above problems some are NP-complete while others are not. Characterizing exactly which types of mappings that gives NP-complete problems for unordered versions would certainly improve the understanding of all of the above problems.
  \item The currently best worst case upper bound on the ordered tree edit distance problem is the algorithm of \cite{Klein1998} using $O(|T_1|^2|T_2|\log |T_2|)$. Conversely, the quadratic lower bound for the longest common subsequence problem \cite{AHU1976} problem is the best general lower bound for the ordered tree edit distance problem. Hence, a large gap in complexity exists which needs to be closed. 
  \item Several meaningful edit operations other than the above may be considered depending on the particular application. Each set of operations yield a new edit distance problem for which we can determine the complexity. Some extensions of the tree edit distance problem have been considered \cite{CRMW1996, CM1997, KTSK2000}.
\end{itemize}

\section{Acknowledgments}
Thanks to Inge Li Gørtz and Anna \"{O}stlin Pagh for proof reading and helpful discussions.


\emptythanks
\chapter{The Tree Inclusion Problem: In Linear Space and Faster}\label{chap:tree2}

\title{The Tree Inclusion Problem: In Linear Space and Faster}

\author{Philip Bille \\ IT University of Copenhagen \\ \texttt{beetle@itu.dk}
\and Inge Li G{\o}rtz\thanks{Part of this work was performed while the author was a PhD student at the IT University of Copenhagen.} \\ Technical University of Denmark \\ {\tt ilg@imm.dtu.dk}}

\date{}
\cleartooddpage

\maketitle

\begin{abstract}
Given two rooted, ordered, and labeled trees $P$ and $T$ the tree
inclusion problem is to determine if $P$ can be obtained from $T$
by deleting nodes in $T$. This problem has recently been
recognized as an important query primitive in XML databases.
Kilpel\"{a}inen and Mannila [\emph{SIAM J. Comput. 1995}] presented the
first polynomial time algorithm using quadratic time and space.
Since then several improved results have been obtained for special
cases when $P$ and $T$ have a small number of leaves or small
depth. However, in the worst case these algorithms still use
quadratic time and space. In this paper we present a new approach
to the problem which leads to an algorithm using linear space and subquadratic running time. Our algorithm improves all previous time and space bounds. Most importantly, the space is improved by a linear factor. This will likely make it possible to query larger XML databases and speed up the query time since more of the computation can be kept in main memory.
\end{abstract}

\section{Introduction}
Let $T$ be a rooted tree. We say that $T$ is \emph{labeled} if
each node is assigned a character from an alphabet
$\Sigma$ and we say that $T$ is \emph{ordered} if a left-to-right
order among siblings in $T$ is given. All trees in this paper are rooted, ordered, and labeled.
A tree $P$ is \emph{included} in $T$, denoted $P\sqsubseteq T$, if $P$ can be
obtained from $T$ by deleting nodes of $T$. Deleting a node $v$ in
$T$ means making the children of $v$ children of the parent of $v$
and then removing $v$. The children are inserted in the place of
$v$ in the left-to-right order among the siblings of $v$. The \emph{tree inclusion problem}
is to determine if $P$ can be included in $T$ and if so report all subtrees of $T$ that include $P$.
\begin{figure}[t]
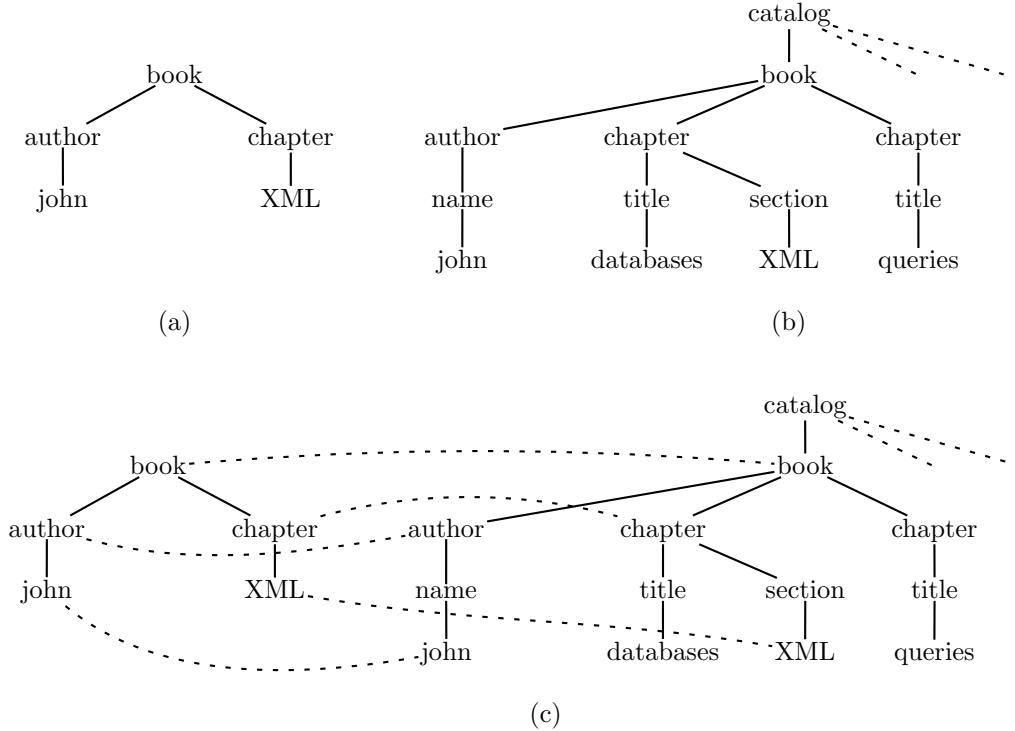

\begin{center}
  \begin{psmatrix}[colsep=0.6cm,rowsep=0.4cm,labelsep=1pt, nodesep=1pt]
  &&&&&&&&[name=cat] catalog \\
  &&[name=book] book &&&&&&[name=book2] book & [name=book3] & [name=book4]\\
  & [name=author] author &&[name=chapter] chapter&&
  [name=author2] author && [name=chapter2] chapter && [name=chapter3] chapter \\
  & [name=john] john && [name=xml] XML &&
  [name=name] name && [name=title] title  & [name=section] section & [name=title2] title \\
  &&&&&  [name=john2] john && [name=DB] databases & [name=xml2] XML & [name=queries] queries \\
  && (a) &&&&&& (b) \\
  \ncline{author}{book} \ncline{chapter}{book}
  \ncline{john}{author} \ncline{xml}{chapter}
  \ncline{cat}{book2}\psset{linestyle=dashed,dash=2pt 4pt}\ncline{cat}{book3} \ncline{cat}{book4}
  \psset{linestyle=solid}
  \ncline{book2}{author2}\ncline{book2}{chapter2}\ncline{book2}{chapter3}
  \ncline{author2}{name}\ncline{chapter2}{title}\ncline{chapter2}{section}\ncline{chapter3}{title2}
  \ncline{name}{john2}\ncline{title}{DB}\ncline{section}{xml2}\ncline{title2}{queries}
  \end{psmatrix}

  \begin{psmatrix}[colsep=0.6cm,rowsep=0.4cm,labelsep=1pt, nodesep=1pt]
  &&&&&&&&[name=cat] catalog \\
  &&[name=book] book &&&&&&[name=book2] book & [name=book3] & [name=book4]\\
  & [name=author] author &&[name=chapter] chapter&&
  [name=author2] author && [name=chapter2] chapter && [name=chapter3] chapter \\
  & [name=john] john && [name=xml] XML &&
  [name=name] name && [name=title] title  & [name=section] section & [name=title2] title \\
  &&&&&  [name=john2] john && [name=DB] databases & [name=xml2] XML & [name=queries] queries \\
  &&&&&&(c)&&&& \\
  \ncline{author}{book} \ncline{chapter}{book}
  \ncline{john}{author} \ncline{xml}{chapter}
  \ncline{cat}{book2}\psset{linestyle=dashed,dash=2pt 4pt} \ncline{cat}{book3} \ncline{cat}{book4}
  \psset{linestyle=solid}
  \ncline{book2}{author2}\ncline{book2}{chapter2}\ncline{book2}{chapter3}
  \ncline{author2}{name}\ncline{chapter2}{title}\ncline{chapter2}{section}\ncline{chapter3}{title2}
  \ncline{name}{john2}\ncline{title}{DB}\ncline{section}{xml2}\ncline{title2}{queries}

  \psset{linestyle=dashed}
  \nccurve[angleA=5,angleB=175]{book}{book2}
  \nccurve[angleA=345,angleB=190]{author}{author2}
  \nccurve[angleA=320,angleB=190]{john}{john2}
  \nccurve[angleA=15,angleB=165]{chapter}{chapter2}
  \nccurve[angleA=350,angleB=170]{xml}{xml2}
  \end{psmatrix}
   \caption{Can the tree (a) be included in the tree (b)? It can and an embedding is given in (c).}
  \label{t2:inclusionexample}
  \end{center}
\end{figure}

Recently, the problem has been recognized as an important query
primitive for XML data and has received considerable attention,
see e.g., \cite{SM2002, YLH2003,YLH2004, ZADR03, SN2000, TRS2002}.
The key idea is that an XML document can be viewed as a tree and queries on the document correspond to a tree
inclusion problem. As an example consider Figure~\ref{t2:inclusionexample}. Suppose that we want to maintain
a catalog of books for a bookstore. A fragment of the tree,
denoted $D$, corresponding to the catalog is shown in (b). In addition to supporting
full-text queries, such as find all documents containing the word
"John", we can also utilize the tree structure of the catalog to
ask more specific queries, such as "find all books written by John
with a chapter that has something to do with XML". We can model
this query by constructing the tree, denoted $Q$, shown in (a) and
solve the tree inclusion problem: is $Q \sqsubseteq D$? The answer
is yes and a possible way to include $Q$ in $D$ is indicated by
the dashed lines in (c). If we delete all the nodes in $D$ not
touched by dashed lines the trees $Q$ and $D$ become isomorphic.
Such a mapping of the nodes from $Q$ to $D$ given by the dashed
lines is called an \emph{embedding} (formally defined in
Section~\ref{t2:sec:recursion}).

The tree inclusion problem was initially introduced by Knuth
\cite[exercise 2.3.2-22]{Knuth1969} who gave a sufficient
condition for testing inclusion. Motivated by applications in
structured databases \cite{KM93, MR90} Kilpel\"{a}inen and Mannila
\cite{KM1995} presented the first polynomial time algorithm using
$O(n_Pn_T)$ time and space, where $n_P$ and $n_T$ is the number of
nodes in $P$ and $T$, respectively. During the last decade
several improvements of the original algorithm of \cite{KM1995}
have been suggested \cite{Kilpelainen1992,AS2001, Richter1997a,
Chen1998}. The previously best known bound is due to Chen
\cite{Chen1998} who presented an algorithm using $O(l_Pn_T)$ time
and $O(l_P \cdot \min\{d_T, l_T\})$ space. Here, $l_S$ and $d_S$ denotes
the number of leaves and the maximum depth of a tree $S$,
respectively. This algorithm is based on an algorithm of
Kilpel\"{a}inen \cite{Kilpelainen1992}. Note that the time and
space is still $\Theta(n_Pn_T)$ for worst-case input trees.

In this paper we present three algorithms which combined improves all of the previously known time and space bounds. To avoid trivial cases we always assume that $1 \leq n_P \leq n_T$. We show the following theorem:
\begin{theorem}\label{t2:thm:main}
For trees $P$ and $T$ the tree inclusion problem can be solved in $O(n_T)$ space with the following running times:
\begin{equation*}
\min
\begin{cases}
      O(l_Pn_T), \\
      O(n_Pl_T\log \log n_T + n_T), \\
      O(\frac{n_Pn_T}{\log n_T} + n_{T}\log n_{T}).
\end{cases}
\end{equation*}
\end{theorem}
Hence, when either $P$ or $T$ has few leaves we obtain fast algorithms. When both trees have many leaves and $n_{P} = \Omega (\log^{2} n_{T})$, we instead improve the previous quadratic time bound by a logarithmic factor. Most importantly, the space used is linear. In the context of XML databases this will likely make it possible to query larger trees and speed up the query time since more of the computation can be kept in main memory.

\subsection{Techniques}
Most of the previous algorithms, including the best one
\cite{Chen1998}, are essentially based on a simple dynamic
programming approach from the original algorithm of \cite{KM1995}. The
main idea behind this algorithm is the following: Let $v$ be a node in $P$ with children $v_1, \ldots, v_i$ and
let $w$ be a node in $T$ with children $w_1, \ldots, w_j$. Consider the subtrees rooted at $v$ and $w$, denoted by $P(v)$ and $T(w)$. To decide if $P(v)$ can be included in $T(w)$ we try to find a sequence of numbers $1 \leq x_1 < x_2 < \cdots < x_i \leq j$ such that $P(v_k)$ can be included in $T(w_{x_k})$ for all $k$, $1 \leq k \leq i$. If we have already determined whether or not $P(v_s) \sqsubseteq T(w_t)$, for all $s$ and $t$, $1\leq s \leq i$, $1\leq t \leq j$, we can efficiently find such a sequence by scanning the children of $v$ from left to
right. Hence, applying this approach in a bottom-up fashion we can determine, if $P(v) \sqsubseteq T(w)$, for all pairs of nodes $v$ in $P$ and $w$ in $T$.

In this paper we take a different approach. The main
idea is to construct a data structure on $T$ supporting a small
number of procedures, called the \emph{set procedures}, on subsets
of nodes of $T$. We show that any such data structure implies an
algorithm for the tree inclusion problem. We consider various
implementations of this data structure which all use linear space.
The first simple implementation gives an algorithm with
$O(l_Pn_T)$ running time. As it turns out, the running time
depends on a well-studied problem known as the \emph{tree color
problem}. We show a direct connection between a data structure
for the tree color problem and the tree inclusion problem.
Plugging in a data structure of Dietz \cite{Die89} we obtain an
algorithm with $O(n_Pl_T\log \log n_T + n_T)$ running time.

Based on the simple algorithms above we show how to improve the
worst-case running time of the set procedures by a logarithmic
factor. The general idea used to achieve this is to divide $T$
into small trees called \emph{clusters} of logarithmic size which overlap with other clusters in at most $2$ nodes. Each cluster is represented by a constant number of nodes in a \emph{macro tree}. The nodes in the
macro tree are then connected according to the overlap of the cluster they represent. We show how to efficiently preprocess the clusters and the macro tree such that the set procedures use constant time for each cluster. Hence, the worst-case quadratic running time is improved by a logarithmic factor.

Throughout the paper we assume a unit-cost RAM model of computation with word size $\Theta(\log n_T)$ and a standard instruction set including bitwise boolean operations, shifts, addition, and multiplication. All space complexities refer to the number of words used by the algorithm.

\subsection{Related Work}
For some applications considering \emph{unordered} trees is more
natural. However, in \cite{MT1992,KM1995} this problem was proved
to be NP-complete. The tree inclusion problem is closely related
to the \emph{tree pattern matching problem} \cite{CO1982,
Kosaraju1989,DGM1990, CHI1999}. The goal is here to find an
injective mapping $f$ from the nodes of $P$ to the nodes of $T$
such that for every node $v$ in $P$ the $i$th child of $v$ is
mapped to the $i$th child of $f(v)$. The tree pattern matching
problem can be solved in $(n_P+n_T) \log^{O(1)} (n_P+n_T)$ time. Another similar problem is the \emph{subtree isomorphism} problem \cite{Chung1987, ST1999}, which is to determine if $T$ has
a subgraph isomorphic to $P$. The subtree isomorphism
problem can be solved efficiently for ordered and unordered trees.
The best algorithms for this problem use $O(\frac{n_P^{1.5}n_T}{\log
n_P} + n_T)$ time for unordered trees and $O(\frac{n_Pn_T}{\log n_P} + n_T)$ time for ordered
trees \cite{Chung1987, ST1999}. Both use $O(n_Pn_T)$ space. The
tree inclusion problem can be considered a special case of the
\emph{tree edit distance problem} \cite{Tai1979, ZS1989,
Klein1998, DMRW2006}. Here one wants to find the minimum sequence of insert,
delete, and relabel operations needed to transform $P$ into $T$.
Currently the best algorithm for this problem uses $O(n_T n_P^2 (1 + \log \frac{n_T}{n_P}))$ time~\cite{DMRW2006}. For more details and references see the survey \cite{Bille2005}.
\subsection{Outline}
In Section~\ref{t2:def} we give notation and definitions used
throughout the paper. In Section~\ref{t2:sec:recursion} a common
framework for our tree inclusion algorithms is given.
Section~\ref{t2:simple} present two simple algorithms and then, based on these result, we show how to get a faster algorithm in Section~\ref{t2:micromacro}.

\section{Notation and Definitions}\label{t2:def}
In this section we define the notation and definitions we will use
throughout the paper. For a graph $G$ we denote the set of nodes
and edges by $V(G)$ and $E(G)$, respectively. Let $T$ be a rooted
tree. The root of $T$ is denoted by $\roots(T)$. The \emph{size}
of $T$, denoted by $n_T$, is $|V(T)|$. The \emph{depth} of a node
$v\in V(T)$, $\depth(v)$, is the number of edges on the path from
$v$ to $\roots(T)$ and the depth of $T$, denoted $d_T$, is the
maximum depth of any node in $T$. The parent of $v$ is denoted $\parent(v)$ and the set of children of $v$ is denoted $\child(v)$. A node with no children is a leaf and otherwise an internal node. The set of leaves of $T$ is
denoted $L(T)$ and we define $l_T = |L(T)|$. We say that $T$ is \emph{labeled} if
each node $v$ is a assigned a character, denoted $\lab(v)$, from an alphabet
$\Sigma$ and we say that $T$ is \emph{ordered} if a left-to-right
order among siblings in $T$ is given. All trees in this paper are rooted, ordered, and labeled.

\paragraph{Ancestors and Descendants}
Let $T(v)$ denote the subtree of $T$ rooted at a node $v \in
V(T)$. If $w\in V(T(v))$ then $v$ is an ancestor of $w$, denoted
$v \preceq w$, and if $w\in V(T(v))\backslash \{v\}$ then $v$ is a
proper ancestor of $w$, denoted $v \prec w$. If $v$ is a (proper)
ancestor of $w$ then $w$ is a (proper) descendant of $v$. A node
$z$ is a common ancestor of $v$ and $w$ if it is an ancestor of
both $v$ and $w$. The nearest common ancestor of $v$ and $w$,
$\nca(v,w)$, is the common ancestor of $v$ and $w$ of greatest
depth. The \emph{first ancestor of $w$ labeled $\alpha$}, denoted
$\fl(w,\alpha)$, is the node $v$ such that $v \preceq w$, $\lab(v)
= \alpha$, and no node on the path between $v$ and $w$ is labeled
$\alpha$. If no such node exists then $\fl(w,\alpha) = \bot$,
where $\bot \not\in V(T)$ is a special \emph{null node}. 

\paragraph{Traversals and Orderings}
Let $T$ be a tree with root $v$ and let $v_1, \ldots ,v_k$ be the
children of $v$ from left-to-right. The \emph{preorder traversal}
of $T$ is obtained by visiting $v$ and then recursively visiting
$T(v_i)$, $1 \leq i \leq k$, in order. Similarly, the
\emph{postorder traversal} is obtained by first visiting $T(v_i)$,
$1 \leq i \leq k$, in order and then $v$. The \emph{preorder
number} and \emph{postorder number} of a node $w \in T(v)$,
denoted by $\pre(w)$ and $\post(w)$, is the number of nodes
preceding $w$ in the preorder and postorder traversal of $T$,
respectively. The nodes to the \emph{left} of $w$ in $T$ is the
set of nodes $u \in V(T)$ such that $\pre(u) < \pre(w)$ and
$\post(u) < \post(w)$. If $u$ is to the left of $w$, denoted by $u
\lhd w$, then $w$ is to the \emph{right} of $u$. If $u \lhd w$, $u
\preceq w$, or $w \prec u$ we write $u \unlhd w$. The null node
$\bot$ is not in the ordering, i.e., $\bot \ntriangleleft v$ for
all nodes $v$.

\begin{figure}[t]
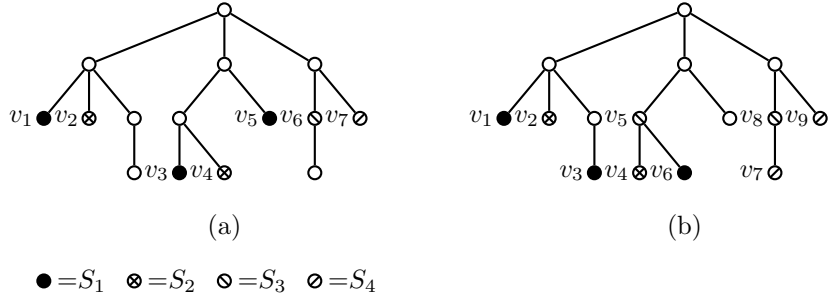

\begin{center}
\begin{psmatrix}[colsep=0.6cm,rowsep=0.3cm,labelsep=1pt]
  &&&& \cnode{.1}{root}\rput(0,.3){}
  \\
  & \cnode{.1}{l}\rput(0,.3){} &&&  \cnode{.1}{c}\rput(-.3,.3){}
  &&
  \cnode{.1}{r}\rput(0,.3){} \\
  \cnode*[fillcolor=black]{.1}{ll}\rput(-.3,0){$v_1$}\rput(0,-.3){} &
  \psset{fillstyle=crosshatch}\cnode{.1}{lc}\rput(-.3,0){$v_2$}\rput(0,-.3){} &
  \cnode{.1}{lr}\rput(-.2,-.3){} &  \cnode{.1}{cl}\rput(0,-.3){}
  & &
  \cnode*[fillcolor=black]{.1}{cr}\rput(0,-.3){}\rput(-.3,0){$v_5$} & \psset{fillstyle=vlines}
  \cnode{.1}{rl}\rput(-.2,-.3){}\rput(-.3,0){$v_6$}
  &
  \psset{fillstyle=hlines}\cnode{.1}{rr}\rput(0,-.3){}\rput(-.3,0){$v_7$}&&&
     \\
  && \cnode{.1}{lrc}\rput(0,-.3){} & \cnode*[fillcolor=black]{.1}{cll}\rput(0,-.3){}\rput(-.3,0){$v_3$}
  & \psset{fillstyle=crosshatch}\cnode{.1}{clr}\rput(0,-.3){}\rput(-.3,0){$v_4$}&&
  \cnode{.1}{rlc}\rput(0,-.3){}\\
  &&&&\rput(0,0){(a)} \\
  \ncline{root}{l}\ncline{root}{c}\ncline{root}{r}
  \ncline{l}{ll}\ncline{l}{lc}\ncline{l}{lr}
  \ncline{c}{cr}\ncline{c}{cl}
  \ncline{r}{rl}\ncline{r}{rr}
  \ncline{lr}{lcr}\ncline{cl}{cll}\ncline{cl}{clr}
  \ncline{rl}{rlc}\ncline{lr}{lrc}
  \cnode*[fillcolor=black]{.1}{1}\rput(.5,0){=$S_1$} &&
  \psset{fillstyle=crosshatch}\cnode{.1}{2}\rput(.5,0){=$S_2$} &&
  \psset{fillstyle=vlines}\cnode{.1}{3}\rput(.5,0){=$S_3$} &&
  \psset{fillstyle=hlines}\cnode{.1}{4}\rput(.5,0){=$S_4$} &&
  \end{psmatrix}
  \begin{psmatrix}[colsep=0.6cm,rowsep=0.3cm,labelsep=1pt]
  &&&& \cnode{.1}{root}\rput(0,.3){}
  \\
  & \cnode{.1}{l}\rput(0,.3){} &&&  \cnode{.1}{c}\rput(-.3,.3){}
  &&
  \cnode{.1}{r}\rput(0,.3){} \\
  \cnode*[fillcolor=black]{.1}{ll}\rput(-.3,0){$v_1$}\rput(0,-.3){} &
  \psset{fillstyle=crosshatch}\cnode{.1}{lc}\rput(-.3,0){$v_2$}\rput(0,-.3){} &
  \cnode{.1}{lr}\rput(-.2,-.3){} &  \psset{fillstyle=vlines}\cnode{.1}{cl}\rput(0,-.3){}\rput(-.3,0){$v_5$}
  &&
  \cnode{.1}{cr}\rput(0,-.3){} & \psset{fillstyle=vlines}
  \cnode{.1}{rl}\rput(-.2,-.3){}\rput(-.3,0){$v_8$}
  &
  \psset{fillstyle=hlines}\cnode{.1}{rr}\rput(0,-.3){}\rput(-.3,0){$v_9$}
     \\
  && \cnode*[fillcolor=black]{.1}{lrc}\rput(0,-.3){}\rput(-.3,0){$v_3$}
  & \psset{fillstyle=crosshatch}\cnode{.1}{cll}\rput(0,-.3){}\rput(-.3,0){$v_4$}
  & \cnode*[fillcolor=black]{.1}{clr}\rput(0,-.3){}\rput(-.3,0){$v_6$}&&
  \psset{fillstyle=hlines}\cnode{.1}{rlc}\rput(0,-.3){}\rput(-.3,0){$v_7$}\\
  &&&&\rput(0,0){(b)} \\
  \ncline{root}{l}\ncline{root}{c}\ncline{root}{r}
  \ncline{l}{ll}\ncline{l}{lc}\ncline{l}{lr}
  \ncline{c}{cr}\ncline{c}{cl}
  \ncline{r}{rl}\ncline{r}{rr}
  \ncline{lr}{lcr}\ncline{cl}{cll}\ncline{cl}{clr}
  \ncline{rl}{rlc}\ncline{lr}{lrc}
  \end{psmatrix}
     \caption{In (a) we have
      $\{(v_1,v_2,v_3,v_6,v_7),(v_1,v_2,v_5,v_6,v_7),
      (v_1,v_4,v_5,v_6,v_7),(v_3,v_4,v_5,v_6,v_7)\}=\Phi(S_1,S_2,S_1,S_3,S_4)$ and
      thus
      $\mop(S_1,S_2,S_1,S_3,S_4)=\{(v_3,v_7)\}$. In  (b) we have
      $\Phi(S_1,S_2,S_1,S_3,S_4)=\{(v_1,v_2,v_3,v_5,v_7),(v_1,v_2,v_6,v_8,v_9),
      (v_1,v_2,v_3,v_8,v_9),(v_1,v_2,v_3,v_5,v_9),(v_1,v_4,v_6,v_8,v_9),
      (v_3,v_4,v_6,v_8,v_9) \}$ and thus
      $\mop(S_1,S_2,S_1,S_3,S_4)=\{(v_1,v_7),(v_3,v_9)\}$.}
  \label{t2:fig:mopexample}
\end{center}
\end{figure}

\paragraph{Minimum Ordered Pairs}
A set of nodes $X \subseteq V(T)$ is \emph{deep} if no node in $X$
is a proper ancestor of another node in $X$. For $k$ deep sets of
nodes $X_1, \ldots, X_k$ let $\Phi(X_1,\ldots,X_k) \subseteq (X_1
\times \cdots \times X_k)$, be the set of tuples such that $(x_1,
\ldots, x_k) \in \Phi(X_1,\ldots,X_k)$ iff $x_1 \lhd \cdots \lhd
x_k$. If $(x_1, \ldots, x_k) \in \Phi(X_1,\ldots,X_k) $ and there
is no $(x_1', \ldots, x_k') \in \Phi(X_1,\ldots,X_k) $, where
either $x_1 \lhd x_1' \lhd x_k' \unlhd x_k$ or $x_1 \unlhd x_1'
\lhd x_k' \lhd x_k$ then the pair $(x_1, x_k)$ is a \emph{minimum
ordered pair}. The set of minimum ordered pairs for $X_1, \ldots,
X_k$ is denoted by $\mop(X_1, \ldots, X_k)$.
Figure~\ref{t2:fig:mopexample} illustrates these concepts on a small
example.
For any set of pairs $Y$, let $\restrict{Y}{1}$ and
$\restrict{Y}{2}$ denote the \emph{projection} of $Y$ to the first
and second coordinate, that is, if $(y_1,y_2) \in Y$ then $y_1 \in
\restrict{Y}{1}$ and $y_2 \in \restrict{Y}{2}$. We say that $Y$ is
\emph{deep} if $\restrict{Y}{1}$ and $\restrict{Y}{2}$ are deep.
The following lemma shows that given deep sets $X_1, \ldots, X_k$
we can compute $\mop(X_1,\ldots,X_k)$ iteratively by first
computing $\mop(X_1,X_2)$ and then
$\mop(\restrict{\mop(X_1,X_2)}{2},X_3)$ and so on.

\begin{lemma}\label{t2:lem:nnsmaller2}
  For any deep sets of nodes $X_1, \ldots, X_k$ we have,
  $(x_1,x_k) \in \mop(X_1, \ldots, X_k)$ iff there exists a
  $x_{k-1}$  such that $(x_1,x_{k-1}) \in \mop(X_1, \ldots,
  X_{k-1})$ and $(x_{k-1},x_k) \in \mop(\restrict{\mop(X_1,
  \ldots, X_{k-1})}{2}, X_k)$.
\end{lemma}

\begin{proof}
  We start by showing that if $(x_1,x_k) \in \mop(X_1, \ldots,
  X_k)$ then there exists  a node $x_{k-1}$  such that $(x_1,x_{k-1}) \in
  \mop(X_1, \ldots, X_{k-1})$ and $(x_{k-1},x_k) \in
  \mop(\restrict{\mop(X_1, \ldots, X_{k-1})}{2}, X_k)$.

  First note that $(z_1,\ldots,z_k) \in
  \Phi(X_1,\ldots,X_k)$ implies $(z_1,\ldots,z_{k-1}) \in
  \Phi(X_1,\ldots,X_{k-1})$. Since $(x_1,x_k) \in
  \mop(X_1,\ldots,X_k)$ there must be a minimum $x_{k-1}$ such that
  the tuple $(x_1,\ldots,x_{k-1})$ is in $\Phi(X_1,\ldots,X_{k-1})$.
  We have $(x_1, x_{k-1}) \in \mop(X_1,\ldots,X_{k-1})$. We need to
  show $(x_{k-1},x_k) \in \mop(\restrict{\mop(X_1, \ldots, X_{k-1})}{2}, X_k)$.
  Since $(x_1,x_k) \in \mop(X_1, \ldots,
  X_k)$ there exists no $z \in X_k$ such that  $x_{k-1} \lhd z
  \lhd x_k$. Assume there exists a $x \in \restrict{\mop(X_1, \ldots,
  X_{k-1})}{2}$ such that  $x_{k-1} \lhd z
  \lhd x_k$. Since $(x,x_{k-1})\in \mop(X_1,\ldots,X_{k-1})$ this
  implies that there is a $z' \rhd x_1$ such that $(z',z) \in
  \mop(X_1,\ldots,X_{k-1})$. But this implies that the tuple
  $(z',\ldots,z,x_k)$ is in $\Phi(X_1,\ldots,X_k)$ contradicting that
  $(x_1,x_k) \in \mop(X_1,\ldots,X_k)$.

  We will now show that if there exists a $x_{k-1}$  such that $(x_1,x_{k-1}) \in
  \mop(X_1, \ldots, X_{k-1})$ and $(x_{k-1},x_k) \in
  \mop(\restrict{\mop(X_1, \ldots, X_{k-1})}{2}, X_k)$
  then the pair $(x_1,x_k) \in \mop(X_1, \ldots, X_k)$.
  Clearly, there exists a tuple $(x_1,\ldots,x_{k-1},x_k)\in \Phi(X_1,\ldots,X_k)$.
  Assume that there exists a tuple $(z_1,\ldots,z_k)\in
  \Phi(X_1,\ldots,X_k)$ such that $x_1 \lhd z_1 \lhd z_k \unlhd
  x_k$. Since $z_{k-1} \unlhd x_{k-1}$ this contradicts that
  $(x_1,x_{k-1}) \in \mop(X_1,\ldots,X_{k-1})$.
  Assume that there exists a tuple $(z_1,\ldots,z_k)\in
  \Phi(X_1,\ldots,X_k)$ such that $x_1 \unlhd z_1 \lhd z_k \lhd
  x_k$. Since $(x_1,x_{k-1})\in \mop(X_1,\ldots,X_{k-1})$ we have
  $x_{k-1} \unlhd z_{k-1}$ and thus $z_k \rhd x_{k-1}$ contradicting
  $(x_{k-1},x_k) \in \mop(\restrict{\mop(X_1, \ldots, X_{k-1})}{2}, X_k)$.
\qed \end{proof}

When we want to specify which tree we mean in the above relations we add a subscript. For instance, $v \prec_{T} w$ indicates that $v$ is an ancestor of $w$ \emph{in $T$}.

\section{Computing Deep Embeddings}\label{t2:sec:recursion}
In this section we present a general framework for answering tree
inclusion queries. As in \cite{KM1995} we solve the equivalent
\emph{tree embedding problem}. Let $P$ and $T$ be rooted labeled
trees. An \emph{embedding} of $P$ in $T$ is an injective function
$f : V(P) \rightarrow V(T)$ such that for all nodes $v,u \in
V(P)$,
\begin{itemize}
  \item[(i)] $\lab(v) = \lab(f(v))$. (label preservation condition)
  \item[(ii)] $v \prec u$  iff $f(v) \prec f(u)$. (ancestor condition)
  \item[(iii)] $v \lhd u$ iff $f(v) \lhd f(u)$. (order condition)
\end{itemize}
An example of an embedding is given in
Figure~\ref{t2:inclusionexample}(c).
\begin{lemma}[Kilpel\"{a}inen and Mannila~\cite{KM1995}]
For any trees $P$ and $T$, $P \sqsubseteq T$ iff there exists an
embedding of $P$ in $T$.
\end{lemma}
We say that the embedding $f$ is \emph{deep} if there is no
embedding $g$ such that $f(\roots(P)) \prec g(\roots(P))$. The
\emph{deep occurrences} of $P$ in $T$, denoted $\emb(P, T)$ is the
set of nodes,
\begin{equation*}
\emb(P,T) = \{f(\roots(P)) \mid \text{$f$ is a deep embedding of $P$ in $T$}\}.
\end{equation*}
By definition the set of ancestors of nodes in $\emb(P,T)$ is exactly the set of nodes $\{u \mid P \sqsubseteq T(u)\}$. Hence, to solve the tree inclusion problem it is sufficient to compute $\emb(P,T)$ and then, using additional $O(n_T)$ time, report all ancestors of this set. Note that the set $\emb(P,T)$ is deep.

\begin{figure}
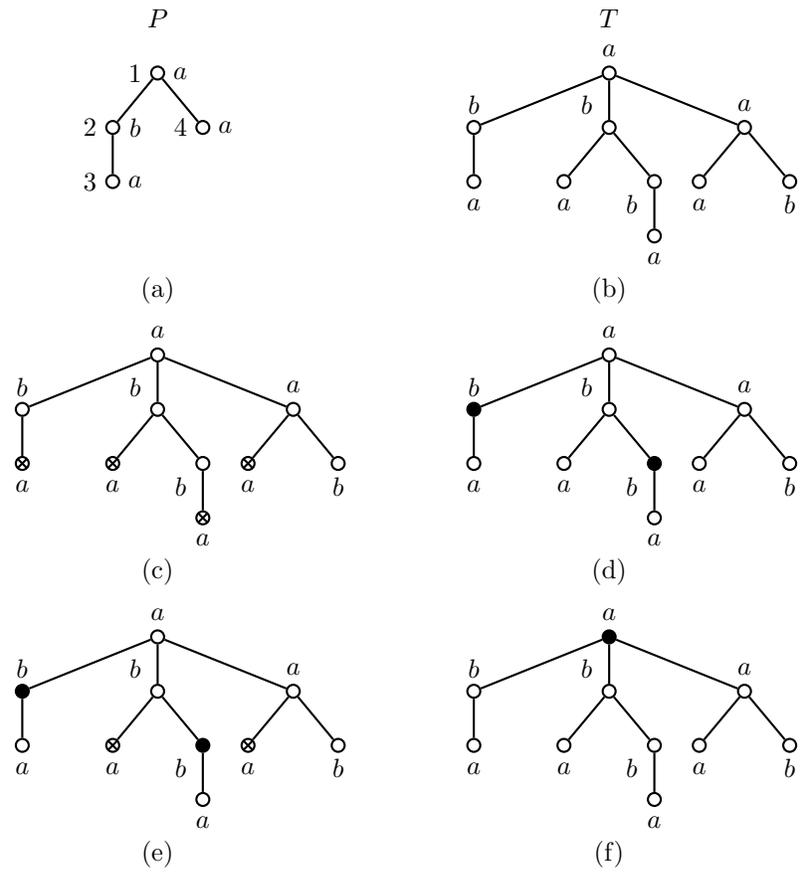

\begin{center}
  \begin{psmatrix}[colsep=0.6cm,rowsep=0.3cm,labelsep=1pt]
  &&&\rput(0,0){$P$} &&&&&&&&&& \rput(0,0){$T$} \\
  &&& \cnode{.1}{root1}\rput(.3,0){$a$}\rput(-.3,0){$1$} &&&&&&&&&&
  \cnode{.1}{root2}\rput(0,.3){$a$}\\
  && \cnode{.1}{l1}\rput(.3,0){$b$}\rput(-.3,0){$2$} &&  \cnode{.1}{r1}\rput(.3,0){$a$}\rput(-.3,0){$4$} &&&&&&
  \cnode{.1}{l2}\rput(0,.3){$b$} &&&  \cnode{.1}{c2}\rput(-.3,.3){$b$} &&&
  \cnode{.1}{r2}\rput(0,.3){$a$} \\
  && \cnode{.1}{ll1}\rput(.3,0){$a$}\rput(-.3,0){$3$} && &&&&&&
  \cnode{.1}{ll2}\rput(0,-.3){$a$} && \cnode{.1}{lc2}\rput(0,-.3){$a$} &&
  \cnode{.1}{rc2}\rput(-.3,-.3){$b$} & \cnode{.1}{lr2}\rput(0,-.3){$a$} &&
  \cnode{.1}{rr2}\rput(0,-.3){$b$} \\
  &&&&&&&&&&&&&&\cnode{.1}{e2}\rput(0,-.3){$a$}\\
  &&&\rput(0,0){(a)} &&&&&&&&&& \rput(0,0){(b)} \\
  \ncline{root1}{l1}\ncline{root1}{r1}
  \ncline{l1}{ll1}
  \ncline{root2}{l2}\ncline{root2}{c2}\ncline{root2}{r2}
  \ncline{l2}{ll2}\ncline{c2}{rc2}\ncline{c2}{lc2}
  \ncline{r2}{lr2}\ncline{r2}{rr2}
  \ncline{rc2}{e2}
  \end{psmatrix}
  \begin{psmatrix}[colsep=0.6cm,rowsep=0.3cm,labelsep=1pt]
  &&& \cnode{.1}{root1}\rput(0,.3){$a$} &&&&&&&&&&
  \cnode{.1}{root2}\rput(0,.3){$a$}\\
  \cnode{.1}{l1}\rput(0,.3){$b$} &&&  \cnode{.1}{c1}\rput(-.3,.3){$b$} &&&
  \cnode{.1}{r1}\rput(0,.3){$a$} &&&&
  \cnode*[fillcolor=black]{.1}{l2}\rput(0,.3){$b$} &&&  \cnode{.1}{c2}\rput(-.3,.3){$b$} &&&
  \cnode{.1}{r2}\rput(0,.3){$a$} \\
  \psset{fillstyle=crosshatch}\cnode{.1}{ll1}\rput(0,-.3){$a$} && \psset{fillstyle=crosshatch}\cnode{.1}{lc1}\rput(0,-.3){$a$} &&
  \cnode{.1}{rc1}\rput(-.3,-.3){$b$} & \psset{fillstyle=crosshatch}\cnode{.1}{lr1}\rput(0,-.3){$a$} &&
  \cnode{.1}{rr1}\rput(0,-.3){$b$} &&&
  \cnode{.1}{ll2}\rput(0,-.3){$a$} && \cnode{.1}{lc2}\rput(0,-.3){$a$} &&
  \cnode*[fillcolor=black]{.1}{rc2}\rput(-.3,-.3){$b$} & \cnode{.1}{lr2}\rput(0,-.3){$a$} &&
  \cnode{.1}{rr2}\rput(0,-.3){$b$} \\
  &&&& \psset{fillstyle=crosshatch}\cnode{.1}{e1}\rput(0,-.3){$a$} &&&&&&&&&&
  \cnode{.1}{e2}\rput(0,-.3){$a$}\\
  &&&\rput(0,0){(c)} &&&&&&&&&& \rput(0,0){(d)} \\
  \ncline{root1}{l1}\ncline{root1}{c1}\ncline{root1}{r1}
  \ncline{l1}{ll1}\ncline{c1}{rc1}\ncline{c1}{lc1}
  \ncline{r1}{lr1}\ncline{r1}{rr1}
  \ncline{rc1}{e1}
  \ncline{root2}{l2}\ncline{root2}{c2}\ncline{root2}{r2}
  \ncline{l2}{ll2}\ncline{c2}{rc2}\ncline{c2}{lc2}
  \ncline{r2}{lr2}\ncline{r2}{rr2}
  \ncline{rc2}{e2}
  \end{psmatrix}
  \begin{psmatrix}[colsep=0.6cm,rowsep=0.3cm,labelsep=1pt]
  &&& \cnode{.1}{root1}\rput(0,.3){$a$} &&&&&&&&&&
  \cnode*[fillcolor=black]{.1}{root2}\rput(0,.3){$a$}\\
  \cnode*[fillcolor=black]{.1}{l1}\rput(0,.3){$b$} &&&  \cnode{.1}{c1}\rput(-.3,.3){$b$} &&&
  \cnode{.1}{r1}\rput(0,.3){$a$} &&&&
  \cnode{.1}{l2}\rput(0,.3){$b$} &&&  \cnode{.1}{c2}\rput(-.3,.3){$b$} &&&
  \cnode{.1}{r2}\rput(0,.3){$a$} \\

  \cnode{.1}{ll1}\rput(0,-.3){$a$} &&\psset{fillstyle=crosshatch}\cnode{.1}{lc1}\rput(0,-.3){$a$} &&
  \cnode*[fillcolor=black]{.1}{rc1}\rput(-.3,-.3){$b$} & \psset{fillstyle=crosshatch}\cnode{.1}{lr1}\rput(0,-.3){$a$} &&
  \cnode{.1}{rr1}\rput(0,-.3){$b$} &&&
  \cnode{.1}{ll2}\rput(0,-.3){$a$} && \cnode{.1}{lc2}\rput(0,-.3){$a$} &&
  \cnode{.1}{rc2}\rput(-.3,-.3){$b$} & \cnode{.1}{lr2}\rput(0,-.3){$a$} &&
  \cnode{.1}{rr2}\rput(0,-.3){$b$} \\
  &&&& \cnode{.1}{e1}\rput(0,-.3){$a$} &&&&&&&&&&
  \cnode{.1}{e2}\rput(0,-.3){$a$}\\
  &&&\rput(0,0){(e)} &&&&&&&&&& \rput(0,0){(f)} \\
  \ncline{root1}{l1}\ncline{root1}{c1}\ncline{root1}{r1}
  \ncline{l1}{ll1}\ncline{c1}{rc1}\ncline{c1}{lc1}
  \ncline{r1}{lr1}\ncline{r1}{rr1}
  \ncline{rc1}{e1}
  \ncline{root2}{l2}\ncline{root2}{c2}\ncline{root2}{r2}
  \ncline{l2}{ll2}\ncline{c2}{rc2}\ncline{c2}{lc2}
  \ncline{r2}{lr2}\ncline{r2}{rr2}
  \ncline{rc2}{e2}
  \end{psmatrix}
   \caption{Computing the deep occurrences of $P$ into $T$ depicted in (a) and (b)
   respectively. The nodes in $P$ are numbered $1$--$4$ for easy reference.
   (c) Case 1 of $\Emb$: The set $\Emb(3)$.
   Since $3$ and $4$ are leaves and $\lab(3) = \lab(4)$ we have $\Emb(3) = \Emb(4)$. (d)
   Case 2 of $\Emb$. The set $\Emb(2)$.
   Note that the middle child of the root of $T$ is not in the set since it is not a
   deep occurrence. (e) Case 3 of $\Emb$: The two minimal ordered pairs of
   (d) and (c). (f) The nearest common ancestors of both pairs in (e) give
   the root node of $T$ which is the only (deep) occurrence of $P$.}
  \label{t2:fig:embexample}
\end{center}
\end{figure}

In the following we show how to compute deep embeddings. The key
idea is to build a data structure for $T$ allowing a fast
implementation of the following procedures. For all $X \subseteq
V(T)$, $Y \subseteq V(T)\times V(T)$, and $\alpha \in \Sigma$
define:
\begin{relate}
    \item[$\Parent(X)$:] Return the set
    $\{\parent(x) \mid x \in X\}$.
    \item[$\Nca(Y)$:]
    Return the set $\{\nca(y_1,y_2) \mid (y_1,y_2) \in Y\}$.
    \item[$\Deep(X)$:] Return the set $\{x \in X\mid \text{there is no }
    z \in X \text{ such that } x \prec z\}$.
    \item[$\Mop(Y,X)$:]
    Return the set of pairs $R$ such that for any pair $(y_1,y_2)
    \in Y$, $(y_1,x) \in R$ iff $(y_2, x) \in
    \mop(\restrict{Y}{2}, X)$.
    \item[$\Fl(X, \alpha)$:]
    Return the set $\{\fl(x, \alpha) \mid x \in X\}$.
\end{relate}
Collectively we call these procedures the \emph{set procedures}.
The procedures $\Parent$, $\Nca$, and $\Fl$ are selfexplanatory.
$\Deep(X)$ returns the set of all nodes in $X$ that have no
descendants in $X$. Hence,  the returned set is always deep.
$\Mop$ is used to iteratively compute minimum ordered pairs. If we
want to specify that a procedure applies to a certain tree $T$ we
add the subscript $T$. With the set procedures we can compute deep
embeddings. The following procedure $\Emb(v)$, $v \in V(P)$,
recursively computes the set of deep occurrences of $P(v)$ in $T$.
Figure~\ref{t2:fig:embexample} illustrates how $\Emb$ works on a
small example.
\begin{relate}
\item[$\Emb(v)$:] Let $v_1, \ldots, v_k$ be the sequence of
children of $v$ ordered from left to right. There are three cases:
    \begin{enumerate}
    \item $k=0$ ($v$ is a leaf). Compute $R := \Deep(\Fl(L(T),\lab(v)))$.
    \item $k=1$. Recursively compute $R_1 := \Emb(v_1)$. 
    
    Compute $R := \Deep(\Fl(\Deep(\Parent(R_1)),\lab(v)))$.
    \item $k > 1$. Compute $R_1 := \Emb(v_1)$ and set $U_1 := \{(r,r) \mid r \in R_1\}$. 
    
    For $i := 2$ to $k$, compute $R_i := \Emb(v_i)$ and $U_i := \Mop(U_{i-1}, R_i)$.  
    
    Finally, compute $R := \Deep(\Fl(\Deep(\Nca(U_k)), \lab(v)))$.
    \end{enumerate}
If $R = \emptyset$ stop and report that there is no deep embedding
of $P(v)$ in $T$. Otherwise return $R$.
\end{relate}
\begin{lemma}
For trees $P$ and $T$ and node $v\in V(P)$, $\Emb(v)$ computes the set of
deep occurrences of $P(v)$ in $T$.
\end{lemma}
\begin{proof}
By induction on the size of the subtree $P(v)$. If $v$ is a leaf
we immediately have that $\emb(v,T) = \Deep(\Fl(L(T),
\lab(v)))$ and thus case 1 follows. Suppose that $v$ is an
internal node with $k\geq1$ children $v_1, \ldots, v_k$. We show
that $\emb(P(v),T) = \Emb(v)$. Consider cases 2 and 3 of the
algorithm.

If $k=1$ we have that $w\in \Emb(v)$ implies that $\lab(w) =
\lab(v)$ and there is a node $w_1 \in \Emb(v_1)$ such that
$\fl(\parent(w_1), \lab(v)) = w$, that is, no node on the path
between $w_1$ and $w$ is labeled $\lab(v)$.  By induction
$\Emb(v_1) = \emb(P(v_1),T)$ and therefore $w$ is the root of an
embedding of $P(v)$ in $T$. Since $\Emb(v)$ is the deep set of
all such nodes it follows that $w \in \emb(P(v),T)$. Conversely,
if $w \in \emb(P(v),T)$ then $\lab(w) = \lab(v)$, there is a node
$w_1 \in \emb(P(v_1),T)$ such that $w \prec w_1$, and no node on
the path between $w$ and $w_1$ is labeled $\lab(v)$, that is,
$\fl(w_1, \lab(v)) = w$. Hence, $w \in \Emb(v)$.

Before considering case 3 we first show that $U_j =
\mop(\Emb(v_1), \ldots, \Emb(v_j))$ by induction on $j$, $2
\leq j \leq k$. For $j=2$ it follows from the definition of
$\Mop$ that $U_2 = \mop(\Emb(v_1), \Emb(v_2))$. Hence,
assume that $j > 2$. We have $U_j = \Mop(U_{j-1}, \Emb(v_j)) =
\Mop(\mop(\Emb(v_1), \ldots, \Emb(v_{j-1})), R_j)$. By
definition of $\Mop$, $U_j$ is the set of pairs such that for
any pair  $(r_1, r_{j-1}) \in \mop(\Emb(v_1), \ldots,
\Emb(v_{j-1}))$, $(r_1, r_j) \in U_j$ iff $(r_{j-1}, r_j) \in
\mop(\restrict{\mop(\Emb(v_1), \ldots, \Emb(v_{j-1}))}{2},
R_j)$. By Lemma~\ref{t2:lem:nnsmaller2} it follows that $(r_1, r_j)
\in U_j$ iff $(r_1, r_j) \in \mop(\Emb(v_1), \ldots,
\Emb(v_j))$.

Next consider the case when $k>1$. If $w \in \Emb(v)$ we have
that $\lab(w) = \lab(v)$ and there are nodes $(w_1, w_k) \in
\mop(\emb(P(v_1),T), \ldots, \emb(P(v_k),T))$ such that
$w=\fl(\nca(w_1, w_k), \lab(v))$. Clearly, $w$ is the root of an
embedding of $P(v)$ in $T$. Assume for contradiction that $w$ is
not a deep embedding, that is, $w \prec u$ for some node $u \in
\emb(P(v),T)$. Since $w=\fl(\nca(w_1, w_k), \lab(v))$ there must
be nodes $u_1 \lhd \cdots \lhd u_k$, such that $u_i \in
\emb(P(v_i),T)$ and $u=\fl(\nca(u_1, u_k), \lab(v))$. However,
this contradicts the fact that $(w_1, w_k) \in
\mop(\emb(P(v_1),T), \ldots, \emb(P(v_k),T))$. If $w \in
\emb(P(v),T)$ a similar argument implies that $w\in \Emb(v)$.
\qed \end{proof}
The set $L(T)$ is deep and in all tree cases of $\Emb(V)$ the returned set is also deep. By induction it follows that the input to $\Parent$, $\Fl$, $\Nca$, and $\Mop$ is always deep. We will use this fact to our advantage in the following algorithms.

\section{A Simple Tree Inclusion Algorithm}\label{t2:simple}
In this section we a present a simple implementation of the set
procedures which leads to an efficient tree inclusion algorithm.
Subsequently, we modify one of the procedures to obtain a family
of tree inclusion algorithms where the complexities depend on the
solution to a well-studied problem known as the \emph{tree color
problem}.

\subsection{Preprocessing}\label{t2:sec:simplepreprocessing}
To compute deep embeddings we require a data structure
for $T$ which allows us, for any $v,w \in V(T)$, to compute
$\nca_T(v,w)$ and determine if $v \prec w$ or $v \lhd w$. In
linear time we can compute $\pre(v)$ and $\post(v)$ for all nodes
$v \in V(T)$, and with these it is straightforward to test the two
conditions. Furthermore,
\begin{lemma}[Harel and Tarjan~\cite{HT1984}]
For any tree $T$ there is a data structure using $O(n_T)$ space
and preprocessing time which supports nearest common ancestor
queries in $O(1)$ time.
\end{lemma}
Hence, our data structure uses linear preprocessing time and space (see also~\cite{BFC2000,AGKR2004} for more recent nearest common ancestor data structures). 

\subsection{Implementation of the Set Procedures}\label{t2:implementationsimple}
To answer tree inclusion queries we give an efficient
implementation of the set procedures. The idea is to represent
sets of nodes and sets of pairs of nodes in a left-to-right order
using linked lists. For this purpose we introduce some helpful
notation. Let $X = [x_1, \ldots, x_k]$ be a linked list of nodes.
The \emph{length} of $X$, denoted $|X|$, is the number of elements
in $X$ and the list with no elements is written $[]$. The $i$th
node of $X$, denoted $X[i]$, is $x_i$. Given any node $y$ the list
obtained by \emph{appending} $y$ to $X$, is the list $X \circ y =
[x_1, \ldots, x_k, y]$. If for all $i$, $1 \leq i \leq |X|-1$,
$X[i] \lhd X[i+1]$ then $X$ is \emph{ordered} and if $X[i] \unlhd
X[i+1]$ then $X$ is \emph{semiordered}. A list $Y = [(x_1, z_k),
\ldots, (x_k, z_k)]$ is a \emph{node pair list}. By analogy, we
define length, append, etc. for $Y$. For a pair $Y[i] = (x_i,
z_i)$ define $Y[i]_1 = x_i$ and $Y[i]_2 = z_i$. If the lists
$[Y[1]_1, \ldots, Y[k]_1]$ and $[Y[1]_2, \ldots, Y[k]_2]$ are both
ordered or semiordered then $Y$ is \emph{ordered} or
\emph{semiordered}, respectively.

The set procedures are implemented using node lists. All lists
used in the procedures are either ordered or semiordered. As noted
in Section~\ref{t2:sec:recursion} we may assume that the input to all
of the procedures, except $\Deep$, represent a deep set, that is,
the corresponding node list or node pair list is ordered. We
assume that the input list given to $\Deep$ is semiordered and the
output, of course, is ordered. Hence, the output of all the other
set procedures must be semiordered. In the following let $X$ be a
node list, $Y$ a node pair list, and $\alpha$ a character in
$\Sigma$. The detailed implementation of the set procedures is
given below. We show the correctness in
Section~\ref{t2:sec:simplecorrectness} and discuss the complexity in
Section~\ref{t2:sec:simplecomplexity}.
\begin{relate}[simple]
\item[$\Parent(X)$:] Return the list $[\parent(X[1]), \ldots,
\parent(X[\norm{X}])]$.

\item[$\Nca(Y)$:] Return the list $[\nca(Y[1]), \ldots, \nca(Y[\norm{Y}])]$.

\item[$\Deep(X)$:] Initially, set $x := X[1]$ and $R := []$.

For $i:=2$ to $|X|$ do:
\begin{itemize}
\item[] Compare $x$ and $X[i]$. There are three cases:
    \begin{enumerate}
    \item 
    $x \lhd X[i]$. Set $R:= R \circ x$ and $x:= X[i]$.

    \item 
    $x \prec X[i]$. Set $x := X[i]$.  

    \item $X[i] \prec x$. Do nothing.
    \end{enumerate}
\end{itemize}
Return $R \circ x$.
\end{relate}
The implementation of procedure \Deep\ takes advantage of the fact
that the input list is semiordered. In case 1 node $X[i]$ to the
right of our "potential output node" $x$. Since any node that is a
descendant of $x$ must be to the right of $X[i]$ it cannot not
appear later in the list $X$ than $X[i]$. We can thus safely add
$x$ to $R$ at this point. In case 2 node $x$ is an ancestor of
$X[i]$ and can thus not be in the output list. In case 3 node
$X[i]$ is an ancestor of $x$ and can thus not be in the output
list.
\begin{relate}[simple]
\item[$\Mop(Y,X)$:]
    Initially, set $R:=[]$.

    Find the smallest $j$ such that $Y[1]_2 \lhd X[j]$ and set
    $y:=Y[1]_1$, $x:= X[j]$, and $h:=j$. If no such $j$ exists stop.

    For $i :=2$ to $\norm{Y}$ do:
    \begin{itemize}
    \item[] Set $h:=h+1$ until $Y[i]_2 \lhd X[h]$ or $h>\norm{X}$.

    If $h> \norm{X}$
    stop and return $R := R \circ (y,x)$.
    Otherwise, compare $X[h]$ and $x$.
    There are two cases:
    \begin{enumerate}
    \item If $x \lhd X[h]$ set
    $R:=R\circ (y,x)$, $y:=Y[i]_1$, and $x:=X[h]$.
    \item If $x=X[h]$
    set $y:=Y[i]_1$.
    \end{enumerate}
    \end{itemize}
    Return $R := R \circ (y,x)$.
\end{relate}
In procedure \Mop\ we have a "potential pair" $(y,x)$ where
$y=Y[i]_1$ for some $i$ and $Y[i]_2 \lhd x$. Let $j$ be the index
such that $y=Y[j]_1$. In case 1 we have $x \lhd X[h]$ and also
$Y[j]_2 \lhd Y[i]_2$ since the input lists are ordered (see
Figure~\ref{t2:fig:mopimplexample}(a)). Therefore, $(y,x)$ is
inserted into $R$. In case 2 we have $x=X[h]$, i.e., $Y[i]_2\lhd
x$, and as before $Y[j]_2 \lhd Y[i]_2$ (see
Figure~\ref{t2:fig:mopimplexample}(b)). Therefore $(y,x)$ cannot be
in the output, and we set $(Y[i]_1,x)$ to be the new potential
pair.
\begin{figure}[t]
\begin{center}
\begin{psmatrix}[colsep=0.7cm,rowsep=0.3cm,labelsep=2pt]
  &&&& \cnode{.1}{root}\rput(0,.3){}
  \\
  & \cnode{.1}{l}\rput(0,.3){} &&&  \cnode{.1}{c}\rput(-.3,.3){}
  &&
  \cnode{.1}{r}\rput(0,.3){} \\
  \cnode{.1}{ll}&
  \cnode*{.1}{lc}\rput(0,-.3){$Y[j]_2$} &
  \cnode{.1}{lr}\rput(-.2,-.3){} &  \cnode*{.1}{cl}\rput(.6,0){$Y[i]_2$}
  & &
  \cnode{.1}{cr}& 
  \cnode{.1}{rl}
  &
  \psset{fillstyle=crosshatch}\cnode{.1}{rr}\rput(0,-.3){$X[h]$}&&&
     \\
  && \cnode{.1}{lrc}& \psset{fillstyle=crosshatch}\cnode{.1}{cll}\rput(0,-.3){$x$}
  & \cnode{.1}{clr}&&
  \cnode{.1}{rlc}\rput(0,-.3){}\\[.2cm]
  &&&&\rput(0,0){(a)}
  \ncline{root}{l}\ncline{root}{c}\ncline{root}{r}
  \ncline{l}{ll}\ncline{l}{lc}\ncline{l}{lr}
  \ncline{c}{cr}\ncline{c}{cl}
  \ncline{r}{rl}\ncline{r}{rr}
  \ncline{lr}{lcr}\ncline{cl}{cll}\ncline{cl}{clr}
  \ncline{rl}{rlc}\ncline{lr}{lrc}
  \end{psmatrix}
  \begin{psmatrix}[colsep=0.7cm,rowsep=0.3cm,labelsep=1pt]
  &&&& \cnode{.1}{root}
  \\
  & \cnode{.1}{l} &&&  \cnode{.1}{c}
  &&
  \cnode{.1}{r} \\
  \cnode{.1}{ll} &
  \cnode*{.1}{lc}\rput(0,-.3){$Y[j]_2$} &
  \cnode{.1}{lr} &  \cnode{.1}{cl}
  &&
  \cnode{.1}{cr} &
  \cnode{.1}{rl}
  &
  \cnode{.1}{rr}
     \\
  && \cnode{.1}{lrc}
  & \cnode{.1}{cll}
  & \cnode*{.1}{clr}\rput(0,-.3){$Y[i]_2$}&&
  \psset{fillstyle=crosshatch}\cnode{.1}{rlc}\rput(.3,-.3){$x=X[h]$}\\[.2cm]
  &&&&\rput(0,0){(b)}
  \ncline{root}{l}\ncline{root}{c}\ncline{root}{r}
  \ncline{l}{ll}\ncline{l}{lc}\ncline{l}{lr}
  \ncline{c}{cr}\ncline{c}{cl}
  \ncline{r}{rl}\ncline{r}{rr}
  \ncline{lr}{lcr}\ncline{cl}{cll}\ncline{cl}{clr}
  \ncline{rl}{rlc}\ncline{lr}{lrc}
  \end{psmatrix}
     \caption{Case 1 and 2 from the implementation of \Mop.
     In (a) we have $Y[i]_2 \ntriangleleft x$.
     In  (b) we have $Y[j]_2 \lhd Y[i]_2 \lhd x = X[h]$.
      }
  \label{t2:fig:mopimplexample}
\end{center}
\end{figure}
\begin{relate}[simple]
\item[$\Fl(X,\alpha)$:]
Initially, set $Z:=X$, $R:=[]$, and
    $S:=[]$.

    Repeat until $Z:=[]$:
    \begin{itemize}
    \item[] For $i:=1$ to $\norm{Z}$ do: If
    $\lab(Z[i])=\alpha$ set $R:=\Insertt(Z[i],R)$. Otherwise set $S:=S
    \circ \parent(Z[i])$.

    \item[] Set $S:=\Deep(S)$, $W:=\Deep^*(S,R)$, and $S:=[]$.
    \end{itemize}
    Return $R$.
\end{relate}
The procedure $\Fl$ calls two auxiliary procedures: $\Insertt(x,R)$
that takes an ordered list $R$ and insert the node $x$ such that
the resulting list is ordered, and $\Deep^*(S,R)$ that takes two
ordered lists and returns the ordered list representing the set
$\Deep(S \cup R) \cap S$, i.e., $\Deep^*(S,R)=[s\in S|\nexists
z\in R:s\prec z]$. Below we describe in more detail how to
implement \Fl\ together with the auxiliary procedures.

We use one doubly linked list to represent all the lists $Z$, $S$,
and $R$. For each element in $Z$ we have pointers \Pred\ and
\Successor\ pointing to the predecessor and successor in the list,
respectively. We also have  at each element a pointer \Next\
pointing to the next element in $Z$. In the beginning
$\Next=\Successor$ for all elements, since all elements in the list are
in $Z$. When going through $Z$ in one iteration we simple follow
the \Next\ pointers. When \Fl\ calls $\Insertt(Z[i],R)$ we set
$\Next(\Pred(Z[i]))$ to $\Next(Z[i])$. That is, all nodes in the
list not in $Z$, i.e., nodes not having a \Next\ pointer pointing
to them, are in $R$. We do not explicitly maintain $S$. Instead we
just set save $\Parent(Z[i])$ at the position in the list instead
of $Z[i]$. Now $\Deep(S)$ can be performed following the \Next\
pointers and removing elements from the doubly linked list
accordingly to procedure \Deep. It remains to show how to
calculate $\Deep^*(S,R)$. This can be done by running through $S$
following the \Next\ pointers. At each node $s$ compare $\Pred(s)$
and $\Successor(s)$ with $s$. If one of them is a descendant of $s$
remove $s$ from the doubly linked list.

Using this linked list implementation $\Deep^*(S,R)$ takes time
$O(\norm{S})$, whereas using \Deep\ to calculate this would have
used time $O(\norm{S}+\norm{R})$.

\subsection{Correctness of the Set Procedures}\label{t2:sec:simplecorrectness}
Clearly, \Parent\ and \Nca\ are correct. The following lemmas show
that $\Deep$, $\Fl$, and $\Mop$ are also correctly implemented.
For notational convenience we write $x\in X$, for a list $X$, if
$x = X[i]$ for some $i$, $1 \leq i \leq \norm{X}$.
\begin{lemma}\label{t2:lem:deep}
    Procedure $\Deep(X)$ is correct.
\end{lemma}

\begin{proof} Let $y$ be an element in $X$.
    We will first prove that if there are no descendants of $y$ in $X$, i.e.,
    $X \cap V(T(y))=\emptyset$, then $y\in R$. Since
    $X \cap V(T(y))=\emptyset$ we must at some point during the procedure have
    $x=y$, and $x$ will not change before $x$ is added to $R$. If
    $y$ occurs several times in $X$ we will have $x=y$ each time
    we meet a copy of $y$ (except the first) and it follows from
    the implementation that $y$ will occur exactly once in $R$.

    We will now prove that if there are any descendants of $y$ in $V$, i.e.,
    $X \cap V(T(y))\neq \emptyset$, then $y \not \in
    R$. Let $z$ be the rightmost and deepest descendant of $y$ in $V$. There
    are two cases:
    \begin{enumerate}
    \item $y$ is before $z$ in $X$.
        Look at the time in the execution of the procedure when we
        look at $z$. There are two cases.
        \begin{enumerate}
        \item $x=y$. Since $y \prec z$ we set $x=z$ and proceed. It
            follows that $y \not \in R$.

        \item $x = x' \neq y$. Since any node to the left of $y$ also is to
            the left of $z$ and $X$ is an semiordered list we must have
            $x' \in V(T(y))$ and thus $y \not \in R$.
        \end{enumerate}

    \item $y$ is after $z$ in $X$. Since $z$ is the rightmost and
        deepest descendant of  $y$ and $V$ is semiordered we must have
        $x=z$ at the time in the procedure where we look at $y$. Therefore
        $y \not \in R$.
    \end{enumerate}
    If $y$ occurs several times in $X$, each copy will be taken
    care of by either case 1 or 2.
\qed \end{proof}

\begin{lemma}\label{t2:lem:nnm}
    Procedure $\Mop(Y,X)$ is correct.
\end{lemma}
\begin{proof}
    We want to show that for any $1\leq l < \norm{Y}$, $1 \leq k <
    \norm{X}$ the pair
    $(Y[l]_1,X[k])$ is in  $R$ if and only if $(Y[l]_2,X[k])
    \in \mop(\restrict{Y}{2},X)$.
    Since $\restrict{Y}{2}$ and $X$ are ordered lists we have
    \begin{equation*}
    (Y[l]_2,X[k]) \in \mop(X|_2,X) \quad \Leftrightarrow \quad
    X[k-1] \unlhd Y[l]_2 \lhd X[k] \unlhd Y[l+1]_2 \;,
    \end{equation*}
    for $k \geq 2$, and
    $$(Y[l]_2,X[1]) \in \mop(X|_2,X) \quad \Leftrightarrow \quad
    Y[l]_2 \lhd X[1] \unlhd Y[l+1]_2 \;,$$
    when $k=1$.

    It follows immediately from the implementation of the procedure, that
    if $Y[j]_2 \lhd X[t]$, $X[t-1] \unlhd Y[j]_2$,
    and $Y[j+1]_2 \unrhd
    X[t]$ then $(Y[j]_1, X[t]) \in R$.

    We will now show that $(Y[l]_1,X[k]) \in R \Rightarrow (Y[l]_2,X[k]) \in
    \mop(\restrict{Y}{2},X)$. 
    That $(Y[l]_1,X[k]) \in R \Rightarrow
    X[k-1] \unlhd Y[l]_2 \lhd X[k]$ follows immediately from the
        implementation of the procedure by induction on $l$.








        It remains to show that $(Y[l]_1,X[k]) \in R
        \Rightarrow  X[k] \unlhd Y[l+1]_2$. Assume for the
        sake of contradiction that $Y[l+1]_2 \lhd X[k]$.
        Consider
        the iteration in the execution of the procedure
        when we look at $Y[l+1]_2$. We have
        $x=X[k]$ and thus set $y:=Y[l+1]_1$
        contradicting $(Y[l]_1,X[k]) \in R$.
\qed \end{proof}
To show that \Fl\ is correct we need the following proposition.
\begin{prop}\label{t2:prop:deeplist}
Let $X$ be an ordered list and let $x$ be an ancestor of $X[i]$
for some $i \in \{1,\ldots,k\}$. If $x$ is an ancestor of some
node in $X$ other than $X[i]$ then $x$ is an ancestor of $X[i-1]$
or $X[i+1]$.
\end{prop}
\begin{proof}
Assume for the sake of contradiction that $x \npreceq X[i-1]$, $x
\npreceq X[i+1]$, and $x \preceq z$, where $z \in X$ and $z \neq
X[i]$. Since $X$ is ordered either $z \lhd X[i-1]$ or $X[i+1] \lhd
z$. Assume $z \lhd X[i-1]$. Since $x \prec X[i]$, $x \npreceq
X[i-1]$, and $X[i-1]$ is to the left of $X[i]$, $X[i-1]$ is to the
left of $x$. Since $z \lhd X[i-1]$ and $X[i-1]\lhd x$ we have $z
\lhd x$ contradicting $x \prec z$. Assume $X[i+1] \lhd z$. Since
$x \prec X[i]$, $x \npreceq X[i+1]$, and $X[i+1]$ is to the right
of $X[i]$, $X[i+1]$ is to the right of $x$. Thus $x \lhd z$
contradicting $x \prec z$.
\qed \end{proof}
Proposition~\ref{t2:prop:deeplist} shows that the doubly linked list
implementation of $\Deep^*$ is correct. Clearly, \Insertt\ is
implemented correct by the doubly linked list representation,
since the nodes in the list remains in the same order throughout
the execution of the procedure.
\begin{lemma}\label{t2:lem:flcorrect}
    Procedure $\Fl(X,\alpha)$ is correct.
\end{lemma}
\begin{proof}
    Let $F=\{\fl(x,\alpha)\mid x \in X\}$.
    It follows immediately from the implementation of the
    procedure  that $\Fl(X,\alpha) \subseteq X$.
    It remains to show that $\Deep(F)\subseteq \Fl(X,\alpha)$.
    Let $x$ be a
    node in $\Deep(F))$, let $z \in X$ be the node such that
    $x=\fl(z,\alpha)$, and let $z=x_1,x_2,\ldots,x_k=x$
    be the nodes on the
    path from $z$ to $x$. In each iteration of the algorithm we have $x_i \in Z$
    for some $i$ unless $x \in R$.
\qed \end{proof}

\subsection{Complexity of the Set Procedures}\label{t2:sec:simplecomplexity}
For the running time of the node list implementation observe that,
given the data structure described in
Section~\ref{t2:sec:simplepreprocessing}, all set procedures, except
$\Fl$, perform a single pass over the input using constant time at
each step. Hence we have,
\begin{lemma}\label{t2:lem:auxprocedures}
For any tree $T$ there is a data structure using $O(n_T)$ space
and preprocessing which supports each of the procedures $\Parent$,
$\Deep$, $\Mop$, and $\Nca$ in linear time (in the size of their
input).
\end{lemma}
The running time of a single call to \Fl\ might take time
$O(n_T)$. Instead we will divide the calls to \Fl\ into groups and
analyze the total time used on such a group of calls. The
intuition behind the division is that for a path in $P$ the calls
made to \Fl\ by \Emb\ is done bottom up on disjoint lists of
nodes in $T$.

\begin{lemma}\label{t2:lem:fl}
For disjoint ordered node lists $X_1, \ldots, X_k$ and labels
$\alpha_1, \ldots, \alpha_k$, such that any node in $X_{i+1}$ is
an ancestor of some node in $\Deep(\Fl_T(X_i, \alpha_i))$, $2 \leq
i < k$, all of $\Fl_T(X_1, \alpha_1), \ldots , \Fl_T(X_k,
\alpha_k)$ can be computed in $O(n_T)$ time.
\end{lemma}
\begin{proof}
Let $Z$, $R$, and $S$ be as in the implementation of the
procedure. Since \Deep\ and $\Deep^*$ takes time $O(\norm{S})$, we
only need to show that the total length of the lists $S$---summed
over all the calls---is $O(n_T)$ to analyze the total time usage
of \Deep\ and $\Deep^*$. We note that in one iteration $\norm{S}
\leq \norm{Z}$. \Insertt\ takes constant time and it is thus enough
to show that any node in $T$ can be in $Z$ at most twice during
all calls to \Fl.

Consider a call to \Fl. Note that $Z$ is ordered at all times.
Except for the first iteration, a node can be in $Z$ only if one
of its children were in $Z$ in the last iteration. Thus in one
call to \Fl\ a node can be in $Z$ only once.

Look at a node $z$ the first time it appears in $Z$. Assume that
this is in the call $\Fl(X_i,\alpha_i)$. If $z \in X$ then $z$
cannot be in $Z$ in any later calls, since no node in $X_j$ where
$j>i$ can be a descendant of a node in $X_i$. If $z \not \in R$ in
this call then $z$ cannot be in $Z$ in any later calls. To see
this look at the time when $z$ removed from $Z$. Since the set $Z
\cup R$ is deep at all times no descendant of $z$ will appear in
$Z$ later in this call to \Fl, and no node in $R$ can be a
descendant of $z$. Since any node in $X_j$, $j>i$, is an ancestor
of some node in $\Deep(\Fl(X_i,\alpha_i))$ neither $z$ or any
descendant of $z$ can be in any $X_j$, $j>i$. Thus $z$ cannot
appear in $Z$ in any later calls to \Fl. Now if $z \in R$ then we
might have $z \in X_{i+1}$. In that case, $z$ will appear in $Z$
in the first iteration of the procedure call
$\Fl(X_{i+1},\alpha_i)$, but not in any later calls since the
lists are disjoint, and since no node in $X_j$ where $j>i+1$ can
be a descendant of a node in $X_{i+1}$. If $z \in R$ and $z \not
\in X_{i+1}$ then clearly $z$ cannot appear in $Z$ in any later
call. Thus a node in $T$ is in $Z$ at most twice during all the
calls.
\qed \end{proof}

\subsection{Complexity of the Tree Inclusion Algorithm}
Using the node list implementation of the set procedures we get:

\begin{theorem}\label{t2:thm:simple}
For trees $P$ and $T$ the tree inclusion problem can be solved in
$O(l_Pn_T)$ time and $O(n_T)$  space.
\end{theorem}
\begin{proof}
By Lemma~\ref{t2:lem:auxprocedures} we can preprocess $T$ in $O(n_T)$
time and space. Let $g(n)$ denote the time used by $\Fl$ on a
list of length $n$. Consider the time used by $\Emb(\roots(P))$.
We bound the contribution for each node $v\in V(P)$. From Lemma
\ref{t2:lem:auxprocedures} it follows that if $v$ is a leaf the cost
of $v$ is at most $O(g(l_T))$. Hence, by
Lemma~\ref{t2:lem:fl}, the total cost of all leaves is $O(l_Pg(l_T)) = O(l_P  n_T)$.
If $v$ has a single child $w$ the cost is $O(g(|\Emb(w)|))$. If
$v$ has more than one child the cost of $\Mop$, $\Nca$, and $\Deep$ is bounded by $\sum_{w \in \child(v)} O(|\Emb(w)|)$. Furthermore, since the length of the output of
$\Mop$ (and thus $\Nca$) is at most $z = \min_{w \in \child(v)}
|\Emb(w)|$ the cost of $\Fl$ is $O(g(z))$. Hence, the total cost
for internal nodes is,
\begin{equation}
\label{t2:eq:internal}  \sum_{v \in V(P)\backslash L(P)}
O\bigg(g(\min_{w \in \child(v)} |\Emb(w)|) + \sum_{w \in
\child(v)} |\Emb(w)|\bigg)\leq \sum_{v \in V(P)}
O(g(|\Emb(v)|)).
\end{equation}
Next we bound (\ref{t2:eq:internal}). For any $w \in \child(v)$ we have that $\Emb(w)$ and $\Emb(v)$ are disjoint ordered lists. Furthermore we have that any node in $\Emb(v)$ must be an ancestor of a node in $\Deep(\Fl(\Emb(w), \lab(v)))$. Hence, by Lemma~\ref{t2:lem:fl}, for any leaf to root path $\delta = v_1, \ldots, v_k$ in $P$, we have that $\sum_{u \in \delta} g(|\Emb(u)|) \leq O(n_T)$. Let
$\Delta$ denote the set of all root to leaf paths in $P$. It
follows that,
\begin{equation*}
\sum_{v \in V(T)} g(|\Emb(v)|) \leq \sum_{p \in \Delta} \sum_{u
\in p} g(|\Emb(u)|) \leq O(l_Pn_T).
\end{equation*}
Since this time dominates the time spent at the leaves the time
bound follows. Next consider the space used by
$\Emb(\roots(P))$. The preprocessing of
Section~\ref{t2:sec:simplepreprocessing} uses only $O(n_T)$ space.
Furthermore, by induction on the size of the subtree $P(v)$ it
follows immediately that at each step in the algorithm at most
$O(\max_{v\in V(P)}|\Emb(v)|)$ space is needed. Since
$\Emb(v)$ is a deep embedding, it follows that $|\Emb(v)| \leq
l_T$.
\qed \end{proof}

\subsection{An Alternative Algorithm}\label{t2:sec:alt}
In this section we present an alternative algorithm. Since the
time complexity of the algorithm in the previous section is
dominated by the time used by $\Fl$, we present an implementation
of this procedure which leads to a different complexity. Define a
\emph{firstlabel data structure} as a data structure supporting
queries of the form $\fl(v, \alpha)$, $v\in V(T)$, $\alpha \in
\Sigma$. Maintaining such a data structure is known as the
\emph{tree color problem}. This is a well-studied problem, see
e.g. \cite{Die89,MM1996, FM1996,AHR1998}. With such a data
structure available we can compute $\Fl$ as follows,
\begin{relate}
\item[$\Fl(X, \alpha)$:] Return the list $R := [\fl(X[1], \alpha),
\ldots, \fl(X[\norm{X}], \alpha)]$.
\end{relate}
\begin{theorem}\label{t2:thm:simple2}
Let $P$ and $T$ be trees. Given a firstlabel data structure using
$s(n_T)$ space, $p(n_T)$ preprocessing time, and $q(n_T)$ time for
queries, the tree inclusion problem can be solved in $O(p(n_T) +
n_Pl_T\cdot q(n_T))$ time and $O(s(n_T) + n_T)$ space.
\end{theorem}
\begin{proof}
Constructing the firstlabel data structures uses $O(s(n_T))$ and
$O(p(n_T))$ time. As in the proof of Theorem~\ref{t2:thm:simple} we
have that the total time used by $\Emb(\roots(P))$ is bounded by
$\sum_{v \in V(P)} g(|\Emb(v)|)$, where $g(n)$ is the time used
by $\Fl$ on a list of length $n$. Since $\Emb(v)$ is a deep
embedding and each $\fl$ takes $q(n_T)$ we have,
\begin{equation*}
\sum_{v \in V(P)} g(|\Emb(v)|) \leq \sum_{v \in V(P)} g(l_T) =
n_P l_T\cdot q(n_T).
\end{equation*}
\qed \end{proof}
Several firstlabel data structures are available, for instance, if we want to maintain linear space we have,
\begin{lemma}[Dietz~\cite{Die89}]\label{t2:lem:dietz}
For any tree $T$ there is a data structure using $O(n_T)$ space,
$O(n_T)$ expected preprocessing time which supports firstlabel
queries in $O(\log \log n_T)$ time.
\end{lemma}
The expectation in the preprocessing time is due to perfect hashing. Since our data structure does not need to support efficient updates we can remove the expectation by using the deterministic dictionary of Hagerup et. al. \cite{HMP2001}. This gives a worst-case preprocessing time of $O(n_T \log n_T)$, however, using a simple two-level approach this can be reduced to $O(n_T)$ (see e.g. \cite{Thorup2003}). Plugging in this data structure we obtain,
\begin{corollary}\label{t2:cor:simple}
For trees $P$ and $T$ the tree inclusion problem can be solved in
$O(n_Pl_T\log\log n_T + n_T)$ time and $O(n_T)$ space.
\end{corollary}

\section{A Faster Tree Inclusion Algorithm}\label{t2:micromacro}
In this section we present a new tree inclusion algorithm which
has a worst-case subquadratic running time. As discussed in the introduction the general idea is to divide $T$ into clusters of logarithmic size which we can efficiently preprocess and then use this to speedup the computation with a logarithmic factor.

\subsection{Clustering}
In this section we describe how to divide $T$ into clusters and
how the macro tree is created. For simplicity in the presentation
we assume that $T$ is a binary tree. If this is not the case it is
straightforward to construct a binary tree $B$, where $n_{B} \leq
2n_T$, and a mapping $g : V(T) \rightarrow V(B)$ such that for any
pair of nodes $v,w \in V(T)$, $\lab(v) = \lab(g(v))$, $v \prec w$
iff $g(v) \prec g(w)$, and $v \lhd w$ iff $g(v) \lhd g(w)$. If the
nodes in the set $U = V(B)\backslash \{g(v) \mid v \in V(T)\}$ is
assigned a special label $\beta \not\in \Sigma$ it follows that
for any tree $P$, $P \sqsubseteq T$ iff $P \sqsubseteq B$.

Let $C$ be a connected subgraph of $T$. A node in $V(C)$ incident
to a node in $V(T)\backslash V(C)$ is a \emph{boundary} node. The
boundary nodes of $C$ are denoted by $\delta C$. A \emph{cluster}
of $C$ is a connected subgraph of $C$ with at most two boundary
nodes. A set of clusters $CS$ is a \emph{cluster partition} of $T$
iff $V(T) = \cup_{C\in CS} V(C)$, $E(T) = \cup_{C\in CS} E(C)$,
and for any $C_1 ,C_2 \in CS$, $E(C_1) \cap E(C_2) = \emptyset$,
$|E(C_1)| \geq 1$, $\roots(T) \in \delta C$ if $\roots(T) \in
V(C)$. If $|\delta C| = 1$ we call $C$ a \emph{leaf cluster} and
otherwise an \emph{internal cluster}.

We use the following recursive procedure $\Cluster_T(v,s)$,
adopted from \cite{AR2002c}, which creates a cluster partition
$CS$ of the tree $T(v)$ with the property that $|CS| = O(s)$ and
$|V(C)| \leq \ceil{n_T/s}$. A similar cluster partitioning
achieving the same result follows from \cite{AHT2000, AHLT1997,
Frederickson1997}.
\begin{relate}
\item[$\Cluster_T(v,s)$:] For each child $u$ of $v$ there are two
cases:
\begin{enumerate}
  \item $|V(T(u))| + 1 \leq \ceil{n_T/s}$. Let the nodes $\{v\} \cup V(T(u))$ be a leaf cluster with boundary node $v$.
  \item $|V(T(u))| > \ceil{n_T/s}$. Pick a node $w \in V(T(u))$ of
  maximum depth such that $|V(T(u))| + 2 - |V(T(w))| \leq \ceil{n_T/s}$.
  Let the nodes $V(T(u)) \backslash V(T(w)) \cup \{v,w\}$ be an internal
  cluster with boundary nodes $v$ and $w$. Recursively, compute $\Cluster_T(w, s)$.
\end{enumerate}
\end{relate}

\begin{lemma}\label{t2:lem:clustering}
Given a tree $T$ with $n_T>1$ nodes, and a parameter $s$, where
$\ceil{n_T/s} \geq 2$, we can build a cluster partition $CS$ in
$O(n_T)$ time, such that $|CS| = O(s)$ and $|V(C)| \leq
\ceil{n_T/s}$ for any $C \in CS$.
\end{lemma}
\begin{proof}
The procedure $\Cluster_T(\roots(T), s)$ clearly creates a cluster
partition of $T$ and it is straightforward to implement in
$O(n_T)$ time. Consider the size of the clusters created. There
are two cases for $u$. In case $1$, $|V(T(u))| + 1 \leq
\ceil{n_T/s}$ and hence the cluster $C = \{v\} \cup V(T(u))$
has size $|V(C)| \leq \ceil{n_T/s}$. In case $2$, $|V(T(u))| + 2 -
|V(T(w))| \leq \ceil{n_T/s}$ and hence the cluster $C = V(T(u))
\backslash V(T(w)) \cup \{v,w\}$ has size $|V(C)| \leq
\ceil{n_T/s}$.

Next consider the size of the cluster partition.  Let $c=
\ceil{n_T/s}$. We say that a cluster $C$ is \emph{bad} if
$\norm{V(C)} \leq c/2$ and \emph{good} otherwise. We will show
that at least a constant fraction of the clusters in the cluster
partition are good. It is easy to verify that the cluster
partition created by procedure \Cluster\ has the following
properties:
\begin{itemize}
\item[(i)] Let $C$ be a bad internal cluster with boundary nodes
$v$ and $w$ ($v \prec
w$). Then $w$ has two children with at least $c/2$ descendants
each.
\item[(ii)] Let $C$ be a bad leaf cluster with boundary node
$v$. Then the boundary node $v$ is contained in  a good cluster.
\end{itemize}
By (ii) the number of bad leaf clusters is no larger than twice
the number of good internal clusters. By (i) each bad internal
cluster $C$ is sharing its lowest boundary node of $C$ with two
other clusters, and each of these two clusters are either internal
clusters or good leaf clusters. This together with (ii) shows that
number of bad clusters is at most a constant fraction of the total
number of clusters. Since a good cluster is of size more than
$c/2$, there can be at most $2s$ good clusters and thus
$\norm{CS}=O(s)$.
\qed \end{proof}

Let $C\in CS$ be an internal cluster $v, w \in \delta C$. The
\emph{spine path} of $C$ is the path between $v,w$
excluding $v$ and $w$. A node on the spine path is a \emph{spine
node}. A node to the left and right of $v$, $w$, or any node on
the spine path is a \emph{left node} and \emph{right node}, respectively.
If $C$ is a leaf cluster with $v \in \delta C$ then any proper
descendant of $v$ is a \emph{leaf node}.

\begin{figure}[t]
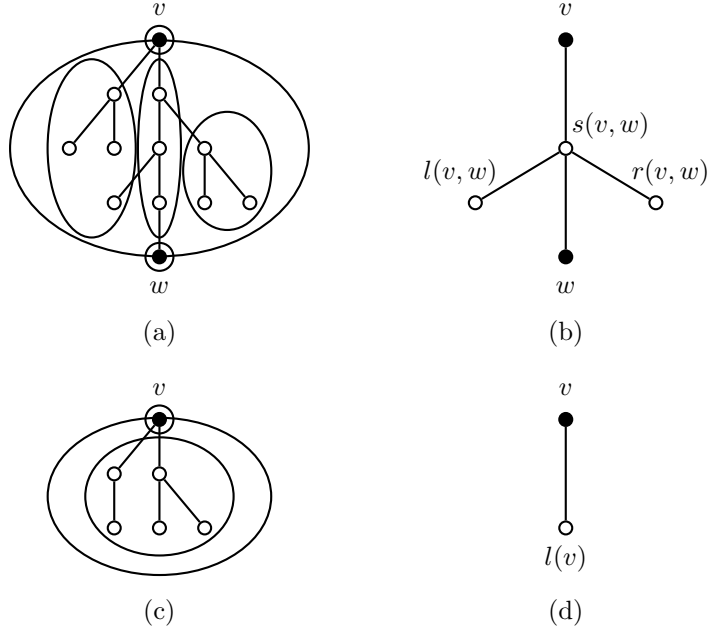

\begin{center}
  \begin{psmatrix}[colsep=0.6cm,rowsep=0.3cm,labelsep=1pt]
  &&& \psellipse(0,0)(.2,.2)\cnode*[fillcolor=black]{.1}{root1}\rput(0,.4){$v$} &&&&&&&&&
  \cnode*[fillcolor=black]{.1}{v}\rput(0,.4){$v$}\\
  && \cnode{.1}{a1} &  \cnode{.1}{a2} &&&&&& \\
  & \cnode{.1}{b1} & \psellipse(-.3,0)(.6,1.2)\cnode{.1}{b2} & \cnode{.1}{b3}\psellipse(0,0)(.3,1.2) \psellipse(0,0)(2,1.45) & \psellipse(.3,-.3)(.6,.8)\cnode{.1}{b4}&&&&&&&& \cnode{.1}{s}\rput(.6,.3){$s(v,w)$} \\
   & & \cnode{.1}{c1} & \cnode{.1}{c2} & \cnode{.1}{c3} & \cnode{.1}{c4} &&&&&
   \cnode{.1}{l}\rput(-.2,.4){$l(v,w)$} &&&& \cnode{.1}{r}\rput(.2,.4){$r(v,w)$}\\
  & && \psellipse(0,0)(.2,.2)\cnode*[fillcolor=black]{.1}{d2}\rput(0,-.4){$w$} &&&&&&&&& \cnode*[fillcolor=black]{.1}{w}\rput(0,-.4){$w$}\\
  &&& \rput(0,-.3){(a)} &&&&&&&&& \rput(0,-.3){(b)} \\\\
  \ncline{root1}{a1}\ncline{root1}{a2}
  \ncline{a1}{b1}\ncline{a1}{b2} \ncline{a2}{b3}\ncline{a2}{b4}
  \ncline{b3}{c1}\ncline{b3}{c2} \ncline{b4}{c3}\ncline{b4}{c4}
  \ncline{c2}{d2}
  \ncline{v}{s}\ncline{w}{s}\ncline{l}{s}\ncline{r}{s}
  &&& \psellipse(0,0)(.2,.2)\cnode*[fillcolor=black]{.1}{root1}\rput(0,.4){$v$} &&&&&&&&&
  \cnode*[fillcolor=black]{.1}{v}\rput(0,.4){$v$}\\
  && \cnode{.1}{a1} &  \psellipse(0,-.3)(1,.8)\psellipse(0,-.3)(1.5,1.06)\cnode{.1}{a2} &&&&&& \\
  &&\cnode{.1}{b2} & \cnode{.1}{b3} & \cnode{.1}{b4}&&&&&&&& \cnode{.1}{s}\rput(0,-.4){$l(v)$} \\
  &&& \rput(0,-.4){(c)} &&&&&&&&& \rput(0,-.4){(d)} \\
  \ncline{root1}{a1}\ncline{root1}{a2}
  \ncline{a1}{b2} \ncline{a2}{b3}\ncline{a2}{b4}
  \ncline{v}{s}
  \end{psmatrix}
   \caption{The clustering and the macro tree. (a) An internal cluster. The black nodes are the
   boundary node and the internal ellipses correspond to the boundary nodes,
   the right and left nodes, and spine path. (b) The macro tree corresponding to
   the cluster in (a). (c) A leaf cluster. The internal ellipses are the boundary node
   and the leaf nodes. (d) The macro tree corresponding to the cluster in (c).}
  \label{t2:clusterexample}
\end{center}
\end{figure}

Let $CS$ be a cluster partition of $T$ as described in Lemma
\ref{t2:lem:clustering}. We define an ordered \emph{macro tree}
$M$. Our definition of $M$ may be viewed as an ''ordered''
version of the macro tree defined in \cite{AR2002c}. The node set $V(M)$ consists of the boundary nodes in $CS$. Additionally, for each internal cluster $C \in CS$, $v,w \in \delta C$, $v \prec w$, we have the nodes $s(v,w)$, $l(v,w)$ and $r(v,w)$ and edges $(v, s(v,w)), (s(v,w), w), (l(v,w),s(v,w))$, and $(r(v,w),s(v,w))$. The nodes are ordered such that $l(v,w) \lhd w \lhd r(v,w)$. For each leaf cluster $C$, $v \in \delta C$, we have the node $l(v)$ and edge $(l(v), v)$.  Since $\roots(T)$ is a boundary node $M$ is rooted at $\roots(T)$. Figure~\ref{t2:clusterexample} illustrates these definitions.

To each node $v \in V(T)$ we associate a unique macro node denoted
$c(v)$. Let $u \in V(C)$, where $C \in CS$.
\begin{equation*}
c(u) =
\begin{cases}
        u & \text{if $u$ is boundary node}, \\
        l(v) & \text{if $u$ is a leaf node and $v \in \delta C$}, \\
        s(v,w) & \text{if $u$ is a spine node, $v,w \in \delta C$, and $v\prec w$}, \\
    l(v,w) & \text{if $u$ is a left node, $v,w \in \delta C$, and $v\prec w$}, \\
    r(v,w) & \text{if $u$ is a right node, $v,w \in \delta C$, and $v\prec w$}.
\end{cases}
\end{equation*}

Conversely, for any macro node $i \in V(M)$ define the
\emph{micro forest}, denoted $C(i)$, as the induced
subgraph of $T$ of the set of nodes $\{v \mid v \in V(T), i =
c(v)\}$. We also assign a \emph{set} of labels to $i$ given by
$\lab(i) = \{\lab(v) \mid v \in V(C(i))\}$. If $i$ is spine node
or a boundary node the unique node in $V(C(i))$ of greatest depth
is denoted by $\first(i)$. Finally, for any set of nodes $\{i_1,
\ldots, i_k\} \subseteq V(M)$ we define $C(i_1, \ldots, i_k)$ as the induced subgraph of the set of nodes $V(C(i_1)) \cup \cdots \cup V(C(i_k))$.

\begin{figure}[t]
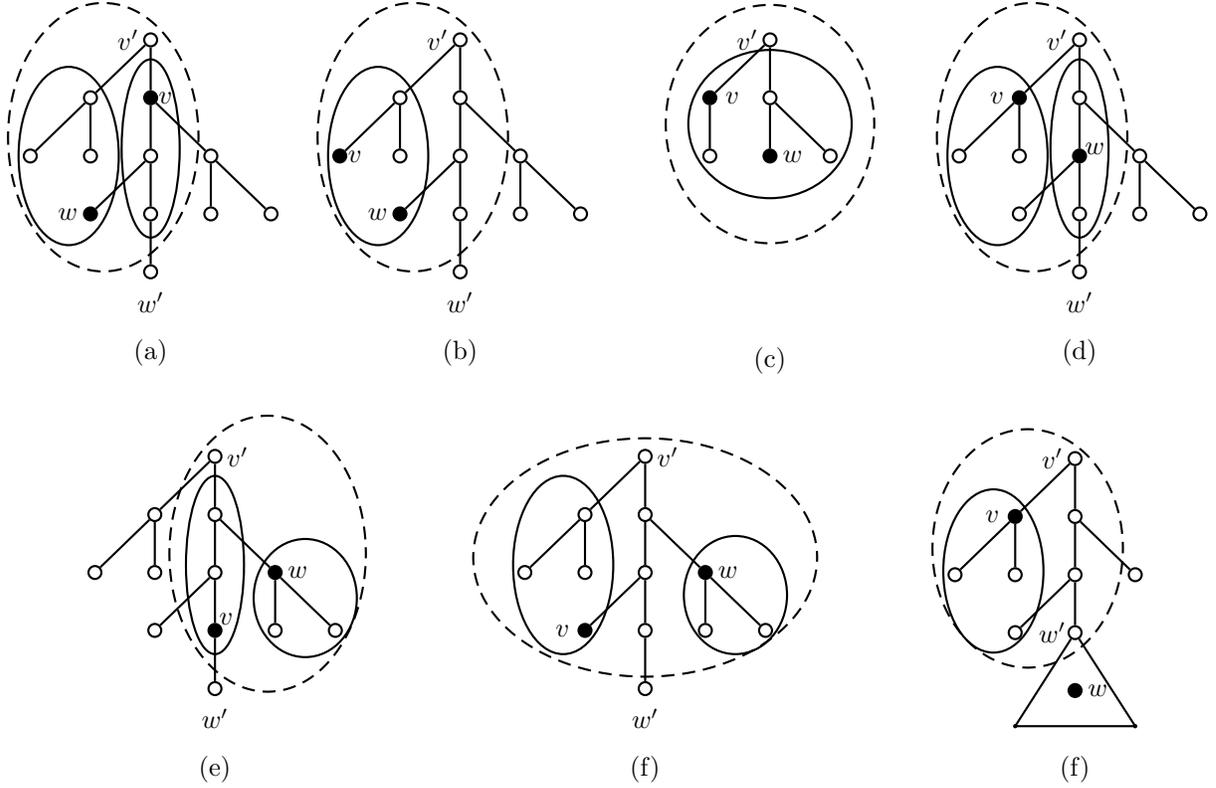

\begin{center}
  \begin{psmatrix}[colsep=0.8cm,rowsep=0.35cm,labelsep=1pt]
  &&& 
  \cnode{.1}{root1}\rput(-.3,0){$v'$}
  \\
  && \cnode{.1}{a1} &  \cnode*{.1}{a2}\rput(.2,0){$v$}
  \\
  & \cnode{.1}{b1} & \psellipse(-.29,0)(.68,1.2) 
  \cnode{.1}{b2} &
  \cnode{.1}{b3} \psellipse(0,.1)(.4,1.2) 
    \psellipse[linestyle=dashed](-.63,.25)(1.28,1.8)
  &
  \cnode{.1}{b4}
 \\
   & & \cnode*{.1}{c1}\rput(-.3,0){$w$} & \cnode{.1}{c2} & \cnode{.1}{c3} & \cnode{.1}{c4} \\
  & && 
  \cnode{.1}{d2}\rput(0,-.4){$w'$}
 \\
  &&&
  \rput(0,-.3){(a)}
  \\\\
  \ncline{root1}{a1}\ncline{root1}{a2}
  \ncline{a1}{b1}\ncline{a1}{b2} \ncline{a2}{b3}\ncline{a2}{b4}
  \ncline{b3}{c1}\ncline{b3}{c2} \ncline{b4}{c3}\ncline{b4}{c4}
  \ncline{c2}{d2}
  \end{psmatrix}
  \begin{psmatrix}[colsep=0.8cm,rowsep=0.35cm,labelsep=1pt]
  &&& 
  \cnode{.1}{root1}\rput(-.3,0){$v'$}
  \\
  && \cnode{.1}{a1} &  \cnode{.1}{a2}
  \\
  & \cnode*{.1}{b1} \rput(.2,0){$v$}& \psellipse(-.29,0)(.68,1.2) 
  \cnode{.1}{b2} &
  \cnode{.1}{b3}
    \psellipse[linestyle=dashed](-.63,.25)(1.28,1.8)
  &
  \cnode{.1}{b4}
 \\
   & & \cnode*{.1}{c1}\rput(-.3,0){$w$} & \cnode{.1}{c2} & \cnode{.1}{c3} & \cnode{.1}{c4} \\
  & && 
  \cnode{.1}{d2}\rput(0,-.4){$w'$}
 \\
  &&&
  \rput(0,-.3){(b)}
  \\\\
  \ncline{root1}{a1}\ncline{root1}{a2}
  \ncline{a1}{b1}\ncline{a1}{b2} \ncline{a2}{b3}\ncline{a2}{b4}
  \ncline{b3}{c1}\ncline{b3}{c2} \ncline{b4}{c3}\ncline{b4}{c4}
  \ncline{c2}{d2}
  \end{psmatrix}
  \begin{psmatrix}[colsep=0.8cm,rowsep=0.35cm,labelsep=1pt]
  &&&
  \cnode{.1}{root1}\rput(-.3,0){$v'$} 
  \\
   &&\cnode*{.1}{a1}\rput(.3,0){$v$} &  \psellipse(0,-.35)(1.1,1)\psellipse[linestyle=dashed](0,-.35)(1.4,1.6)\cnode{.1}{a2} \\ 
  &&\cnode{.1}{b2} & \cnode*{.1}{b3}\rput(.3,0){$w$} & \cnode{.1}{b4} &
  \\\\\\
  &&& \rput(0,-.4){(c)}    \\\\
  \ncline{root1}{a1}\ncline{root1}{a2}
  \ncline{a1}{b2} \ncline{a2}{b3}\ncline{a2}{b4}
  \end{psmatrix}
  \begin{psmatrix}[colsep=0.8cm,rowsep=0.35cm,labelsep=1pt]
  &&& 
  \cnode{.1}{root1}\rput(-.3,0){$v'$}
  \\
  && \cnode*{.1}{a1}\rput(-.3,0){$v$} &  \cnode{.1}{a2}
  \\
  & \cnode{.1}{b1} & \psellipse(-.29,0)(.68,1.2) 
  \cnode{.1}{b2} &
  \cnode*{.1}{b3}\rput(.2,.1){$w$} \psellipse(0,.1)(.4,1.2) 
    \psellipse[linestyle=dashed](-.63,.25)(1.28,1.8)
  &
  \cnode{.1}{b4}
 \\
   & & \cnode{.1}{c1} & \cnode{.1}{c2} & \cnode{.1}{c3} & \cnode{.1}{c4} \\
  & && 
  \cnode{.1}{d2}\rput(0,-.4){$w'$}
 \\
  &&&
  \rput(0,-.3){(d)}
  \\\\
  \ncline{root1}{a1}\ncline{root1}{a2}
  \ncline{a1}{b1}\ncline{a1}{b2} \ncline{a2}{b3}\ncline{a2}{b4}
  \ncline{b3}{c1}\ncline{b3}{c2} \ncline{b4}{c3}\ncline{b4}{c4}
  \ncline{c2}{d2}
  \end{psmatrix}
  \begin{psmatrix}[colsep=0.8cm,rowsep=0.35cm,labelsep=1pt]
  &&& 
  \cnode{.1}{root1}\rput(.3,0){$v'$}
  \\
  && \cnode{.1}{a1} &  \cnode{.1}{a2}
  \\
  & \cnode{.1}{b1} & 
  \cnode{.1}{b2} &
  \cnode{.1}{b3} \psellipse(0,.1)(.4,1.2) 
    \psellipse[linestyle=dashed](.7,.25)(1.32,1.85)
  &
  \psellipse(.4,-.34)(.7,.8) 
  \cnode*{.1}{b4}\rput(.3,0){$w$}
 \\
   & & \cnode{.1}{c1} & \cnode*{.1}{c2}\rput(.15,.15){$v$} & \cnode{.1}{c3} & \cnode{.1}{c4} && \\
  & && 
  \cnode{.1}{d2}\rput(0,-.4){$w'$}
 \\
  &&&
  \rput(0,-.3){(e)}
  \\
  \ncline{root1}{a1}\ncline{root1}{a2}
  \ncline{a1}{b1}\ncline{a1}{b2} \ncline{a2}{b3}\ncline{a2}{b4}
  \ncline{b3}{c1}\ncline{b3}{c2} \ncline{b4}{c3}\ncline{b4}{c4}
  \ncline{c2}{d2}
  \end{psmatrix}
  \begin{psmatrix}[colsep=0.8cm,rowsep=0.35cm,labelsep=1pt]
  &&& 
  \cnode{.1}{root1}\rput(.3,0){$v'$}
  \\
  && \cnode{.1}{a1} &  \cnode{.1}{a2}
  \\
  & \cnode{.1}{b1} & \psellipse(-.29,0.1)(.68,1.2) 
  \cnode{.1}{b2} &
  \cnode{.1}{b3} 
    \psellipse[linestyle=dashed](0,0.2)(2.3,1.6)
  &
  \psellipse(.4,-.3)(.7,.8) 
  \cnode*{.1}{b4}\rput(.3,0){$w$}
 \\
   & & \cnode*{.1}{c1}\rput(-.3,0){$v$}& \cnode{.1}{c2} & \cnode{.1}{c3} & \cnode{.1}{c4} &&\\
  & && 
  \cnode{.1}{d2}\rput(0,-0.4){$w'$}
 \\
  &&&
  \rput(0,-.3){(f)}
  \\
  \ncline{root1}{a1}\ncline{root1}{a2}
  \ncline{a1}{b1}\ncline{a1}{b2} \ncline{a2}{b3}\ncline{a2}{b4}
  \ncline{b3}{c1}\ncline{b3}{c2} \ncline{b4}{c3}\ncline{b4}{c4}
  \ncline{c2}{d2}
  \end{psmatrix}
   \begin{psmatrix}[colsep=0.8cm,rowsep=0.35cm,labelsep=1pt]
  &&& 
  \cnode{.1}{root1}\rput(-.3,0){$v'$}
  \\
  && \cnode*{.1}{a1}\rput(-.3,0){$v$} &  \cnode{.1}{a2}
  \\
  & \cnode{.1}{b1} & \psellipse(-.29,0.05)(.68,1.1) 
  \cnode{.1}{b2} &
  \cnode{.1}{b3} 
    \psellipse[linestyle=dashed](-.63,.35)(1.28,1.6)
  &
  \cnode{.1}{b4}
 \\
   & & \cnode{.1}{c1} & \cnode{.1}{c2}\rput(-.3,0){$w'$} 
   \\
  & & & 
  \cnode*{.1}{d3}\rput(.3,0){$w$}
 \\[-.3cm]
  && \cnode{0}{d2} && \cnode{0}{d4}
  \\[-.5cm]
  &&&
  \rput(0,-.3){(f)}
  \\
  \ncline{root1}{a1}\ncline{root1}{a2}
  \ncline{a1}{b1}\ncline{a1}{b2} \ncline{a2}{b3}\ncline{a2}{b4}
  \ncline{b3}{c1}\ncline{b3}{c2} 
  \ncline{c2}{d2}\ncline{c2}{d4}\ncline{d2}{d4}
  \end{psmatrix}
   \caption{Examples from the propositions. In all cases $v'$ and $w'$ are top and bottom
   boundary nodes of the cluster, respectively. (a)
   Proposition~\ref{t2:lem:ancestorlemma}(ii).
   Here $c(v)=s(v',w')$ and $c(w)=l(v',w')$ (solid ellipses). The dashed ellipse corresponds to
   $C(c(v),s(v',w'),v')$.
   (b) Proposition~\ref{t2:lem:orderlemma}(i) and~\ref{t2:lem:ncalemma}(ii). Here $c(v)=c(w)=l(v',w')$ (solid ellipse). 
   The dashed ellipse corresponds to
   $C(c(v),s(v',w'),v')$. 
   (c) Proposition~\ref{t2:lem:orderlemma}(ii) and~\ref{t2:lem:ncalemma}(i). Here $c(v)=c(w)=l(v')$ (solid ellipse). 
   The dashed ellipse corresponds to   $C(c(v),v')$. 
   (d)
    Proposition~\ref{t2:lem:orderlemma}(iii). Here $c(v)=l(v',w')$ and $c(w)=s(v',w')$ (solid ellipses). The dashed ellipse corresponds to
   $C(c(v),c(w),v')$.
   (e) Proposition~\ref{t2:lem:orderlemma}(iv). Here $c(v)=s(v',w')$ and $c(w)=r(v',w')$ (solid ellipses). The dashed ellipse corresponds to
   $C(c(v),c(w),v')$. (f) Proposition~\ref{t2:lem:ncalemma}(iv). Here $c(v)=r(v',w')$ and $c(w)=l(v',w')$ (solid ellipses). The dashed ellipse corresponds to $C(c(v),c(w),s(v',w'),v')$.  (g) Proposition~\ref{t2:lem:ncalemma}(v). Here $c(v)=r(v',w')$ (solid ellipse) and $w' \preceq_M c(w)$. The dashed ellipse corresponds to $C(c(v),s(v',w'),v',w'))$.}
  \label{t2:fig:propositions}
\end{center}
\end{figure}

The following propositions states useful properties of ancestors,
nearest common ancestor, and the left-to-right ordering in the
micro forests and in $T$. The propositions follows directly
from the definition of the clustering. See also
Figure~\ref{t2:fig:propositions}.
\begin{prop}[Ancestor relations]\label{t2:lem:ancestorlemma}
For any pair of nodes $v, w \in V(T)$, the following hold
\begin{itemize}
  \item[(i)] If $c(v) = c(w)$ then $v \prec_T w$ iff $v \prec_{C(c(v))} w$.
  \item[(ii)] If $c(v) \neq c(w)$, $c(v) \in \{s(v',w'), v'\}$, and $c(w)
  \in \{l(v', w'), r(v',w')\}$ then we have $v \prec_T w$ iff $v \prec_{C(c(v),s(v',w'), v')} w$.
  \item[(iii)] In all other cases, $w \prec_T v$ iff $c(w) \prec_{M} c(v)$.
\end{itemize}
\end{prop}
Case (i) says that if $v$ and $w$ belongs to the same macro node
then $v$ is an ancestor of $w$ iff $v$ is an ancestor of $w$ in
the micro forest for that macro node. Case (ii) says that if $v$
is a spine node or a top boundary node and
$w$ is a left or right node in the same cluster then
$v$ is an ancestor of $w$ iff $v$ is an ancestor of $w$ in the
micro tree induced by that cluster (Figure~\ref{t2:fig:propositions}(a)). Case (iii) says that in all
other cases $v$ is an ancestor of $w$ iff the macro node $v$
belongs to is an ancestor of the macro node $w$ belongs to in the
macro tree.

\begin{prop}[Left-of relations]\label{t2:lem:orderlemma}
For any pair of nodes $v, w \in V(T)$, the following hold
\begin{itemize}
  \item[(i)] If $c(v) = c(w) \in \{r(v',w'),l(v',w')\}$ then $v \lhd w$
     iff $v \lhd_{C(c(v),v',s(v',w'))}
    w$.
  \item[(ii)] If $c(v) = c(w)=l(v')$ then $v \lhd w$ iff $v \lhd_{C(c(v),v')} w$.
  \item[(iii)] If $c(v) = l(v', w')$ and $c(w)=s(v',w')$  then
  $v \lhd w$ iff $v \lhd_{C(c(v), c(w), v')} w$.
  \item[(iv)] If  $c(v)=s(v',w')$  and $c(w) = r(v', w')$ then
  $v \lhd w$ iff $v \lhd_{C(c(v), c(w), v')} w$.
  \item[(v)] In all other cases, $v \lhd w$ iff $c(v) \lhd_{M} c(w)$.
\end{itemize}
\end{prop}
Case (i) says that if $v$ and $w$  are both either left or right nodes in the same cluster then $v$ is to the left of $w$ iff
$v$ is to the left of $w$ in the micro tree induced by their macro
node together with the spine and top boundary node of the cluster (Figure~\ref{t2:fig:propositions}(b)).
Case (ii) says that if $v$ and $w$ are both leaf nodes in the same cluster then $v$ is to the left of $w$ iff $v$ is to the
left of $w$ in the micro tree induced by that leaf cluster (Figure~\ref{t2:fig:propositions}(c)). Case (iii) says that if $v$ is a left node and $w$ is a spine node in the
same cluster then $v$ is to the left of $w$ iff $v$ is to the left
of $w$ in the micro tree induced by their two macro nodes and the
top boundary node of the cluster (Figure~\ref{t2:fig:propositions}(d)). Case (iv) says that if $v$ is a spine node and $w$  is a  right node in the same cluster then $v$ is to the left of $w$ iff $v$ is to the left
of $w$ in the micro tree induced by their two macro nodes  and the top boundary node of the cluster (Figure~\ref{t2:fig:propositions}(e)). In all other cases $v$ is to the
left of $w$ if the macro node $v$ belongs to is to the left of the
macro node of $w$ in the macro tree (Case (v)).

\begin{prop}[Nca relations]\label{t2:lem:ncalemma}
    For any pair of nodes $v, w \in V(T)$, the following hold
    \begin{itemize}
    \item[(i)] If $c(v) = c(w)=l(v')$
        then $\nca_T(v,w)=\nca_{C(c(v),v')}(v,w)$.
    \item[(ii)] If $c(v) = c(w)\in\{l(v',w'),r(v',w')\}$
        then $\nca_T(v,w)=\nca_{C(c(v), s(v',w'), v')}(v,w)$.
    \item[(iii)] If $c(v) = c(w)=s(v',w')$
        then $\nca_T(v,w)=\nca_{C(c(v))}(v,w)$.
    \item[(iv)] If $c(v) \neq c(w)$ and $c(v),c(w)\in\{l(v',w'),r(v',w'),s(v',w')\}$
        then \\ $\nca_T(v,w)=\nca_{C(c(v),c(w),s(v',w'),v')}(v,w)$.
    \item[(v)] If $c(v) \neq c(w)$,
        $c(v)\in\{l(v',w'),r(v',w'),s(v',w')\}$, and $w' \preceq_{M}
        c(w)$ then \\ $\nca_T(v,w)=\nca_{C(c(v),s(v',w'),v',w')}(v,w')$.
    \item[(vi)] If $c(v) \neq c(w)$,
        $c(w)\in\{l(v',w'),r(v',w'),s(v',w')\}$, and $w' \preceq_{M}
        c(v)$ then \\ $\nca_T(v,w)=\nca_{C(c(w'),s(v',w'),v',w')}(w,w')$.
    \item[(vii)] In all other cases, $\nca_T(v,w)=\nca_{M}(c(v),c(w))$.
    \end{itemize}
\end{prop}
Case (i) says that if $v$ and $w$ are leaf nodes in the same cluster
then the nearest common ancestor of $v$ and $w$ is the nearest
common ancestor of $v$ and $w$ in the micro tree induced by that
leaf cluster (Figure~\ref{t2:fig:propositions}(c)). Case (ii) says
that if $v$ and $w$ are both either left nodes
or right nodes then the nearest common ancestor of $v$ and $w$ is
the nearest common ancestor in the micro tree induced by their
macro node together with the spine and top boundary node of the
cluster (Figure~\ref{t2:fig:propositions}(b)). Case (iii) says that if $v$ and $w$ are both
spine nodes in the same cluster then the nearest common ancestor of $v$ and $w$ is the
nearest common ancestor of $v$ and $w$ in the micro tree induced
by their macro node. Case (iv) says that if $v$ and $w$ are in different macro nodes but are right, left, or spine nodes in the same cluster then the
nearest common ancestor of $v$ and $w$ is the nearest common
ancestor of $v$ and $w$ in the micro tree induced by that cluster
(we can omit the bottom boundary node) (Figure~\ref{t2:fig:propositions}(f)). Case (v) says that if $v$
is a left, right, or spine node, and the bottom boundary
node $w'$ of $v$'s cluster is an ancestor in the macro tree of the
macro node containing $w$, then the nearest common ancestor of $v$
and $w$ is the nearest common ancestor of $v$ and $w'$ in the
micro tree induced by the macro node of $v$, the spine node, and the top and bottom boundary nodes of $v$'s cluster (Figure~\ref{t2:fig:propositions}(g)). Case (vi) is the same as
case (v) with $v$ and $w$ interchanged. In all other cases the
nearest common ancestor of $v$ and $w$ is the nearest common
ancestor of their macro nodes in the macro tree (Case (vii)).

\subsection{Preprocessing}\label{t2:sec:preprocessing}
In this section we describe how to preprocess $T$. First build a
cluster partition $CS$ of the tree $T$ with clusters of size $s$,
to be fixed later, and the corresponding macro tree $M$ in
$O(n_T)$ time. The macro tree is preprocessed as in Section~\ref{t2:sec:simplepreprocessing}. However, since nodes in $M$ contain a set of labels, we now store a dictionary for $\lab(v)$ for each node $v \in V(M)$. Using the deterministic using the deterministic dictionary of Hagerup et. al. \cite{HMP2001} all these dictionaries can be constructed in $O(n_{T}\log n_{T})$ time and $O(n_{T})$ space.
Furthermore, we extend the definition of $\fl$ such that $\fl_{M}(v, \alpha)$ is the nearest ancestor $w$ of $v$ such that $\alpha \in \lab(w)$.

Next we show how to preprocess the micro forests. For any cluster $C\in CS$, deep sets $X, Y, Z\subseteq V(C)$, $i \in \mathbb{N}$, and $\alpha \in \Sigma$ define the following procedures on cluster $C$.

\begin{relate}[clusterprocedures]

\item[$\leftn_C(i,X)$:] Return the leftmost $i$ nodes in $X$.

\item[$\rn_C(i,X)$:] Return the rightmost $i$ nodes in $X$.

\item[$\leftof_C(X,Y)$:] Return all nodes of $X$ to the left of the leftmost node in $Y$.


\item[$\match_C(X,Y,Z)$,] where $X=\{m_1 \lhd \cdots \lhd m_k\}$,
$Y=\{v_1 \lhd \cdots \lhd v_k\}$, and $Z \subseteq Y$.
Return $R:=\{m_{j} \mid v_j \in Z \}$.

\item[$\mopc_{C}(X,Y)$] Return
the pair $(R_1,R_2)$. Where
$R_1=\restrict{\mop(M,N)}{1}$ and $R_2=\restrict{\mop(M,N)}{2}$.

\end{relate}
In addition to these procedures we also define the set procedures on clusters, that is, $\Parent_C$, $\Nca_C$, $\Deep_C$, 
and $\Fl_C$, as in Section~\ref{t2:sec:recursion}. Collectively, we will call these the \emph{cluster procedures}.  We represent the input and outputs set in the 
procedures as bit strings indexed by preorder numbers. Specifically, a subset $X$ in a cluster $C$ is given by a bit string $b_{1} \ldots b_{s}$, such that $b_{i} = 1$ iff the $i$th node in a preorder traversal of $C$ is in $X$. If $C$ contains fewer than $s$ nodes we leave the remaining values undefined.

The procedures $\leftn_C(i,X)$  then corresponds to setting
 all bits in $X$ larger than the $i$th set bit to zero. Similarly, $\rn_C(i,X)$ corresponds to setting all bits smaller than the $i$th largest set bit to zero. Similarly, the procedures $\leftof_C(X,Y)$, $\Rightof_C(X,Y)$, and $\match_C(X,Y,Z)$ only depends on the preorder of the nodes and thus only on the bit string not any other information about the cluster. 
We can thus ommit the subscript $C$ from these five procedures.

Next we show how to implement the cluster procedures efficiently. We precompute the value of all procedures, except $\Fl_{C}$, for all possible inputs and clusters. By definition, these procedures do not depend on any specific labeling of the nodes in the cluster. Hence, it suffices to precompute the value for all rooted, ordered trees with at most $s$ nodes. The total number of these is less than $2^{2s}$ (consider e.g. an encoding using balanced parenthesis). Furthermore, the number of possible input sets is at most $2^{s}$. Since at most $3$ sets are given as input to a cluster procedure, it follows that we can tabulate all solutions using less than $2^{3s}\cdot 2^{2s} = 2^{5s}$ bits of memory. Hence, choosing $s \leq 1/10\log n$ we use $O(2^{\frac{1}{2}\log n}) = O(\sqrt{n})$ bits. Using standard bit wise operations each solution is easily implemented in $O(s)$ time giving a total time of $O(\sqrt{n}\log n)$. 

Since the procedure $\Fl_{C}$ depends on the alphabet, which may be of size $n_{T}$, we cannot efficiently apply the same trick as above. Instead define for any cluster $C \in CS$, $X \subseteq V(C)$, and $\alpha \in \Sigma$:
\begin{relate}[simple]
\item[$\Ancestor_{C}(X)$:] Return the set $\{x \mid \text{$x$ is an ancestor of a node in $X$}\}$.
\item[$\Eq_C(\alpha)$:] Return the set $\{x \mid x \in V(C), \lab(x) = \alpha\}$.
\end{relate}
Clearly, $\Ancestor_{C}$ can be implemented as above. For $\Eq_{C}$ note that the
 total number of distinct labels in $C$ is at most $s$. Hence, $\Eq_{C}$ can be stored in a dictionary with at most $s$ entries each of which is a bit string of length $s$. Thus, (using again the result of \cite{HMP2001}) the total time to build all such dictionaries is $O(n_{T}\log n_{T})$.

By the definition of $\Fl_{C}$ we have that,
\begin{equation*}
\Fl_C(X,\alpha) = \Deep_C(\Ancestor_C(X) \cap \Eq_C(\alpha)).
\end{equation*}
Since intersection can be implemented using a binary \emph{and}-operation, $\Fl_{C}(X, \alpha)$ can be computed in constant time. Later, we will also need to compute union of bit strings and we note that this can be done using a binary \emph{or}-operation.

To implement the set procedures in the following section we often need to ``restrict'' the cluster procedures to work on a subtree of a cluster. Specifically, for any set of macro nodes $\{i_{1}, \ldots, i_{k}\}$ in the \emph{same} cluster $C$ (hence, $k \leq 5$), we will replace the subscript $C$ with $C(i_{1}, \ldots, i_{k})$. For instance, $\Parent_{C(s(v,w), l(v,w))}(X) = \{\parent(x) \mid x\in X \cap V(C(s(v,w), l(v,w))\} \cap V(C(s(v,w), l(v,w))$. To implement all restricted versions of the cluster procedures, we compute for each cluster $C \in CS$ a bit string representing the set of nodes in each micro forest. Clearly, this can be done in $O(n_{T})$ time. Since there are at most $5$ micro forests in each cluster it follows that we can compute any restricted version using an additional constant number of and-operations.

Note that the total preprocessing time and space is dominated by the construction of deterministic dictionaries which use $O(n_{T}\log n_{T})$ time and $O(n_T)$ space.


\subsection{Implementation of the Set Procedures}
Using the preprocessing from the previous section we show how to implement the set procedures in sublinear time. First we define a compact representation of node sets. Let $T$ be a tree with macro tree $M$. For simplicity, we identify nodes in $M$ with their preorder number. Let $S \subseteq V(T)$ be any subset of nodes of $T$.
A \emph{micro-macro node array} (abbreviated node array) $X$ representing $S$ is an array of size $n_{M}$. The $i$th entry, denoted $X[i]$, represents the subset of nodes in $C(i)$, that is, $X[i] =  V(C(i)) \cap S$. The set $X[i]$ is encoded using the same bit representation as in Section~\ref{t2:sec:preprocessing}. By our choice of parameter in the clustering the space used for this representation is $O(n_{T}/\log n_{T})$.

We now present the detailed implementation of the set procedures on node arrays. Let $X$ be a node array.
\begin{relate}
\item[$\Parent(X)$:] Initialize a node array $R$ of size $n_{M}$ and set $i:=1$.

Repeat until $i>n_M$:
\begin{itemize}
\item[] Set $i:=i+1$ until $X[i]\neq \emptyset$.

There are three cases depending on the type of $i$:
\begin{enumerate}
  \item $i\in \{l(v,w), r(v,w)\}$. Compute $N := \Parent_{C(i, s(v,w), v)}(X[i])$. For each $j \in \{i, s(v,w), v\}$, set $R[j] := R[j] \cup (N \cap V(C(j)))$.
  \item $i = l(v)$. Compute $N := \Parent_{C(i, v)}(X[i])$. For each $j \in \{i, v\}$, set $R[j] :=R[j] \cup (N \cap V(C(j)))$.
  \item $i \not\in \{l(v,w), r(v,w), l(v)\}$. Compute $N := \Parent_{C(i)}(X[i])$. If $N \neq \emptyset$ set $R[i]:=R[i] \cup N$. Otherwise, if $j := \parent_{M}(i) \neq \bot$ set
    $R[j] := R[j] \cup \{\first(j)\}$.
\end{enumerate}
Set $i:=i+1$.
\end{itemize}
Return $R$.
\end{relate}
To see the correctness of the implementation of procedure  $\Parent$
consider the three cases of the procedure. Case 1 handles
the fact that  left or right nodes may have a node on a spine or
boundary node as parent. Since no left or right nodes can have a
parent outside their cluster there is no need to compute parents
in the macro tree. Case 2 handles the fact that a leaf node may have the boundary node as parent. Since no leaf node can have a parent outside its cluster
there is no need to compute parents in the macro tree. Case 3
handles boundary and spine nodes. In this case there is either a parent within the micro forest or we can use
the macro tree to compute the parent of the root of the micro
tree.
Since the input to \Parent\ is deep we only need to do one of the two things. If the computation of parent in the micro tree returns a node $j$, this will either be a spine node or a boundary node. To take care of the case where $j$ is a spine node, we add the lowest node ($\first(j)$) in $j$ to the output. Procedure \Parent\ thus correctly computes parent for all kinds of macro nodes.

We now give the implementation of procedure \Nca.
The input to procedure \Nca\ is two node arrays $X$ and $Y$ representing two subsets $\mc{X}, \mc{Y} \subseteq V(T)$,  $\norm{\mc{X}}=\norm{\mc{Y}}=k$. 
 The output is a node array $R$ representing the set $\{\nca(\mc{X}_i,\mc{Y}_i) \mid 1\leq i\leq k\}$, where $\mc{X}_i$ and $\mc{Y}_i$ is the $i$th element of  $\mc{X}$ and $\mc{Y}$, wrt.\ to their preorder number in the tree, respectively. We also assume that we have $\mc{X}_i \prec \mc{Y}_i$ for all $i$ (since $\Nca$ is always called on a set of minimum ordered pairs).
\begin{relate}
\item[$\Nca(X,Y)$:] Initialize a node array $R$ of size $n_{M}$, set $i:=1$ and $j:=1$.

    Repeat until $i>n_M$ or $j>n_M$:
    \begin{itemize}
    \item[] Until $X[i] \neq \emptyset$ set $i:=i+1$. Until  $Y[j] \neq \emptyset$ set $j:=j+1$.

    Compare $i$ and $j$.  There are two cases:
    \begin{enumerate}
    \item $i=j$. There are two subcases:
 	\begin{enumerate}
   	\item $i$ is a boundary node.
	
    	Set $R[i]:=X[i]$, $i:=i+1$ and $j:=j+1$.

        \item $i$ is not a boundary node. 
        
        	Compare the sizes of $X[i]$ and $Y[j]$. There are two cases:
             \begin{itemize}
             \item $\norm{X[i]} > \norm{Y[j]}$. Set 
             $X_i:= \leftn (\norm{Y[j]},X[i])$,
             \item $\norm{X[i]} = \norm{Y[j]}$. Set $X_i=X[i]$.
             \end{itemize}
        Set
        \begin{equation*}
        S:=
            \begin{cases}
            C(i,v), &  \textrm{if } i=l(v),\\
            C(i, s(v,w), v), & \textrm{if } i\in\{l(v,w),r(v,w)\}, \\
            C(i), & \textrm{if } i=s(v,w).
            \end{cases}
            \end{equation*}
            Compute $N:=\Nca_S(X_i,Y_j)$.

        For each macronode $h$ in $S$ set 
        $R[h]:=R[h] \cup (N\cap V(C(h)))$.

        Set $X[i]:=X[i]\setminus X_i$ and $j:=j+1$.
	\end{enumerate}
    \item $i \neq j$. 
	    	Compare the sizes of $X[i]$ and $Y[j]$. There are three cases:
         	\begin{itemize}
		\item $\norm{X[i]} > \norm{Y[j]}$. 
			Set $X_i:=\leftn(\norm{Y[j]},X[i])$ 
			and $Y_j:=Y[j]$, 
		
		\item $\norm{X[i]} < \norm{Y[j]}$. Set 
		$X_i:=X[i]$ and $Y_j :=\leftn(\norm{X[i]},Y[j])$,
		
		\item $\norm{X[i]} = \norm{Y[j]}$. Set 
		$X_i:=X[i]$ and $Y_j:=Y[j]$.
		\end{itemize}
	    Compute $h:=\Nca_M(i,j)$. There are two subcases:
            \begin{enumerate}
            \item $h$ is a boundary node. Set $R[h]:=1$.
            \item  $h$ is a spine node $s(v,w)$. There are three cases:
                \begin{enumerate}
                \item $i \in \{l(v,w),s(v,w)\}$ and $j\in \{s(v,w),r(v,w)\}$. 
                
                Compute $N:=\Nca_{C(i,j,h,v)}(X_i,Y_j)$.
                \item $i=l(v,w)$ and $w \preceq j$. 
                
                Compute $N:=\Nca_{C(i,h,v,w)}(\rn(1,X_i),w)$.
                \item $j=r(v,w)$ and $w \preceq i$. 
                
                Compute $N:=\Nca_{C(j,h,w,v)}(w,\leftn(1,Y_j))$.
                \end{enumerate}
                Set $R[h]:=R[h]\cup (N\cap V(C(h)))$ and 
                $R[v]:=R[v]\cup (N\cap V(C(v)))$.
            \end{enumerate}
            Set $X[i]:=X[i]\setminus X_i$ and $Y[j]:=Y[j]\setminus Y_j$.
    \end{enumerate}
    \end{itemize}
Return $R$.
\end{relate}
In procedure \Nca\ we first find the next non-empty entries in the node arrays $X[i]$ and $Y[j]$. Then we have two cases depending on whether $i=j$ or not.
If $i=j$ (Case 1) we have two subcases. If $i$ is a boundary node (Case 1(a)) then $C(i)$ only consists of one node $v= X[i]=Y[j]$ and therefore $\nca(v,v)=v=X[i]$.
If $i$ is not a boundary node (Case 1(b)) we compare the sizes of the subsets represented by $X[i]$ and $Y[i]$. If $\norm{X[i]} > \norm{Y[j]}$  we compute nearest common ancestors of the first/leftmost $\norm{Y[j]}$ nodes in $X[i]$ and the nodes in $Y[j]$. Due to the assumption on the input ($\mc{X}_i \prec \mc{Y}_i$) we either have $\norm{X[i]} > \norm{Y[j]}$ or  $\norm{X[i]} = \norm{Y[j]}$. If $\norm{X[i]} > \norm{Y[j]}$ we must compute nearest common ancestors of the first/leftmost $\norm{Y[j]}$ nodes in $X[i]$ and the nodes in $Y[j]$. If $\norm{X[i]} = \norm{Y[j]}$ we must compute nearest common ancestors of all nodes in $X[i]$ and $Y[j]$. We now compute nearest common ancestors of the described nodes in a cluster $S$ depending on what kind of node $i$ is. If $i$ is a leaf node then the nearest common ancestors of the nodes in $X[i]$ and $Y[j]$ is either in $i$ or in the boundary node (Proposition~\ref{t2:lem:ncalemma}(i)). If $i$ is a left or right node then the nearest common ancestors must be in $i$ on the spine or in the top boundary node (Proposition~\ref{t2:lem:ncalemma}(ii)). If $i$ is a spine node then the nearest common ancestors must be on the spine or in the top boundary node (Proposition~\ref{t2:lem:ncalemma}(iii)). We update the output node array, remove from  $X[i]$  the nodes we have just computed nearest common ancestors of, and increment $j$ since we have now computed nearest common ancestors for all nodes in $Y[j]$.

Now consider the case where $i \neq j$. First we
 compare the sizes of the subsets represented by $X[i]$ and $Y[i]$. If $\norm{X[i]} > \norm{Y[j]}$  we should compute nearest common ancestors of the first/leftmost $\norm{Y[j]}$ nodes in $X[i]$ and the nodes in $Y[j]$ as in Case 1(b). If $\norm{X[i]} < \norm{Y[j]}$ we must compute nearest common ancestors of the first/leftmost $\norm{X[i]}$ nodes in $Y[j]$ and the nodes in $X[i]$. Otherwise $\norm{X[i]} = \norm{Y[j]}$ and we  compute nearest common ancestors of the all nodes  in $X[i]$ and $Y[j]$. We now compute the nearest common ancestor of $i$ and $j$ in the macro tree. This must either be a boundary node or a spine node due to the structure of the macro tree. If it is a boundary node then the nearest common ancestor of all nodes in $i$ and $j$ is this boundary node. If it is a spine node we have three different cases depending on the types of $i$ and $j$. If $i$ is a left or spine node and $j$ is a spine or right node in the same cluster then we compute nearest common ancestors in that cluster (Proposition~\ref{t2:lem:ncalemma}(iv)). If $i$ is a left node and $j$ is a descendant of the bottom boundary node in $i$'s cluster then we compute the nearest common ancestor of the rightmost node in $X_i$ and $w$  in $i$'s cluster(Proposition~\ref{t2:lem:ncalemma}(v)). That we can restrict the computation to only the rightmost node in $X_i$ and $w$ is due to the fact that we always run \Deep\ on the output from \Nca\ before using it in any other computations. In the last case $j$ is a right node and $i$ is a descendant of the bottom boundary node of $j'$s cluster. Then we compute the nearest common ancestor of the leftmost node in $Y_j$ and $w$ (Proposition~\ref{t2:lem:ncalemma}(vi)) in $j$'s cluster. The argument for restricting the computation to the leftmost node of $Y_j$ and $w$ is the same as in the previous case. Due to the assumption on the input ($\mc{X}_i \prec \mc{Y}_i$) the rest of the cases from Proposition~\ref{t2:lem:ncalemma}(iv)--(vi) cannot happen. Therefore, we have now argued that the procedure correctly takes care of all cases from Proposition~\ref{t2:lem:ncalemma}. Finally, we update the output node array and remove from $X[i]$ and $Y[j]$ the nodes we have just computed nearest common ancestors of.

The correctness of the procedure follows from the above and induction on the rank of the elements. 
 
\ignore{**************Case 1(b) handles the cases
(i), (ii), and (iii) from Proposition~\ref{t2:lem:ncalemma} and Case 1(a) handles part of case (vii). Case 2
handles the cases (iv), (v), (vi) and (vii) from
Proposition~\ref{t2:lem:ncalemma}. In all cases we make sure that the number of nodes in the microtrees/forrests that we work on are the same by comparing the sizes of $X[i]$ and $Y[j]$. In Case 1(a) $i=j$ is a boundary node and thus only contains one node. In Case 1(b) it is possible that there are more nodes in the set represented by $X[i]$ than in the one represented by $Y[j]$ (but not vice versa due to the assumptions on the input). In that case we extract the leftmost $\norm{Y[j]}$ nodes from $X[i]$ and use these for the computation. Before iterating we delete these nodes from $X[i]$. In Case 2 we have three cases since the set represented by $Y[j]$ now can be larger than the set represented by $X[i]$. Again we extract the relevant nodes, compute $\nca$ on these and update $X[i]$ and $Y[j]$ before iterating. 
}

\begin{relate}
\item[$\Deep(X)$:] Initialize a node array $R$ of size $n_{M}$ and set $j := 1$.

Repeat until $i>n_M$:
\begin{itemize}
\item[] Set $i:=i+1$ until $X[i]\neq \emptyset$.

Compare $j$ and $i$. There are three cases:
\begin{enumerate}
  \item $j \lhd i$. Set \begin{equation*}
        S:=
            \begin{cases}
            C(j,v), &  \textrm{if } j=l(v),\\
            C(j, s(v,w), v), & \textrm{if } j\in\{l(v,w),r(v,w)\}, \\
            C(j), & \textrm{otherwise}.
            \end{cases}
            \end{equation*}
  Set $R[j] := \Deep_S (X[j])$ and $j := i$.
  \item $j \prec i$. If $i \in \{l(v,w), r(v,w)\}$ and $j=s(v,w)$ compute
  $N := \Deep_{C(i, s(v,w), v)}(X[i] \cup X[j])$, and set $R[j]:=R[j]\cap N$.

  Set $j:=i$.
\end{enumerate}
Set $i:=i+1$.
\end{itemize}
  Set $R[j] := \Deep_S (X[j])$, where $S$ is set as in Case 1.
  
Return $R$.
\end{relate}
The above $\Deep$ procedure resembles the previous $\Deep$
procedure implemented on the macro tree in the two first cases. The third case from the previous implementation can be omitted since the input list is now in preorder.
In case 1 node $i$ is to the
right of our "potential output node" $j$. Since any node $l$ that is a
descendant of $j$ must be to the left of $i$ ($l<i$) it cannot not
appear later in the list $X$ than $i$. We can thus safely add
$\Deep_S(X[j])$ to $R$ at this point. To ensure that the cluster we compute \Deep\ on is a tree we include the top boundary node if $j$ is a leaf node and the top and spine node if $j$ is a left or right node. In case 2 node $j$ is an ancestor of
$i$ and can therefore not be in the output list unless $j$ is a spine node and $i$ is the corresponding left or right node. If this is the case we first compute \Deep\ of $X[j]$ in the cluster containing $i$ and $j$ and add the result to the output before setting $i$ to be our new potential node.
After scanning the whole node array $X$ we add the last potential node $j$ to the output after computing \Deep\ of it as in case 1.

That the procedure is correct follows by the proof of Lemma~\ref{t2:lem:deep} and the above.
\\\\
We now give the implementation of procedure \Mop.
Procedure $\Mop$ takes a pair of node arrays $(X,Y)$ and another node array $Z$ as input.  The pair $(X,Y)$ represents a set of minimum ordered pairs, where the first coordinates are in $X$ and the second coordinates are in $Y$.
To simplify the implementation of procedure $\Mop$ it calls two auxiliary procedures $\Mopsim$ and $\Match$ defined below. Procedure $\Mopsim$ computes $\mop$ of $Y$ and $Z$, and procedure $\Match$ takes care of finding the first-coordinates from $X$ corresponding to the first coordinates from the minimum ordered pairs from $M$.

\begin{relate}
\item[$\Mop((X,Y),Z)$] Compute $M:=\Mopsim(Y,Z)$. Compute $R:=\Match(X,Y,\restrict{M}{1})$. Return $(R,\restrict{M}{2})$.
\end{relate}
Procedure $\Mopsim$ takes two node arrays as input and computes $\mop$ of these.
\begin{relate}
\item[$\Mopsim(X,Y)$] 
Initialize two node arrays $R$ and $S$ of size $n_M$, set $i:=1$,
$j:=1$, $h:=1$,  $(r_1,r_2):=(0,\emptyset)$, $(s_1,s_2):=(0,\emptyset)$.
Repeat the following until $i>n_M$ or $j>n_M$:
\begin{itemize}
\item[] Set $i:=i+1$ until $X[i] \neq \emptyset$. There are three cases:
\begin{enumerate}
\item If $i=l(v,w)$ for some $v, w$ set $j:=j+1$ until $Y[j]\neq \emptyset$ and either $i \lhd j$, $i=j$, or $j=s(v,w)$.
\item If $i=s(v,w)$ for some $v, w$ set $j:=j+1$ until $Y[j]\neq \emptyset$ and either $i \lhd j$ or $j=r(v,w)$.
\item If $i \in \{r(v,w), l(v)\}$ for some $v, w$ set $j:=j+1$ until $Y[j]\neq \emptyset$ and either $i \lhd j$ or  $i=j$.
\item Otherwise ($i$ is a boundary node) set $j:=j+1$ until $Y[j]\neq \emptyset$ and $i \lhd j$.
\end{enumerate}
Compare $i$ and $j$. There are two cases:
    \begin{enumerate}
    \item $i\lhd j$: Compare $s_1$ and $j$. If $s_1 \lhd j$ set
        $R[r_1]:=R[r_1] \cup r_2$, $S[s_1]:=S[s_1] \cup s_2$, and
        $(s_1,s_2):= (j, \leftn_{C(j)}(1,Y[j]))$. 
        
        Set
        $(r_1,r_2):=(i, \rn_{C(i)}(1,X[i]))$ and $i=i+1$.
    \item Otherwise compute $(r,s):=\mopc_{C(i,j,v)}(X[i],Y[j])$, where $v$ is the top boundary node in the cluster $i$ and $j$ belongs to.
    
	If $r\neq \emptyset$ do:
            \begin{itemize}
            \item Compare $s_1$ and $j$. If  $s_1 \lhd j$ or if $s_1=j$
             and $\leftof_{C(i,j)}(X[i],s_2)=\emptyset$  then set
                $R[r_1]:=R[r_1] \cup r_2$, $S[s_1]:=S[s_1] \cup s_2$.

            \item  Set $(r_1,r_2):=(i,r)$ and $(s_1,s_2):=(j,s)$.

            \end{itemize}
		There are two subcases:
        \begin{enumerate}
        \item $i=j$ or $i=l(v,w)$ and $j=s(v,w)$. 
        Set  $X[i]:=\rn_{C(i)}(X[i])\setminus r_2$ and 
        $j:=j+1$.

        \item  $i=s(v,w)$ and $j=r(v,w)$. If $r_2=\emptyset$ set 
        $j:=j+1$ otherwise set $i:=i+1$.

        \end{enumerate}            
    \end{enumerate}
\end{itemize}
Set $R[r_1]:=R[r_1] \cup r_2$ and $S[s_1]:=S[s_1] \cup s_2$. Return $(R,S)$.
\end{relate}
Procedure $\Mopsim$ is somewhat similar to the previous implementation of the procedure $\Mop$ from Section~\ref{t2:implementationsimple}. We again have a "potential pair" $((r_1,r_2), (s_1,s_2))$ but we need more cases to take care of the different kinds of macro nodes.

We first find the next non-empty macro node $i$. We then have 4 cases depending on which kind of node $i$ is. In Case 1 $i$ is a left node. Due to Proposition~\ref{t2:lem:orderlemma} we can have $\mop$ in $i$ (case (i)), in the spine (case (iii)), or in a node to the left of $i$ (case(v)). In Case 2 $i$ is a spine node. Due to Proposition~\ref{t2:lem:orderlemma} we can have $\mop$  in the right node (case (iv)) or in a node to the left of $i$ (case(v)). In Case 3 $i$ is a right node or a leaf node. Due to Proposition~\ref{t2:lem:orderlemma} we can have $\mop$ in $i$ (case (i) and (ii)) or in a node to the left of $i$ (case(v)). In the last case (Case 5) $i$ must be a boundary node and $\mop$ must be in a node to the left of $i$.

We then compare $i$ and $j$. The case were $i \lhd j$ is similar to the previous implementation of the procedure. We compare $j$ with our potential pair. If $s_1 \lhd j$ then we can insert $r_2$ and $s_2$ into our output node arrays $R$ and $S$, respectively. We also set $s_1$ to $j$ and $s_2$ to the leftmost node in $Y[j]$. Then---both if $s_1 \lhd j$ or $s_1=j$---we set $r_1$ to $i$ and $r_2$ to the rightmost node in $X[i]$. We have thus updated $((r_1,r_2), (s_1,s_2))$ to be our new potential pair. That we only need the rightmost node in $X[i]$ and the leftmost node in $Y[j]$ follows directly from the definition of $\mop$.

Case 2 ($i \ntriangleleft j$) is more complicated. In this case we need to compute $\mop$ in the cluster $i$ and $j$ belongs to. If this results in any minimum ordered pairs ($r \neq \emptyset$) we must update our potential pair.  As in the previous case we compare $s_1$ and $j$, but this time we must also add $r_1$ and $s_1$ to the output if $s_1 =j$ and no nodes in $X[i]$ are to the left of the leftmost node in $s_2$. To see this first note that since $r_1 \lhd i$ (the input is deep) we must have $r_1 \neq s_1$ and thus $s_2$ contains only one node $s'$. If $s'$ is to the  left of all nodes in $X[i]$ then no node in $X[i]$ can be in a minimum ordered pair with $s'$ and we can safely add our potential pair to the output. We then update our potential pair. Finally, we need to update $X[i]$, $i$, and $j$.  This update depends on which kind of macro nodes we have been working on. In Case (a) we either have $i=j$  or $i$ is a left node and $j$ is a spine node. In both cases we can have nodes in $X[i]$ that are to not to the left of any node in $Y[j]$. The rightmost of these nodes can be in a minimum ordered pair with a node from another macro node and we thus update $X[i]$ to contain this node only (if it exists).  Now all nodes in $Y[j]$ must be to the left of all nodes in $X[i]$ in the next iteration and  thus we increment $j$. In Case (b) $i$ is a spine node and $j$ is a right node. If $r_2= \emptyset$ then no node in $Y[j]$ is to the right of the node in $X[i]$. Since the input arrays are deep, no node later in the array $X$ can be to the left of any node in $Y[j]$ and we therefore increment $j$. If $r_2 \neq \emptyset$ then the single node in $X[i]$ is in the potential pair and we increment $i$. We do not increment $j$ as there could be nodes in $X[j]$ to the left of the nodes in $Y[j]$.
When reaching the end of one of the arrays we add our potential pair to the output and return.

The correctness of the procedure follows from the proof of Lemma~\ref{t2:lem:nnm} and the above.
\\\\
Procedure \Match\ takes three node arrays $X$, $Y$, and $Y'$ representing deep sets $\mc{X}$, $\mc{Y}$, and $\mc{Y}'$, where 
$\norm{\mc{X}}=\norm{\mc{Y}}$, and $\mc{Y}' \subseteq \mc{Y}$. The output is a node array representing the set $\{\mc{X}_j \mid \mc{Y}_j \in \mc{Y'}\}$.
\begin{relate}
\item[$\Match(X,Y,Y')$]
Initialize a node array $R$  of size $n_M$, set $X_L:=\emptyset$, $Y_L:=\emptyset$, $Y_L':=\emptyset$, $x:=0$, $y:=0$, $i:=1$ and $j:=1$.

Repeat until $i>n_M$ or $j>n_M$:
    \begin{itemize}
    \item[] Until $X[i] \neq \emptyset$ set $i:=i+1$. Set $x:=\norm{X[i]}$.

    Until  $Y[j] \neq \emptyset$ set $j:=j+1$. Set $y:=\norm{Y[j]}$.

    Compare $Y[j]$ and $Y'[j]$. There are two cases:
        \begin{enumerate}
        \item $Y[j]=Y'[j]$. Compare $x$ and $y$. There are three cases:
            \begin{enumerate}
            \item $x=y$. Set $R[i]:=R[i] \cup X[i]$, $i:=i+1$, and $j:=j+1$.
            \item $x < y$. Set $R[i]:=R[i] \cup X[i]$, $
            Y[j]:=Y[j]\setminus \leftn(x,Y[j])$, $Y'[j]:=Y[j]$,  and $i:=i+1$.
            \item $x > y$. Set $X_L:=\leftn(y,X[i])$, $R[i]:=R[i] \cup X_L$, $X[i]:=X[i]\setminus X_L$, and $j:=j+1$.
            \end{enumerate}

        \item $Y[j]\neq Y'[j]$. Compare $x$ and $y$. There are three cases:
            \begin{enumerate}
            \item $x=y$. Set $R[i]:=R[i] \cup \match(X[i],Y[j],Y'[j])$, $i:=i+1$, and $j:=j+1$.
            \item $x < y$. Set $Y_L:=\leftn(x,Y[j])$, $Y'_L:=Y'[j] \cap Y_L$,
            
             $R[i]:=R[i] \cup \match(X[i],Y_L,Y'_L)$, $Y[j]:=Y[j]\setminus Y_L$, $Y'[j]:=Y'[j] \setminus Y'_L$,  and $i:=i+1$.

            \item $x > y$. Set $X_L:=\leftn(y,X[i])$, $R[i]:=R[i] \cup \match(X_L,Y[j],Y'[j])$, $X[i]:=X[i]\setminus X_L$, and $j:=j+1$.
            \end{enumerate}
        \end{enumerate}
    \end{itemize}
    Return $R$.
\end{relate}
Procedure $\Match$ proceeds as follows. First we find the next non-empty entries in the two node arrays $X[i]$ and $Y[j]$. We then compare $Y[j]$ and $Y'[j]$. 

If they are equal we keep all nodes in $X$ with the same rank as the nodes in $Y[j]$. We do this by splitting into three cases. If there are the same number of nodes $X[i]$ and $Y[j]$ we add all nodes in $X[i]$ to the output and increment $i$ and $j$. If there are more nodes in $Y[j]$ than in $X[i]$ we add all nodes in $X[i]$ to the output and update $Y[j]$ to contain only the $y-x$ lefmost nodes in $Y[j]$. We then increment $i$ and iterate. If there are more nodes in $X[i]$ than in $Y[j]$ we add the first $y$ nodes in $X[i]$ to the output, increment $j$,  and update $X[i]$ to contain only the nodes we did not add to the output. 

If $Y[j] \neq Y'[j]$ we call the cluster procedure \match. Again we split into three cases depending on the number of nodes in $X[i]$ and $Y[j]$. If they have the same number of nodes we can just call \match\ on $X[i]$, $Y[j]$, and $Y'[j]$ and increment $i$ and $j$. If $\norm{Y[j]} > \norm{X[i]}$ we call match with $X[i]$ the leftmost $\norm{X[i]}$ nodes of $Y[j]$ and with the part of $Y'[j]$ that are a subset of these leftmost $\norm{X[i]}$ nodes of $Y[j]$. We then update $Y[j]$ and $Y'[j]$ to contain only the nodes we did not use in the call to \match\ and increment $i$. If $\norm{Y[j]} < \norm{X[i]}$ we call \match\ with the leftmost $\norm{Y[j]}$ nodes of $X[i]$, $Y[j]$, and $Y'[j]$. We then update $X[i]$ to contain only the nodes we did not use in the call to \match\ and increment $j$. 

It follows by induction on the rank of the elements that the procedure is correct.

\begin{relate}
\item[$\Fl(X, \alpha)$:] Initialize a node array $R$ of size $n_M$ and two node lists $L$ and $S$.

Repeat until $i>n_M$:
\begin{itemize}
\item[] Until $X[i] \neq \emptyset$ set $i:=i+1$. 

There are three cases depending on the type of $i$:
\begin{enumerate}
  \item $i \in \{l(v,w), r(v,w)\}$. Compute $N := \Fl_{C(i, s(v,w), v)}(X[i], \alpha)$.
  
  If $N \neq \emptyset$ for each $j \in \{i, s(v,w), v\}$ set $R[j] = R[j] \cup (N \cap V(C(j)))$. 
  
  Otherwise,
  set $L := L \circ \parent_{{M}}(v)$.
  \item  $i = l(v)$. Compute $N := \Fl_{C(i, v)}(X[i])$.
  
   If $N \neq \emptyset$ for each $j \in \{i, v\}$, set $R[j] :=R[j] \cup (N \cap V(C(j)))$. 
   
   Otherwise,
  set $L := L \circ \parent_{{M}}(v)$.
  \item $i \not\in \{l(v,w), r(v,w),l(v)\}$. Compute $N := \Fl_{C(i)}(X[i], \alpha)$.
  
  If $N \neq \emptyset$ set $R[i] := R[i] \cup N$. 
  
  Otherwise set $L := L \circ \parent_{{M}}(i)$.
\end{enumerate}
\end{itemize}
Subsequently, compute the list $S := \Fl_{M}(L, \alpha)$. For each node $i \in S$ set $R[i] := R[i] \cup \Fl_{C(S[i])}(\first(S[i]), \alpha))$. Return $R$.
\end{relate}
The $\Fl$ procedure is similar to $\Parent$. The cases 1, 2 and 3
compute $\Fl$ on a micro forest. If the result is within the micro
tree we add it to $R$ and otherwise we store the node in the
macro tree which contains the parent of the root of the micro forest in
a node list $L$. Since we always call \Deep\
on the output from $\Fl(X,\alpha)$ there is no need to
compute \Fl\ in the macro tree if $N$ is nonempty. We then compute $\Fl$ in the macro tree on the
list $L$, store the results in a list $S$, and use this to compute the final result.

Consider the cases of procedure \Fl.  In Case 1 $i$ is a left or right node. Due to Proposition~\ref{t2:lem:ancestorlemma} case (i) and (ii) $\fl$ of a node in $i$ can be in $i$ on the spine or in the top boundary node. If this is not the case it can be found  by a computation of $\Fl$ of the parent of the top boundary node of the $i$'s cluster in the macro tree (Proposition~\ref{t2:lem:ancestorlemma} case (iii)).
In Case 2 $i$ is a leaf node. Then $\fl$ of a node in $i$ must either be in $i$, in the top boundary node, or can be found  by a computation of $\Fl$ of the parent of the top boundary node of the $i$'s cluster in the macro tree. If $i$ is a spine node or a boundary node   $\fl$ of a node in $i$ is either in $i$ or can be found by a computation of $\Fl$ of the parent of $i$ in the macro tree. 
 
The correctness of the procedure follows from Proposition~\ref{t2:lem:ancestorlemma}, the
above, and the correctness of procedure $\Fl_{M}$.

\subsection{Complexity of the Tree Inclusion Algorithm}
To analyse the complexity of the node array implementation we first bound the running time of the above implementation of the set procedures. All procedures scan the input from left-to-right while gradually producing the output. In addition to this procedure $\Fl$ needs a call to a node list implementation of $\Fl$ on the macro tree. Given the data structure described in Section~\ref{t2:sec:preprocessing} it is easy to check that each step in the scan can be performed in $O(1)$ time giving a total of $O(n_{T}/\log n_T)$ time. Since the number of nodes in the macro tree is $O(n_{T}/\log n_T)$ the call to the node list implementation of $\Fl$ is easily done within the same time. Hence, we have the following lemma.
\begin{lemma}\label{t2:lem:auxmacro}
For any tree $T$ there is a data structure using $O(n_T)$ space
and $O(n_T\log n_{T})$ preprocessing time which supports all of the
set procedures in $O(n_T/\log n_T)$ time.
\end{lemma}
Next consider computing the deep occurrences of $P$ in $T$ using
the procedure $\Emb$ of Section~\ref{t2:sec:recursion} and
Lemma~\ref{t2:lem:auxmacro}. Since each node $v \in V(P)$ contributes
at most a constant number of calls to set procedures it follows
immediately that,
\begin{theorem}\label{t2:thm:complex}
For trees $P$ and $T$ the tree inclusion problem can be solved in
$O(n_Pn_T/\log n_T + n_{T}\log n_{T})$ time and $O(n_T)$ space.
\end{theorem}
Combining the results in Theorems~\ref{t2:thm:simple},
\ref{t2:thm:complex} and Corollary~\ref{t2:cor:simple} we
have the main result of Theorem~\ref{t2:thm:main}.



\emptythanks
\chapter{Matching Subsequences in Trees}\label{chap:tree3}

\title{Matching Subsequences in Trees}

\author{Philip Bille \\ IT University of Copenhagen \\ \texttt{beetle@itu.dk}
\and Inge Li G{\o}rtz\thanks{This work was performed while the author was a PhD student at the IT University of Copenhagen.} \\ Technical University of Denmark \\ {\tt ilg@imm.dtu.dk}}

\date{}
\cleartooddpage

\maketitle
\begin{abstract}
Given two rooted, labeled trees $P$ and $T$ the tree path
subsequence problem is to determine which paths in $P$ are
subsequences of which paths in $T$. Here a path begins at the root and
ends at a leaf. In this paper we propose this problem as a useful
query primitive for XML data, and provide new algorithms improving
the previously best known time and space bounds.
\end{abstract}

\section{Introduction}
We say that a tree is \emph{labeled} if each node is assigned a
character from an alphabet $\Sigma$. Given two sequences of
labeled nodes $p$ and $t$, we say that $p$ is a \emph{subsequence}
of $t$, denoted $p \sqsubseteq t$, if $p$ can be obtained by
removing nodes from $t$. Given two rooted, labeled trees $P$ and
$T$ the \emph{tree path subsequence problem} (TPS) is to determine
which paths in $P$ are subsequences of which paths in $T$. Here a path
begins at the root and ends at a leaf. That is, for each path $p$ 
in $P$ we must report all paths $t$ in $T$ such that $p \sqsubseteq t$.

This problem was introduced by Chen~\cite{Chen2000} who gave an
algorithm using $O(\min(l_{P}n_{T} + n_P, n_{P}l_{T}+n_T))$ time and
$O(l_{P}d_{T} + n_P + n_T)$ space. Here, $n_{S}$, $l_{S}$, and $d_{S}$ denotes
the number of nodes, number of leaves, and depth, respectively, of
a tree $S$. Note that in the worst-case this is quadratic time and
space. In this paper we present improved algorithms giving the following result:
\begin{theorem}\label{t3:main}
For trees $P$ and $T$ the tree path subsequence problem can be 
solved in $O(n_{P} + n_{T})$ space with the following running times:
\begin{equation*}
\min
\begin{cases}
    O(l_{P}n_{T} + n_P) , \\
     O(n_{P}l_{T}+n_T) , \\
     O(\frac{n_{P}n_{T}}{\log n_{T}}+ n_T + n_P \log n_P).
\end{cases}
\end{equation*}
\end{theorem}
The first two bounds in Theorem~\ref{t3:main} match the previous time bounds while improving the space to linear. This is achieved using a algorithm that resembles the algorithm of Chen~\cite{Chen2000}. At a high level, the algorithms are essentially identical and therefore the bounds should be regarded as an improved analysis of Chen's algorithm. The latter bound is obtained by using an entirely new algorithm that improves the worst-case quadratic time. Specifically, whenever $\log n_{P} = O(n_{T}/ \log n_{T})$ the running time is improved by a logarithmic factor. Note that -- in the worst-case -- the number of pairs consisting of a path from $P$ and a path $T$ is $\Omega(n_{P}n_{T})$, and therefore we need at least as many bits to report the solution to TPS. Hence, on a RAM with logarithmic word size our worst-case bound is optimal. Most importantly, all our algorithms use linear space. For practical applications this will likely make it possible to solve TPS on large trees and improve running time since more of the computation can be kept in main memory.

\subsection{Applications}\label{t3:applications} We propose TPS as a useful query
primitive for XML data. The key idea is that an XML document $D$
may be viewed as a rooted, labeled tree.
\begin{figure}[t]
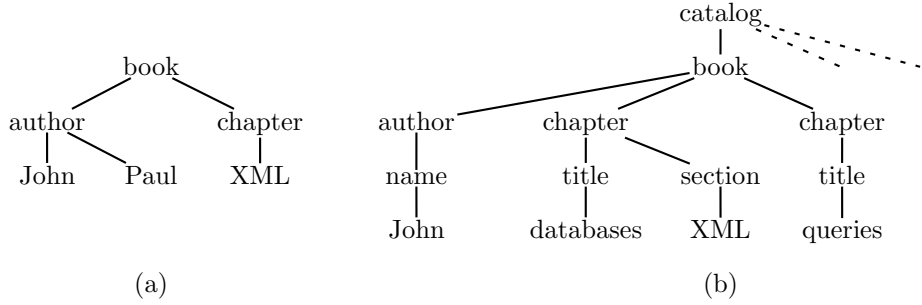

\begin{center}
  \begin{psmatrix}[colsep=0.5cm,rowsep=0.3cm,labelsep=1pt, nodesep=1pt]
  &&&&&&&&[name=cat] catalog \\
  &&[name=book] book &&&&&&[name=book2] book & [name=book3] & [name=book4]\\
  & [name=author] author &&[name=chapter] chapter&&
  [name=author2] author && [name=chapter2] chapter && [name=chapter3] chapter \\
  & [name=john] John & [name=paul] Paul & [name=xml] XML &&
  [name=name] name && [name=title] title  & [name=section] section & [name=title2] title \\
  &&&&&  [name=john2] John && [name=DB] databases & [name=xml2] XML & [name=queries] queries \\
  && (a) &&&&&& (b) 
  \ncline{author}{book} \ncline{chapter}{book}
  \ncline{john}{author} \ncline{paul}{author} \ncline{xml}{chapter}
  \ncline{cat}{book2}\psset{linestyle=dashed,dash=2pt 4pt}\ncline{cat}{book3} \ncline{cat}{book4}
  \psset{linestyle=solid}
  \ncline{book2}{author2}\ncline{book2}{chapter2}\ncline{book2}{chapter3}
  \ncline{author2}{name}\ncline{chapter2}{title}\ncline{chapter2}{section}\ncline{chapter3}{title2}
  \ncline{name}{john2}\ncline{title}{DB}\ncline{section}{xml2}\ncline{title2}{queries}
  \end{psmatrix}
   \caption{(a) The trie of queries 1,2,3, or the tree for query 4. (b)
   A fragment of a catalog of books.}
  \label{t3:xmlexample}
  \end{center}
\end{figure}
For example, suppose that we want to maintain a catalog of books
for a bookstore. A fragment of a possible XML tree, denoted $D$,
corresponding to the catalog is shown in Fig.~\ref{t3:xmlexample}(b).
In addition to supporting full-text queries, such as find all
documents containing the word ``John'', we can also use the tree
structure of the catalog to ask more specific queries, such as the
following examples:
\begin{enumerate}
  \item Find all books written by John,
  \item find all books written by Paul,
  \item find all books with a chapter that has something to do with XML, or
  \item find all books written by John and Paul with a chapter that has something to do with XML.
\end{enumerate}
The queries 1,2, and 3 correspond to a \emph{path query} on $D$,
that is, compute which paths in $D$ that contains a specific path
as a subsequence. For instance, computing the paths in $D$ that
contain the path of three nodes labeled ``book'', ``chapter'', and
``XML'', respectively, effectively answers query 3. Most XML-query
languages, such as XPath \cite{CD1999}, support such queries.

Using a depth-first traversal of $D$ a path query can be solved in linear time.
More precisely, if $q$ is a path consisting of $n_{q}$ nodes, answering the path query on $D$ takes $O(n_{q} + n_{D})$ time. Hence, if we are given path queries $q_{1}, \ldots, q_{k}$ we can
answer them in $O(n_{q_{1}} + \cdots + n_{q_{k}} + kn_{D})$ time. However, we can do better by constructing the \emph{trie}, $Q$, of $q_{1}, \ldots, q_{k}$. Answering all paths queries now correspond to solving TPS on $Q$ and $D$. As an example the queries 1,2, and 3 form the trie shown in Fig.~\ref{t3:xmlexample}(a). As $l_Q \leq k$, Theorem~\ref{t3:main} gives us an algorithm with running time 
\begin{equation}
\label{t3:bound}
O\left(n_{q_{1}} + \cdots + n_{q_{k}} + \min\left(kn_D + n_Q, n_Ql_D + n_D, \frac{n_Qn_D}{\log n_D} + n_D + n_Q\log n_Q\right)\right). 
\end{equation}
Since $n_Q \leq n_{q_{1}} + \cdots + n_{q_{k}}$ this is at least as good as answering the queries individually and better in many cases. If many paths share a prefix, i.e., queries 1 and 2 share "book" and "author", 
the size of $n_Q$ can much smaller than $n_{q_{1}} + \cdots + n_{q_{k}}$. Using our solution to TPS we can efficiently take advantage of this situation since the latter two terms in \eqref{t3:bound} depend on $n_Q$ and not on $n_{q_{1}} + \cdots + n_{q_{k}}$.  

Next consider query 4. This query cannot be answered by solving a
TPS problem but is an instance of the \emph{tree inclusion
problem} (TI). Here we want to decide if $P$ is \emph{included} in
$T$, that is, if $P$ can be obtained from $T$ by \emph{deleting}
nodes of $T$. Deleting a node $y$ in $T$ means making the children
of $y$ children of the parent of $y$ and then removing $y$. It is
straightforward to check that we can answer query 4 by deciding if
the tree in Fig.~\ref{t3:xmlexample}(a) can be included in the tree
in Fig.~\ref{t3:xmlexample}(b).

Recently, TI has been recognized as an important XML query
primitive and has recieved considerable attention, see e.g.,
\cite{SM2002, YLH2003,YLH2004, ZADR03, SN2000, TRS2002}.
Unfortunately, TI is NP-complete in general 
\cite{KM1995} and therefore the existing algorithms are based on
heuristics. Observe that a necessary condition for $P$ to included
in $T$ is that all paths in $P$ are subsequences of paths in $T$.
Hence, we can use TPS to quickly identify trees or parts of trees that cannot be
included $T$. We believe that in this way TPS can be used as an
effective "filter" for many tree inclusion problems that occur in
practice.


\subsection{Technical Overview} Given two strings (or labeled
paths) $a$ and $b$, it is straightforward to determine if $a$ is a
subsequence of $b$ by scanning the character from left to right in
$b$. This uses $O(|a| + |b|)$ time. We can solve TPS by applying
this algorithm to each of the pair of paths in $P$ and $T$,
however, this may use as much as $O(n_Pn_T(n_P + n_T))$ time.
Alternatively, Baeza-Yates \cite{BaezaYates1991} showed how to
preprocess $b$ in $O(|b|\log |b|)$ time such that testing whether
$a$ is a subsequence of $b$ can be done in $O(|a|\log |b|)$ time.
Using this data structure on each path in $T$ we can solve the TPS
problem, however, this may take as much as $O(n^2_{T}\log n_{T} +
n_{P}^{2}\log n_{T})$. Hence, none of the availiable subsequence algorithms
on strings provide an immediate efficient solution to TPS. 

Inspired by the work of Chen~\cite{Chen2000} we take another
approach. We provide a framework for solving TPS. The main
idea is to traverse $T$ while maintaining a subset of nodes in
$P$, called the \emph{state}. When reaching a leaf $z$ in $T$ the
state represents the paths in $P$ that are a subsequences of the
path from the root to $z$. At each step the state is updated using a simple
procedure defined on subset of nodes. The result of
Theorem~\ref{t3:main} is obtained by taking the best of two
algorithms based on our framework: The first one uses a simple
data structure to maintain the state. This leads to an algorithm
using $O(\min(l_{P}n_{T} + n_P, n_{P}l_{T}+n_T))$ time. At a high level this
algorithm resembles the algorithm of Chen~\cite{Chen2000} and
achieves the same running time. However, we improve the analysis
of the algorithm and show a space bound of  $O(n_P +
n_T)$. This should be compared to the worst-case quadratic space
bound of $O(l_{P}d_{T} + n_P + n_T)$ given by Chen~\cite{Chen2000}. Our second
algorithm takes a different approach combining several 
techniques. Starting with a simple quadratic time and space
algorithm, we show how to reduce the space to $O(n_P \log n_T)$
using a decomposition of $T$ into disjoint paths. We then divide
$P$ into small subtrees of logarithmic size called \emph{micro
trees}. The micro trees are then  preprocessed such that subsets of
nodes in a micro tree can be maintained in constant time and
space. Intuitively, this leads to a logarithmic improvement of the time and space bound.



\subsection{Notation and Definitions} In this section we define
the notation and definitions we will use throughout the paper. For
a graph $G$ we denote the set of nodes and edges by $V(G)$ and
$E(G)$, respectively. Let $T$ be a rooted tree. The root of $T$ is
denoted by $\roots(T)$. The \emph{size} of $T$, denoted by $n_T$,
is $|V(T)|$. The \emph{depth} of a node $y\in V(T)$, $\depth(y)$,
is the number of edges on the path from $y$ to $\roots(T)$ and the
depth of $T$, denoted $d_T$, is the maximum depth of any node in
$T$. The parent of $y$ is denoted $\parent(y)$. A node with no
children is a leaf and otherwise it is an internal node. The
number of leaves in $T$ is denoted $l_T$. Let $T(y)$ denote the
subtree of $T$ rooted at a node $y \in V(T)$. If $z\in V(T(y))$
then $y$ is an ancestor of $z$ and if $z\in V(T(y))\backslash
\{y\}$ then $y$ is a proper ancestor of $z$. If $y$ is a (proper)
ancestor of $z$ then $z$ is a (proper) descendant of $y$. We say
that $T$ is \emph{labeled} if each node $y$ is assigned a
character, denoted $\lab(y)$, from an alphabet $\Sigma$. The path
from $y$ to $\roots(T)$, of nodes $\roots(T) = y_1, \ldots, y_k =
y$ is denoted $\path(y)$. Hence, we can formally state TPS as
follows: Given two rooted tree $P$ and $T$ with leaves $x_1,
\ldots, x_r$ and $y_1, \ldots, y_s$, respectively, determine all
pairs $(i,j)$ such that $\path(x_{i}) \sqsubseteq \path(y_{j})$.
For simplicity we will assume that leaves in $P$ and $T$ are
always numbered as above and we identify each of the paths by the
number of the corresponding leaf.

Throughout the paper we assume a unit-cost RAM model of computation
with word size $\Theta(\log n_T)$ and a standard instruction set including bitwise boolean operations, shifts, addition and multiplication. All space complexities refer to the number of words used by the algorithm.

\section{A Framework for solving TPS}
In this section we present a simple general
algorithm for the tree path subsequence problem. The key
ingredient in our algorithm is the following procedure. For any $X
\subseteq V(P)$ and $y \in V(T)$ define:
\begin{relate}
\item[$\Down(X,y)$:] Return the set
    $\Child(\{x \in X \mid \lab(x) = \lab(y)\})
    \cup \{x \in X \mid \lab(x) \neq \lab(y)\}$.
\end{relate}
The notation $\Child(X)$ denotes the set of children of $X$. 
Hence, $\Down(X,y)$ is the set consisting of nodes in $X$ with a
different label than $y$ and the children of the nodes $X$ with
the same label as $y$. We will now show how to solve TPS using
this procedure.

First assign a unique number in the range $\{1,\ldots,l_{P}\}$ to
each leaf in $P$. Then, for each $i$, $1\leq i \leq l_{P}$, add a
\emph{pseudo-leaf} $\bot_{i}$ as the single child of the $i$th
leaf. All pseudo-leaves are assigned a special label $\beta
\not\in \Sigma$. The algorithm traverses $T$ in a depth first
order and computes at each node $y$ the set $X_{y}$. We call this
set the \emph{state} at $y$. Initially, the state consists of
$\{\roots(P)\}$. For $z \in \child(y)$, the state $X_z$ can be
computed from state $X_y$ as 
\begin{equation*}
X_{z} =   \Down(X_{y}, z).
\end{equation*}
If $z$ is a leaf we report the number of each pseudo-leaf in $X_z$
as the paths in $P$ that are subsequences of $\path(z)$. See
Figure~\ref{t3:fig:framework}  for an example. To show
the correctness of this approach we need the following lemma.

\begin{figure}[tb]
\begin{center}
\begin{psmatrix}[colsep=0.6cm,rowsep=0.3cm,labelsep=3pt]
  && \cnode{.2}{rootp}\rput(0,0){$a$}\rput(0,.4){$\roots(P)$} &&&&&&
  \cnode{.2}{roott}\rput(0,0){$a$}\rput(0,.4){$\roots(T)$}
  \\
  &\cnode{.2}{v1}\rput(0,0){$c$}\rput(.5,0){$x_1$} & &
  \cnode{.2}{v2}\rput(0,0){$b$}\rput(.5,0){$x_2$}
  &&&&& \cnode{.2}{w1}\rput(0,0){$c$}\rput(.4,0){$1$}\\
 &\cnode{.2}{v3}\rput(0,0){$a$}\rput(.5,0){$x_3$} & &
  [name=b2]\psframebox{}\rput(.3,0){$\bot_2$}
  &&&&\cnode{.2}{w2}\rput(0,0){$a$}\rput(.4,0){$2$} & &
  \cnode{.2}{w3}\rput(0,0){$b$}\rput(.4,0){$4$}
  \\
  &[name=b1] \psframebox{}\rput(.3,0){$\bot_1$}
  &&&&&&\cnode{.2}{w4}\rput(0,0){$b$}\rput(.4,0){$3$} & &
  \cnode{.2}{w5}\rput(0,0){$b$}\rput(.4,0){$5$}
  \\
  && $P$ &&&&&& $T$

  \ncline{rootp}{v1}\ncline{rootp}{v2}
  \ncline{v1}{v3}\ncline{v2}{b2}\ncline{v3}{b1}
  \ncline{roott}{w1}
  \ncline{w1}{w3}\ncline{w1}{w2}\ncline{w2}{w4}
    \ncline{w3}{w5}
  \end{psmatrix}
  \caption{The letters inside the nodes are the labels, and the
  identifier of each node is written outside the node. Initially we have $X=\{\roots(P)\}$.
  Since $\lab(\roots(P))=a=\lab(\roots(T))$ we replace $\roots(P)$ with is children and
  get
  $X_{\roots(T)}=\{x_1,x_2\}$. Since
  $\lab(1)=\lab(x_1)\neq\lab(x_2)$ we get
  $X_1=\{x_3,x_2\}$. Continuing this way we get
  $X_2=\{\bot_1,x_2\}$, $X_3=\{\bot_1,\bot_2\}$, $X_4=\{x_3,\bot_2\}$,
  and $X_5=\{x_3,\bot_2\}$. The nodes $3$ and $5$ are leaves of $T$ and we thus
  report paths $1$ and $2$ after computing $X_3$ and path $2$ after computing
  $X_5$.}\label{t3:fig:framework}
  \end{center}
\end{figure}
\begin{lemma}\label{t3:lem:invariant}
For any node $y \in V(T)$ the state $X_{y}$ satisfies the
following property: 
$$x \in X_{y} \Rightarrow \path(\parent(x)) \sqsubseteq \path(y)\;.$$
\end{lemma}
\begin{proof}
By induction on the number of iterations of the procedure.
Initially, $X = \{\roots(P)\}$ satisfies the property
 since $\roots(P)$ has no parent. Suppose that
$X_{y}$ is the current state and $z\in \child(y)$ is the next node
in the depth first traversal of $T$. By the induction hypothesis
$X_{y}$ satisfies the property, that is, for
any $x \in X_{y}$, $\path(\parent(x)) \sqsubseteq \path(y))$. Then, 
\begin{equation*}
X_{z} = \Down(X_{y},z) =  \Child(\{x \in X_{y} \mid \lab(x) =
\lab(z)\}) \cup \{x \in X_{y} \mid \lab(x) \neq \lab(z)\}\;.
\end{equation*}
Let $x$ be a node in $X_y$. There are two cases. If $\lab(x) =
\lab(z)$ then $\path(x) \sqsubseteq \path(z)$ since
$\path(\parent(x)) \sqsubseteq \path(y)$. Hence, for any child
$x'$ of $x$ we have $\path(\parent(x')) \sqsubseteq \path(z)$. On
the other hand, if $\lab(x) \neq \lab(z)$ then $x \in X_z$. Since
$y=\parent(z)$ we have $\path(y) \sqsubseteq \path(z)$, and hence
$\path(\parent(x)) \sqsubseteq \path(y) \sqsubseteq \path(z)$.
\qed \end{proof}
By the above lemma all paths reported at a leaf $z
\in V(T)$ are subsequences of $\path(z)$. The following lemma shows that the paths reported at a leaf $z \in
V(T)$ are \emph{exactly} the paths in $P$ that are subsequences of $\path(z)$.
\begin{lemma}\label{t3:lem:correct}
Let $z$ be a leaf in $T$ and let $\bot_i$ be a pseudo-leaf in $P$.
Then,
\begin{equation*}
\bot_i \in X_{z} \Leftrightarrow
\path(\parent(\bot_i)) \sqsubseteq \path(z)\;.
\end{equation*}
\end{lemma}
\begin{proof}
It follows immediately from Lemma~\ref{t3:lem:invariant} that $\bot_i
\in X_{z} \Rightarrow \path(\parent(\bot_i)) \sqsubseteq
\path(z)$. It remains to show that $\path(\parent(\bot_i))
\sqsubseteq \path(z) \Rightarrow \bot_i \in X_{z}$. Let
$\path(z)=z_1,\ldots,z_k$, where $z_1=\roots(T)$ and $z_k=z$, and
let $\path(\parent(\bot_i))=y_1,\ldots,y_\ell$, where
$y_1=\roots(P)$ and $y_\ell=\parent(\bot_i)$. Since
$\path(\parent(\bot_i)) \sqsubseteq \path(z)$ there are nodes
$z_{j_i}=y_i$ for $1\leq i\leq k$, such that (i) $j_i< j_{i+1}$
and (ii) there exists no node $z_j$ with $\lab(z_j)=\lab(y_i)$,
where $j_{i-1}<j<j_i$. Initially, $X=\{\roots(P)\}$. We have
$\roots(P) \in X_{z_j}$ for all $j<j_{1}$, since $z_{j_1}$ is the
first node on $\path(z)$ with label $\lab(\roots(P))$. When we get
to $z_{j_1}$, $\roots(P)$ is removed from the state and $y_2$ is
inserted. Similarly, $y_i$ is in all states $X_{z_j}$ for $j_{i-1}
\leq j <j_i$. It follows that $\bot_i$ is in all states $X_{z_j}$
where $j\geq j_\ell$ and thus $\bot_i \in X_{z_k}=X_z$.
\qed \end{proof}
The next lemma can be used to give an upper bound on the number of
nodes in a state.
\begin{lemma}\label{t3:lem:statesize}
For any $y\in V(T)$ the state $X_y$ has the following property:
Let $x \in X_y$. Then no ancestor of $x$ is in $X_y$.
\end{lemma}
\begin{proof}
 By induction on the length of $\path(y)$.
Initially, the state only contains $\roots(P)$. Let $z$ be the
parent of $y$, and thus $X_y$ is computed from $X_z$. First we
note that for all nodes $x\in X_y$ either $x\in X_z$ or
$\parent(x)\in X_z$. If $x\in X_z$ it follows from the induction
hypothesis that no ancestor of $x$ is in $X_z$, and thus no
ancestors of $x$ can be in $X_y$. If $\parent(x)\in X_z$ then due
to the definition of \Down\ we must have $\lab(x)=\lab(y)$. It
follows from the definition of \Down\ that $\parent(x)\not \in
X_y$.
\qed \end{proof}
It follows from Lemma~\ref{t3:lem:statesize} that $|X_y|\leq l_P$ for
any $y\in V(T)$. If we store the
state in an unordered linked list each step of the depth-first
traversal takes time $O(l_{P})$ giving a total $O(l_{P}n_{T} +n_{P})$
time algorithm.
Since each state is of size at most $l_P$ the space used is
$O(n_{P} + l_P n_{T})$. In the following sections we show how to improve these bounds.

\section{A Simple Algorithm}
In this section we consider a simple
implementation of the above algorithm, which has running time
$O\left(\min(l_Pn_T +n_P, n_Pl_T + n_T)\right)$ and uses $O(n_P + n_T)$ space. We assume
that the size of the alphabet is $n_T + n_P$ and each character in
$\Sigma$ is represented by an integer in the range $\{1,\ldots,n_T
+ n_P\}$. If this is not the case we can sort all characters in
$V(P) \cup V(T)$ and replace each label by its rank in the sorted
order. This does not change the solution to the problem, and
assuming at least a logarithmic number of leaves in both trees it
does not affect the running time. To get the space usage down to
linear we will avoid saving all states. For this purpose we
introduce the procedure \Up, which reconstructs the state $X_z$
from the state $X_y$, where $z=\parent(y)$. We can thus save space
as we only need to save the current state.

We use the following data structure to represent the current state $X_y$: A
\emph{node dictionary} consists of two dictionaries denoted
$X^{c}$ and $X^{p}$.  The dictionary $X^c$ represents the node set
corresponding to $X_y$, and the dictionary $X^p$ represents the
node set corresponding to the set $\{x \in X_z \mid x \not\in X_y
\text{ and } z \text{ is an ancestor of } y\}$.  That is, $X^c$
represents the nodes in the current state, and $X^p$ represents
the nodes that is in a state $X_z$, where $z$ is an ancestor of
$y$ in $T$, but not in $X_y$. We will use $X^p$ to reconstruct
previous states. The dictionary $X^c$ is indexed by $\Sigma$ and
$X^p$ is indexed by $V(T)$.
The subsets stored at each entry are represented by doubly-linked
lists. Furthermore, each node in $X^{c}$ maintains a pointer to
its parent in $X^{p}$ and each node $x'$ in $X^{p}$ stores a
linked list of pointers to its children in $X^{p}$.
With this representation the total size of the node dictionary is
$O(n_{P}+n_T)$.

%
Next we show how to solve the tree path subsequence problem in our
framework using the node dictionary representation. For
simplicity, we add a node $\top$ to $P$ as a the parent of
$\roots(P)$. Initially, the $X^{p}$ represents $\top$ and $X^{c}$
represents $\roots(P)$. The $\Down$ and $\Up$ procedures are
implemented as follows:
\begin{relate}
\item[$\Down((X^{p}, X^{c}), y)$:]\begin{enumerate}
\item Set
$X:=X^c[\lab(y)]$ and $X^c[\lab(y)]:=\emptyset$.

\item For each $x\in X$ do:
\begin{enumerate}
\item Set 
$X^p[y] := X^p[y] \cup \{x\}$. 
\item For each $x' \in \child(x)$ do:
\begin{enumerate}
  \item Set $X^c[\lab(x')] := X^c[\lab(x')] \cup \{x\}$.
  \item Create pointers between $x'$ and $x$.
\end{enumerate}
\end{enumerate}
\item Return $(X^{p}, X^{c})$.
\end{enumerate}
\item[$\Up((X^{p}, X^{c}), y)$:]
\begin{enumerate}
\item Set $X:=X^p[y]$ and $X^p[y]:=\emptyset$.

\item For each $x \in X$ do:
\begin{enumerate}
\item Set
$X^c[\lab(x)] := X^c[\lab(x)] \cup \{x\}$.
\item For each $x' \in \child(x)$ do:
\begin{enumerate}
\item Remove pointers between $x'$ and $x$.
\item Set $X^c[\lab(x')] := X^c[\lab(x')] \setminus \{x'\}$.
\end{enumerate}
\end{enumerate}
\item Return $(X^{p}, X^{c})$.
\end{enumerate}
\end{relate}
The next lemma shows that 
\Up\ correctly reconstructs
the former state.
\begin{lemma}\label{t3:lem:updown}
Let $X_z=(X^c,X^p)$ be a state computed at a node $z\in V(T)$, and
let $y$ be a child of $z$. Then, 
\begin{equation*}
X_z=\Up(\Down(X_z,y),y)\;.
\end{equation*}
\end{lemma}
\begin{proof}
Let $(X^c_1,X^p_1)= \Down(X_z,y)$ and $(X^c_2,X^p_2)=
\Up((X^c_1,X^p_1),y)$.
We will first show that $x \in X_z \Rightarrow x \in
\Up(\Down(X_z,y),y)$.

Let $x$ be a node in $X^c$. There are two cases. If $x \in
X^c[\lab(y)]$, then it follows  from the implementation of \Down\
that $x \in X^p_1[y]$. By the implementation of \Up, $x \in
X^p_1[y]$ implies $x \in X^c_2$. If $x \not\in X^c[\lab(y)]$ then
$x\in X^c_1$. We need to show $\parent(x) \not\in X^p_1[y]$. This
will imply $x\in X^c_2$, since the only nodes removed from $X^c_1$
when computing $X^c_2$ are the nodes with a parent in $X^p_1[y]$.
Since $y$ is unique it follows from the implementation of \Down\
that $\parent(x)\in X^p_1$ implies $x\in X^c[\lab(y)]$.

Let $x$ be a node in $X^p$. Since $y$ is unique we have $x \in
X^p[y']$ for some $y'\neq y$.  It follows immediately from the
implementation of \Up\ and \Down\ that
$X^p[y']=X^p_1[y']=X^p_2[y']$, when $y'\neq y$, and thus
$X^p=X^p_2$.

We will now show $x \in \Up(\Down(X_z,y),y) \Rightarrow x \in
X_z$.
Let $x$ be a node in $X^c_2$.  There are two cases. If $x \not \in
X^c_1$ then it follows from the implementation of \Up\ that $x \in
X^p_1[y]$. By the implementation of \Down, $x \in X^p_1[y]$
implies $x \in X^c[\lab(y)]$, i.e., $x\in X^c$. If $x\in X_1^c$
then by the implementation of \Up, $x \in X^c_2$ implies
$\parent(x) \not\in x^p_1[y]$. It follows from the implementation
of \Down\ that $x\in X^c$. Finally, let $x$ be a node in $X^p_2$.
As argued above $X^p=X^p_2$, and thus $x \in X^p$.
\qed \end{proof}
From the current state $X_{y}=(X^c,X^p)$ the next state $X_{z}$ is
computed as follows:
\begin{equation*}
X_{z} =
\begin{cases}
   \Down(X_{y}, z)   & \text{if $y = \parent(z)$}, \\
   \Up(X_{y},y)  & \text{if $z = \parent(y)$}.
\end{cases}
\end{equation*}
The correctness of the algorithm follows from
Lemma~\ref{t3:lem:correct} and Lemma~\ref{t3:lem:updown}. 
We will now analyze the running time of the algorithm. The procedures
\Down\ and \Up\ uses time linear in the size of the current state
and the state computed.
By Lemma~\ref{t3:lem:statesize} the size of each state is $O(l_P)$.
Each step in the depth-first traversal thus takes time $O(l_P)$,
which gives a total running time of $O(l_P n_T+n_P)$. On the other
hand consider a path $t$ in $T$. We will argue that the
computation of all the states along the path takes total time
$O(n_P+n_t)$, where $n_T$ is the number of nodes in $t$. 
To show this we need the following lemma.
\begin{lemma}\label{t3:lem:Tpath}
Let $t$ be a path in $T$. During the computation of the states
along the path $t$, any node $x\in V(P)$ is inserted into $X^c$ at
most once.
\end{lemma}
\begin{proof}
Since $t$ is a path we only need to consider the \Down\
computations. The only way a node $x\in V(P)$ can be inserted into
$X^c$ is if $\parent(x)\in X^c$. It thus follows from
Lemma~\ref{t3:lem:statesize} that $x$ can be inserted into $X^c$ at
most once. 
\qed \end{proof}
It follows from Lemma~\ref{t3:lem:Tpath} that 
the computations of the all states when $T$ is a path takes time
$O(n_P+n_T)$. Consider a path-decomposition of $T$. A
path-decomposition of $T$ is a decomposition of $T$ into disjoint
paths. We can make such a path-decomposition of the tree $T$
consisting of $l_T$ paths. Since the running time of \Up\ and
\Down\ both are linear in the size of the current and computed
state it follows from Lemma~\ref{t3:lem:updown} that we only need to
consider the total cost of the \Down\ computations on the paths in
the path-decompostion. Thus, the algorithm uses time at most
$\sum_{t\in T}O(n_p + n_t)=O(n_P l_T+ n_T)$.

Next we consider the space used by the algorithm. Lemma~\ref{t3:lem:statesize}
implies that $|X^c|\leq l_P$. Now consider the size of $X^p$. A
node is inserted into $X^p$ when it is removed from $X^c$. It is
removed again when inserted into $X^c$ again. Thus
Lemma~\ref{t3:lem:Tpath} implies $|X^p| \leq n_P$ at any time. The
total space usage is thus $O(n_P+n_T)$.
To summarize we have shown,
\begin{theorem}\label{t3:simple} For trees $P$ and $T$ the tree path subsequence
problem can be solved in $O(n_P + n_T)$ space and $O\left(\min(l_Pn_T + n_P, n_Pl_T+n_T)\right)$ time.
\end{theorem}

\section{A Worst-Case Efficient Algorithm}\label{t3:worstcase}
In this section we consider the worst-case
complexity of TPS and present an algorithm using subquadratic
running time and linear space. The new algorithm works within our framework but does not use the $\Up$
procedure or the node dictionaries from the previous section.

Recall that using a simple linked list to represent the states we
immediately get an algorithm using $O(n_{P}n_{T})$ time and space.
We first show how to modify the traversal of $T$ and discard
states along the way such that at most $O(\log n_{T})$ states are
stored at any step in the traversal. This improves the space to
$O(n_{P}\log n_{T})$. Secondly, we decompose $P$ into small
subtrees, called \emph{micro trees}, of size $O(\log n_{T})$. Each micro tree can be represented
in a single word of memory and therefore a state uses only
$O(\ceil{\frac{n_{P}}{\log n_{T}}})$ space. In total the space used to
represent the $O(\log n_{T})$ states is $O(\ceil{\frac{n_{P}}{\log
n_{T}}} \cdot \log n_{T}) = O(n_{P} + \log n_T)$. Finally, we show how to
preprocess $P$ in linear time and space such that computing the
new state can be done in constant time per micro tree.
Intuitively, this achieves the $O(\log n_{T})$ speedup.
\subsection{Heavy Path Traversal}\label{t3:heavy} 
In this section we present the
modified traversal of $T$. We first partition $T$ into disjoint
paths as follows. For each node $y\in V(T)$ let $\size(y) =
|V(T(y))|$. We classify each node as either \emph{heavy} or
\emph{light} as follows. The root is light. For each internal node
$y$ we pick a child $z$ of $y$ of maximum size among the children
of $y$ and classify $z$ as heavy. The remaining children are
light. An edge to a light child is a \emph{light edge}, and an
edge to a heavy child is a \emph{heavy edge}. The heavy child of a
node $y$ is denoted $\heavy(y)$. Let $\ldepth(y)$ denote the
number of light edges on the path from $y$ to $\roots(T)$.
\begin{lemma}[Harel and Tarjan \cite{HT1984}]\label{t3:lightdepth}
For any tree $T$ and node $y\in V(T)$,  $\ldepth(y) \leq \log
n_{T} + O(1)$.
\end{lemma}
Removing the light edges, $T$ is partitioned into \emph{heavy
paths}. We traverse $T$ according to the heavy paths using the
following procedure. For  node $y \in V(T)$ define:
\bigskip
\begin{relate}
  \item[$\Visit(y)$:]  
  \begin{enumerate}
  \item If $y$ is a leaf report all leaves in $X_y$ and return.
  \item Else let $y_{1}, \ldots, y_{k}$ be the light children of $y$ and let $z = \heavy(y)$.
  \item For $i:= 1$ to $k$ do:
  \begin{enumerate}
    \item Compute $X_{y_{i}} := \Down(X_{y}, y_{i})$
    \item Compute $\Visit(y_{i})$.
  \end{enumerate}
  \item Compute $X_{z} := \Down(X_{y}, z)$.
  \item Discard $X_{y}$ and compute $\Visit(z)$.
\end{enumerate}
\end{relate}
The procedure is called on the root node of $T$ with the initial
state $\{\roots(P)\}$. The traversal resembles a depth first
traversal, however, at each step the light children are visited
before the heavy child. We therefore call this a \emph{heavy path
traversal}. Furthermore, after the heavy child (and therefore all
children) has been visited we discard $X_{y}$. At any step we have
that before calling $\Visit(y)$ the state $X_{y}$ is availiable,
and therefore the procedure is correct. We have the following
property:
\begin{lemma}\label{t3:traversal}
For any tree $T$ the heavy path traversal stores at most $\log
n_{T} + O(1)$ states.
\end{lemma}
\begin{proof}
At any node $y \in V(T)$ we store at most one state for each of
the light nodes on the path from $y$ to $\roots(T)$. Hence, by
Lemma~\ref{t3:lightdepth} the result follows. 
\qed \end{proof}
Using the heavy-path traversal immediately gives an $O(n_Pn_T)$ time and $O(n_P\log n_T)$ space algorithm. In the following section we improve the time and space by an additional $O(\log n_T)$ factor.

\subsection{Micro Tree Decomposition} 
In this section we present
the decomposition of $P$ into small subtrees. A \emph{micro tree}
is a connected subgraph of $P$. A set of micro trees $MS$ is a
\emph{micro tree decomposition} iff $V(P) = \cup_{M \in MS} V(M)$
and for any $M, M' \in MS$, $(V(M) \backslash \{\roots(M)\}) \cap
(V(M') \backslash \{\roots(M')\}) = \emptyset$.
 Hence, two micro trees in a decomposition share at most one node and this node must be the root in at least one of the micro trees. If $\roots(M') \in V(M)$ then $M$ is the \emph{parent} of $M'$ and $M'$ is the \emph{child} of $M$. A micro tree with no children is a \emph{leaf} and a micro tree  with no parent is a \emph{root}. Note that we may have several root micro trees since they can overlap at the node $\roots(P)$. We decompose $P$ according to the following classic result:
\begin{lemma}[Gabow and Tarjan \cite{GT1989}]\label{t3:clustering}
For any tree $P$ and parameter $s > 1$, it is possible to build  a
micro tree decomposition $MS$ of $P$ in linear time such that
$|MS| = O(\ceil{n_P/s})$ and $|V(M)| \leq s$ for any $M \in MS$
\end{lemma}

\subsection{Implementing the Algorithm} In this section we show
how to implement the $\Down$ procedure using the micro tree
decomposition. First decompose $P$ according to
Lemma~\ref{t3:clustering} for a parameter $s$ to be chosen later.
Hence, each micro tree has at most $s$ nodes and $|MS| =
O(\ceil{n_P/s})$. We represent the state $X$ compactly using a bit vector
for each micro tree. Specifically, for any micro tree $M$ we store
a bit vector $X_M = [b_{1}, \ldots, b_{s}]$, such that $X_M[i] = 1$
iff the $i$th node in a preorder traversal of $M$ is in $X$. If
$|V(M)| < s$ we leave the remaining values undefined. Later we
choose $s= \Theta(\log n_{T})$ such that each bit vector can be
represented in a single word.

Next we define a $\Down_{M}$ procedure on each micro tree $M\in
MS$. Due to the overlap between micro trees the $\Down_{M}$
procedure takes a bit $b$ which will be used to propagate
information between micro trees. For each micro tree $M \in MS$,
bit vector $X_M$, bit $b$, and $y\in V(T)$ define:
\begin{relate}
\item[$\Down_{M}(X_M, b, y)$:] Compute the state  $X'_M := \Child(\{x \in X_M \mid \lab(x) = \lab(y)\}) \cup \{x \in X_M \mid \lab(x) \neq \lab(y)\}$. If $b=0$, return $X_M'$, else return $X_M' \cup \{\roots(M)\}$.
\end{relate}
Later we will show how to implemenent $\Down_{M}$ in constant time
for $s = \Theta(\log n_{T})$. First we show how to use $\Down_M$ to
simulate $\Down$ on $P$. We define a recursive procedure $\Down$
which traverse the hiearchy of micro trees. For micro tree
$M$, state $X$, bit $b$, and $y \in V(T)$ define:
\begin{relate}
\item[$\Down(X,M,b,y)$:] Let $M_1, \ldots, M_k$ be the children of $M$.
\begin{enumerate}
\item Compute  $X_M := \Down_{M}(X_M, b, y)$.
\item For $i:=1$ to $k$ do:
\begin{enumerate}
\item Compute $\Down(X, M_i, b_{i}, y)$, where $b_{i} = 1$
iff

$\roots(M_i) \in X_M$.
\end{enumerate}
\end{enumerate}
\end{relate}
Intuitively, the $\Down$ procedure works in a top-down fashion
using the $b$ bit to propagate the new state of the root of micro
tree. To solve the problem within our framework we initially
construct the state representing $\{\roots(P)\}$. Then, at each
step we call $\Down(R_{j}, 0, y)$ on each root micro tree $R_{j}$.
We formally show that this is correct:
\begin{lemma}
The above algorithm correctly simulates the $\Down$ procedure on
$P$.
\end{lemma}
\begin{proof}
Let $X$ be the state and let $X' :=\Down(X, y)$. For
simplicity, assume that there is only one root micro tree $R$.
Since the root micro trees can only overlap at $\roots(P)$ it is
straightforward to generalize the result to any number of roots.
We show that if $X$ is represented by bit vectors at each micro
tree then calling $\Down(R, 0, y)$ correctly produces the new
state $X'$.

If $R$ is the only micro tree then only line 1 is executed. Since
$b = 0$ this produces the correct state by definition of
$\Down_{M}$. Otherwise, consider a micro tree $M$ with children
$M_{1}, \ldots, M_{k}$ and assume that $b = 1$ iff $\roots(M) \in
X'$. Line 1 computes and stores the new state returned by
$\Down_{M}$. If $b=0$ the correctness follows immediately. If
$b=1$ observe that $\Down_{M}$ first computes the new state and
then adds $\roots(M)$. Hence, in both cases the state of $M$ is
correctly computed. Line 2 recursively computes the new state of
the children of $M$. 
\qed \end{proof}

If each micro tree has size at most $s$ and $\Down_{M}$ can be
computed in constant time it follows that the above algorithm
solves TPS in $O(\ceil{n_{P}/s})$ time. In the following section we show
how to do this for $s = \Theta(\log n_{T})$, while maintaining
linear space.

\subsection{Representing Micro Trees} In this section we show
how to preprocess all micro trees $M \in MS$ such that
$\Down_{M}$ can be computed in constant time. This preprocessing
may be viewed as a ``Four Russian Technique''~\cite{ADKF1970}. To
achieve this in linear space we need the following auxiliary
procedures on micro trees. For each micro tree $M$, bit vector
$X_M$, and $\alpha \in \Sigma$ define:
\begin{relate}
\item[$\Child_{M}(X_M)$:] Return the bit vector of nodes in $M$ that are children of nodes in $X_M$.
\item[$\Eq_{M}(\alpha)$:] Return the bit vector of nodes in  $M$ labeled $\alpha$.
\end{relate}
By definition it follows that:
\begin{equation*}
\begin{aligned}
\Down_{M}(X_M,b, y) &=
\begin{cases}
\Child_{M}(X_M \cap \Eq_M(\lab(y)))\; \cup \\
\quad (X_M \backslash (X_M \cap  \Eq_M(\lab(y)))      & \text{if $b = 0$}, \\
\Child_{M}(X_M \cap \Eq_M(\lab(y))) \; \cup \\
\quad (X_M \backslash (X_M \cap \Eq_M(\lab(y))) \cup \{\roots(M)\}
& \text{if $b=1$}.
\end{cases} \\
\end{aligned}
\end{equation*}
Recall that the bit vectors are represented in a single word. Hence,
given $\Child_{M}$ and $\Eq_{M}$ we can compute $\Down_M$ using
standard bit-operations in constant time.

Next we show how to efficiently implement the operations. For each
micro tree $M \in MS$ we store the value $\Eq_{M}(\alpha)$ in a
hash table indexed by $\alpha$. Since the total number of
different characters in any $M\in MS$ is at most $s$, the hash
table $\Eq_{M}$ contains at most $s$ entries. Hence, the total
number of entries in all hash tables is $O(n_{P})$. Using perfect
hashing we can thus represent $\Eq_{M}$ for all micro trees, $M\in
MS$, in $O(n_{P})$ space and $O(1)$ worst-case lookup time. The
preprocessing time is expected $O(n_{P})$ w.h.p.. To get a worst-case bound we 
use the deterministic dictionary of Hagerup et. al.
\cite{HMP2001} with $O((n_{P})\log (n_{P}))$ worst-case preprocessing
time. 

Next consider implementing $\Child_{M}$. Since this
procedure is independent of the labeling of $M$ it suffices to
precompute it for all \emph{topologically} different rooted trees
of size at most $s$. The total number of such trees is less than
$2^{2s}$ and the number of different states in each tree is at
most $2^{s}$. Therefore $\Child_{M}$ has to be computed for a
total of $2^{2s}\cdot 2^{s} = 2^{3s}$ different inputs. For any
given tree and any given state, the value of $\Child_{M}$ can be
computed and encoded in $O(s)$ time. In total we can precompute
all values of $\Child_{M}$ in $O(s2^{3s})$ time. Choosing the
largest $s$ such that $3s + \log s \leq n_{T}$ (hence $s =
\Theta(\log n_{T})$) this uses $O(n_{T})$ time and space. Each of
the inputs to $\Child_{M}$ are encoded in a single word such that
we can look them up in constant time.

Finally, note that we also need to report the leaves of a state efficiently since this is needed in line 1 in the $\Visit$-procedure. To do this compute the state $L$ corresponding to all leaves in $P$. Clearly, the leaves of a state $X$ can be computed by performing a bitwise AND of each pair of bit vectors in $L$ and $X$. Computing $L$ uses $O(n_{P})$ time and the bitwise AND operation uses $O(\ceil{n_{P}/s})$ time.

Combining the results, we decompose $P$, for $s$ as described
above, and compute all values of $\Eq_{M}$ and $\Child_{M}$. Then, we solve TPS using the heavy-path
traversal. Since $s = \Theta(\log n_{T})$ and from Lemmas~\ref{t3:traversal} and \ref{t3:clustering} we have the following
theorem:
\begin{theorem}\label{t3:faster} For trees $P$ and $T$ the tree path subsequence problem can be solved in $O(n_P + n_T)$ space and $O(\frac{n_Pn_T}{\log n_T} +n_T+ n_P\log n_P)$ time.
\end{theorem}
Combining the results of Theorems~\ref{t3:simple} and \ref{t3:faster} this proves Theorem~\ref{t3:main}.

\section{Acknowledgments}
The authors would like to thank Anna {\"O}stlin Pagh many helpful comments.



\emptythanks
\chapter{Fast and Compact Regular Expression Matching}\label{chap:string1}

\title{Fast and Compact Regular Expression Matching}

\author{
    Philip Bille \\ IT University of Copenhagen \\ \texttt{beetle@itu.dk}
    \and
    Martin~Farach-Colton \\
    Rutgers University \\
    \texttt{farach@cs.rutgers.edu}
}

\date{}
\cleartooddpage


\maketitle

\begin{abstract}
The use of word operations has led to fast algorithms for classic
problems such as shortest paths and sorting.  Many classic problems in
stringology, notably regular expression matching and its variants, as well as
edit distance computation, also have transdichotomous algorithms.
Some of these algorithms have alphabet restrictions or require a large
amount of space.  In this paper, we improve on several of the keys
results by providing algorithms that improve on known time/space
bounds, or algorithms that remove restrictions on the alphabet
size.
\end{abstract}

\section{Introduction}\label{s1:sec:intro}
Transdichotomous algorithms~\cite{FredmanWillardFussion,FredmanWillardSP-n-MST} allow
logarithmic-sized words to be manipulated in constant time.  Many
classic problems, such as MST~\cite{FredmanWillardSP-n-MST}, Shortest
Paths~\cite{Thorup1999} and Sorting~\cite{HT2002}, have fast
transdichotomous algorithms.  Many classic stringology problems also
have transdichotomous solutions, though some of these, such as Myers
algorithm for regular expression matching~\cite{Myers1992} uses a lot of
space, whereas others, such as the algorithm by Masek and Paterson~\cite{MP1980} for edit distance computation requires that the alphabet be of constant size.

In this paper, we give improved algorithms for several such classic
problems. In particular:

\paragraph{Regular Expression Matching}
Given a regular expression $R$ and a string $Q$, the \textsc{Regular
  Expression Matching} problem is to determine if $Q$ is a member of
  the language denoted by $R$. This problem occurs in several text
  processing applications, such as in editors like
  Emacs~\cite{Stallman1981} or in the \texttt{Grep}
  utilities~\cite{WM1992,Navarro2001}.  It is also used in the lexical
  analysis phase of compilers and interpreters, regular expressions
  are commonly used to match tokens for the syntax analysis phase, and
  more recently for querying and validating XML databases, see
  e.g.,~\cite{HP2001,LM2001,Murata2001,BMLMA2004}. The standard
  textbook solution to the problem, due to Thompson~\cite{Thomp1968},
  constructs a non-deterministic finite automaton (NFA) for $R$ and
  simulates it on the string $Q$. For $R$ and $Q$ of sizes $m$ and
  $n$, respectively, this algorithm uses $O(mn)$ time and $O(m)$
  space. If the NFA is converted into a deterministic finite automaton
  (DFA), the DFA needs $O(\frac{m}{w} 2^{2m}\sigma)$ words, where
  $\sigma$ is the size of the alphabet $\Sigma$ and $w$ is the word
  size.  Using clever representations of the DFA the space can be reduced to $O(\frac{m}{w}(2^m +
  \sigma))$~\cite{WM1992b,NR2004}.

Normally, it is reported that the running time of traversing the DFA
is $O(n)$, but this complexity analysis ignores the word size. Since
nodes in the DFA may need $\Omega(m)$ bits to be addressed, we may need
$\Omega(m/w + 1)$ time to identify the next node in the
traversal. Therefore the running time becomes $O(mn/w + n + m)$ with a
potential exponential blowup in the space. Hence, in the
transdichotomous model, where $w$ is $\Theta(\log (n + m))$, using
worst-case exponential preprocessing time improves the query time by a
log factor.  The fastest known algorithm is due to
Myers~\cite{Myers1992}, who showed how to achieve $O(mn/k + m2^k + (n
+ m)\log m)$ time and $O(2^km)$ space, for any $k \leq w$. In
particular, for $k = \log (n/ \log n)$ this gives an algorithm using
$O(mn/\log n + (n+ m)\log m)$ time and $O(mn/\log n)$ space.

In Section \ref{s1:sec:regex}, we present an algorithm for \textsc{Regular
Expression Matching} that takes time $O(nm/k + n + m\log m)$ time and
uses $O(2^k + m)$ space, for any $k\leq w$. In particular, if we pick $k=\log n$, we
are (at least) as fast as the algorithm of Myers, while achieving $O(n+m)$ space. 

\paragraph{Approximate Regular Expression Matching}
Motivated by applications in computational biology, Myers and
Miller~\cite{MM1989} studied the \textsc{Approximate Regular
  Expression Matching} problem. Here, we want to determine if
$Q$ is within \emph{edit distance} $d$ to any string in the language
given by $R$. The edit distance between two strings is the minimum
number of insertions, deletions, and substitutions needed to transform
one string into the other. Myers and Miller~\cite{MM1989} gave an
$O(mn)$ time and $O(m)$ space dynamic programming
algorithm. Subsequently, assuming as a constant sized alphabet, Wu, Manber and Myers~\cite{WMM1995} gave an $O(\frac{mn\log(d+2)}{\log n} + n + m)$ time and $O(\frac{m\sqrt{n}
  \log(d+2)}{\log n} + n + m)$ space algorithm. Recently, an exponential space solution based on DFAs for the problem has been proposed by Navarro~\cite{Navarro2004}. 
  
  In Section \ref{s1:sec:appregexmatching}, we extend our results of Section \ref{s1:sec:regex} and  
  give an algorithm, without any assumption on the alphabet size, using $O(\frac{mn\log(d+2)}{k} + n + m\log m)$ time
and $O(2^k + m)$ space, for any $k\leq w$.

\paragraph{Subsequence Indexing}
We also consider a special case of regular expression matching. Given
text $T$, the \textsc{Subsequence Indexing} problem is to preprocess
$T$ to allow queries of the form ``is $Q$ a subsequence of $T$?''
Baeza-Yates~\cite{BaezaYates1991} showed that this problem can be
solved with $O(n)$ preprocessing time and space, and query time
$O(m\log n)$, where $Q$ has length $m$ and $T$ has length $n$.
Conversely, one can achieve queries of time $O(m)$ with $O(n\sigma)$
preprocessing time and space.  As before, $\sigma$ is the size of the
alphabet. 

In Section \ref{s1:sec:subseq}, we give an algorithm that improves the former
results to $O(m\log\log\sigma)$ query time or the latter result to
$O(n\sigma^{\epsilon})$ preprocessing time and space.

\paragraph{String Edit Distance} We conclude by giving a simple way to
improve the complexity of the \textsc{String Edit Distance} problem,
which is defined as that of computing the minimum number of edit
operations needed to transform given string $S$ of length $m$ into
given string $T$ of length $n$.  The standard dynamic programming
solution to this problem uses $O(mn)$ time and $O(\min(m,n))$
space. The fastest algorithm for this problem, due to Masek and
Paterson~\cite{MP1980}, achieves $O(mn/k^2 +m + n)$ time and $O(2^k +
\min(n,m))$ space for any $k \leq w$. However, this algorithm assumes
a constant size alphabet. 

In Section \ref{s1:sec:stringedit}, we show how to
achieve $O(nm\log k/k^2 + m + n)$ time and $O(2^k + \min(n,m))$ space
for any $k \leq w$ for an arbitrary alphabet. Hence, we remove the
dependency of the alphabet at the cost of a $\log k$ factor to the
running time.

\section{Regular Expression Matching}\label{s1:sec:regex} Given an string $Q$ and a
regular expression $R$ the \textsc{Regular Expression Matching}
problem is to determine if $Q$ is in the language given by $R$. Let
$n$ and $m$ be the sizes of $Q$ and $R$, respectively. In this section
we show that \textsc{Regular Expression Matching} can be solved in
$O(mn/k + n + m\log m)$ time and $O(2^{k}+m)$ space, for $k \leq w$.

\subsection{Regular Expressions and NFAs} We briefly review
Thompson's construction and the standard node set simulation.  The set
of \emph{regular expressions} over $\Sigma$ is defined recursively as
follows:
\begin{itemize}
\item A character $\alpha \in \Sigma$ is a regular expression.
\item If $S$ and $T$ are regular expressions then so is the
  \emph{catenation}, $(S)\cdot(T)$, the \emph{union}, $(S)|(T)$, and
  the \emph{star}, $(S)^*$.
\end{itemize}
Unnecessary parentheses can be removed by observing that $\cdot$ and
$|$ are associative and by using the standard precedence of the
operators, that is $*$ precedes $\cdot$, which in turn precedes $|$. Furthermore, we will often remove the $\cdot$ when writing regular expressions.  The \emph{language} $L(R)$ generated by $R$ is the set of all strings
matching $R$. The \emph{parse tree} $T(R)$ of $R$ is the rooted and ordered
tree representing the hiearchical structure of $R$. All leaves are represented by a character in $\Sigma$ and all internal nodes are labeled $\cdot$, $|$, or $^*$. We assume that parse trees are binary and constructed such that they are in one-to-one correspondance with the regular expressions. An example parse tree of the
regular expression $ac|a^*b$ is shown in Fig. \ref{s1:fig:clustering}(a).

A \emph{finite automaton} $A$ is a tuple $A = (G, \Sigma, \theta, \Phi)$ such that,
\begin{itemize}
  \item $G$ is a directed graph,
  \item Each edge $e \in E(G)$ is labeled with a character $\alpha \in \Sigma$ or $\epsilon$,
  \item $\theta \in V(G)$ is a \emph{start node},
  \item $\Phi \subseteq V(G)$ is the set of \emph{accepting nodes}.
\end{itemize} 
$A$ is a \emph{deterministic finite automaton} (DFA) if $A$ does not contain any $\epsilon$-edges, and for each node $v \in V(G)$ all outcoming edges have different labels. Otherwise, $A$ is a \emph{non-deterministic automaton} (NFA). We say that $A$ \emph{accepts} a string $Q$ if there is a path from $\theta$ to a node in $\Phi$ which spells out $Q$. 

\begin{figure}[t] 
  \centering \includegraphics[scale=.5]{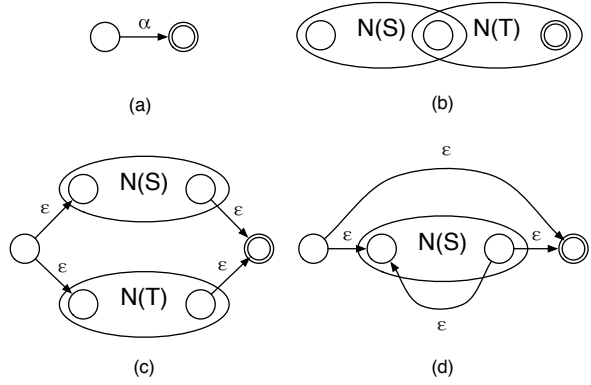}
  \caption{Thompson's NFA construction. The regular expression for a character $\alpha \in \Sigma$ correspond to NFA $(a)$. If $S$ and $T$ are regular expression then $N(ST)$, $N(S|T)$, and
    $N(S^*)$ correspond to NFAs $(a)$, $(b)$, and $(c)$, respectively.
    Accepting nodes are marked with a double circle.}
  \label{s1:fig:thompson}
\end{figure}

Using Thompson's method \cite{Thomp1968} we can recursively construct an NFA $N(R)$ accepting all strings in $L(R)$. The set of rules is presented below and illustrated in
Fig.~\ref{s1:fig:thompson}. 

\begin{itemize}
\item $N(\alpha)$ is the automaton consisting of a start node
  $\theta_\alpha$, accepting node $\phi_\alpha$, and an $\alpha$-edge
  from $\theta_\alpha$ to $\phi_\alpha$.
\item Let $N(S)$ and $N(T)$ be automata for regular expression $S$ and
  $T$ with start and accepting nodes $\theta_S$, $\theta_T$, $\phi_S$,
  and $\phi_T$, respectively. Then, NFAs for $N(S\cdot T)$, $N(S|T)$,
  and $N(S^*)$ are constructed as follows:
  \begin{relate}
  \item[$N(ST)$:] Merge the nodes $\phi_S$ and $\theta_T$ into a
    single node. The new start node is $\theta_S$ and the new
    accepting node is $\phi_T$.
  \item[$N(S|T)$:] Add a new start node $\theta_{S|T}$ and new
    accepting node $\phi_{S|T}$. Then, add $\epsilon$ edges from
    $\theta_{S|T}$ to $\theta_S$ and $\theta_T$, and from $\phi_S$ and
    $\phi_T$ to $\phi_{S|T}$.
  \item [$N(S^*)$:] Add a new start node $\theta_{S^*}$ and new
    accepting node $\phi_{S^*}$. Then, add $\epsilon$ edges from
    $\theta_{S^*}$ to $\theta_S$ and $\phi_{S^*}$, and from $\phi_S$
    to $\phi_{S^*}$ and $\theta_S$.
\end{relate}
\end{itemize}
By construction, $N(R)$ has a single start and accepting node, denoted
$\theta$ and $\phi$, respectively. $\theta$ has no incoming edges and $\phi$ has no outcoming edges. 
The total number of nodes is at most $2m$ and since each node has at most $2$ outgoing edges that the
total number of edges is less than $4m$.  Furthermore, all incoming
edges have the same label, and we denote a node with incoming
$\alpha$-edges an \emph{$\alpha$-node}. Note that the star
construction in Fig. \ref{s1:fig:thompson}(d) introduces an edge from the
accepting node of $N(S)$ to the start node of $N(S)$. All such edges
in $N(R)$ are called \emph{back edges} and all other edges are
\emph{forward edges}. We need the following important property of
$N(R)$.
\begin{lemma}[Myers \cite{Myers1992}]\label{s1:lem:cyclepath}
  Any cycle-free path in $N(R)$ contains at most one back edge.
\end{lemma}
For a string $Q$ of length $n$ the standard node-set simulation of
$N(R)$ on $Q$ produces a sequence of node-sets $S_0, \ldots, S_n$. A node $v$ is in $S_i$ iff there is a path from $\theta$ to $v$ that spells out the $i$th prefix of $Q$.
The simulation can be implemented with the following simple
operations. Let $S$ be a node-set in $N(R)$ and let $\alpha$ be a
character in $\Sigma$.
\begin{relate}
\item[$\Move(S,\alpha)$:] Compute and return the set of nodes
  reachable from $S$ via a single $\alpha$-edge.
\item[$\Close(S)$:] Compute and return the set of nodes reachable from
  $S$ via $0$ or more $\epsilon$-edges.
\end{relate}
The number of nodes and edges in $N(R)$ is $O(m)$, and both operations
are implementable in $O(m)$ time.  The simulation proceed as follows:
Initially, $S_0 := Close(\{\theta\})$. If $Q[j]=\alpha$, $1\leq j \leq
n$, then $S_j := \Close(\Move(S_{j-1}, \alpha))$. Finally, $Q \in
L(R)$ iff $\phi \in S_n$. Since each node-set $S_j$ only depends on
$S_{j-1}$ this algorithm uses $O(mn)$ time $O(m)$ space. 

\subsection{Outline of Algorithm}
The algorithm presented in the following section resembles the one by Myers~\cite{Myers1992}. The key to improving the space is the use of compact data structures and an efficient encoding of small automatons. We first present a clustering of $T(R)$ in Section~\ref{s1:sec:clustering}. This leads to a decomposition of $N(R)$ into small subautomata. In Section~\ref{s1:sec:simulation} we define appropiate $\Move$ and $\Close$ operations on the subautomata. With these we show how to simulate the node-set algorithm on $N(R)$. Finally, in Section~\ref{s1:sec:representation} we give a compact representation for the $\Move$ and $\Close$ operations on subautomata of size $\Theta(k)$. The representation allows constant time simulation of each subautomata leading to the speedup.

\subsection{Decomposing the NFA}\label{s1:sec:clustering} 
\begin{figure}[ht]
  \centering \includegraphics[scale=.5]{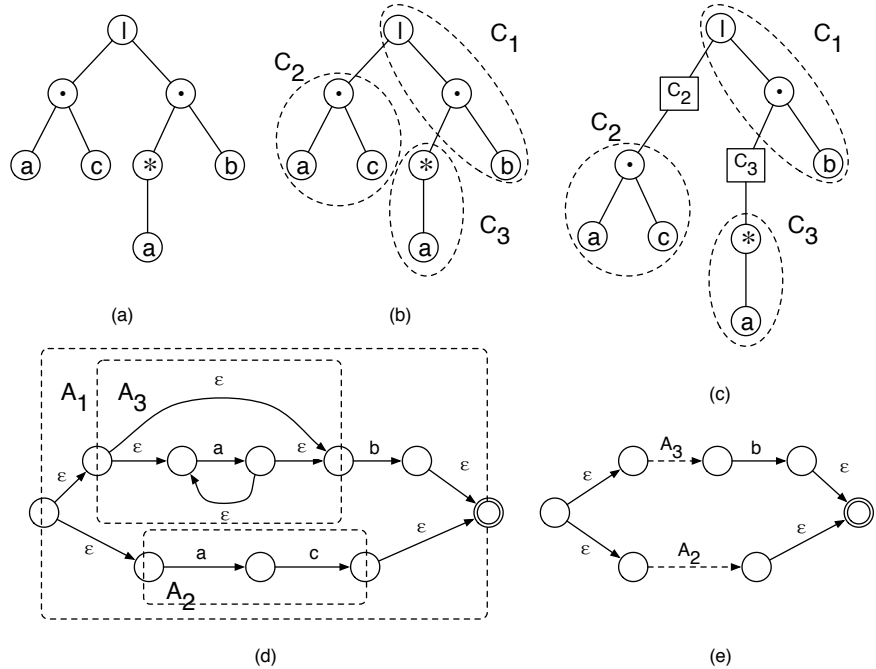}
   \caption{(a) The parse tree for the regular expression
     $ac|a^*b$. (b) A clustering of $(a)$ into
     node-disjoint connected subtrees $C_1$, $C_2$, and $C_3$.  Here,
     $x=3$. (c) The clustering from (b) extended with pseudo-nodes.
     (d) The automaton for the parse tree divided into subautomata
     corresponding to the clustering. (e) The subautomaton $A_1$ with
     pseudo-edges corresponding to the child automata.}
   \label{s1:fig:clustering}
\end{figure}

In this section we show
how to decompose $N(R)$ into small subautomata. In the final algorithm
transitions through these subautomata will be simulated in constant
time.  The decomposition is based on a clustering of the parse tree
$T(R)$. Our decomposition is similar to the one given in
\cite{Myers1992, WMM1995}.  A \emph{cluster} $C$ is a connected
subgraph of $T(R)$. A \emph{cluster partition} $CS$ is a partition of
the nodes of $T(R)$ into node-disjoint clusters. Since $T(R)$ is a
binary tree, a bottom-up procedure yields the following lemma.
\begin{lemma}\label{s1:lem:clustering}
  For any regular expression $R$ of size $m$ and a parameter $x$, it
  is possible to build a cluster partition $CS$ of $T(R)$, such that
  $|CS| = O(m/x)$ and for any $C\in CS$ the number of nodes in $C$ is
  at most $x$.
\end{lemma}
An example clustering of a parse tree is shown in
Fig.~\ref{s1:fig:clustering}(b).

Before proceding, we need some definitions.  Assume that $CS$ is a
cluster partition of $T(R)$ for a some yet-to-be-determined parameter
$x$. Edges adjacent to two clusters are \emph{external edges} and all
other edges are \emph{internal edges}. Contracting all internal edges
induces a \emph{macro tree}, where each cluster is represented by a
single \emph{macro node}. Let $C_v$ and $C_w$ be two clusters with
corresponding macro nodes $v$ and $w$. We say that $C_v$ is a
\emph{parent cluster} (resp. \emph{child cluster}) of $C_w$ if $v$ is
the parent (resp. child) of $w$ in the macro tree. The \emph{root
  cluster and leaf clusters} are the clusters corresponding to the
root and the leaves of the macro tree.

Next we show how to decompose $N(R)$ into small subautomata. Each
cluster $C$ will correspond to a subautomaton $A$ and we use the terms
child, parent, root, and leaf for subautomata in the same way we do
with clusters. For a cluster $C$, we insert a special
\emph{pseudo-node} $p_i$ for each child cluster $C_1, \ldots, C_l$ in
the middle of the external edge connecting $C$ and $C_i$. Now, $C$'s
subautomaton $A$ is the automaton corresponding to the parse tree
induced by the set of nodes $V(C) \cup \{p_1, \ldots, p_l\}$. The
pseudo-nodes are alphabet placeholders, since the leaves of a
well-formed parse tree must be characters.

In $A$, child automaton $A_i$ is represented by its start and
accepting node $\theta_{A_i}$ and $\phi_{A_i}$ and a
\emph{pseudo-edge} connecting them. An example of these definitions is
given in Fig.~\ref{s1:fig:clustering}. Any cluster $C$ of size at most
$x$ has less than $2x$ pseudo-children and therefore the size of the
corresponding subautomaton is at most $6x$. Note, therefore, that
automata derived from regular expressions can be thus decomposed into
$O(m/z)$ subautomata each of size at most $z$, by
Lemma~\ref{s1:lem:clustering} and the above construction.

\subsection{Simulating the NFA}\label{s1:sec:simulation}

In this section we show how to do a node-set simulation of $N(R)$
using the subautomata. Recall that each subautomaton has size less
than $z$. Topologically sort all nodes in each subautomaton $A$
ignoring back edges. This can be done for all subautomata in total
$O(m)$ time. We represent the current node-set $S$ of $N(R)$ compactly using a bitvector for each subautomaton. Specifically, for each subautomaton $A$ we store a \emph{characteristic bitvector} $\vec{B} = [b_1,
\ldots, b_z]$, where nodes in $\vec{B}$ are indexed by the their
topological order, such that $\vec{B}[i] = 1$ iff the $i$th node is in $S$. 
If $A$ contains fewer than $z$ nodes we leave the remaining values undefined. For simplicity, we will refer to the \emph{state} of $A$ as the node-set represented by the characteristic vector stored at $A$. Similarly, the state of $N(R)$ is the set of characteristic vectors representing $S$. The state of a node is the bit indicating if the node is in $S$. Since any child $A'$ of $A$ overlap at the nodes $\theta_{A'}$ and $\phi_{A'}$ we will insure that the state of $\theta_{A'}$ and $\phi_{A'}$ is the same in the characteristic vectors of both $A$ and $A'$.

Below we present appropiate move and $\epsilon$-closure operations defined on subautomata. Due to the overlap between parent and child nodes these operations take a bit $b$ which will use to propagate the new state of the start node. For each subautomaton $A$, characteristic vector $\vec{B}$, bit $b$, and character $\alpha \in \Sigma$
define:
\begin{relate}
\item[$\Move^A(\vec{B}, b, \alpha)$:] Compute the state $\vec{B}'$ of all nodes in
  $A$ reachable via a single $\alpha$ edge from $\vec{B}$. If $b=0$, return
  $\vec{B}'$, else return $\vec{B}' \cup \{\theta_A\}$.
\item[$\Close^A(\vec{B}, b)$:] Return the set $\vec{B}'$ of all nodes in $A$
  reachable via a path of $0$ or more $\epsilon$-edges from $\vec{B}$, if
  $b=0$, or reachable from $\vec{B} \cup \{\theta_A\}$, if $b = 1$.
\end{relate}
We will later show how to implement these operations in constant time and
total $2^{O(k)}$ space when $z = \Theta(k)$. Before doing so we show
how to use these operations to perform the node-set simulation of
$N(R)$. Assume that the current node-set of $N(R)$ is represented by
its characteristic vector for each subautomaton. The following $\Move$ and $\Close$ operations recursively traverse the hiearchy of subautomata top-down. At each subautomata the current state of $N(R)$ is modified using primarily  $\Move^A$ and  $\Close^A$. For any subautomaton $A$, bit $b$, and character $\alpha \in \Sigma$ define:
\begin{relate}
\item[$\Move(A, b, \alpha)$:] Let $\vec{B}$ be the current state of $A$
  and let $A_1, \ldots, A_l$ be children of $A$ in topological order
  of their start node.
\begin{enumerate}
\item Compute $\vec{B}':= \Move^A(\vec{B}, b, \alpha)$.
\item For each $A_i$, $1 \leq i \leq l$, 
\begin{enumerate}
\item Compute $f_i := \Move(A_i, b_i, \alpha)$, where $b_i = 1$ iff $\theta_{A_i} \in \vec{B}'$.
\item If $f_i = 1$ set $\vec{B}' := \vec{B}' \cup \{\phi_{A_i}\}$.
\end{enumerate}
\item Store $\vec{B}'$ and return the value $1$ if $\phi_{A} \in \vec{B}'$ and $0$ otherwise.
\end{enumerate}
\item[$\Close(A, b)$:] Let $\vec{B}$ be the current state of $A$ and let
  $A_1, \ldots, A_l$ be children of $A$ in topological order of their
  start node.
\begin{enumerate}
\item Compute $\vec{B}' := \Close^A(\vec{B}, b)$.
\item For each child automaton $A_i$, $1 \leq i \leq l$,
\begin{enumerate}
\item Compute $f_i := \Close(A_i, b_i)$, where $b_i = 1$ if
  $\theta_{A_i} \in \vec{B}'$.
\item If $f_i = 1$ set $\vec{B}' := \vec{B}' \cup \{\phi_{A_i}\}$.
\item $\vec{B}' := \Close^A(\vec{B}, b)$.
\end{enumerate}
\item Store $\vec{B}'$ and return the value $1$ if $\phi_{A} \in \vec{B}'$ and
  $0$ otherwise.
\end{enumerate}
\end{relate}
The ``store'' in line 3 of both operations updates the state of the subautomaton. The node-set simulation of $N(R)$ on string $Q$ of length $n$ produces the states $S_0, \ldots, S_n$ as follows.  Let $A_r$ be the root automaton. Initialize the state of $N(R)$ to be empty, i.e., set all bitvectors to $0$. $S_0$ is computed by calling $\Close(A_r, 1)$ twice. Assume that $S_{j-1}$, $1 \leq j \leq n$, is the current state of $N(R)$ and let $\alpha = Q[j]$.  Compute $S_j$ by calling $\Move(A_r, 0, \alpha)$ and then calling $\Close(A_r, 0)$ twice.  Finally, $Q \in L(R)$ iff $\phi \in S_n$.

We argue that the above algorithm is correct. To do this we need to show that the call to the $\Move$ operation and the two calls to the $\Close$ operation simulates the standard $\Move$ and $\Close$ operations.

First consider the $\Move$ operation. Let $S$ be the state of $N(R)$ and let $S'$ be the state after a call to $\Move(A_r, 0, \alpha)$. Consider any subautomaton $A$ and let $\vec{B}$ and $\vec{B}'$ be the bitvectors of $A$ corresponding to states $S$ and $S'$, respectively. We first show by induction that after $\Move(A, 0,  \alpha)$ the new state $\vec{B}'$ is the set of nodes reachable from $\vec{B}$ via a single $\alpha$-edge in $N(R)$. For $\Move(A, 1, \alpha)$ a similar argument shows that new state is the union of the set of nodes reachable from $\vec{B}$ via a single $\alpha$-edge and $\{\theta_A\}$. 

Initially, we compute $\vec{B}' := \Move^A(\vec{B}, 0, \alpha)$. Thus $\vec{B}'$ contains the set of nodes reachable via a single $\alpha$-edge in $A$. If $A$ is a leaf automaton then $\vec{B}'$ satisfies the property and the algorithm returns. Otherwise, there may be an $\alpha$-edge to some accepting node $\phi_{A_i}$ of a child automaton $A_i$. Since this edge is not contained $A$, $\phi_{A_i}$ is not initially in $\vec{B}'$. However, since each child is handled recursively in topological order and the new state of start and accepting nodes are propagated, it follows that $\phi_{A_i}$ is ultimately added to $\vec{B}'$. Note that since a single node can be the accepting node of a child $A_i$ and the start node of child $A_{i+1}$, the topological order is needed to ensure a consistent update of the state.

It now follows that the state $S'$ of $N(R)$ after $\Move(A_r, 0, \alpha)$, consists of all nodes reachable via a single $\alpha$-edge from $S$. Hence, $\Move(A_r, 0, \alpha)$ correctly simulates a standard $\Move$ operation.

Next consider the two calls to the $\Close$ operation. Let $S$ be the state of $N(R)$ and let $S'$ be the state after  the first call to $\Close(A_r,0)$. As above consider any subautomaton $A$ and let $\vec{B}$ and $\vec{B}'$ be the bitvectors of $A$ corresponding to $S$ and $S'$, respectively. We show by induction that after $\Close(A,0)$ the state $\vec{B}'$ \emph{contains} the set of nodes in $N(R)$ reachable via a path of $0$ or more \emph{forward} $\epsilon$-edges from $\vec{B}$. Initially, $\vec{B}' := \Close^A(\vec{B}, 0)$, and hence $\vec{B}'$ contains  all nodes reachable via a path of $0$ or more $\epsilon$-edges from $\vec{B}$, where the path consists solely of edges in $A$. If $A$ is a leaf automaton, the result immediately holds. Otherwise, there may be a path of $\epsilon$-edges to a node $v$  going through the children of $A$. As above, the recursive topological processing of the children ensures that $v$ is added to $\vec{B}'$. 

Hence, after the first call to $\Close(A_r, 0)$ the state $S'$ contains all nodes reachable from $S$ via a path of $0$ or more forward $\epsilon$-edges. By a similar argument it follows that the second call to $\Close(A_r, 0)$ produces the state $S''$ that contains all the nodes reachable from $S$ via a path of $0$ or more forward $\epsilon$-edge and  $1$ back edge. However, by Lemma~\ref{s1:lem:cyclepath} this is exactly the set of nodes reachable via a path of $0$ or more $\epsilon$-edges. Furthermore, since $\Close(A_r, 0)$ never produces a state with nodes that are not reachable through $\epsilon$-edges, it follows that the two calls to $\Close(A_r, 0)$ correctly simulates a standard $\Close$ operation. 

Finally, note that if we start with a state with no nodes, we can compute the state $S_0$ in the node-set simulation by calling $\Close(A_r, 1)$ twice. Hence, the above algorithm correctly solves \textsc{Regular Expression Matching}.

If the subautomata have size at most $z$ and $\Move^A$ and $\Close^A$ can be computed in constant time the above algorithm computes a step in the node-set simulation in $O(m/z)$ time. In the following section
we show how to do this in $O(2^{k})$ space for $z = \Theta(k)$. Note that computing the clustering uses an additional $O(m)$ time and space.

\subsection{Representing Subautomata}\label{s1:sec:representation} To efficiently
represent $\Move^A$ and $\Close^A$ we apply a Four Russians trick.
Consider a straightforward code for $\Move^A$: Precompute the value of
$\Move^A$ for all $\vec{B}$, both values of $b$, and all characters
$\alpha$. Since the number of different bitvectors is $2^{z}$ and the
size of the alphabet is $\sigma$, this table has $2^{z+1}\sigma$
entries.  Each entry can be stored in a single word, so the table also uses
a total of $2^{z+1}\sigma$ space. The total number of subautomata is
$O(m/z)$, and therefore the total size of these tables is an unacceptable
$O(\frac{m}{z} \cdot 2^{z} \sigma)$.  

To improve this we use a more elaborate approach. First we factor out
the dependency on the alphabet, as follows. For all subautomata $A$ and all
characters $\alpha \in \Sigma$ define:
\begin{relate}
\item[$\Succ^A(\vec{B})$:] Return the set of all nodes in $A$ reachable
  from $\vec{B}$ by a single edge.
\item[$\Equal^A(\alpha)$:] Return the set of all $\alpha$-nodes in $A$.
\end{relate}
Since all incoming edges to a node are labeled with the same character
it follows that,
\begin{equation*}
\Move^A(\vec{B}, b, \alpha) =
\begin{cases}
  \Succ^A(\vec{B}) \cap \Equal^A(\alpha) &\text{if $b=0$}, \\
  (\Succ^A(\vec{B}) \cap \Equal^A(\alpha)) \cup \{\theta_A\} & \text{if $b=1$}.
\end{cases} 
\end{equation*}
Hence, given $\Succ^A$ and $\Equal^A$ we can implement $\Move^A$ in
constant time using bit operations. To efficiently represent $\Equal^{A}$,
for each subautomaton $A$, store the value of $\Equal^A(\alpha)$ in a
hash table. Since the total number of different characters in $A$ is
at most $z$ the hash table $\Equal^A$ contains at most $z$ entries.
Hence, we can represent $\Equal^{A}$ for all subautomata is $O(m)$ space and
constant worst-case lookup time. The preprocessing time is $O(m)$
w.h.p.. To get a worst-case preprocessing bound we use the deterministic dictionary
of~\cite{HMP2001} with $O(m \log m)$ worst-case preprocessing time.

We note that the idea of using $\Equal^A(\alpha)$ to represent the $\alpha$-nodes is not new and has been used in several string matching algorithms, for instance, in the classical Shift-Or algorithm~\cite{BYG1992} and in the recent optimized DFA construction for regular expression matching~\cite{NR2004}.

To represent $\Succ$ compactly we proceed as follows. Let $\hat{A}$ be
the automaton obtained by removing the labels from edges in $A$.
$\Succ^{A_1}$ and $\Succ^{A_2}$ compute the same function if
$\hat{A_1} = \hat{A_2}$. Hence, to represent $\Succ$ it suffices to
precompute $\Succ$ on all possible subautomata $\hat{A}$. By the
one-to-one correspondance of parse trees and automata we have that
each subautomata $\hat{A}$ corresponds to a parse tree with leaf
labels removed. Each such parse tree has at most
$x$ internal nodes and $2x$ leaves. The number of rooted, ordered,
binary trees with at most $3x$ nodes is less than $2^{6x+1}$, and for
each such tree each internal node can have one of $3$ different labels.
Hence, the total number of distinct subautomata is less than $2^{6x +
  1}3^x$. Each subautomaton has at most $6x$ nodes and therefore the
result of $\Succ^A$ has to be computed for each of the $2^{6x}$
different values for $\vec{B}$ using $O(x2^{6x})$ time. Therefore we can
precompute all values of $\Succ$ in $O(x2^{12x + 1}3^x)$ time.
Choosing $x$ such that $x+\frac{\log x}{12 + \log 3} \leq \frac{k -
  1}{12 + \log 3}$ gives us $O(2^k)$ space and preprocessing time.

Using an analogous argument, it follows that $\Close^A$ can be
precomputed for all distinct subautomata within the same complexity.
By our discussion in the previous sections and since $x = \Theta(k)$ we have shown the following theorem:
\begin{theorem}\label{s1:thm:regex}
  For regular expression $R$ of length $m$, string $Q$ of length
  $n$, and $k \leq w$,  \textsc{Regular Expression Matching} can be solved in $O(mn/k +  n + m\log m)$ time and $O(2^k + m)$ space.
\end{theorem}

\section{Approximate Regular Expression Matching}\label{s1:sec:appregexmatching}
Given a string $Q$, a regular expression $R$, and an integer $d \geq
0$, the \textsc{Approximate Regular Expression Matching} problem is to
determine if $Q$ is within edit distance $d$ to a string in
$L(R)$. In this section we extend our solution for \textsc{Regular Expression
  Matching} to \textsc{Approximate Regular Expression Matching}.
Specifically, we show that the problem can be solved in $O(\frac{mn\log(d+2)}{k} + n + m\log m)$ time and $O(2^k +m)$ space, for any $k\leq w$. 

\subsection{Dynamic Programming Recurrence} Our algorithm is
based on a dynamic programming recurrence due to Myers and
Miller~\cite{MM1989}, which we describe below. Let $\Delta(v,i)$
denote the minimum over all paths ${\cal P}$ between $\theta$ and $v$
of the edit distance between ${\cal P}$ and the $i$th prefix of $Q$. The recurrence avoids cyclic dependencies from the back
edges by splitting the recurrence into two passes. Intuitively, the
first pass handles forward edges and the second pass propagates values
from back edges.  The \emph{pass-1 value} of $v$ is denoted
$\Delta_1(v,i)$, and the \emph{pass-2 value} is $\Delta_2(v,i)$. For a
given $i$, the \emph{pass-1 (resp. pass-2) value of $N(R)$} is the set of
pass-1 (resp. pass-2) values of all nodes of $N(R)$.  For all
$v$ and $i$, we set $\Delta(v,i) = \Delta_2(v,i)$.

The set of \emph{predecessors} of $v$ is the set of nodes $\Pre(v) =
\{w \mid \text{ $(w, v)$ is an edge}\}$.  We define
$\overline{\Pre}(v) = \{w \mid \text{ $(w,v)$ is a forward edge}\}$.
For notational convenience, we extend the definitions of $\Delta_1$
and $\Delta_2$ to apply to sets, as follows: $\Delta_1(\Pre(v),i) =
\min_{w\in \Pre(v)} \Delta_1(w, i)$ and
$\Delta_1(\overline{\Pre}(v),i) = \min_{w\in \overline{\Pre}(v)}
\Delta_1(w, i)$, and analogously for $\Delta_2$. The pass-1 and pass-2
values satisfy the following recurrence:
 
\begin{align*}
  \Delta_2(\theta, i) &= \Delta_1(\theta, i) = i \qquad \text{$0 \leq i \leq n$}. \\
  \Delta_2(v,0) &= \Delta_1(v, 0) = \min
\begin{cases}
  \Delta_2(\overline{\Pre}(v), 0) + 1 & \text{if $v$ is a $\Sigma$-node}, \\
  \Delta_2(\overline{\Pre}(v), 0) & \text{if $v \neq \theta$ is an
    $\epsilon$-node}.
\end{cases}  \\
\intertext{$\qquad$ For $1\leq i \leq n$,}
\Delta_1(v, i) &=
\begin{cases}
  \min(\Delta_2(v, i-1) + 1, \Delta_2(\Pre(v), i) + \lambda(v,Q[i]),
  \Delta_1(\overline{\Pre}(v), i) + 1)&
  \text{if $v$ is a $\Sigma$-node}, \\
  \Delta_1(\overline{\Pre}(v), i) & \text{if $v\neq \theta$ is an $\epsilon$-node},
\end{cases} \\
\intertext{$\qquad$ where $\lambda(v,Q[i]) = 1$ if $v$ is a
  $Q[i]$-node and $0$ otherwise,}  
  \Delta_2(v,i) &=
\begin{cases}
  \min(\Delta_1(\Pre(v), i), \Delta_2(\overline{\Pre}(v), i)) + 1
  & \text{if $v$ is a $\Sigma$-node}, \\
  \min(\Delta_1(\Pre(v), i), \Delta_2(\overline{\Pre}(v), i))
  & \text{if $v$ is a $\epsilon$-node}. \\
\end{cases} 
\end{align*}
A full proof of the correctness of the above recurrence can be found
in~\cite{MM1989, WMM1995}. Intuitively, the first pass handles forward
edges as follows: For $\Sigma$-nodes the recurrence handles
insertions, substitution/matches, and deletions (in this order). For
$\epsilon$-nodes the values computed so far are propagated.
Subsequently, the second pass handles the back edges. For our problem
we want to determine if $Q$ is within edit distance $d$. Hence, we can
replace all values exceeding $d$ by $d+1$.  

\subsection{Simulating the Recurrence} Our algorithm now
proceeds analogously to the case with $d=0$ above.  We will decompose
the automaton into subautomata, and we will compute the above dynamic
program on an appropriate encoding of the subautomata, leading to a
small-space speedup.

As before, we decompose $N(R)$ into subautomata of size less than $z$.
For a subautomaton $A$ we define operations $\Nextt^A_1$ and
$\Nextt^A_2$ which we use to compute the pass-1 and pass-2 values of
$A$, respectively. However, the new (pass-1 or pass-2) value of $A$
depends on pseudo-edges in a more complicated way than before: If $A'$
is a child of $A$, then all nodes preceding $\phi_{A'}$ depend on
the value of $\phi_{A'}$. Hence, we need the value of $\phi_{A'}$
before we can compute values of the nodes preceding $\phi_{A'}$. To
address this problem we partition the nodes of a subautomaton as
described below.

For each subautomaton $A$ topologically sort the nodes (ignoring back
edges) with the requirement that for each child $A'$ the start and
accepting nodes $\theta_{A'}$ and $\phi_{A'}$ are consecutive in the
order. Contracting all pseudo-edges in $A$ this can be done for all
subautomata in $O(m)$ time. Let $A_1, \ldots, A_l$ be the children of
$A$ in this order. We partition the nodes in $A$, except $\{\theta_A\}
\cup \{\phi_{A_1}, \ldots, \phi_{A_l}\}$ , into $l+1$ \emph{chunks}.
The first chunk is the nodes in the interval $[\theta_A + 1,
\theta_{A_1}]$. If we let $\phi_{A_{l+1}} = \phi_A$, then the $i$th
chunk, $1\leq l \leq l+1$, is the set of nodes in the interval
$[\phi_{A_{i-1}}+1, \theta_{A_i}]$. A leaf automaton has a single
chunk consisting of all nodes except the start node. We represent the
$i$th chunk in $A$ by a characteristic vector $\vec{L_i}$ identifying
the nodes in the chunks, that is, $\vec{L_i}[j] = 1$ if node $j$ is in the $i$th chunk and $0$ otherwise. From the topological order we can compute all chunks and their corresponding characteristic vectors in total $O(m)$
time.

The value of $A$ is represented by a vector $\vec{B} = [b_1, \ldots,
b_z]$, such that $b_i \in [0, d+1]$. Hence, the total number of bits
used to encode $\vec{B}$ is $z\ceil{\log d+2}$ bits.  For an
automaton $A$, characteristic vectors $\vec{B}$ and $\vec{L}$, and a
character $\alpha \in \Sigma$ define the operations
$\Nextt^A_1(\vec{B}, \vec{L}, b, \alpha)$ and
$\Nextt^A_2(\vec{B},\vec{L}, b)$ as the vectors $\vec{B}_1$ and
$\vec{B}_2$, respectively, given by:
\begin{align*}
  \vec{B}_1[v] &= B[v]  \qquad\quad \text{if $v \not\in \vec{L}$} \\
  \vec{B}_1[v] &=
\begin{cases}
  \min(\vec{B}[v] + 1, \vec{B}[\Pre(v)] + \lambda(v, \alpha),  \vec{B}_1[\overline{\Pre}(v)]  + 1) & \text{if $v \in\vec{L}$ is a $\Sigma$-node}, \\
  \vec{B}_1[\Pre(v)] & \text{if $v \in \vec{L}$ is an $\epsilon$-node}
\end{cases} \\
\vec{B}_2[v] &= B[v] \qquad\quad \text{if $v \not\in \vec{L}$}  \\
\vec{B}_2[v] &=
\begin{cases}
  \min(\vec{B}[\Pre(v)], \vec{B}_2[\overline{\Pre}(v)] + 1) & \text{if $v \in\vec{L}$ is a $\Sigma$-node}, \\
  \min(\vec{B}[\Pre(v)], \vec{B}_2[\overline{\Pre}(v)]) & \text{if $v
    \not\in\vec{L}$ is an $\epsilon$-node}
\end{cases} \\
\end{align*}
Importantly, note that the operations only affect the nodes in the
chunk specified by $\vec{L}$. We will use this below to
compute new values of $A$ by advancing one chunk at each step. We use
the following recursive operations: For subautomaton $A$,  integer
$b$, and character $\alpha$ define:
\begin{relate}
\item[$\Nextt_1(A, b, \alpha)$:] Let $\vec{B}$ be the current value of
  $A$ and let $A_1, \ldots, A_l$ be children of $A$ in topological
  order of their start node.
\begin{enumerate}
\item Set $\vec{B}_1 := \vec{B}$ and $\vec{B}_1[\theta_{A}] := b$.
\item For each chunk $L_i$, $1 \leq i \leq l$,
        \begin{enumerate}
        \item Compute $\vec{B}_1 := \Nextt^{A}_1(\vec{B}_1, \vec{L_i},
          \alpha)$.
        \item Compute $f_i := \Nextt_1(A_i, \vec{B}_1[\theta_{A_i}],
          \alpha)$.
        \item Set $\vec{B}_1[\phi_{A_i}] := f_i$.
\end{enumerate}
\item Compute $\vec{B}_1 := \Nextt^{A}_1(\vec{B}_1, \vec{L}_{l+1},
  \alpha)$.
\item Return $\vec{B}_1[\phi_A]$.
\end{enumerate}
\item[$\Nextt_2(A, b)$:] Let $\vec{B}$ be the current value of $A$ and
  let $A_1, \ldots, A_l$ be children of $A$ in topological order of
  their start node.
\begin{enumerate}
\item Set $\vec{B}_2 := \vec{B}$ and $\vec{B}_2[\theta_{A}] := b$.
\item For each chunk $L_i$, $1 \leq i \leq l$,
        \begin{enumerate}
        \item Compute $\vec{B}_2 := \Nextt^{A}_2(\vec{B}_2,
          \vec{L_i})$.
        \item Compute $f_i := \Nextt_2(A_i, \vec{B}_2[\theta_{A_i}])$.
        \item Set $\vec{B}_2[\phi_{A_i}] := f_i$.
\end{enumerate}
\item Compute $\vec{B}_2 := \Nextt^{A}_2(\vec{B}_2, \vec{L}_{l+1})$.
\item Return $\vec{B}_2[\phi_A]$.
\end{enumerate}
\end{relate}
The simulation of the dynamic programming recurrence on a string $Q$
of length $n$ proceeds as follows: First encode the initial values of
the all nodes in $N(R)$ using the recurrence. Let $A_r$ be the root
automaton, let $S_{j-1}$ be the current value of $N(R)$, and let
$\alpha = Q[j]$. Compute the next value $S_j$ by calling $\Nextt_1(A_r,
j, \alpha)$ and then $\Nextt_2(A_r, j, \alpha)$. Finally, if the
value of $\phi$ in the pass-2 value of $S_n$ is less than $d$, report a match. 

To see the correctness, we need to show that the calls $\Nextt_1$ and $\Nextt_2$ operations correctly compute the pass-1 and pass-2 values of $N(R)$. First consider $\Nextt_1$, and let $A$ be any subautomaton. The key property is that if $p_1$ is the pass-1 value of $\theta_A$ then after a call to $\Nextt_1(A, p_1, \alpha)$, the value of $A$ is correctly updated to the pass-1 value. This follows by a straightforward induction similar to the exact case. Since the pass-1 value of $\theta$ after reading the $j$th prefix of $Q$ is $j$, the correctness of the call to $\Nextt_1$ follows. For $\Nextt_2$ the result follows by an analogous argument.

Next we show how to efficiently represent $\Nextt^A_1$ and $\Nextt^A_2$.
First consider $\Nextt^A_1$. Note that again the alphabet size is a
problem. Since the $\vec{B}_1$ value of a node in $A$ depends on other
$\vec{B}_1$ values in $A$ we cannot ``split'' the computation of
$\Nextt^A_1$ as before. However, the alphabet character only affects
the value of $\lambda(v,\alpha)$, which is $1$ if $v$ is an
$\alpha$-node and $0$ otherwise. Hence, we can represent $\lambda(v,
\alpha)$ for all nodes in $A$ with $\Equal^A(\alpha)$ from the previous
section.  Recall that $\Equal^A(\alpha)$ can be represented for all
subautomata in total $O(m)$ space. With this representation the total
number of possible inputs to $\Nextt^A_1$ can be represented using
$(d+2)^{z} + 2^{2z}$ bits. Note that for $z = \frac{k}{\log(d+2)}$ we have
that $(d+2)^z = 2^k$.  Furthermore, since $\Nextt^A_1$ is now alphabet
independent we can apply the same trick as before and only precompute
it for all possible parse trees with leaf labels removed. It follows
that we can choose $z = \Theta(\frac{k}{\log(d+2})$ such that $\Nextt^A_1$
can precomputed in total $O(2^k)$ time and space. An analogous argument
applies to $\Nextt^A_2$.  Hence, by our discussion in the previous
sections we have shown that,
\begin{theorem}\label{s1:thm:approxregex}
  For regular expression $R$ of length $m$, string $Q$ of length $n$,
  and integer $d \geq 0$ \textsc{Approximate Regular Expression
    Matching} can be solved in $O(\frac{mn \log(d+2)}{k} + n + m\log m)$ time and $O(2^k+m)$
  space, for any $k\leq w$.
\end{theorem}

\section{Subsequence Indexing}\label{s1:sec:subseq}
The \textsc{Subsequence Indexing} problem is to preprocess a string $T$ to build a data structure supporting queries of the form:``is $Q$ a subsequence of $T$?'' for any string $Q$. This problem was considered by Baeza-Yates~\cite{BaezaYates1991} who showed the trade-offs listed in Table~\ref{s1:subseqcomplexity}.  We assume throughout the section that $T$ and $Q$ have lenght $n$ and $m$, respectively. For properties of automata accepting subsequences of string and generalizations of the problem see the recent survey~\cite{CMT2003}.
\begin{table}[t]
  \centering 
  \begin{tabular}{|c|c|c|}
\hline
Space   			& Preprocessing 	& Query \\ \hline
$O(n\sigma)$  		& $O(n\sigma)$ 		& $O(m)$ \\\hline
$O(n\log \sigma)$  	& $O(n \log \sigma)$& $O(m\log \sigma)$ \\ \hline
$O(n)$				& $O(n)$			& $O(m\log n)$ \\ \hline
\end{tabular}
  \caption{Trade-offs for \textsc{Subsequence Indexing}.} \label{s1:subseqcomplexity}
\end{table}

Using recent data structures and a few observations we improve all previous bounds. As a notational shorthand, we
will say that a data structure with preprocessing time and space
$f(n,\sigma)$ and query time $g(m,n,\sigma)$ has complexity
\pqtime{f(n,\sigma)}{g(m,n,\sigma)}

Let us consider the simplest algorithm for \textsc{Subsequence
  Indexing}. One can build a DFA of size $O(n\sigma)$
for recognizing all subsequences of $T$.  To do so, create an
accepting node for each character of $T$, and for node $v_i$,
  corresponding to character $T[i]$, create an edge to
  $v_j$ on character $\alpha$ if $T[j]$ is the first $\alpha$ after
  position $i$.  The start node has edges to the first
  occurence of each character.  Such an automaton yields an algorithm
  with complexity \pqtime{O(n\sigma)}{O(m)}.
  
  An alternative is to build, for each character $\alpha$, a data
  structure $D_{\alpha}$ with the positions of $\alpha$ in $T$.
  $D_{\alpha}$ should support fast successor queries.  The
  $D_{\alpha}$'s can all be built in a total of linear time and space
  using, for instance, van Emde Boas trees and perfect hashing~\cite{Boas1977, BKZ1977, MN1990}.  These
  trees have query time $O(\log\log n)$.  We use these vEB trees to simulate the above automaton-based algorithm: whenever we are in state $v_{i}$, and the next character to be read from $P$ is $\alpha$, we look up the successor of $i$ in   $D_{\alpha}$ in $O(\log\log n)$ time.  The complexity of this algorithm is \pqtime{O(n)}{O(m\log\log n}.
  
  We combine these two data structures as follows: Consider an automaton consisting of nodes $u_1,   \ldots, u_{n/\sigma}$, where node $u_i$ corresponds to characters   $T[\sigma(i-1), \ldots, \sigma i - 1]$, that is, each node $u_i$
  corresponds to $\sigma$ nodes in $T$. Within each such node, apply
  the vEB based data structure.  Between such nodes, apply the full
  automaton data structure.  That is, for node $w_i$, compute the
  first occurrence of each character $\alpha$ after $T[\sigma i -1]$.
  Call these \emph{long jumps}.  A edge takes you to a node
  $u_j$, and as many characters of $P$ are consumed with $u_j$ as
  possible.  When no valid edge is possible within $w_j$, take a
  long jump. The automaton uses $O(\frac{n}{\sigma} \cdot \sigma) = O(n)$ space and preprocessing time. The total size of the vEB data structures is $O(n)$. Since each $u_{i}$ consist of at most $\sigma$ nodes, the query time is improved to $O(\log \log \sigma)$. Hence, the complexity of this algorithm is \pqtime{O(n)}{O(m\log \log \sigma)}. To get a trade-off we can replace the vEB data structures by a recent data structure of Thorup~\cite[Thm. 2]{Thorup2003}. This data structure supports successor queries of $x$ integers in the range $[1,X]$ using $O(xX^{1/2^l})$ preprocessing time and space with query time $O(l+1)$, for $0\leq l \leq \log \log X$. Since each of the $n/\sigma$ groups of nodes contain at most $\sigma$ nodes, this implies the following result:
    
\begin{theorem}\label{s1:subseqthm}
  \textsc{Subsequence Indexing} can be solved in
  \pqtime{O(n\sigma^{1/2^l})}{O(m(l+1))}, for $0\leq l \leq\log\log\sigma$.
\end{theorem}

\begin{corollary} 
  \textsc{Subsequence Indexing} can be solved in
  \pqtime{O(n\sigma^{\epsilon})}{O(m)} or \pqtime{O(n)}{O(m\log\log\sigma)}.
\end{corollary}
\begin{proof} We set $l$ to be a constant or $\log\log\sigma$, respectively.\qed
\end{proof}

\section{String Edit Distance}\label{s1:sec:stringedit}
The \textsc{String Edit Distance} problem is to compute the minimum number of edit operations needed to transform a string $S$ into a string $T$. Let $m$ and $n$ be the size of $S$ and $T$, respectively. The classical solution to this problem, due to Wagner and Fischer~\cite{WF1974}, fills in the entries of an $m + 1 \times n+1$ matrix $D$. The entry $D_{i,j}$ is the edit distance between $S[1..i]$ and $T[1..j]$, and can be computed using the following recursion:
\begin{align*}
    D_{i,0} &= i   \\
    D_{0,j} &= j  \\
    D_{i,j} &= \min\{D_{i-1, j-1} + \lambda(i,j), D_{i-1, j} + 1, D_{i, j-1} + 1\}
\end{align*}
where $\lambda(i,j) = 0$ if $S[i] = T[j]$ and $1$ otherwise. The edit distance between $S$ and $T$ is the entry $D_{m,n}$. Using dynamic programming the problem can be solved in $O(mn)$ time. When filling out the matrix we only need to store the previous row or column and hence the space used is $O(\min(m,n))$. For further details, see the book by Gusfield~\cite[Chap. 11]{Gusfield1997}. 

The best algorithm for this problem, due to Masek and Paterson~\cite{MP1980}, improves the time to $O(\frac{mn}{k^{2}} +m+ n)$ time and $O(2^k + \min(m,n))$ space, for any $k \leq w$. This algorithm, however, assumes that the alphabet size is constant. In this section we give an algorithm using $O(\frac{mn\log k}{k^{2}} +m+ n)$ time and $O(2^k + \min(m,n))$ space, for any $k \leq w$, that works for any alphabet. Hence, we remove the dependency of the alphabet at the cost of a $\log k$ factor.

We first describe the algorithm by Masek and Paterson~\cite{MP1980}, and then modify it to handle arbitrary alphabets. The algorithm uses a Four Russian Trick. The matrix $D$ is divided into \emph{cells} of size $x \times x$ and all possible inputs of a cell is then precomputed and stored in a table. From the above recursion it follows that the values inside each cell $C$ depend on the corresponding substrings in $S$ and $T$, denoted $S_{C}$ and $T_{C}$, and on the values in the top row and the leftmost colunm in $C$. The number of different strings of length $x$ is $\sigma^{x}$ and hence there are $\sigma^{2x}$ possible choices for $S_{C}$ and $T_{C}$. Masek and Paterson~\cite{MP1980} showed that adjacent entries in $D$ differ by at most one, and therefore if we know the value of an entry there are exactly three choices for each adjacent entry. Since there are at most $m$ different values for the top left corner of a cell it follows that the number of different inputs for the top row and the leftmost column is $m3^{2x}$. In total, there are at $m(\sigma3)^{2x}$ different inputs to a cell. Assuming that the alphabet has constant size, we can choose $x = \Theta(k)$ such that all cells can be precomputed in $O(2^k)$ time and space. The input of each cell is stored in a single machine word and therefore all values in a cell can be computed in constant time. The total number of cells in the matrix is $O(\frac{mn}{k^{2}})$ and hence this implies an algorithm using $O(\frac{mn}{k^{2}} +m +n)$ time and $O(2^k + \min(m,n))$ space. 

We show how to generalize this to arbitrary alphabets. The first observation, similar to the idea in Section~\ref{s1:sec:appregexmatching}, is that the values inside a cell $C$ does not depend on the actual characters of $S_{C}$ and $T_{C}$, but only on the $\lambda$ function on $S_C$ and $T_C$. Hence, we only need to encode whether or not $S_{C}[i] = T_{C}[j]$ for all $1 \leq i,j \leq x$. To do this we assign a code $c(\alpha)$ to each character $\alpha$ that appears in $T_C$ or $S_C$ as follows. If $\alpha$ only appears in only one of $S_C$ or $T_C$ then $c(\alpha) = 0$.  Otherwise, $c(\alpha)$ is the rank of $\alpha$ in the sorted list of characters that appears in both $S_C$ and $T_C$. The representation is given by two vectors $\vec{S}_{C}$ and $\vec{T}_{C}$ of size $x$, where $\vec{S}_C[i] = c(S_C[i])$ and $\vec{T}_C[i] = c(T_C[i])$, for all $i$, $1\leq i \leq x$.  Clearly, $S_C[i] = T_C[j]$ iff $\vec{S}_{C}[i] = \vec{T}_{C}[j]$ and $\vec{S}_{C}[i] > 0$ and $\vec{T}_{C}[j] >0$ and hence $\vec{S}_C$ and $\vec{T}_C$ suffices to represent $\lambda$ on $C$. 

The number of characters appearing in both $T_{C}$ and $S_{C}$ is at most $x$ and hence each entry of the vectors is assigned an integer value in the range $[1, x]$. Thus, the total number of bits needed for both vectors is $2x\ceil{\log x + 1}$. Hence, we can choose $x = \Theta(\frac{k}{\log k})$ such that the vectors for a cell can be represented in a single machine word. It follows that if all vectors have been precomputed we get an algorithm for \textsc{String Edit Distance} using $O(\frac{mn\log k}{k^{2}} + m + n)$ time and $O(2^k + \min(m,n))$ space.

Next we show how to compute vectors efficiently. Given any cell $C$, we can identify the characters appearing in both $S_C$ and $T_C$ by sorting $S_C$ and then for each index $i$ in $T_C$ use a binary search to see if $T_C[i]$ appears in $S_C$. Next we sort the characters appearing in both substrings and insert their ranks into the corresponding positions in $\vec{S}_C$ and $\vec{T}_C$. All other positions in the vectors are given the value $0$.  This algorithm uses $O(x\log x)$ time for each cell. However, since the number of cells is $O(\frac{nm}{x^{2}})$ the total time becomes $O(\frac{nm\log x}{x})$, which for our choice of $x$ is $O(\frac{nm (\log k)^{2}}{k})$. To improve this we group the cells into \emph{macro cells} of $y \times y$ cells. We then compute the vector representation for each of these macro cells. The vector representation for a cell $C$ is now the corresponding subvectors of the macro cell containing $C$. Hence, each vector entry is now in the range $[0, \ldots, xy]$ and thus uses $\ceil{\log(xy + 1)}$ bits. Computing the vector representation uses $O(xy \log (xy))$ time for each macro cell and since the number of macro cells is $O(\frac{nm}{(xy)^{2}})$ the total time to compute it is $O(\frac{nm\log(xy)}{xy} +m +n)$. It follows that we can choose $y = k\log k$ and $x = \Theta(\frac{k}{\log k})$ such that vectors for a cell can be represented in a single word. Furthermore, with this choice of $x$ and $y$ all vectors are computed in $O(\frac{nm\log k}{k^{2}} + m + n)$ time. Combined with the time used to compute the distance we have shown:
\begin{theorem}\label{s1:thm:stringeditdistance}
For strings $S$ and $T$ of length $n$ and $m$, respectively, \textsc{String Edit Distance} can be solved in $O(\frac{mn\log k}{k^{2}} + m + n)$ time and $O(2^k + \min(m,n))$ space.
\end{theorem} 



\emptythanks
\chapter{New Algorithms for Regular Expression Matching}\label{chap:string2}

\title{New Algorithms for Regular Expression Matching}

\author{Philip Bille \\ IT University of Copenhagen \\ \texttt{beetle@itu.dk}}
\date{}

\cleartooddpage

\maketitle

\begin{abstract}
In this paper we revisit the classical regular expression matching problem, namely, given a regular expression $R$  and a string $Q$, decide if $Q$ matches one of the strings specified by $R$. Let $m$ and $n$ be the length of $R$ and $Q$, respectively. On a standard unit-cost RAM with word length $w \geq \log n$, we show that the problem can be solved in $O(m)$ space with the following running times:
\begin{equation*}
\begin{cases}
      O(n\frac{m \log w}{w} + m \log w) & \text{ if $m > w$} \\
      O(n\log m + m\log m) & \text{ if $\sqrt{w} < m \leq  w$} \\
      O(\min(n+ m^2, n\log m + m\log m)) & \text{ if $m \leq \sqrt{w}$.}
\end{cases}
\end{equation*} 
This improves the best known time bound among algorithms using $O(m)$ space. Whenever $w \geq \log^2 n$ it improves all known time bounds regardless of how much space is used. 
\end{abstract}

\section{Introduction} 
Regular expressions are a powerful and simple way to describe a set of strings. For this reason, they are often chosen as the input language for text processing applications. For instance, in the lexical analysis phase of compilers, regular expressions are often used to specify and distinguish tokens to be passed to the syntax analysis phase. Utilities such as Grep, the programming language Perl, and most modern text editors provide mechanisms for handling regular expressions. These applications all need to solve the classical \textsc{Regular Expression Matching} problem, namely, given a regular expression $R$ and a string $Q$, decide if $Q$ matches one of the strings specified by $R$. 

The standard textbook solution, proposed by Thompson~\cite{Thomp1968} in 1968, constructs a \emph{non-deterministic finite automaton} (NFA) accepting all strings matching $R$. Subsequently, a state-set simulation checks if the NFA accepts $Q$. This leads to a simple $O(nm)$ time and $O(m)$ space algorithm, where $m$ and $n$ are the number of symbols in $R$ and $Q$, respectively. The full details are reviewed later in Sec.~\ref{s2:sec:regex} and can found in most textbooks on compilers (e.g. Aho et. al. \cite{ASU1986}). Despite the importance of the problem, it took 24 years before the $O(nm)$ time bound was improved by Myers~\cite{Myers1992} in 1992, who achieved $O(\frac{nm}{\log n} + (n+m)\log n)$ time and $O(\frac{nm}{\log n})$ space. For most values of $m$ and $n$ this improves the $O(nm)$ algorithm by a $O(\log n)$ factor. Currently, this is the fastest known algorithm. Recently, Bille and Farach-Colton~\cite{BFC2005} showed how to reduce the space of Myers' solution to $O(n)$. Alternatively, they showed how to achieve a speedup of $O(\log m)$ over Thompson's algorithm while using $O(m)$ space. These results are all valid on a unit-cost RAM with $w$-bit words and a standard instruction set including addition, bitwise boolean operations, shifts, and multiplication. Each word is capable of holding a character of $Q$ and hence $w \geq \log n$. The space complexities refer to the number of words used by the algorithm, not counting the input which is assumed to be read-only. All results presented here assume the same model. In this paper we present new algorithms achieving the following complexities:
\begin{theorem}\label{s2:thm:main}
Given a regular expression $R$ and a string $Q$ of lengths $m$ and $n$, respectively, \textsc{Regular Expression Matching} can be solved using $O(m)$ space with the following running times:
\begin{equation*}
\begin{cases}
      O(n\frac{m \log w}{w} + m \log w) & \text{ if $m > w$} \\
      O(n\log m + m\log m) & \text{ if $\sqrt{w} < m \leq  w$} \\
      O(\min(n+ m^2, n\log m + m\log m)) & \text{ if $m \leq \sqrt{w}$.}
\end{cases}
\end{equation*} 
\end{theorem}
This represents the best known time bound among algorithms using $O(m)$ space. To compare these with previous results, consider a conservative word length of $w = \log n$. When the regular expression is "large", e.g.,  $m > \log n$, we achieve an $O(\frac{\log n}{\log \log n})$ factor speedup over Thompson's algorithm using $O(m)$ space. Hence, we simultaneously match the best known time and space bounds for the problem, with the exception of an $O(\log \log n)$ factor in time. More interestingly, consider the case when the regular expression is "small", e.g., $m = O(\log n)$. This is usually the case in most applications. To beat the $O(n\log n)$ time of Thompson's algorithm, the fast algorithms~\cite{Myers1992, BFC2005} essentially convert the NFA mentioned above into a \emph{deterministic finite automaton} (DFA) and then simulate this instead. Constructing and storing the DFA incurs an additional exponential time and space cost in $m$, i.e., $O(2^m) = O(n)$ (see~\cite{WM1992b,NR2004} for compact DFA representations). However, the DFA can now be simulated in $O(n)$ time, leading to an $O(n)$ time and space algorithm. Surprisingly, our result shows that this exponential blow-up in $m$ can be avoided with very little loss of efficiency. More precisely, we get an algorithm using $O(n\log \log n)$ time and $O(\log n)$ space. Hence, the space is improved exponentially at the cost of an $O(\log \log n)$ factor in time. In the case of an even smaller regular expression, e.g., $m = O(\sqrt{\log n})$, the slowdown can be eliminated and we achieve optimal $O(n)$ time. For larger word lengths our time bounds improve. In particular, when $w > \log n \log \log n$ the bound is better in all cases, except for $\sqrt{w} \leq m \leq w$, and when $w > \log^2n$ it improves all known time bounds regardless of how much space is used.

The key to obtain our results is to avoid explicitly converting small NFAs into DFAs. Instead we show how to effectively simulate them directly using the parallelism available at the word-level of the machine model. The kind of idea is not new and has been applied to many other string matching problems, most famously, the Shift-Or algorithm~\cite{BYG1992}, and the approximate string matching algorithm by Myers \cite{Myers1999}. However, none of these algorithms can be easily extended to \textsc{Regular Expression Matching}. The main problem is the complicated dependencies between states in an NFA. Intuitively, a state may have long paths of $\epsilon$-transitions to a large number of other states, all of which have to be traversed in parallel in the state-set simulation. To overcome this problem we develop several new techniques ultimately leading to Theorem~\ref{s2:thm:main}. For instance, we introduce a new hierarchical decomposition of NFAs suitable for a parallel state-set simulation. We also show how state-set simulations of large NFAs efficiently reduces to simulating small NFAs.

The results presented in this paper are primarily of theoretical interest. However, we believe that most of the ideas are useful in practice. The previous algorithms require large tables for storing DFAs, and perform a long series of lookups in these tables. As the tables become large we can expect a high number of cache-misses during the lookups, thus limiting the speedup in practice. Since we avoid these tables, our algorithms do not suffer from this defect. 

The paper is organized as follows. In Sec.~\ref{s2:sec:regex} we review Thompson's NFA construction, and in Sec.~\ref{s2:sec:simul} we present the above mentioned reduction. In Sec.~\ref{s2:sec:simple} we present our first simple algorithm for the problem which is then improved in Sec.~\ref{s2:sec:notsimple}. Combining these algorithms with our reduction leads to Theorem~\ref{s2:thm:main}. We conclude with a couple of remarks and open problems in Sec.~\ref{s2:sec:remarks}.

\section{Regular Expressions and Finite Automata}\label{s2:sec:regex}
In this section we briefly review Thompson's construction and the standard
state-set simulation.  The set of \emph{regular expressions} over an alphabet 
$\Sigma$ are defined recursively as follows: 
\begin{itemize}
\item A character $\alpha \in \Sigma$ is a regular expression.
\item If $S$ and $T$ are regular expressions then so is the
  \emph{concatenation}, $(S)\cdot(T)$, the \emph{union}, $(S)|(T)$, and
  the \emph{star}, $(S)^*$.
\end{itemize}
Unnecessary parentheses can be removed by observing that $\cdot$ and
$|$ is associative and by using the standard precedence of the
operators, that is $*$ precedes $\cdot$, which in turn precedes $|$. We often remove the $\cdot$ when writing regular expressions. 

\begin{figure}[t] 
  \centering \includegraphics[scale=.5]{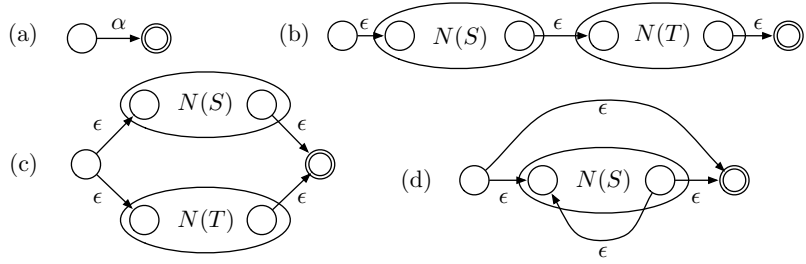}
  \caption{Thompson's NFA construction. The regular expression for a character $\alpha \in \Sigma$ corresponds to NFA $(a)$. If $S$ and $T$ are regular expressions then $N(ST)$, $N(S|T)$, and
    $N(S^*)$ correspond to NFAs $(b)$, $(c)$, and $(d)$, respectively.
    Accepting nodes are marked with a double circle.}
  \label{s2:fig:thompson}
\end{figure}

The \emph{language} $L(R)$ generated by $R$ is the set of all strings matching $R$. The \emph{parse tree} $T(R)$ of $R$ is the binary rooted tree representing the hiearchical structure of $R$. Each leaf is labeled by a character in $\Sigma$ and each internal node is labeled either $\cdot$, $|$, or $*$. A \emph{finite automaton} is a tuple $A = (V, E, \delta, \theta, \phi)$, where 
\begin{itemize}
  \item $V$ is a set of nodes called \emph{states},
  \item $E$ is set of directed edges between states called \emph{transitions}, 
  \item $\delta : E \rightarrow \Sigma \cup \{\epsilon\}$ is a function assigning labels to transitions, and
  \item $\theta, \phi \in V$ are distinguished states called the \emph{start state} and \emph{accepting state}, respectively\footnote{Sometimes NFAs are allowed a \emph{set} of accepting states, but this is not necessary for our purposes.}.
\end{itemize} 
Intuitively, $A$ is an edge-labeled directed graph with special start and accepting nodes. $A$ is a \emph{deterministic finite automaton} (DFA) if $A$ does not contain any $\epsilon$-transitions, and all outgoing transitions of any state have different labels. Otherwise, $A$ is a \emph{non-deterministic automaton} (NFA). We say that $A$ \emph{accepts} a string $Q$ if there is a path from $\theta$ to $\phi$ such that the concatenation of labels on the path spells out $Q$. 
Thompson~\cite{Thomp1968} showed how to recursively construct a NFA $N(R)$ accepting all strings in $L(R)$. 
The rules are presented below and illustrated in Fig.~\ref{s2:fig:thompson}.

\begin{itemize}
\item $N(\alpha)$ is the automaton consisting of states $\theta_\alpha$, $\phi_\alpha$, and an $\alpha$-transition
  from $\theta_\alpha$ to $\phi_\alpha$.
\item Let $N(S)$ and $N(T)$ be automata for regular expressions $S$ and
  $T$ with start and accepting states $\theta_S$, $\theta_T$, $\phi_S$,
  and $\phi_T$, respectively. Then, NFAs $N(S\cdot T)$, $N(S|T)$,
  and $N(S^*)$ are constructed as follows:
  \begin{relate}
  \item[$N(ST)$:] Add start state $\theta_{ST}$ and accepting state $\phi_{ST}$, and $\epsilon$-transitions $(\theta_{ST}, \theta_{S})$, $(\phi_{S}, \theta_T)$, and $(\phi_T, \phi_{ST})$.
  \item[$N(S|T)$:] Add start state $\theta_{S|T}$ and accepting state $\phi_{S|T}$, and add $\epsilon$-transitions $(\theta_{S|T}, \theta_S)$, $(\theta_{S|T}, \theta_T)$, $(\phi_S, \phi_{S|T})$, and $(\phi_T, \phi_{S|T})$.
  \item [$N(S^*)$:] Add a new start state $\theta_{S^*}$ and
    accepting state $\phi_{S^*}$, and $\epsilon$-transitions $(\theta_{S^*}, \theta_S)$, $(\theta_{S^*}, \phi_{S^*})$, $(\phi_S, \phi_{S^*})$, and $(\phi_S, \theta_S)$.
\end{relate}
\end{itemize}

Readers familiar with Thompson's construction will notice that $N(ST)$ is slightly different from the usual construction. 
This is done to simplify our later presentation and does not affect the worst case complexity of the problem. Any automaton produced by these rules we call a \emph{Thompson-NFA} (TNFA). By construction, $N(R)$ has a single start and accepting state, denoted $\theta$ and $\phi$, respectively. $\theta$ has no incoming transitions and $\phi$ has no outgoing transitions. The total number of states is $2m$ and since each state has at most $2$ outgoing transitions that the total number of transitions is at most $4m$.  Furthermore, all incoming
transitions have the same label, and we denote a state with incoming $\alpha$-transitions an \emph{$\alpha$-state}. Note that the star construction in Fig. \ref{s2:fig:thompson}(d) introduces a transition from the accepting state of $N(S)$ to the start state of $N(S)$. All such transitions are called \emph{back transitions} and all other transitions are \emph{forward transitions}. We need the following property. 
\begin{lemma}[Myers \cite{Myers1992}]\label{s2:lem:cyclefree} Any cycle-free path in a TNFA contains at most one back transition.
\end{lemma}
For a string $Q$ of length $n$ the standard state-set simulation of
$N(R)$ on $Q$ produces a sequence of state-sets $S_0, \ldots, S_n$. The
$i$th set $S_i$, $0\leq i \leq n$, consists of all states in $N(R)$ for which there is a path from $\theta$ that spells out the $i$th prefix of $Q$. The simulation can be implemented with the following simple
operations. For a state-set $S$ and a character $\alpha \in \Sigma$, define
\begin{relate}
\item[$\Move(S,\alpha)$:] Return the set of states reachable from $S$ via a single $\alpha$-transition.
\item[$\Close(S)$:] Return the set of states reachable from $S$ via $0$ or more $\epsilon$-transitions.
\end{relate}
Since the number of states and transitions in $N(R)$ is $O(m)$, both operations can be easily implemented in $O(m)$ time. The $\Close$ operation is often called an \emph{$\epsilon$-closure}. The simulation proceeds as follows: Initially, $S_0 := \Close(\{\theta\})$. If $Q[j]=\alpha$, $1\leq j \leq
n$, then $S_j := \Close(\Move(S_{j-1}, \alpha))$. Finally, $Q \in L(R)$ iff $\phi \in S_n$. Since each state-set $S_j$ only depends on $S_{j-1}$ this algorithm uses $O(mn)$ time and $O(m)$ space. 

\section{From Large to Small TNFAs}\label{s2:sec:simul}
In this section we show how to simulate $N(R)$ by simulating a number of smaller TNFAs. We will use this to achieve our bounds when $R$ is large. 

\subsection{Clustering Parse Trees and Decomposing TNFAs}\label{s2:sec:clustering}
\begin{figure}[t]
  \centering \includegraphics[scale=.5]{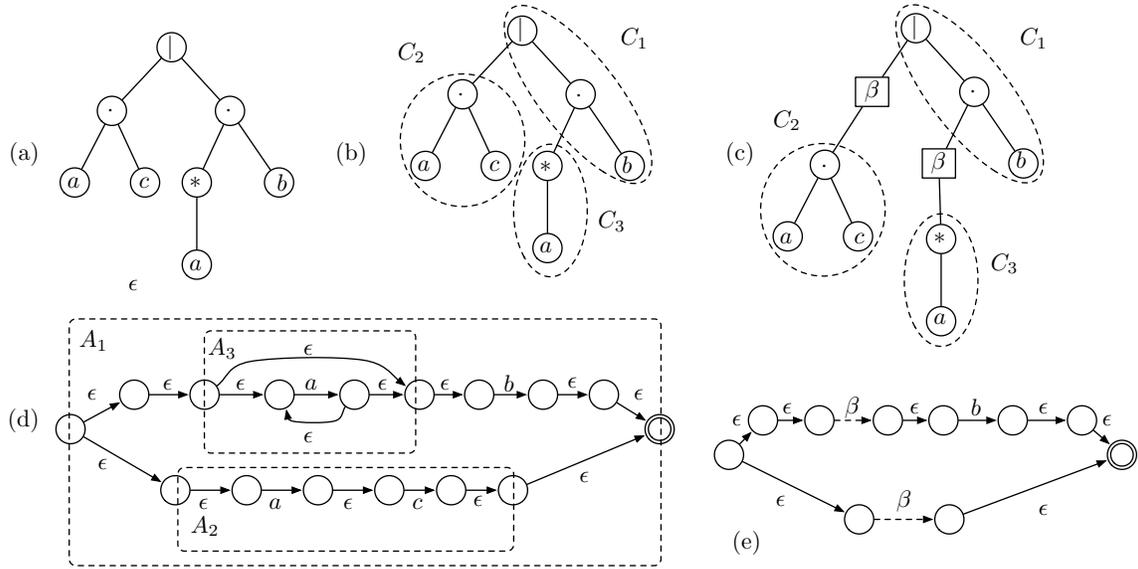}
   \caption{(a) The parse tree for the regular expression
     $ac|a^*b$. (b) A clustering of $(a)$ into
     node-disjoint connected subtrees $C_1$, $C_2$, and $C_3$, each with at most $3$ nodes. (c) The clustering from (b) extended with pseudo-nodes. (d) The nested decomposition of $N(ac|a^*b)$. (e) The TNFA corresponding to $C_1$.}
   \label{s2:fig:clustering}
\end{figure}

Let $R$ be a regular expression of length $m$. We first show how to decompose $N(R)$ into smaller TNFAs. This decomposition is based on a simple clustering of the parse tree $T(R)$. A \emph{cluster} $C$ is a connected subgraph of $T(R)$ and a \emph{cluster partition} $CS$ is a partition of the nodes of $T(R)$ into node-disjoint clusters. Since $T(R)$ is a binary tree with $O(m)$ nodes, a simple top-down procedure provides the following result (see e.g. \cite{Myers1992}):
\begin{lemma}\label{s2:lem:clustering}
Given a regular expression $R$ of length $m$ and a parameter $x$, a cluster partition $CS$ of $T(R)$ can be constructed in $O(m)$ time such that $|CS| = O(\ceil{m/x})$, and for any $C\in CS$, the number of nodes in $C$ is
  at most $x$.
\end{lemma}
For a cluster partition $CS$, edges adjacent to two clusters are \emph{external edges} and all other edges are \emph{internal edges}. Contracting all internal edges in $CS$ induces a \emph{macro tree}, where each cluster is represented by a single \emph{macro node}. Let $C_v$ and $C_w$ be two clusters with corresponding macro nodes $v$ and $w$. We say that $C_v$ is the \emph{parent cluster} (resp. \emph{child cluster}) of $C_w$ if $v$ is the parent (resp. child) of $w$ in the macro tree. The \emph{root cluster and leaf clusters} are the clusters corresponding to the root and the leaves of the macro tree. An example clustering of a parse tree is shown in Fig.~\ref{s2:fig:clustering}(b).
Given a cluster partition $CS$ of $T(R)$ we show how to divide $N(R)$ into a set of small nested TNFAs. Each cluster $C \in CS$ will correspond to a TNFA $A$, and we use the terms child, parent, root, and leaf for the TNFAs in the same way we do with clusters. For a cluster $C \in CS$ with children $C_1, \ldots, C_l$, insert  a special \emph{pseudo-node} $p_i$, $1 \leq i \leq l$, in the middle of the external edge connecting $C$ with $C_i$. We label each pseudo-node by a special character $\beta \not \in \Sigma$. Let $T_C$ be the tree induced by the set of nodes in $C$ and $\{p_1, \ldots, p_l\}$. Each leaf in $T_C$ is labeled with a character from $\Sigma \cup \{\beta\}$, and hence $T_C$ is a well-formed parse tree for some regular expression $R_C$ over $\Sigma \cup \{\beta\}$. Now, the TNFA $A$ corresponding to $C$ is $N(R_C)$. In $A$, child TNFA $A_i$ is represented by its start and accepting state $\theta_{A_i}$ and $\phi_{A_i}$ and a \emph{pseudo-transition} labeled $\beta$ connecting them. An example of these definitions is given in Fig.~\ref{s2:fig:clustering}. We call any set of TNFAs obtained from a cluster partition as above a \emph{nested decomposition} $AS$ of $N(R)$. 
\begin{lemma}\label{s2:lem:decomp}
  Given a regular expression $R$ of length $m$ and a parameter $x$, a nested decomposition $AS$ of $N(R)$ can be constructed in $O(m)$ time such that $|AS| = O(\ceil{m/x})$, and for any $A\in AS$, the number of states in $A$ is at most $x$.
\end{lemma}
\begin{proof}
Construct the parse tree $T(R)$ for $R$ and build a cluster partition $CS$ according to Lemma~\ref{s2:lem:clustering} with parameter $y = \frac{x}{4} - \frac{1}{2}$. From $CS$ build a nested decomposition $AS$ as described above. Each $C \in CS$ corresponds to a TNFA $A \in AS$ and hence $|AS| = O(\ceil{m/y}) = O(\ceil{m/x})$. Furthermore, if $|V(C)| \leq y$ we have $|V(T_C)| \leq 2y + 1$. Each node in $T_C$ contributes two states to the corresponding TNFA $A$, and hence the total number of states in $A$ is at most $4y + 2 = x$. Since the parse tree, the cluster partition, and the nested decomposition can be constructed in $O(m)$ time the result follows.\qed
\end{proof}

\subsection{Simulating Large Automata}
We now show how $N(R)$ can be simulated using the TNFAs in a nested decomposition. For this purpose we define a simple data structure to dynamically maintain the TNFAs. Let $AS$ be a nested decomposition of $N(R)$ according to Lemma~\ref{s2:lem:decomp}, for some parameter $x$. Let $A \in AS$ be a TNFA, let $S_A$ be a state-set of $A$, let $s$ be a state in $A$, and let $\alpha \in \Sigma$.  A \emph{simulation data structure} supports the $4$ operations: $\Move_A(S_A, \alpha)$, $\Close_A(S_A)$, $\Member_A(S_A, s)$, and $\Insert_A(S_A, s)$. Here, the operations $\Move_A$ and $\Close_A$ are defined exactly as in Sec.~\ref{s2:sec:regex}, with the modification that they only work on $A$ and not $N(R)$. The operation $\Member_A(S_A, s)$ returns yes if $s \in S_A$ and no otherwise and $\Insert_A(S_A, s)$ returns the set $S_A \cup \{s\}$.

In the following sections we consider various efficient implementations of simulation data structures. For now assume that we have a black-box data structure for each $A \in AS$. To simulate $N(R)$ we proceed as follows. First, fix an ordering of  the TNFAs in the nested decomposition $AS$, e.g., by a preorder traversal of the tree represented given by the parent/child relationship of the TNFAs. The collection of state-sets for each TNFA in $AS$ are represented in a \emph{state-set array} $X$ of length $|AS|$. The state-set array is indexed by the above numbering, that is, $X[i]$ is the state-set of the $i$th TNFA in $AS$. For notational convenience we write $X[A]$ to denote the entry in $X$ corresponding to $A$. Note that a parent TNFA share two states with each child, and therefore a state may be represented more than once in $X$. To avoid complications we will always assure that $X$ is \emph{consistent}, meaning that if a state $s$ is included in the state-set of some TNFA, then it is also included in the state-sets of all other TNFAs that share $s$. If $S =  \bigcup_{A\in AS} X[A]$ we say that $X$ \emph{models} the state-set $S$ and write $S \equiv X$. 

Next we show how to do a state-set simulation of $N(R)$ using the operations $\Move_{AS}$ and $\Close_{AS}$, which we define below. These operations recursively update a state-set array using the simulation data structures. For any $A\in AS$, state-set array $X$, and $\alpha \in \Sigma$ define
\begin{relate}
\item[$\Move_{AS}(A, X, \alpha)$:] 
\begin{enumerate}
\item $X[A] :=  \Move_{A}(X[A], \alpha)$
\item For each child $A_i$ of $A$ in topological order do
\begin{enumerate}
\item $X := \Move_{AS}(A_i, X, \alpha)$
\item If $\phi_{A_i} \in X[A_i]$ then $X[A] := \Insert_A(X[A],\phi_{A_i})$
\end{enumerate}
\item Return $X$
\end{enumerate}
\item[$\Close_{AS}(A, X)$:] 
\begin{enumerate}
\item $X[A] := \Close_A(X[A])$
\item For each child $A_i$ of $A$ in topological order do
\begin{enumerate}
\item If $\theta_{A_i} \in X[A]$ then $X[A_i] := \Insert_{A_i}(X[A_i], \theta_{A_i})$
\item X := $\Close_{AS}(A_i, X)$
\item If $\phi_{A_i} \in X[A_i]$ then $X[A] := \Insert_A(X[A],\phi_{A_i})$
\item $X[A] := \Close_A(X[A])$
\end{enumerate}
\item Return $X$
\end{enumerate}
\end{relate}
The $\Move_{AS}$ and $\Close_{AS}$ operations recursively traverses the nested decomposition top-down processing the children in topological order. At each child the shared start and accepting states are propagated in the state-set array. For simplicity, we have written $\Member_A$ using the symbol $\in$.

The state-set simulation of $N(R)$ on a string $Q$ of length $n$ produces the sequence of state-set arrays $X_0, \ldots, X_n$ as follows: Let $A_r$ be the root automaton and let $X$ be an empty state-set array (all entries in $X$ are $\emptyset$). Initially, set $X[A_r] := \Insert_{A_r}(X[A_r], \theta_{A_r})$ and compute $X_0 := \Close_{AS}(A_r, \Close_{AS}(A_r, X))$. For $i>0$ we compute $X_{i}$ from $X_{i-1}$ as follows: 
\begin{equation*}
X_i := \Close_{AS}(A_r, \Close_{AS}(A_r, \Move_{AS}(A_r, X_{i-1}, Q[i])))
\end{equation*}
Finally, we output $Q \in L(R)$ iff $\phi_{A_r} \in X_n[A_r]$. To see that this algorithm correctly solves \textsc{Regular Expression Matching} it suffices to show that for any $i$, $0\leq i \leq n$, $X_i$ correctly models the $i$th state-set $S_i$ in the standard state-set simulation.  We need the following lemma. 
\begin{lemma}\label{s2:lem:statearray}
Let $X$ be a state-set array and let $A_r$ be the root TNFA in a nested decomposition $AS$. If $S$ is the state-set modeled by $X$, then  
\begin{itemize}
  \item $\Move(S,\alpha) \equiv \Move_{AS}(A_r, X, \alpha)$ and 
  \item $\Close(S) \equiv \Close_{AS}(A_r, \Close_{AS}(A_r, X))$.
\end{itemize}
\end{lemma}
\begin{proof}
First consider the $\Move_{AS}$ operation. Let $\overline{A}$ be the TNFA induced by all states in $A$ and descendants of $A$ in the nested decomposition,  i.e., $\overline{A}$ is obtained by recursively "unfolding" the pseudo-states and pseudo-transitions in $A$, replacing them by the TNFAs they represent.  We show by induction that the state-array $X_A' := \Move_{AS}(A, X, \alpha)$ models $\Move(S,\alpha)$ on $\overline{A}$. In particular, plugging in $A = A_r$, we have that $\Move_{AS}(A_r, X, \alpha)$ models $\Move(S,\alpha)$ as required. 

Initially, line $1$ updates $X[A]$ to be the set of states reachable from a single $\alpha$-transition in $A$.
If $A$ is a leaf, line $2$ is completely bypassed and the result follows immediately. Otherwise, let $A_1, \ldots, A_l$ be the children of $A$ in topological order. Any incoming transition to a state $\theta_{A_i}$ or outgoing transition from a state $\phi_{A_i}$ is an $\epsilon$-transition by Thompson's construction. Hence, no endpoint of an $\alpha$-transition in $A$ can be shared with any of the children $A_1, \ldots, A_l$. It follows that after line $1$ the updated $X[A]$ is the desired state-set, except for the shared states, which have not been handled yet. By induction, the recursive calls in line $2$(a) handle the children. Among the shared states only the accepting ones, $\phi_{A_1}, \ldots, \phi_{A_l}$, may be the endpoint of an $\alpha$-transition and therefore line $2$(b) computes the correct state-set.

The $\Close_{AS}$ operation proceeds in a similar, though slightly more complicated fashion. Let $\widetilde{X}_A$ be the state-array modeling the set of states reachable via a path of \emph{forward} $\epsilon$-transitions in $\overline{A}$, and let $\widehat{X}_A$ be the state array modelling $\Close(S)$ in $\overline{A}$. We show by induction that if $X_A'' := \Close_{AS}(A, X)$ then
\begin{equation*}
\widetilde{X}_A \subseteq X_A'' \subseteq \widehat{X}_A,
\end{equation*}
where the inclusion refers to the underlying state-sets modeled by the state-set arrays. Initially, line $1$ updates $X[A] := \Close_{A}(X[A])$. If $A$ is a leaf then clearly $X_A'' = \widehat{X}_A$. Otherwise, let $A_1, \ldots, A_l$ be the children of $A$ in topological order. Line $2$ recursively update the children and propagate the start and accepting states in (a) and (c). Following each recursive call we again update $X[A] := \Close_{A}(X[A])$ in (d). No state is included in $X_A''$ if there is no $\epsilon$-path in $A$ or through any child of $A$. Furthermore, since the children are processed in topological order it is straightforward to verify that the sequence of updates in line $2$ ensure that $X_A''$ contain all states reachable via a path of forward $\epsilon$-transitions in $A$ or through a child of $A$. Hence, by induction we have $\widetilde{X}_A \subseteq X_A'' \subseteq \widehat{X}_A$ as desired. 

A similar induction shows that the state-set array $\Close_{AS}(A_r, X'')$ models the set of states reachable from $X''$ using a path consisting of forward $\epsilon$-transitions and at most $1$ back transition. However, by Lemma~\ref{s2:lem:cyclefree} this is exactly the set of states reachable by a path of $\epsilon$-transitions. Hence, $\Close_{AS}(A_r, X'')$ models $\Close(S)$ and the result follows.\qed
\end{proof}

By Lemma~\ref{s2:lem:statearray} the state-set simulation can be done using the $\Close_{AS}$ and $\Move_{AS}$ operations and the complexity now directly depends on the complexities of the simulation data structure. Putting it all together the following reduction easily follows:
\begin{lemma}\label{s2:lem:simulation}
Let $R$ be a regular expression  of length $m$ over alphabet $\Sigma$ and let $Q$ a string of length $n$. Given a simulation data structure for TNFAs with $x < m$ states over alphabet $\Sigma \cup \{\beta\}$, where $\beta \not\in \Sigma$, that supports all operations in $O(t(x))$ time, using $O(s(x))$ space, and $O(p(x))$ preprocessing time, \textsc{Regular Expression Matching} for $R$ and $Q$ can be solved in $O(\frac{nm \cdot t(x)}{x} + \frac{m\cdot p(x)}{x})$ time using $O(\frac{m \cdot s(x)}{x})$ space. 
\end{lemma}
\begin{proof}
Given $R$ first compute a nested decomposition $AS$ of $N(R)$ using Lemma~\ref{s2:lem:decomp} for parameter $x$. For each TNFA $A \in AS$ sort $A$'s children to topologically and keep pointers to start and accepting states. By Lemma~\ref{s2:lem:decomp} and since topological sort can be done in $O(m)$ time this step uses $O(m)$ time. The total space to represent the decomposition is $O(m)$. Each $A \in AS$ is a TNFA over the alphabet $\Sigma \cup \{\beta\}$ with at most $x$ states and $|AS| = O(\frac{m}{x})$. Hence, constructing simulation data structures for all $A \in AS$ uses $O(\frac{m p(x)}{x})$ time and $O(\frac{m s(x)}{x})$ space. With the above algorithm the state-set simulation of $N(R)$ can now be done in $O(\frac{nm \cdot t(x)}{x})$ time, yielding the desired complexity.\qed
\end{proof}

The idea of decomposing TNFAs is also present in Myers' paper~\cite{Myers1992}, though he does not give a "black-box" reduction as in Lemma~\ref{s2:lem:simulation}. We believe that the framework provided by Lemma~\ref{s2:lem:simulation} helps to simplify the presentation of the algorithms significantly. We can restate Myers' result in our setting as the existence of a simulation data structure with $O(1)$ query time that uses $O(x\cdot 2^x)$ space and preprocessing time. For $x \leq \log (n/\log n)$ this achieves the result mentioned in the introduction. The key idea is to encode and tabulate the results of all queries (such an approach is frequently referred to as the "Four Russian Technique"~\cite{ADKF1970}). Bille and Farach~\cite{BFC2005} give a more space-efficient encoding that does not use Lemma~\ref{s2:lem:simulation} as above. Instead they show how to encode \emph{all possible} simulation data structures in total $O(2^x + m)$ time and space while maintaining $O(1)$ query time.  

In the following sections we show how to efficiently avoid the large tables needed in the previous approaches. Instead we implement the operations of simulation data structures using the word-level parallelism of the machine model. 

\section{A Simple Algorithm}\label{s2:sec:simple}
In this section we present a simple simulation data structure for TNFAs, and develop some of the ideas for the improved result of the next section. Let $A$ be a TNFA with $m = O(\sqrt{w})$ states. We will show how to support all operations in $O(1)$ time using $O(m)$ space and $O(m^2)$ preprocessing time.

To build our simulation data structure for $A$, first sort all states in $A$ in topological order ignoring the back transitions.  We require that the endpoints of an $\alpha$-transition are consecutive in this order. This is automatically guaranteed using a standard $O(m)$ time algorithm for topological sorting (see e.g. \cite{CLRS2001}). We will refer to states in $A$ by their rank in this order. A state-set of $A$ is represented using a bitstring $S = s_1s_2\ldots s_m$ defined such that $s_i = 1$ iff node $i$ is in the state-set. The simulation data structure consists of the following bitstrings:
\begin{itemize}
\item For each $\alpha \in \Sigma$,  a string $D_\alpha = d_1 \ldots d_m$ such that $d_i = 1$ iff $i$ is an $\alpha$-state. 
\item A string $E = 0e_{1,1}e_{1,2}\ldots e_{1,m}0e_{2,1}e_{2,2}\ldots e_{2,m}0 \ldots 0 e_{m,1}e_{m,2}\ldots e_{m,m}$, where $e_{i,j} = 1$ iff $i$ is $\epsilon$-reachable from $j$. The zeros are \emph{test bits} needed for the algorithm.
\item Three constants $I = (10^m)^m$, $X = 1(0^m1)^{m-1}$, and $C = 1(0^{m-1}1)^{m-1}$. Note that $I$ has a $1$ in each test bit position\footnote{We use exponentiation to denote repetition, i.e., $1^30 = 1110$.}.
\end{itemize}
The strings $E$, $I$, $X$, and $C$ are easily computed in $O(m^2)$ time and use $O(m^2)$ bits. Since $m = O(\sqrt{w})$ only $O(1)$ space is needed to store these strings. We store $D_\alpha$ in a hashtable indexed by $\alpha$. Since the total number of different characters in $A$ can be at most $m$, the hashtable contains at most $m$ entries. Using perfect hashing $D_\alpha$ can be represented in $O(m)$ space with $O(1)$ worst-case lookup time. The preprocessing time is expected $O(m)$ w.h.p.. To get a worst-case bound we use the deterministic dictionary of Hagerup et. al. \cite{HMP2001} with $O(m\log m)$ worst-case preprocessing time. In total the data structure requires $O(m)$ space and $O(m^2)$ preprocessing time. 

Next we show how to support each of the operations on $A$. Suppose $S = s_1 \ldots s_m$ is a bitstring representing a state-set of $A$ and $\alpha \in \Sigma$. The result of $\Move_A(S,\alpha)$ is given by
\begin{equation*}
S' := (S >> 1) \: \& \: D_\alpha.
\end{equation*}
This should be understood as C notation, where the right-shift is unsigned. Readers familiar with the Shift-Or algorithm~\cite{BYG1992} will notice the similarity. To see the correctness, observe that state $i$ is put in $S'$ iff state $(i-1)$ is in $S$ and the $i$th state is an $\alpha$-state. Since the endpoints of $\alpha$-transitions  are consecutive in the topological order it follows that $S'$ is correct. Here, state $(i-1)$ can only influence state $i$, and this makes the operation easy to implement in parallel. However, this is not the case for $\Close_A$. Here, any state can potentially affect a large number of states reachable through long $\epsilon$-paths. To deal with this we use the following steps.
\begin{align*}
Y &:= (S \times X) \:\&\: E \\
Z &:= ((Y \: |\: I) - (I >> m)) \:\&\: I \\
S' &:= ((Z \times C) << w-m(m+1)) >> w - m
\end{align*}
We describe in detail why this, at first glance somewhat cryptic sequence, correctly computes $S'$ as the result of $\Close_A(S)$. The variables $Y$ and $Z$ are simply temporary variables inserted to increase the readability of the computation. Let $S = s_1 \ldots s_m$. Initially, $S \times X$ concatenates $m$ copies of $S$ with a zero bit between each copy, that is, 
\begin{equation*}
S \times X = s_1\ldots s_m \times 1(0^m1)^{m-1} = (0s_1\ldots s_m)^m.
\end{equation*}
The bitwise $\&$ with $E$ gives 
\begin{equation*}
Y = 0y_{1,1}y_{1,2}\ldots y_{1,m}0y_{2,1}y_{2,2}\ldots y_{2,m}0 \ldots 0 y_{m,1}y_{m,2}\ldots y_{m,m},
\end{equation*} 
where $y_{i,j} = 1$ iff state $j$ is in $S$ and state $i$ is $\epsilon$-reachable from $j$. In other words, the substring $Y_i = y_{i,1} \ldots y_{i,m}$ indicates the set of states in $S$ that have a path of $\epsilon$-transitions to $i$. Hence, state $i$ should be included in $\Close_A(S)$ precisely if at least one of the bits in $Y_i$ is $1$. This is determined next. First $(Y \: | \: I) - (I >> m)$ sets all test bits to $1$ and subtracts the test bits shifted right by $m$ positions. This ensures that if all positions in $Y_i$ are $0$, the $i$th test bit in the result is $0$ and otherwise $1$. The test bits are then extracted with a bitwise $\&$ with $I$, producing the string $Z = z_10^mz_20^m\ldots z_m0^m$. This is almost what we want since $z_i = 1$ iff state $i$ is in $\Close_A(S)$. 
The final computation \emph{compresses} the $Z$ into the desired format. The multiplication produces the following length $2m^2$ string:
\begin{equation*}
\begin{split}
Z \times C &= z_10^mz_20^m\ldots z_m0^m \times 1(0^{m-1}1)^{m-1}  \\
&= z_10^{m-1} z_1z_20^{m-2} \cdots z_1\ldots z_{k}0^{m-k} \cdots z_1\ldots z_{m-1}0 z_1\ldots z_m  
0z_2\ldots z_m \cdots 0^{k}z_{k+1}\ldots z_m \cdots 0^{m-1} z_m 0^m  
\end{split} 
\end{equation*}
In particular, positions $m(m-1)+1$ through $m^2$ (from the left) contain the test bits compressed into a string of length $m$. The two shifts zeroes all other bits and moves this substring to the rightmost position in the word, producing the final result. Since $m = O(\sqrt{w})$ all of the above operations can be done in constant time.

Finally, observe that $\Insert_A$ and $\Member_A$ are trivially implemented in constant time. Thus, 
\begin{lemma}\label{s2:lem:datastruct1}
For any TNFA with $m = O(\sqrt{w})$ states there is a simulation data structure using $O(m)$ space and $O(m^2)$ preprocessing time which supports all operations in $O(1)$ time. 
\end{lemma}   
The main bottleneck in the above data structure is the string $E$ that represents all $\epsilon$-paths. On a TNFA with $m$ states $E$ requires at least $m^2$ bits and hence this approach only works for $m = O(\sqrt{w})$. In the next section we show how to use the structure of TNFAs to do better.

\section{Overcoming the $\epsilon$-closure Bottleneck}\label{s2:sec:notsimple}
In this section we show how to compute an $\epsilon$-closure on a TNFA with $m = O(w)$ states in $O(\log m)$ time. Compared with the result of the previous section we quadratically increase the size of the TNFA  at the expense of using logarithmic time. The algorithm is easily extended to an efficient simulation data structure. The key idea is a new hierarchical decomposition of TNFAs described below.

\subsection{Partial-TNFAs and Separator Trees} 
\begin{figure}[t]
  \centering \includegraphics[scale=.5]{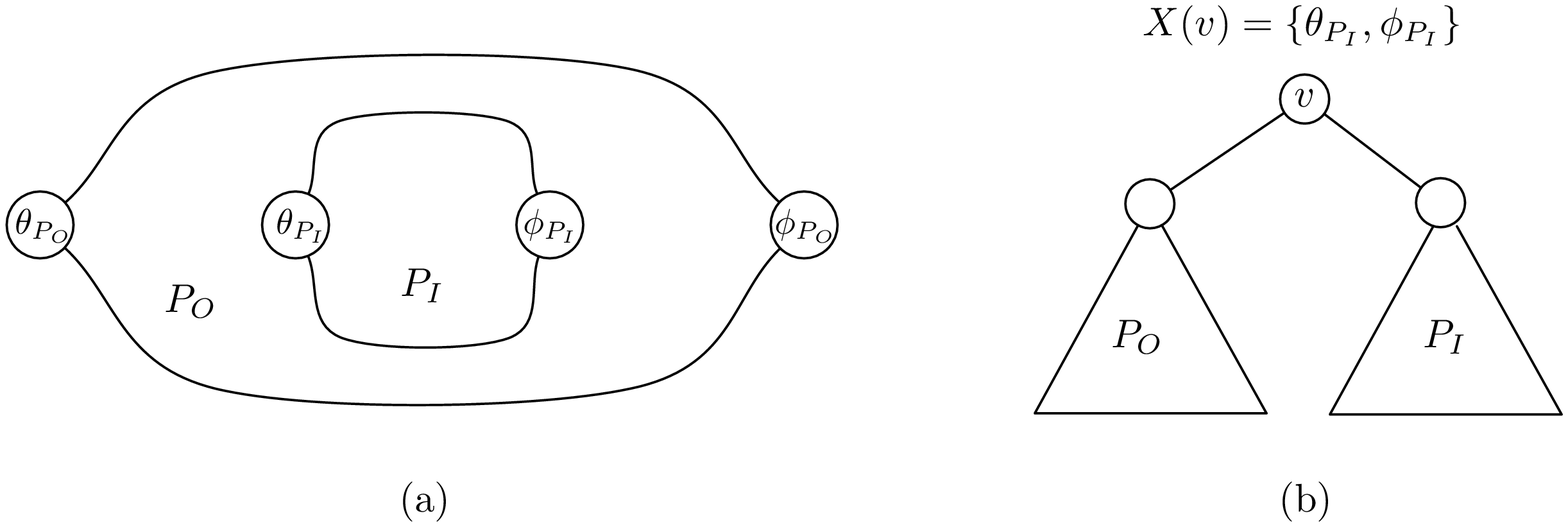}
   \caption{(a) Inner and outer pTNFAs. (b) The corresponding separator tree construction.}
   \label{s2:fig:separator}
\end{figure}

First we need some definitions. Let $A$ be a TNFA with parse tree $T$. Each node $v$ in $T$ uniquely corresponds to two states in $A$, namely, the start and accepting states $\theta_{A'}$ and $\phi_{A'}$ of the TNFA $A'$ with the parse tree consisting of $v$ and all descendants of $v$. We say $v$ \emph{associates} the states $S(v) = \{\theta_{A'}, \phi_{A'}\}$. In general, if $C$ is a cluster of $T$, i.e., any connected subgraph of $T$, we say $C$ associates the \emph{set} of states $S(C) = \cup_{v \in C} S(v)$. We define the \emph{partial-TNFA} (pTNFA) for $C$, as the directed, labeled subgraph of $A$ induced by the set of states $S(C)$. In particular, $A$ is a pTNFA since it is induced by $S(T)$. The two states associated by the root node of $C$ are defined to be the start and accepting state of the corresponding pTNFA. We need the following result. 
\begin{lemma}\label{s2:lem:separator}
For any pTNFA $P$ with $m>2$ states there exists a partitioning of $P$ into two subgraphs $P_O$ and $P_I$ such that
\begin{itemize}
  \item[(i)] $P_O$ and $P_I$ are pTNFAs with at most $2/3m + 2$ states each,
  \item[(ii)] any transition from $P_O$ to $P_I$ ends in $\theta_{P_I}$ and any transition from $P_I$ to $P_O$ starts in $\phi_{P_I}$, and
  \item[(iii)] the partitioning can be computed in $O(m)$ time. 
\end{itemize}
\end{lemma}
\begin{proof}
Let $P$ be pTNFA with $m > 2$ states and let $C$ be the corresponding cluster with $t$ nodes. Since $C$ is a binary tree with more than $1$ node, Jordan's classical result~\cite{Jordan1869} establishes that we can find in $O(t)$ time an edge $e$ in $C$ whose removal splits $C$ into two clusters each with at most $2/3t + 1$ nodes. These two clusters correspond to two pTNFAs, $P_O$ and $P_I$, and since $m = 2t$ each of these have at most $2/3m + 2$ states. Hence, (i) and (iii) follows. For (ii) assume w.l.o.g. that $P_O$ is the pTNFA containing the start and accepting state of $P$, i.e., $\theta_{P_O} = \theta_P$ and $\phi_{P_O} = \phi_P$. Then, $P_O$ is the pTNFA obtained from $P$ by removing all states of $P_I$. From Thompson's construction it is easy to check that any transition from $P_O$ to $P_I$ ends in $\theta_{P_I}$ and any transition from $P_I$ to $P_O$ must start in $\phi_{P_I}$.  \qed
\end{proof}
Intuitively, if we draw $P$, $P_I$ is "surrounded" by $P_O$, and therefore we will often refer to $P_I$ and $P_O$ as the \emph{inner pTNFA}  and the \emph{outer pTNFA}, respectively (see Fig.~\ref{s2:fig:separator}(a)). 
Applying Lemma~\ref{s2:lem:separator} recursively gives the following essential data structure. Let $P$ be a pTNFA with $m$ states. The \emph{separator tree} for $P$ is a binary, rooted  tree $B$ defined as follows: If $m=2$, i.e., $P$ is a trivial pTNFA consisting of two states $\theta_P$ and $\phi_P$, then $B$ is a single leaf node $v$ that stores the set $X(v) = \{\theta_P, \phi_P\}$. Otherwise ($m > 2$), compute $P_O$ and $P_I$ according to Lemma~\ref{s2:lem:separator}. The root $v$ of $B$ stores the set $X(v) = \{\theta_{P_I}, \phi_{P_I}\}$, and the children of $v$ are roots of separator trees for $P_O$ and $P_I$, respectively (see Fig.~\ref{s2:fig:separator}(b)). 

With the above construction each node in the separator tree naturally correspond to a pTNFA, e.g., the root corresponds to $P$, the children to $P_I$ and $P_O$, and so on. We denote the pTNFA corresponding to node $v$ in $B$ by $P(v)$. A simple induction combined with Lemma~\ref{s2:lem:separator}(i) shows that if $v$ is a node of depth $k$ then $P(v)$ contains at most $(\frac{2}{3})^km+6$ states. Hence, the depth of $B$ is at most $d = \log_{3/2} m + O(1)$. By Lemma~\ref{s2:lem:separator}(iii) each level of $B$ can be computed in $O(m)$ time and thus $B$ can be computed in $O(m\log m)$ total time.

\subsection{A Recursive $\epsilon$-Closure Algorithm}
We now present a simple $\epsilon$-closure algorithm for a pTNFA, which recursively traverses the separator tree $B$. We first give the high level idea and then show how it can be implemented in $O(1)$ time for each level of $B$. Since the depth of $B$ is $O(\log m)$ this leads to the desired result. For a pTNFA $P$ with $m$ states, a separator tree $B$ for $P$, and a node $v$ in $B$ define
\begin{relate}
\item[$\Close_{P(v)}(S)$:] 
\begin{enumerate}
\item Compute the set $Z \subseteq X(v)$ of states in $X(v)$ that are $\epsilon$-reachable from $S$ in $P(v)$.
\item If $v$ is a leaf return $S' := Z$, else let $u$ and $w$ be the children of $v$, respectively:
\begin{enumerate}
\item Compute the set $G \subseteq V(P(v))$ of states in $P(v)$ that are $\epsilon$-reachable from $Z$.
\item Return $S' := \Close_{P(u)}((S \cup G) \cap V(P(u))) \cup \Close_{P(w)}((S \cup G)\cap V(P(w)))$.
\end{enumerate}
\end{enumerate}
\end{relate}
\begin{lemma}\label{s2:lem:correctness}
For any node $v$ in the separator tree of a pTNFA $P$, $\Close_{P(v)}(S)$ computes the set of states in $P(v)$ reachable via a path of $\epsilon$-transitions. 
\end{lemma}
\begin{proof}
Let $\widehat{S}$ be the set of states in $P(v)$ reachable via a path of $\epsilon$-transitions. We need to show that $\widehat{S} = S'$. It is easy to check that any state in $S'$ is reachable via a path of $\epsilon$-transitions and hence $S' \subseteq \widehat{S}$. We show the other direction by induction on the separator tree. If $v$ is leaf then the set of states in $P(v)$ is exactly $X(v)$. Since $S' = Z$ the claim follows. Otherwise, let $u$ and $w$ be the children of $v$, and assume w.l.o.g. that $X(v) =\{\theta_{P(u)}, \phi_{P(u)}\}$. Consider a path $p$ of $\epsilon$-transitions from state $s$ to state $s'$. There are two cases to consider:
\begin{description}
  \item[Case 1:] $s' \in V(P(u))$. If $p$ consists entirely of states in $P(u)$ then by induction it follows that $s' \in \Close_{P(u)}(S \cap V(P(u)))$. Otherwise, $p$ contain a state from $P(w)$. However, by Lemma~\ref{s2:lem:separator}(ii) $\theta_{P(u)}$ is on $p$ and hence $\theta_{P(u)} \in Z$. It follows that $s' \in G$ and therefore $s' \in \Close_{P(u)}(G \cap V(P(u)))$.   
  \item[Case 2:] $s'\in V(P(w))$. As above, with the exception that $\phi_{P(u)}$ is now the state in $Z$. 
\end{description}
In all cases $s' \in S'$ and the result follows. \qed 
\end{proof}

\subsection{Implementing the Algorithm}
Next we show how to efficiently implement the above algorithm in parallel. The key ingredient is a compact mapping of states into positions in bitstrings. Suppose $B$ is the separator tree of depth $d$ for a pTNFA $P$ with $m$ states. The \emph{separator mapping} $M$ maps the states of $P$ into an interval of integers $[1, l]$, where $l = 3 \cdot 2^{d}$. The mapping is defined recursively according to the separator tree. Let $v$ be the root of $B$. If $v$ is a leaf node the interval is  $[1, 3]$. The two states of $P$, $\theta_{P}$ and $\phi_{P}$, are mapped to positions $2$ and $3$, respectively, while position $1$ is left intentionally unmapped. Otherwise, let $u$ and $w$ be the children of $v$. Recursively, map $P(u)$ to the interval $[1, l/2]$ and $P(w)$ to the interval $[l/2+1, l]$. Since the separator tree contains at most $2^d$ leaves and each contribute $3$ positions the mapping is well-defined. The size of the interval for $P$ is $l = 3 \cdot 2^{\log_{3/2} m + O(1)} = O(m)$. We will use the unmapped positions as test bits in our algorithm.

The separator mapping compactly maps all pTNFAs represented in $B$ into small intervals. Specifically, if $v$ is a node at depth $k$ in $B$, then $P(v)$ is mapped to an interval of size $l/2^k$ of the form $[(i-1)\cdot \frac{l}{2^k} + 1, i \cdot \frac{l}{2^k}]$, for some $1\leq i \leq 2^k$. The intervals that correspond to a pTNFA  $P(v)$ are \emph{mapped} and all other intervals are \emph{unmapped}. We will refer to a state $s$ of $P$ by its mapped position $M(s)$. A state-set of $P$ is represented by a bitstring $S$ such that, for all mapped positions $i$, $S[i] = 1$ iff the $i$ is in the state-set. Since $m = O(w)$, state-sets are represented in a constant number of words. 
 
To implement the algorithm we define a simple data structure consisting of four length $l$ bitstrings $X^\theta_k$, $X^\phi_k$, $E^\theta_k$, and $E^\phi_k$ for each level $k$ of the separator tree. For notational convenience, we will consider the strings at level $k$ as two-dimensional arrays consisting of $2^k$ intervals of length $l/2^k$, i.e., $X^\theta_k[i,j]$ is position $j$ in the $i$th interval of $X^\theta_k$. If the $i$th interval at level $k$ is unmapped then all positions in this interval are $0$ in all four strings. Otherwise, suppose that the interval corresponds to a pTNFA $P(v)$ and let $X(v) = \{\theta_v, \phi_v\}$. The strings are defined as follows: 
\begin{align*}
    X^\theta_k[i,j] =1 & \text{ iff $\theta_v$ is $\epsilon$-reachable in $P(v)$ from state $j$},  \\
    E^\theta_k[i,j] = 1 & \text{ iff state $j$ is $\epsilon$-reachable  in $P(v)$ from $\theta_v$}, \\
    X^\phi_k[i,j] = 1 & \text{ iff $\phi_v$ is $\epsilon$-reachable in $P(v)$ from state $j$},  \\  
   E^\phi_k[i,j] = 1 & \text{ iff state $j$ is $\epsilon$-reachable  in $P(v)$ from $\phi_v$}. 
\end{align*}
In addtion to these, we also store a string $I_k$ containing a test bit for each interval, that is, $I_k[i,j] = 1$ iff $j = 1$. Since the depth of $B$ is $O(\log m)$ the strings use $O(\log m)$ words. With a simple depth-first search they can all be computed in $O(m\log m)$ time.

Let $S$ be a bitstring representing a state-set of $A$. We implement the operation $\Close_A(S)$ by computing a sequence of intermediate strings $S_0, \ldots, S_d$ each corresponding to a level in the above recursive algorithm. Initially, $S_0 := S$ and the final string $S_d$ is the result of $\Close_A(S)$. At level $k$, $0 \leq k < d$,  we compute $S_{k+1}$ from $S_k$ as follows. Let $t = l/2^k - 1$.

\begin{align*}
Y^\theta &:= S_{k}\: \&\: X^\theta_k \\
Z^\theta &:= ((Y^\theta \: |\: I_k) - (I_k >> t)) \: \&\: I_k\\
F^\theta &:= Z^\theta - (Z^\theta >> t) \\
G^\theta &:= F^\theta \:\&\: E^\theta_k \\
Y^\phi &:= S_{k}\: \&\: X^\phi_k \\
Z^\phi &:= ((Y^\phi \: |\: I_k) - (I_k >> t)) \: \&\: I_k\\
F^\phi &:= Z^\phi - (Z^\phi >> t)\\
G^\phi &:= F^\phi \:\&\: E^\phi_k \\
S_{k+1} &:= S_k \:|\: G^\theta \:|\: G^\phi 
\end{align*}
We argue that the computation correctly simulates (in parallel) a level of the recursive algorithm. Assume that at the beginning of level $k$ the string $S_k$ represents the state-set corresponding the recursive algorithm after $k$ levels. We interpret $S_k$ as divided into $r = l/2^k$ intervals of length $t = l/2^k - 1$, each prefixed with a test bit, i.e., 
\begin{equation*}
S_k = 0s_{1,1}s_{1,2}\ldots s_{1,t}0s_{2,1}s_{2,2}\ldots s_{2,t}0 \ldots 0 s_{r,1}s_{r,2}\ldots s_{r,t} 
\end{equation*}
Assume first that all these intervals are mapped intervals corresponding to pTNFAs $P(v_1), \ldots, P(v_r)$, and let $X(v_i) = \{\theta_{v_i}, \phi_{v_i}\}$, $1\leq i \leq r$. Initially, $S_k\: \&\: X^\theta_k$ produces the string 
$$
Y^\theta = 0y_{1,1}y_{1,2}\ldots y_{1,t}0y_{2,1}y_{2,2}\ldots y_{2,t}0 \ldots 0 y_{r,1}y_{r,2}\ldots y_{r,t}, 
$$
where $y_{i,j} = 1$ iff $\theta_{v_i}$ is $\epsilon$-reachable in $P(v_i)$ from state $j$ and $j$ is in $S_k$. Then, similar to the second line in the simple algorithm, $(Y^\theta \: |\: I_k) - (I_k >> t) \: \&\: I_k$ produces a string of test bits $Z^\theta = z_10^tz_20^t \ldots z_r0^t$, where $z_i = 1$ iff at least one of $y_{i,1}\ldots y_{i,t}$ is $1$. In other words, $z_i = 1$ iff $\theta_{v_i}$ is $\epsilon$-reachable in $P(v_i)$ from any state in $S_k \cap V(P(v_i))$. Intuitively, the $Z^\theta$ corresponds to the "$\theta$-part" of the of $Z$-set in the recursive algorithm. Next we "copy" the test bits to get the string $F^\theta = Z^\theta - (Z^\theta >> t) = 0z_1^t0z_2^t\ldots 0z_r^t$. The bitwise $\&$ with $E^\theta_k$ gives
$$
G^\theta = 0g_{1,1}g_{1,2}\ldots g_{1,t}0g_{2,1}g_{2,2}\ldots g_{2,t}0 \ldots 0 g_{r,1}g_{r,2}\ldots g_{r,t}.
$$
By definition, $g_{i,j} = 1$ iff state $j$ is $\epsilon$-reachable in $P(v_i)$ from $\theta_{v_i}$ and $z_i = 1$. In other words, $G^\theta$ represents, for $1\leq i \leq r$, the states in $P(v_i)$ that are $\epsilon$-reachable from $S_k \cap V(P(v_i))$ through $\theta_{v_i}$. Again, notice the correspondance with the $G$-set in the recursive algorithm. The next $4$ lines are identical to first $4$ with the exception that $\theta$ is exchanged by $\phi$. Hence, $G^\phi$ represents the states that $\epsilon$-reachable through $\phi_{v_1}, \ldots, \phi_{v_r}$. 

Finally, $S_k \:|\: G^\theta \:|\: G^\phi$ computes the union of the states in $S_k$, $G^\theta$, and $G^\phi$ producing the desired state-set $S_{k+1}$ for the next level of the recursion. In the above, we assumed that all intervals were mapped. If this is not the case it is easy to check that the algorithm is still correct since the string in our data structure contain $0$s in all unmapped intervals. The algorithm uses constant time for each of the $d = O(\log m)$ levels and hence the total time is $O(\log m)$. 

\subsection{The Simulation Data Structure}
Next we show how to get a full simulation data structure. First, note that in the separator mapping the endpoints of the $\alpha$-transitions are consecutive (as in Sec.~\ref{s2:sec:simple}). It follows that we can use the same algorithm as in the previous section to compute $\Move_A$ in $O(1)$ time. This requires a dictionary of bitstrings, $D_\alpha$, using additional $O(m)$ space and $O(m\log m)$ preprocessing time. The $\Insert_A$, and $\Member_A$ operations are trivially implemented in $O(1)$. Putting it all together we have:
\begin{lemma}\label{s2:lem:datastruct2}
For a TNFA with $m = O(w)$ states there is a simulation data structure using $O(m)$ space and $O(m\log m)$ preprocessing time which supports all operations in $O(\log m)$ time.
\end{lemma}
Combining the simulation data structures from Lemmas~\ref{s2:lem:datastruct1} and \ref{s2:lem:datastruct2} with the reduction from Lemma~\ref{s2:lem:simulation} and taking the best result gives Theorem~\ref{s2:thm:main}. Note that the simple simulation data structure is the fastest when $m = O(\sqrt{w})$ and $n$ is sufficiently large compared to $m$.

\section{Remarks and Open Problems}\label{s2:sec:remarks}
The presented algorithms assume a unit-cost multiplication operation. Since this operation is not in $AC^0$ (the class of circuits of polynomial size (in $w$), constant depth, and unbounded fan-in) it is interesting to reconsider what happens with our results if we remove multiplication from our machine model. The simulation data structure from Sec.~\ref{s2:sec:simple} uses multiplication to compute $\Close_A$ and also for the constant time hashing to access $D_\alpha$. On the other hand, the algorithm of Sec.~\ref{s2:sec:notsimple} only uses multiplication for the hashing. However, Lemma~\ref{s2:lem:datastruct2} still holds since we can simply replace the hashing by binary search tree, which uses $O(\log m)$ time. It follows that Theorem~\ref{s2:thm:main} still holds except for the $O(n + m^2)$ bound in the last line. 

Another interestring point is to compare our results with the classical Shift-Or algorithm by Baeza-Yates and Gonnet~\cite{BYG1992} for exact pattern matching. Like ours, their algorithm simulates a NFA with $m$ states using word-level parallelism. The structure of this NFA permits a very efficient simulation with an $O(w)$ speedup of the simple $O(nm)$ time simulation. Our results generalize this to regular expressions with a slightly worse speedup of $O(w/\log w)$. We wonder if it is possible to remove the $O(\log w)$ factor separating these bounds. 

From a practical viewpoint, the simple algorithm of Sec.~\ref{s2:sec:simple} seems very promising since only about $15$ instructions are needed to carry out a step in the state-set simulation. Combined with ideas from~\cite{NR2004} we believe that this could lead to a practical improvement over previous algorithms.

\section{Acknowledgments}
The author wishes to thank Rasmus Pagh and Inge Li G{\o}rtz for many comments and interesting discussions.


\emptythanks
\chapter{Improved Approximate String Matching and Regular Expression Matching on Ziv-Lempel Compressed Texts}\label{chap:string3}

\title{Improved Approximate String Matching and Regular Expression Matching on Ziv-Lempel Compressed Texts}

\author{Philip Bille \\ IT University of Copenhagen \\ \texttt{beetle@itu.dk} \and Rolf Fagerberg \\ University of Southern Denmark \\ \texttt{rolf@imada.sdu.dk} \and Inge Li G{\o}rtz \\ Technical University of Denmark \\ \texttt{ilg@imm.dtu.dk}}

\date{}

\cleartooddpage

\maketitle

\begin{abstract}
We study the approximate string matching and regular expression matching problem for the case when the text to be searched is compressed with the Ziv-Lempel adaptive dictionary compression schemes. We present a time-space trade-off that leads to algorithms improving the previously known complexities for both problems. In particular, we significantly improve the space bounds, which in  practical applications are likely to be a bottleneck.
\end{abstract}

\section{Introduction} 
Modern text databases, e.g. for biological and World Wide Web data, are huge. To save time and space, it is desireable if  data can be kept in compressed form and still allow efficient searching. Motivated by this Amir and Benson~\cite{AB1992a, AB1992} initiated the study of \emph{compressed pattern matching problems}, that is, given a text string $Q$ in compressed form $Z$ and a specified (uncompressed) pattern $P$, find all occurrences of $P$ in $Q$ without decompressing $Z$. The goal is to search more efficiently than the na{\"i}ve approach of decompressing $Z$ into $Q$ and then searching for $P$ in $Q$. Various compressed pattern matching algorithms have been proposed depending on the type of pattern and compression method, see e.g.,~\cite{AB1992, FT1998, KTSMA1998, KNU2003, Navarro2003, MUN2003}. For instance, given a string $Q$ of length $u$ compressed with the Ziv-Lempel-Welch scheme~\cite{Welch1984} into a string of length $n$, Amir et al.~\cite{ABF1996} gave an algorithm for finding all exact occurrences of a pattern string of length $m$ in $O(n + m^2)$ time and space. 

In this paper we study the classical approximate string matching and regular expression matching problems in the context of compressed texts. As in previous work on these problems~\cite{KNU2003, Navarro2003} we focus on the popular \zla\  and \zlw\  adaptive dictionary compression schemes~\cite{ZL1978, Welch1984}. We present a new technique that gives a general time-space trade-off. The resulting algorithms improve all previously known complexities for both problems. In particular, we significantly improve the space bounds. When searching large text databases, space is likely to be a bottleneck and therefore this is of crucial importance.

\subsection{Approximate String Matching}
Given strings $P$ and $Q$ and an
\emph{error threshold} $k$, the classical \emph{approximate string matching problem} is to find all ending positions of substrings of $Q$ whose \emph{edit distance} to $P$ is at most $k$. The edit distance between two strings is the minimum number of insertions, deletions, and substitutions needed to convert one string to the other. The classical dynamic programming solution due to Sellers~\cite{Sellers1980} solves the problem in $O(um)$ time and $O(m)$ space, where $u$ and $m$ are the length of $Q$ and $P$, respectively. Several improvements of this result are known, see e.g., the survey by Navarro~\cite{Navarro2001a}. For this paper we are particularly interested in the fast solution for small values of $k$, namely, the $O(uk)$ time algorithm by Landau and Vishkin~\cite{LV1989} and the more recent $O(uk^4/m + u)$ time algorithm due to Cole and Hariharan~\cite{CH2002} (we assume w.l.o.g. that $k < m$). Both of these can be implemented in $O(m)$ space.

Recently, K{\"a}rkk{\"a}inen et al.~\cite{KNU2003} studied this problem for text compressed with the \zla/\zlw\  compression schemes. If $n$ is the length of the compressed text, their algorithm achieves $O(nmk + \occ)$ time and $O(nmk)$ space, where $\occ$ is the number of occurrences of the pattern. Currently, this is the only non-trivial worst-case bound for the general problem on compressed texts. For special cases and restricted versions, other algorithms have been proposed~\cite{MKTSA2000, NR1998}.  An experimental study of
the problem and an optimized practical implementation can be found in~\cite{NKTSA01}.

In this paper, we show that the problem is closely connected to the uncompressed problem and we achieve a simple time-space trade-off. More precisely, let $t(m,u, k)$ and $s(m,u,k)$ denote the time and space, respectively, needed by any algorithm to solve the (uncompressed) approximate string matching problem with error threshold $k$ for pattern and text of length $m$ and $u$, respectively. We show the following result. 
\begin{theorem}\label{s3:thm:approx}
Let $Q$ be a string compressed using {\zla } into a string $Z$ of length $n$ and let $P$ be a pattern of length $m$. Given $Z$, $P$, and a parameter $\tau \geq 1$, we can find all approximate occurrences of $P$ in $Q$ with at most $k$ errors in $O(n(\tau + m + t(m, 2m+2k,k)) + \occ)$ expected time and $O(n/\tau + m + s(m,2m+2k,k) + \occ)$ space.
\end{theorem}
The expectation is due to hashing and can be removed at an additional $O(n)$ space cost. In this case the bound also hold for \zlw\  compressed strings. We assume that the algorithm for the uncompressed problem produces the matches in sorted order (as is the case for all algorithms that we are aware of). Otherwise, additional time for sorting must be included in the bounds. To compare Theorem~\ref{s3:thm:approx} with the result of Karkkainen et al.~\cite{KNU2003}, plug in the Landau-Vishkin algorithm and set $\tau = mk$. This gives an algorithm using $O(nmk + \occ)$ time and $O(n/mk + m + \occ)$ space. This matches the best known time bound while improving the space by a factor $\Theta(m^2k^2)$. Alternatively, if we plug in the Cole-Hariharan algorithm and set $\tau = k^4 + m$ we get an algorithm using $O(nk^4 + nm + \occ)$ time and $O(n/(k^4 + m) + m + \occ)$ space. Whenever $k = O(m^{1/4})$ this is $O(nm + \occ)$ time and $O(n/m + m + \occ)$ space. 

To the best of our knowledge, all previous non-trivial compressed pattern matching algorithms for \zla/\zlw\  compressed text, with the exception of a very slow algorithm for exact string matching by Amir et al.~\cite{ABF1996}, use $\Omega(n)$ space. This is because the algorithms explicitly construct the dictionary trie of the compressed texts. Surprisingly, our results show that for the \zla\  compression schemes this is not needed to get an efficient algorithm. Conversely, if very little space is available our trade-off shows that it is still possible to solve the problem without decompressing the text.

\subsection{Regular Expression Matching}

Given a regular expression $R$ and a string $Q$, the \emph{regular expression matching problem} is to find all ending position of substrings in $Q$ that matches a string in the language denoted by $R$. The classic textbook solution to this problem due to Thompson~\cite{Thomp1968} solves the problem in $O(um)$ time and $O(m)$ space, where $u$ and $m$ are the length of $Q$ and $R$, respectively. Improvements based on the Four Russian Technique or word-level parallelism are given in~\cite{Myers1992, BFC2005, Bille06}. 

The only solution to the compressed problem is due to Navarro~\cite{Navarro2003}. His solution depends on word RAM techniques to encode small sets into memory words, thereby allowing constant time set operations. On a unit-cost RAM with $w$-bit words this technique can be used to improve an algorithm by at most a factor $O(w)$. For $w = O(\log u)$ a similar improvement is straightforward to obtain for our algorithm and we will therefore, for the sake of exposition, ignore this factor in the bounds presented below. With this simplification Navarro's algorithm uses $O(nm^2 + \occ \cdot m\log m)$ time and $O(nm^2)$ space, where $n$ is the length of the compressed string. In this paper we show the following time-space trade-off:

\begin{theorem}\label{s3:thm:regularex}
Let $Q$ be a string compressed using {\zla } or {\zlw } into a string $Z$ of length $n$ and let $R$ be a regular expression of length $m$. Given $Z$, $R$, and a parameter $\tau \geq 1$, we can find all occurrences of substrings matching $R$ in $Q$ in $O(nm(m + \tau) + \occ\cdot m \log m)$ time and $O(nm^2/\tau + nm)$ space.
\end{theorem}
If we choose $\tau = m$ we obtain an algorithm using $O(nm^2 + \occ\cdot m \log m)$ time and $O(nm)$ space. This matches the best known time bound while improving the space by a factor $\Theta(m)$. With word-parallel techniques these bounds can be improved slightly. The full details are given in Section~\ref{s3:sec:wordparallel}.

\subsection{Techniques}
If pattern matching algorithms for \zla\ or \zlw\ compressed texts use $\Omega(n)$ working space they can explicitly store the dictionary trie for the compressed text and apply any linear space data structure to it. This has proven to be very useful for compressed pattern matching. However, as noted by Amir et al.~\cite{ABF1996}, $\Omega(n)$ working space may not be feasible for large texts and therefore more space-efficient algorithms are needed. Our main technical contribution is a simple $o(n)$ data structure for \zla\ compressed texts. The data structure gives a way to compactly represent a subset of the trie which combined with the compressed text enables algorithms to quickly access relevant parts of the trie. This provides a general approach to solve compressed pattern matching problems in $o(n)$ space, which combined with several other techniques leads to the above results.

\section{The Ziv-Lempel Compression Schemes}\label{s3:zlc}
Let $\Sigma$ be an \emph{alphabet} containing $\sigma = |\Sigma|$ \emph{characters}.  A \emph{string} $Q$ is a sequence of characters from $\Sigma$. The \emph{length} of $Q$ is $u = |Q|$ and the unique string of length $0$ is denoted $\epsilon$. The $i$th character of $Q$ is denoted $Q[i]$ and the substring beginning at position $i$ of length $j-i+1$ is denoted $Q[i,j]$. The Ziv-Lempel algorithm from 1978~\cite{ZL1978} provides a simple and natural way to represent strings, which we describe below. Define a \emph{\zla\ compressed string} (abbreviated \emph{compressed string} in the remainder of the paper) to be a string of the form 
$$
Z = z_1 \cdots z_n = (r_1, \alpha_1)(r_2, \alpha_2) \ldots (r_n, \alpha_n), 
$$
where $r_i \in \{0, \ldots, i-1\}$ and $\alpha_i \in \Sigma$. Each pair $z_i = (r_i, \alpha_i)$ is a \emph{compression element}, and $r_i$ and $\alpha_i$ are the \emph{reference} and \emph{label} of $z_i$, denoted by $\reference(z_i)$ and $\lab(z_i)$, respectively. Each compression element \emph{represents} a string, called a \emph{phrase}. The phrase for $z_i$, denoted $\phrase(z_i)$, is given by the following recursion.
\begin{equation*}
\phrase(z_i) = 
\begin{cases}
  \lab(z_i)    & \text{if $\reference(z_i) = 0$}, \\
  \phrase(\reference(z_i))\cdot \lab(z_i)   & \text{otherwise}.
\end{cases}
\end{equation*}
The $\cdot$ denotes concatenation of strings. The compressed string $Z$ \emph{represents} the concatenation of the phrases, i.e., the string $\phrase(z_1)\cdots \phrase(z_n)$.

Let $Q$ be a string of length $u$. In \zla, the compressed string representing $Q$ is obtained by greedily parsing $Q$ from left-to-right with the help of a dictionary $D$. For simplicity in the presentation we assume the existence of an initial compression element $z_0 = (0, \epsilon)$ where $\phrase(z_0) = \epsilon$. Initially, let $z_0 = (0, \epsilon)$ and let  $D = \{\epsilon\}$. After step $i$ we have computed a compressed string $z_0 z_1 \cdots z_i$ representing $Q[1, j]$ and $D = \{\phrase(z_0), \ldots, \phrase(z_i)\}$. We then find the longest prefix of $Q[j+1, u-1]$ that matches a string in $D$, say $\phrase(z_k)$, and let $\phrase(z_{i+1}) = \phrase(z_k) \cdot Q[j+1 + |\phrase(z_k)|]$. Set $D = D \cup \{\phrase(z_{i+1})\}$ and let $z_{i+1} = (k, Q[j + 1 + |\phrase(z_{i+1})|])$. The compressed string $z_0 z_1\ldots z_{i+1}$ now represents the string $Q[1,j + |\phrase(z_{i+1})|])$ and $D = \{\phrase(z_0), \ldots, \phrase(z_{i+1})\}$. We repeat this process until all of $Q$ has been read.

\begin{figure}[t] 
  \centering \includegraphics[scale=.6]{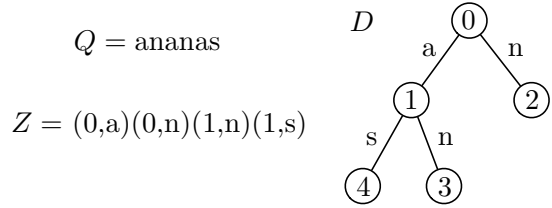}
  \caption{The compressed string $Z$ representing $Q$ and the corresponding dictionary trie $D$. Taken from~\cite{Navarro2003}.}
  \label{s3:fig:lz78}
\end{figure}

Since each phrase is the concatenation of a previous phrase and a single character, the dictionary $D$ is prefix-closed, i.e., any prefix of a phrase is a also a phrase. Hence, we can represent it compactly as a trie where each node $i$ corresponds to a compression element $z_i$ and $\phrase(z_i)$ is the concatenation of the labels on the path from $z_i$ to node $i$. Due to greediness, the phrases are unique and therefore the number of nodes in $D$ for a compressed string $Z$ of length $n$ is $n+1$. An example of a string and the corresponding compressed string is given in Fig.~\ref{s3:fig:lz78}.  

Throughout the paper we will identify compression elements with nodes in the trie $D$, and therefore we use standard tree terminology, briefly summed up here: The \emph{distance} between two elements is the number of edges on the unique simple path between them. The \emph{depth} of element $z$ is the distance from $z$ to $z_0$ (the root of the trie). An element $x$ is an \emph{ancestor} of an element $z$ if $\phrase(x)$ is a prefix of $\phrase(z)$. If also $|\phrase(x)| = |\phrase(z)| - 1$ then $x$ is the \emph{parent} of $z$.  If $x$ is ancestor of $z$ then $z$ is a \emph{descendant} of $x$ and if $x$ is the parent of $z$ then $z$ is the \emph{child} of $x$.The \emph{length} of a path $p$ is the number of edges on the path, and is denoted $|p|$. The \emph{label} of a path is the concatenation of the labels on these edges. 

Note that for a compression element $z$, $\reference(z)$ is a pointer to the parent of $z$ and $\lab(z)$ is the label of the edge to the parent of $z$. Thus, given $z$ we can use the compressed text $Z$ directly to decode the label of the path from $z$ towards the root in constant time per element. We will use this important property in many of our results.

If the dictionary $D$ is implemented as a trie it is straightforward to compress $Q$ or decompress $Z$ in $O(u)$ time. Furthermore, if we do not want to explicitly decompress $Z$ we can compute the trie in $O(n)$ time, and as mentioned above, this is done in almost all previous compressed pattern matching algorithm on Ziv-Lempel compression schemes. However, this requires at least $\Omega(n)$ space which is insufficient to achieve our bounds.  In the next section we show how to partially represent the trie in less space.

\subsection{Selecting Compression Elements}\label{s3:selectingcompression}
Let $Z = z_0\ldots z_n$ be a compressed string. For our results we need an algorithm to select a compact subset of the compression elements such that the distance from any element to an element in the subset is no larger than a given threshold. More precisely, we show the following lemma.
\begin{lemma}\label{s3:lem:special}
Let $Z$ be a compressed string of length $n$ and let $1 \leq \tau \leq n$ be parameter. There is a set of compression elements $C$ of $Z$, computable in $O(n\tau)$ expected time and $O(n/\tau)$ space with the following properties:
\begin{itemize}
  \item[(i)] $|C| = O(n/\tau)$.
  \item[(ii)] For any compression element $z_i$ in $Z$,  the minimum distance to any compression element in $C$ is at most $2\tau$. 
\end{itemize}
\end{lemma}
\begin{proof}
Let $1 \leq \tau \leq n$ be a given parameter. We build $C$ incrementally in a left-to-right scan of $Z$. The set is maintained as a dynamic dictionary using dynamic perfect hashing~\cite{DKMMRT1994}, i.e., constant time worst-case access and constant time amortized expected update. Initially, we set $C = \{z_0\}$. Suppose that we have read $z_0, \ldots, z_i$. To process $z_{i+1}$ we follow the path $p$ of references until we encounter an element $y$ such that $y \in C$.  We call $y$ the \emph{nearest special element} of $z_{i+1}$. Let $l$ be the number of elements in $p$ including $z_{i+1}$ and $y$. Since each lookup in $C$ takes constant time the time to find the nearest special element is $O(l)$. If $l < 2\cdot \tau$ we are done. Otherwise, if $l = 2\cdot \tau$, we find the $\tau$th element $y'$ in the reference path and set $C := C \cup \{y'\}$. As the trie grows under addition of leaves condition (ii) follows. Moreover, any element chosen to be in $C$ has at least $\tau$ descendants of distance at most $\tau$ that are not in $C$ and therefore condition (i) follows. The time for each step is $O(\tau)$ amortized expected and therefore the total time is $O(n\tau)$ expected. The space is proportional to the size of $C$ hence the result follows. \qed
\end{proof}

\subsection{Other Ziv-Lempel Compression Schemes}

A popular variant of \zla\  is the \zlw\ compression scheme~\cite{Welch1984}. Here, the label of compression elements are not explicitly encoded, but are defined to be the first character of the next phrase. Hence, \zlw\ does not offer an asymptotically better compression ratio over \zla\ but gives a better practical performance. The \zlw\ scheme is implemented in the UNIX program \texttt{compress}. From an algorithmic viewpoint \zlw\ is more difficult to handle in a space-efficient manner since labels are not explicitly stored with the compression elements as in \zla. However, if $\Omega(n)$ space is available then we can simply construct the dictionary trie. This gives constant time access to the label of a compression elements and therefore \zla\ and \zlw\ become "equivalent". This is the reason why Theorem~\ref{s3:thm:approx} holds only for \zla\ when space is $o(n)$ but for both when the space is $\Omega(n)$. 

Another well-known variant is the \zlb\ compression scheme~\cite{ZL1977}. Unlike \zla\ and \zlw\ phrases in the \zlb\ scheme can be any substring of text that has already been processed. This makes searching much more difficult and none of the known techniques  for \zla\ and \zlw\ seems to be applicable. The only known algorithm for pattern matching on \zlb\ compressed text is due to Farach and Thorup~\cite{FT1998} who gave an algorithm for the exact string matching problem.

\section{Approximate String Matching}\label{s3:approx}
In this section we consider the compressed approximate string matching problem. 
Before presenting our algorithm we need a few definitions and properties of approximate string matching. 

Let $A$ and $B$ be strings. Define the \emph{edit distance} between $A$ and $B$, $\gamma(A,B)$, to be the minimum number of insertions, deletions, and substitutions needed to transform $A$ to $B$. We say that $j \in [1, |S|]$ is a \emph{match with error at most $k$} of $A$ in a string $S$ if there is an $i \in [1, j]$ such that $\gamma(A, S[i,j]) \leq k$. Whenever $k$ is clear from the context we simply call $j$ a \emph{match}. All positions $i$ satisfying the above property are called a \emph{start} of the match $j$. The set of all matches of $A$ in $S$ is denoted $\Gamma(A, S)$. 
We need the following well-known property of approximate matches.
\begin{prop}\label{s3:prop:match}
Any match $j$ of $A$ in $S$ with at most $k$ errors must start in the interval $[\max(1, j-|A|+1-k), \min(|S|, j-|A|+1+k)]$.
\end{prop}

\begin{proof}
Let $l$ be the length of a substring $B$ matching $A$ and ending at $j$. If the match starts outside the interval then either $l < |A| - k$ or $l > |A| + k$. In these cases, more than $k$ deletions or $k$ insertions, respectively, are needed to transform $B$ to $A$. \qed
\end{proof}

\subsection{Searching for Matches}

Let $P$ be a string of length $m$ and let $k$ be an error threshold. To avoid trivial cases we assume that $k < m$. 
Given a compressed string $Z = z_0z_1\ldots z_n$ representing a string $Q$ of length $u$ we show how to find $\Gamma(P, Q)$ efficiently. 

\begin{figure}[t] 
  \centering \includegraphics[scale=.5]{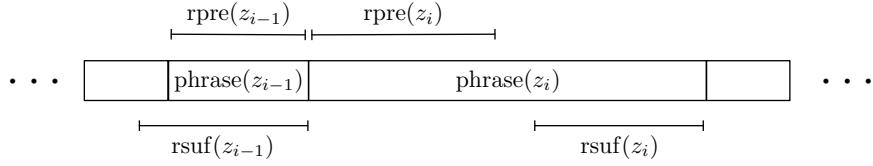}
  \caption{The relevant prefix and the relevant suffix of two phrases in $Q$. Here, $|\phrase(z_{i-1})| < m+k$ and therefore $\rsuf(z_{i-1})$ overlaps with previous phrases.}
  \label{s3:fig:relevant}
\end{figure}

Let $l_i = |\phrase(z_i)|$, let $u_0 = 1$, and let $u_i = u_{i-1} + l_{i-1}$, for $1\leq i \leq n$, i.e., $l_i$ is the length of the $i$th phrase and $u_i$ is the starting position in $Q$ of the $i$th phrase. We process $Z$ from left-to-right and at the $i$th step we find all matches in $[u_i, u_i + l_i-1]$. Matches in this interval can be either \emph{internal} or \emph{overlapping} (or both). A match $j$ in $[u_i, u_i + l_i-1]$ is internal if it has a starting point in $[u_i, u_i + l_i-1]$ and overlapping if it has a starting point in $[1, u_i -1]$. To find all matches we will compute the following information for $z_i$.  
\begin{itemize}
  \item The start position, $u_i$, and length, $l_i$, of $\phrase(z_i)$.
  \item The \emph{relevant prefix}, $\rpre(z_i)$, and the \emph{relevant suffix}, $\rsuf(z_i)$, where 
  \begin{align*}
\rpre(z_i) &= Q[u_i, \min(u_i + m + k-1, u_i + l_i - 1)]\;,   \\ 
\rsuf(z_i) &= Q[\max(1, u_i +l_i - m - k ), u_i + l_i-1]\;.    
\end{align*}
In other words, $\rpre(z_i)$ is the largest prefix of length at most $m+k$ of $\phrase(z_i)$ and $\rsuf(z_i)$ is the substring of length $m+k$ ending at $u_i + l_i -1$. For an example see Fig.~\ref{s3:fig:relevant}.
\item The \emph{match sets} $M_I(z_i)$ and $M_O(z_i)$, where 
\begin{align*}
    M_I(z_i) &= \Gamma(P, \phrase(z_i))\;, \\
    M_O(z_i) &= \Gamma(P, \rsuf(z_{i-1}) \cdot \rpre(z_i))\;. 
\end{align*}
We assume that both sets are represented as sorted lists in increasing order.
\end{itemize}

We call the above information the \emph{description} of $z_i$. In the next section we show how to efficiently compute descriptions. For now, assume that we are given the description of $z_i$. Then, the set of matches in $[u_i, u_i + l_i-1]$ is reported as the set
\begin{eqnarray*}
M(z_i) &=& \{j+u_i - 1 \mid j\in M_I(z_i)\}  \cup \\
&& \{j+u_i-1-|\rsuf(z_{i-1})| \mid j\in M_O(z_i) \cap [u_i, u_i + l_i-1]\}\;.
\end{eqnarray*}

We argue that this is the correct set. Since $\phrase(z_i) = Q[u_i, u_i + l_i-1]$ we have that $$j \in M_I(z_i) \Leftrightarrow   j + u_i - 1 \in \Gamma(P, Q[u_i, u_i + l_i-1]\;.$$ Hence, the set $\{j+u_i - 1\mid j\in M_I(z_i)\}$ is the set of all internal matches. Similarly, $\rsuf(z_{i-1}) \cdot \rpre(z_i) = Q[u_i - |\rsuf(z_{i-1})|, u_i + |\rpre(z_i)| - 1]$ and therefore 
$$j \in M_O(z_i) \Leftrightarrow   j + u_i - 1 - |\rsuf(z_{i-1})| \in \Gamma(P, Q[u_i - |\rsuf(z_{i-1})|, u_i + 1 + |\rpre(z_i)|])\;.$$
By Proposition~\ref{s3:prop:match} any overlapping match must start at a position within the interval $[\max(1, u_i-m+1-k), u_i]$. Hence, $\{j + u_i -1- |\rsuf(z_{i-1})| \mid j\in M_O(z_i)\}$ includes all overlapping matches in $[u_i, u_i + l_i-1]$. Taking the intersection with $[u_i, u_i + l_i-1]$ and the union with the internal matches it follows that the 
set $M(z_i)$ is precisely the set of  matches in $[u_i, u_i + l_i-1]$. 
For an example see Fig.~\ref{s3:fig:approx-description}. 

\begin{figure}[t]
\begin{center}
$$Q=\textrm{ananasbananer}, \quad P=\textrm{base}, \quad Z=\textrm{(0,a)(0,n)(1,n)(1,s)(0,b)(3,a)(2,e)(0,r)}$$
{\bf Descriptions}

$
\begin{array}{l  @{\hspace{5pt}} | @{\hspace{5pt}} l @{\hspace{5pt}} l @{\hspace{5pt}} l @{\hspace{5pt}} l @{\hspace{5pt}} l @{\hspace{5pt}} l @{\hspace{5pt}} l @{\hspace{5pt}} l}
 \hline & z_0 & z_1 & z_2 & z_3 & z_4 & z_5 & z_6 & z_7 \\ \hline
  u_i & 1 & 2& 3 & 5& 7& 8 & 11 & 13 \\
  l_i & 1 & 1 & 2 & 2 & 1& 3 & 2 & 1 \\
 \textrm{rpre}(z_i) & \textrm{a} & \textrm{n} & \textrm{an} & \textrm{as} & \textrm{b} & \textrm{ana} & \textrm{ne} & \textrm{r} \\
 \textrm{rsuf}(z_i) & \textrm{a} & \textrm{an} & \textrm{anas} & \textrm{ananas} & \textrm{nanasb} & \textrm{asbana} & \textrm{banane} & \textrm{ananer} \\
 M_I(z_i) & \emptyset & \emptyset & \emptyset &\{2\}& \emptyset & \emptyset & \emptyset & \emptyset \\
 M_O(z_i)  & \emptyset & \emptyset & \emptyset &\{6\} & \{6,7\} & \{5,6,7,8\} & \{2,3,4,5,6\} & \{2,3,4,6\} \\
 M(z_i) & \emptyset & \emptyset & \emptyset & \{6\} & \{7\} & \{8,9,10\} & \{12\} & \emptyset
\end{array}$
\caption{Example of descriptions. $Z$ is the compressed string representing $Q$. We are looking for all matches of the pattern  $P$ with error threshold $k=2$ in $Z$.  The set of matches is $\{6,7,8,9,10,12\}$.}\label{s3:fig:approx-description}
\end{center}
\end{figure}
Next we consider the complexity of computing the matches. To do this we first bound the size of the $M_I$ and $M_O$ sets. Since the length of any relevant suffix and relevant prefix is at most $m+k$, we have that $|M_O(z_i)|  \leq 2(m+k)<4m$, and therefore the total size of the $M_O$ sets is at most $O(nm)$. Each element in the sets $M_I(z_0), \ldots, M_I(z_n)$ corresponds to a unique match. Thus, the total size of the $M_I$ sets is at most $\occ$, where $\occ$ is the total number of matches. Since both sets are represented as sorted lists the total time to compute the matches for all compression elements is $O(nm + \occ)$.

\subsection{Computing Descriptions}

Next we show how to efficiently compute the descriptions. Let $1\leq \tau \leq n$ be a parameter. Initially, we compute a subset $C$ of the elements in $Z$ according to Lemma~\ref{s3:lem:special} with parameter $\tau$. For each element $z_j \in C$ we store $l_j$, that is, the length of $\phrase(z_j)$. If $l_j > m+k$ we also store the index of the ancestor $x$ of $z_j$ of depth $m+k$. This information can easily be computed while constructing $C$ within the same time and space bounds, i.e., using $O(n\tau)$ time and $O(n/\tau)$ space.

Descriptions are computed from left-to-right as follows. Initially, set $l_0 = 0$, $u_0 = 0$, $\rpre(z_0) = \epsilon$, $\rsuf(z_0) = \epsilon$, $M_I(z_0) = \emptyset$, and $M_O(z_0) = \emptyset$. To compute the description of $z_i$, $1\leq i \leq n$, first
 follow the path $p$
 of references until we encounter an element $z_j \in C$. Using the information stored at $z_j$ we set $l_i := |p| + l_j$ and $u_i = u_{i-1} + l_{i-1}$. By Lemma~\ref{s3:lem:special}(ii) the distance to $z_j$ is at most $2\tau$ and therefore $l_i$ and $u_i$ can be computed in $O(\tau)$ time given the description of $z_{i-1}$.

To compute $\rpre(z_i)$ we 
compute the label of the path from $z_0$ towards $z_i$ of length $\min(m+k, l_i)$. There are two cases to consider: 
If $l_i \leq m+k$ we simply compute the label of the path from $z_i$ to $z_0$ and let $\rpre(z_i)$ be the reverse of this string. Otherwise ($l_i > m+k$), we use the "shortcut" stored at $z_j$ to find the ancestor $z_h$ of distance $m+k$ to $z_0$. The reverse of the label of the path from $z_h$ to $z_0$ is then $\rpre(z_i)$. Hence, $\rpre(z_i)$ is computed in $O(m+k + \tau) = O(m + \tau)$ time.

The string $\rsuf(z_i)$ may be the divided over several phrases and we therefore recursively follow paths towards the root until we have computed the entire string. It is easy to see that the following algorithm correctly decodes the desired substring of length $\min(m+k, u_i)$ ending at position $u_i+l_i-1$.
\begin{enumerate}
  \item Initially, set $l := \min(m+k, u_i + l_i-1)$, $t:=i$, and $s := \epsilon$. 
  \item Compute the path $p$ of references from $z_t$ of length $r = \min(l, \depth(z_t))$ and set $s := s \cdot \lab(p)$.	
  \item If $r < l$ set $l := l-r$, $t := t - 1$, and repeat step $2$. 
  \item Return $\rsuf(z_i)$ as the reverse of $s$. 
\end{enumerate}
Since the length of $\rsuf(z_i)$ is at most $m+k$, the algorithm finds it in $O(m + k) = O(m)$ time.

The match sets $M_I$ and $M_O$ are computed as follows. Let $t(m,u,k)$ and $s(m,u,k)$ denote the time and space to compute $\Gamma(A,B)$ with error threshold $k$ for strings $A$ and $B$ of lengths $m$ and $u$, respectively. Since $|\rsuf(z_{i-1})\cdot \rpre(z_i)| \leq 2m+2k$ it follows that $M_O(z_i)$ can be computed in $t(m, 2m+2k,k)$ time and $s(m, 2m+2k,k)$ space. Since $M_I(z_i) = \Gamma(P, \phrase(z_i))$ we have that $j \in M_I(z_i)$ if and only if $j \in M_I(\reference(z_i))$ or $j =l_i$. By Proposition~\ref{s3:prop:match} any match ending in $l_i$ must start within $[\max(1, l_i-m+1-k), \min(l_i, l_i-m+1+k)]$. Hence, there is a match ending in $l_i$ if and only if $l_i \in \Gamma(P, \rsuf'(z_i))$ where $\rsuf'(z_i)$ is the suffix of $\phrase(z_i)$ of length $\min(m+k, l_i)$. Note that $\rsuf'(z_i)$ is a suffix of $\rsuf(z_i)$ and we can therefore trivially compute it in $O(m+k)$ time. Thus, 
\begin{equation*}
M_I(z_i) = M_I(\reference(z_i)) \cup \{l_i \mid l_i \in  \Gamma(P, \rsuf'(z_i))\}\;.
\end{equation*}
Computing $\Gamma(P, \rsuf'(z_i))$ uses $t(m,m+k,k)$ time and $s(m, m+k, k)$ space. Subsequently, constructing $M_I(z_i)$ takes $O(|M_I(z_i)|)$ time and space. Recall that the elements in the $M_I$ sets correspond uniquely to matches in $Q$ and therefore the total size of the sets is $\occ$. Therefore, using dynamic perfect hashing~\cite{DKMMRT1994} on pointers to non-empty $M_I$ sets we can store these using $O(\occ)$ space in total.

\subsection{Analysis}
Finally, we can put the pieces together to obtain the final algorithm. The preprocessing uses $O(n\tau)$ expected time and $O(n/\tau)$ space. The total time to compute all descriptions and report occurrences is expected $O(n(\tau + m + t(m, 2m+2k,k)) + \occ)$. The description for $z_i$, except for $M_I(z_i)$, depends solely on the description of $z_{i-1}$. Hence, we can discard the description of $z_{i-1}$, except for $M_I(z_{i-1})$, after processing $z_i$ and reuse the space. It follows that the total space used is $O(n/\tau + m + s(m,2m+2k, k) + \occ)$. This completes the proof of Theorem~\ref{s3:thm:approx}. Note that if we use $\Omega(n)$ space we can explicitly construct the dictionary. In this case hashing is not needed and the bounds also hold for the \zlw\ compression scheme.

\section{Regular Expression Matching}\label{s3:regular}

\subsection{Regular Expressions and Finite Automata}
First we briefly review the classical concepts used in the paper. For more details see, e.g., Aho et al.~\cite{ASU1986}. The set of \emph{regular expressions} over $\Sigma$ are defined recursively as follows: A character $\alpha \in \Sigma$ is a regular expression, and if $S$ and $T$ are regular expressions then so is the
  \emph{concatenation}, $(S)\cdot(T)$, the \emph{union}, $(S)|(T)$, and the \emph{star}, $(S)^*$. The \emph{language} $L(R)$ generated by $R$ is defined as follows: $L(\alpha) = \{\alpha\}$, $L(S \cdot T) = L(S)\cdot L(T)$, that is, any string formed by the concatenation of a string in $L(S)$ with a string in $L(T)$, $L(S)|L(T) = L(S) \cup L(T)$, and $L(S^*) = \bigcup_{i \geq 0} L(S)^i$, where $L(S)^0 = \{\epsilon\}$ and $L(S)^i = L(S)^{i-1} \cdot L(S)$, for $i > 0$.     
  
A \emph{finite automaton} is a tuple $A = (V, E, \Sigma, \theta, \Phi)$, where $V$ is a set of nodes called \emph{states}, $E$ is set of directed edges between states called \emph{transitions} each labeled by a character from $\Sigma \cup \{\epsilon\}$, $\theta \in V$ is a \emph{start state}, and $\Phi \subseteq V$ is a set of \emph{final states}. In short, $A$ is an edge-labeled directed graph with a special start node and a set of accepting nodes. 
$A$ is a \emph{deterministic finite automaton} (DFA) if $A$ does not contain any $\epsilon$-transitions, and all outgoing transitions of any state have different labels. Otherwise, $A$ is a \emph{non-deterministic automaton} (NFA). 

The \emph{label} of a path $p$ in $A$ is the concatenation of labels on the transitions in $p$. For a subset $S$ of states in  $A$ and character $\alpha \in \Sigma \cup \{\epsilon\}$, define the \emph{transition map}, $\delta(S, \alpha)$, as the set of states reachable from $S$ via a path labeled $\alpha$. Computing the set $\delta(S, \alpha)$ is called a \emph{state-set transition}. We extend $\delta$ to strings by defining $\delta(S, \alpha \cdot B) = \delta(\delta(S, \alpha), B)$, for any string $B$ and character $\alpha \in \Sigma$. We say that $A$ \emph{accepts} the string $B$ if $\delta(\{\theta\}, B) \cap \Phi \neq \emptyset$. Otherwise $A$ \emph{rejects} $Q$. As in the previous section, we say that $j \in [1, |B|]$ is a \emph{match} iff there is an $i \in [1, j]$ such that $A$ accepts $B[i,j]$. The set of all matches is denoted $\Delta(A, B)$.

Given a regular expression $R$, an NFA $A$ accepting precisely the strings in $L(R)$ can be obtained by several classic methods~\cite{MY1960, Glushkov1961, Thomp1968}. In particular, Thompson~\cite{Thomp1968} gave a simple well-known construction which we will refer to as a \emph{Thompson NFA} (TNFA). A TNFA $A$ for $R$ has at most $2m$ states, at most $4m$ transitions, and can be computed in $O(m)$ time. Hence, a state-set transition can be computed in $O(m)$ time using a breadth-first search of $A$ and therefore we can test acceptance of $Q$ in $O(um)$ time and $O(m)$ space. This solution is easily adapted to find all matches in the same complexity by adding the start state to each of the computed state-sets immediately before computing the next. Formally, $\deltab(S, \alpha \cdot B) = \deltab(\delta(S \cup \{\theta\}, \alpha), B)$, for any string $B$ and character $\alpha \in \Sigma$. 
A match then occurs at position $j$ if $\deltab(\{\theta\}, Q[1,j]) \cap \Phi \neq \emptyset$.

\subsection{Searching for Matches}

Let $A = (V, E, \Sigma, \theta, \Phi)$ be a TNFA with $m$ states. Given a compressed string $Z = z_1\ldots z_n$ representing a string $Q$ of length $u$ we show how to find $\Delta(A, Q)$ efficiently. As in the previous section let $l_i$ and $u_i$, $0\leq i \leq n$ be the length and start position of $\phrase(z_i)$. We process $Z$ from left-to-right and compute a description for $z_i$ consisting of the following information.
\begin{itemize}
  \item The integers $l_i$ and $u_i$.
  \item The state-set $S_{u_i} = \deltab(\{\theta\}, Q[1,u_i]+l_i-1)$.
  \item For each state $s$ of $A$ the compression element $\lastmatch(s, z_i) = x$, where $x$ is the ancestor of $z_i$ of maximum depth such that $\deltab(\{s\}, \phrase(x)) \cap \Phi \neq \emptyset$. If there is no ancestor that satisfies this, then $\lastmatch(s,z_i) = \bot$. 
\end{itemize}

\begin{figure}[t] 
  \centering \includegraphics[scale=.6]{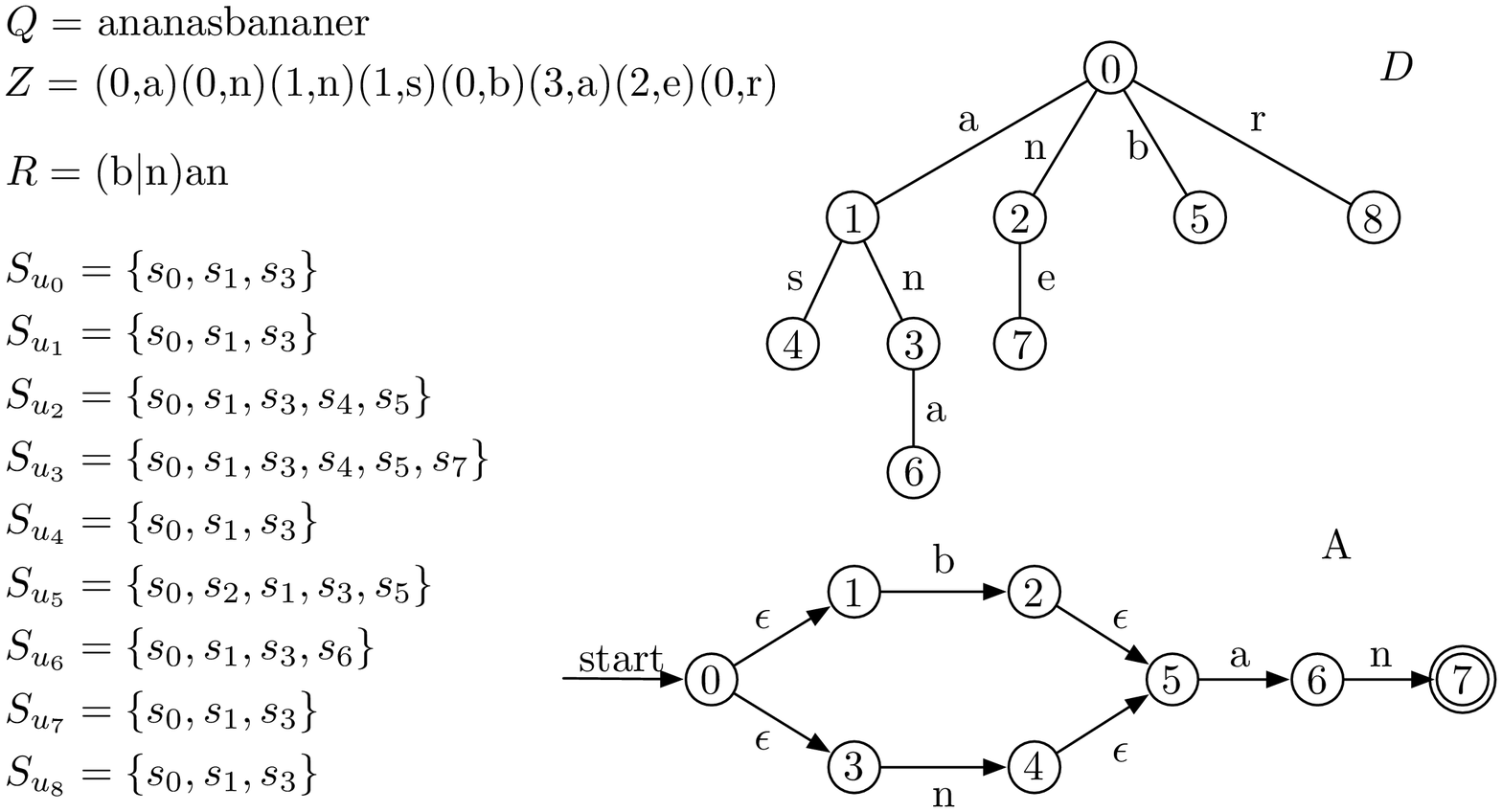}
  \caption{The compressed string $Z$ representing $Q$ and the corresponding dictionary trie $D$. The TNFA $A$ for the regular expression $R$ and the corresponding state-sets $S_{u_i}$ are given.  The lastmatch pointers are as follows: $\lastmatch(s_7,z_i)=\{z_0\}$ for $i=0,1,\ldots,8$, $\lastmatch(s_2,z_i)=\lastmatch(s_4,z_i)= \lastmatch(s_5,z_i)= \{z_3\}$ for $i=3,6$, and $\lastmatch(s_6,z_i)=  \{z_2\}$ for $i=2,7$. All other lastmatch pointers are $\bot$.
  Using the description we can find the matches: Since $s_2 \in S_{u_5}$ the element $z_3 \in M(s_2,z_6)$ represents the match $u_6+\depth(z_3)-1=9$. The other matches can be found similarly.}
  \label{s3:fig:regex}
\end{figure}
An example description is shown in Fig.~\ref{s3:fig:regex}. The total size of the description for $z_i$ is $O(m)$ and therefore the space for all descriptions is $O(nm)$. In the next section we will show how to compute the descriptions. Assume for now that we have processed $z_0, \ldots, z_{i-1}$. We show how to find the matches within $[u_{i}, u_{i} + l_i - 1]$. Given a state $s$ define $M(s, z_i) = \{x_1, \ldots, x_k\}$, where $x_1 = \lastmatch(s, z_i)$, $x_j = \lastmatch(s, \parent(x_{j-1}))$, $1 < j \leq k$, and $\lastmatch(s, x_{k}) = \bot$, i.e., $x_1, \ldots, x_k$ is the sequence of ancestors of $z_i$ obtained by recursively following $\lastmatch$ pointers. By the definition of $\lastmatch$ and $M(s,z_i)$ it follows that $M(s,z_i)$ is the set of ancestors $x$ of $s$ such that $\deltab(s, x) \cap \Phi \neq \emptyset$. Hence, if $s \in S_{u_{i-1}}$ then each element $x \in M(s, z_i)$ represents a match, namely, $u_i + \depth(x)-1$. Each match may occur for each of the $|S_{u_{i-1}}|$ states and to avoid reporting duplicate matches we use a priority queue to merge the sets $M(s,z_i)$ for all $s \in S_{u_{i-1}}$, while generating these sets in parallel. A similar approach is used in~\cite{Navarro2003}. This takes $O(\log m)$ time per match. Since each match can be duplicated at most $|S_{u_{i-1}}| = O(m)$ times the total time for reporting matches is $O(\occ \cdot m\log m)$.

\subsection{Computing Descriptions}

Next we show how to compute descriptions efficiently. Let $1 \leq \tau \leq n$ be a parameter. Initially, compute a set $C$ of compression elements according to Lemma~\ref{s3:lem:special} with parameter $\tau$. For each element $z_j \in C$ we store $l_j$ and the \emph{transition sets} $\deltab(s, \phrase(z_j))$ for each state $s$ in $A$. Each transition set uses $O(m)$ space and therefore the total space used for $z_j$ is $O(m^2)$. During the construction of $C$ we compute each of the transition sets by following the path of references to the nearest element $y\in C$ and computing state-set transitions from $y$ to $z_j$. By Lemma~\ref{s3:lem:special}(ii) the distance to $y$ is at most $2\tau$ and therefore all of the $m$ transition sets can be computed in $O(\tau m^2)$ time.  Since, $|C| = O(n/\tau)$ the total preprocessing time is $O(n/\tau \cdot \tau m^2 ) = O(nm^2)$ and the total space is $O(n/\tau \cdot m^2)$.

The descriptions can now be computed as follows. The integers $l_i$ and $u_i$ can be computed as before in $O(\tau)$ time. All $\lastmatch$ pointers for all compression elements can easily be obtained while computing the transitions sets. Hence, we only show how to compute the state-set values. First, let $S_{u_0} := \{\theta\}$. To compute $S_{u_i}$ from $S_{u_{i-1}}$ we compute the path $p$ to $z_i$ from the nearest element $y \in C$. Let $p'$ be the path from $z_0$ to $y$. Since $\phrase(z_i) = \lab(p') \cdot \lab(p)$ we can compute $S_{u_i} = \deltab(S_{u_{i-1}}, \phrase(z_i))$ in two steps as follows. First compute the set 
\begin{equation}
\label{s3:eq:union}
S' = \bigcup_{s \in S_{u_{i-1}}} \deltab(s, \phrase(y))\;.
\end{equation}
Since $y \in C$ we know the transition sets $\deltab(s, \phrase(y))$ and we can therefore compute the union in $O(m^2)$ time. Secondly, we compute $S_{u_i}$ as the set $\delta(S', \lab(p))$. Since the distance to $y$ is at most $\tau$ this step uses $O(\tau m)$ time. Hence, all the state-sets $S_{u_0}, \ldots, S_{u_n}$ can be computed in $O(nm(m + \tau))$ time.

\subsection{Analysis}
Combining it all, we have an algorithm using $O(nm(m + \tau) + \occ \cdot m \log m)$ time and $O(nm + nm^2/\tau)$ space. Note that since we are using $\Omega(n)$ space, hashing is not needed and the algorithm works for \zlw\ as well. In summary, this completes the proof of Theorem~\ref{s3:thm:regularex}.

\subsection{Exploiting Word-level Parallelism}\label{s3:sec:wordparallel}
If we use the word-parallelism inherent in the word-RAM model, the algorithm of Navarro~\cite{Navarro2003} uses $O(\ceil{m/w}(2^m + nm) + \occ\cdot m \log m)$ time and $O(\ceil{m/w}(2^m + nm))$ space, where $w$ is the number of bits in a word of memory and space is counted as the number of words used. The key idea in Navarro's algorithm is to compactly encode state-sets in bit strings stored in $O(\ceil{m/w})$ words. Using a DFA based on a Glushkov automaton~\cite{Glushkov1961} to  quickly compute state-set transitions, and bitwise OR and AND operations to compute unions and intersections among state-sets, it is possible to obtain the above result. The $O(\ceil{m/w}2^m)$ term in the above bounds is the time and space used to construct the DFA.

A similar idea can be used to improve Theorem~\ref{s3:thm:regularex}. However, since our solution is based on Thompson's automaton we do not need to construct a DFA. More precisely, using the state-set encoding of TNFAs given in~\cite{Myers1992, BFC2005} a state-set transition can be computed in $O(\ceil{m/\log n})$ time after $O(n)$  time and space preprocessing. Since state-sets are encoded as bit strings each transition set uses $\ceil{m/\log n}$ space and the union in \eqref{s3:eq:union} can be computed in $O(m\ceil{m/\log n})$ time using a bitwise OR operation. As $n \geq \sqrt{u}$ in \zla\  and \zlw, we have that $\log n \geq \frac{1}{2}\log u$ and therefore Theorem~\ref{s3:thm:regularex} can be improved by roughly a factor $\log u$. Specifically, we get an algorithm using $O(n\ceil{m/\log u}(m + \tau)+ \occ\cdot m \log m)$ time and $O(nm\ceil{m/\log u}/\tau + nm)$ space.


\bibliography{dissertation}

\end{document}